\documentclass[final,3p,times]{else}

\usepackage{graphicx}

\usepackage{amssymb}

\usepackage[english,francais]{babel}

\def\og{\leavevmode\raise.3ex\hbox{$\scriptscriptstyle\langle\!\langle$~}}
\def\fg{\leavevmode\raise.3ex\hbox{~$\!\scriptscriptstyle\,\rangle\!\rangle$}}

\usepackage[utf8]{inputenc}
\usepackage[T1]{fontenc}
\usepackage{amsmath}
\usepackage{color}
\usepackage{hyperref}
\hypersetup{breaklinks=true}
\hypersetup{colorlinks=true}
\hypersetup{urlcolor=blue}
\hypersetup{citecolor=black}
\hypersetup{linkcolor=black}
\usepackage{cleveref}
\newcommand{\bea}{\begin{eqnarray}}
\newcommand{\eea}{\end{eqnarray}}
\newcommand{\be}{\begin{equation}}
\newcommand{\ee}{\end{equation}}
\newcommand{\rr}{\mathbf{r}}
\newcommand{\kk}{\mathbf{k}}

\newcommand{\qq}{\mathbf{q}}

\newcommand{\ii}{\textrm{i}}
\newcommand{\eee}{\textrm{e}}
\newcommand{\dd}{\mathrm{d}}

\DeclareMathOperator\asin{asin}
\DeclareMathOperator\acos{acos}
\DeclareMathOperator\atan{atan}
\DeclareMathOperator\argth{argth}

\DeclareMathOperator\argsh{argsh}
\DeclareMathOperator\thf{th}
\DeclareMathOperator\ch{ch}
\DeclareMathOperator\sh{sh}
\DeclareMathOperator\re{Re}
\DeclareMathOperator\im{Im}
\usepackage{float}

\begin{document}

\begin{frontmatter}


\selectlanguage{francais}
\title{Collective excitation branch in the continuum of pair-condensed Fermi gases: analytical study and scaling laws}


\author{Yvan Castin}

\address{Laboratoire Kastler Brossel, ENS-Universit\'e PSL, CNRS, Universit\'e de la Sorbonne et Coll\`ege de France, 24 rue Lhomond, 75231 Paris, France}

\author{Hadrien Kurkjian}
\address{TQC, Universiteit Antwerpen, Universiteitsplein 1, B-2610 Antwerp, Belgium}


\begin{abstract}
\selectlanguage{english}
The pair-condensed unpolarized spin-$1/2$ Fermi gases have a collective excitation branch in their pair-breaking continuum (V.A. Andrianov, V.N. Popov, 1976). We study it at zero temperature, with the eigenenergy equation deduced from the linearized time-dependent BCS theory and extended analytically to the lower half complex plane through its branch cut, calculating both the dispersion relation and the spectral weights (quasiparticle residues) of the branch. In the case of BCS superconductors, so called because the effect of the ion lattice is replaced by a short-range electron-electron interaction, we also include the Coulomb interaction and we restrict ourselves to the weak coupling limit $\Delta/\mu\to 0^+$ ($\Delta$ is the order parameter, $\mu $ the chemical potential) and to wavenumbers $q=O(1/\xi)$ where $\xi$ is the size of a pair; when the complex energy $z_q$ is expressed in units of $\Delta$ and $q$ in units of $1/\xi$, the branch follows a universal law insensitive to the Coulomb interaction. In the case of cold atoms in the BEC-BCS crossover, only a contact interaction remains, but the coupling strength $\Delta/\mu$ can take arbitrary values, and we study the branch at any wave number. At weak coupling, we predict three scales, that already mentioned $q\approx 1/\xi$, that $q\approx(\Delta/\mu)^{-1/3}/\xi$ where the real part of the dispersion relation has a minimum and that $q\approx(\mu/\Delta)/\xi\approx k_{\rm F}$ ($k_{\rm F}$ is the Fermi wave number) where the branch reaches the edge of its existence domain. Near the point where the chemical potential vanishes on the BCS side, $\mu/\Delta\to 0^+$, where $\xi\approx k_{\rm F}$, we find two scales $q\approx(\mu/\Delta)^{1/2}/\xi$ and $q\approx 1/\xi$. In all cases, the branch has a limit $2\Delta$ and a quadratic start at $q=0$. These results were obtained for $\mu>0$, where the eigenenergy equation admits at least two branch points $\epsilon_a(q)$ and $\epsilon_b(q)$ on the positive real axis, and for an analytic continuation through the interval $[\epsilon_a(q),\epsilon_b(q)] $. We find new continuum branches by performing the analytic continuation through $[\epsilon_b(q),+\infty[$ or even, for $q$ low enough, where there is a third real positive branch point $\epsilon_c(q)$, through $[\epsilon_b(q),\epsilon_c(q)]$ and $[\epsilon_c(q),+\infty[$. On the BEC side $\mu<0$ not previously studied, where there is only one real positive branch point $ \epsilon_a(q)$, we also find new collective excitation branches under the branch cut $[\epsilon_a (q),+\infty[$. For $\mu>0$, some of these new branches have a low-wavenumber exotic hypoacoustic $z_q\approx q^{3/2}$ or hyperacoustic $z_q\approx q^{4/5}$ behavior. For $\mu<0$, we find a hyperacoustic branch and a nonhypoacoustic branch, with a limit $2\Delta$ and a purely real quadratic start at $q=0$ for $\Delta/|\mu|<0.222$.
\vskip 0.25\baselineskip
\end{abstract} 
\end{frontmatter}


\selectlanguage{english}


\section{Introduction and position of the problem}

We first consider a three-dimensional gas of neutral spin $1/2$ fermions with short-range $s$ wave attractive interactions, spatially homogeneous and taken in the thermodynamic limit, that is nonzero density and infinite size. The gas is unpolarized, that is, it has the same number of particles in each spin state $ \uparrow $ and $ \downarrow $, and is prepared at zero temperature. As predicted by BCS theory, all fermions due to attraction between $ \uparrow $ and $ \downarrow $ form bound pairs $ \uparrow \downarrow $ of zero orbital moment, these pairs constituting moreover a condensate (with an infinite pair coherence length) and a superfluid.

We are interested here in collective excitation modes of the system and in their dispersion relation according to their wave vector $ \qq $. We mean by collective excitation a global oscillation of the gas bringing into play collective variables such as the density $ \rho $ or the complex order parameter $ \Delta $ of the gas, these functions acquiring a spatio-temporal variation $ \rho (\rr, t) $ and $ \Delta (\rr, t) $ at wave vector $ \qq $ and at the real angular frequency $ \omega_\qq $ or complex $ z_\qq/\hbar $ (if the mode is damped). These are linear modes, 
associated with a small deviation of $ \rho $ and $ \Delta $ from their value in the ground state or, in practice, to a linear response of the gas to a weak external perturbation. We exclude any excitation which does not conserve the number of fermions in each spin state, of the coherent Rabi coupling type between $ \uparrow $ and $ \downarrow $. 

To better specify the object of our study, let us recall that the excitations of a gas of paired fermions can fall roughly into two categories. Low energy excitations have a small quantum $ \hbar \omega_\qq $ compared to the binding energy $ E_0 $ of a pair. They then only affect the motion of the center of mass of the pairs, the pairs start to oscillate while keeping their integrity: in this case it is an acoustic excitation of the gas, the modes being part of a sound branch with a linear start, $ \omega_\qq \sim cq $ when $ q \to 0 $, $ c $ being the speed of sound. This type of excitation is shared by all the superfluids subjected to short range interaction, independently of the quantum statistics of the constituent particles. But only the start of the dispersion relation, at best $ q \xi \lesssim 1 $ where $ \xi $ is the size of a bound pair, is universal in our system; the non-universal part was the subject of a very detailed study in reference \cite{CKS}. 
High energy excitations have a quantum $ \hbar \omega_\qq $ greater than $ E_0 $: as nothing prevents it energetically, the pairs will be broken into two free fermions $ \uparrow $ and $ \downarrow $, one of wave vector $ \qq/2+\kk $ and the other of wave vector $ \qq/2-\kk $ since the excitation deposits a quantum of well defined momentum $ \hbar \qq $; on the other hand, the relative wave vector (or internal to the pair) $ \kk $ is not constrained. \footnote{In reality, pair breaking at fixed $ \qq $ starts from the energy threshold $ \inf_\kk E_\kk^{(\qq)} $, which is none other than the edge of the continuum defined below.} The energy cost of such excitation by pair breaking is therefore, in BCS theory, $ E_\kk^{(\qq)} = \epsilon_{\qq/2+\kk}+\epsilon_{\qq/2-\kk} $ where $ k \mapsto \epsilon_{\kk} $ is the dispersion relation of the BCS quasiparticles, which has a band gap of width $ E_0/2 $ by definition of $ E_0 $. As $ \kk $ can span $ \mathbb{R}^3 $, the excitation spectrum $ E_\kk^{(\qq)} $ spans a continuum at $ \qq $ fixed, whose lower edge is $ \geq E_0 $ and which extends up to $+\infty $. 

One could believe that such a continuum does not contain any collective mode and that the intensity of the response functions of the system at the angular frequency $ \omega> E_0/\hbar $ exhibits only broad variations with $ \omega $, excluding narrow structures like a Lorentzian resonance $ \propto | \hbar \omega-z_\qq |^{-2} $ associated with a complex energy mode $ z_\qq $. Since the work of Andrianov and Popov \cite{AndrianovPopov}, in the weak interaction and low wave number regime, and its recent generalization to the arbitrary interaction and wave number regime \cite{PRL2019}, we know that the opposite occurs: as long as the chemical potential $ \mu $ of the gas is $> 0 $, therefore that the minimum $ E_0/2 $ of $ k \mapsto \epsilon_\kk $ is reached in BCS theory at a nonzero wave number $ k_0> 0 $, the fermion gas exhibits indeed, in its broken pair continuum, a branch of collective excitation $ q \mapsto z_\qq $ spanning the entire interval $] 0,2k_0 [$. As the collective continuum mode can be damped by the emission of broken pairs, its energy is complex, $ \im z_\qq <0 $; at low wave number, $ q \to 0 $, $ z_\qq $ tends to $ E_0 $ quadratically in $ q $, the mode acquires an angular eigenfrequency better and better defined and induces a lorentzian peak more and more narrow in the response functions of the system, which guarantees its observability \cite{PRL2019}.

The continuum collective branch $ q \mapsto z_\qq $ of our fermion gas has not been the subject of in-depth analytical studies, apart from the low wave number regime $ q \to 0 $. For example, we do not know, for any interaction strength, to what value of $ q $ extends the law of quadratic variation of $ z_\qq $ previously mentioned. Or, the numerical study of reference \cite{PRL2019} suggests that, in the weak interaction regime, the eigenfrequency $ \re z_\qq/\hbar $ exhibit on $] 0,2 k_0 [$ a minimum at $ q = q_{\rm min} $, but nothing is said about the variation of $ q_{\rm min} $ with the interaction strength. At a fixed wave number $ q $ $> 0 $, we also do not know if the dispersion relation has a simple limiting law when the interaction strength tends to zero. 
In the strong interaction regime, near the point of zero chemical potential ($ \mu \to0^+$) where the continuum collective branch disappears as a whole \cite{PRL2019}, we do not know analytically the form of the dispersion relation, nor the wave number scales which structure it. In addition, the continuum collective branch $ q \mapsto z_\qq $ in references \cite{AndrianovPopov, PRL2019} was obtained by carrying out on the eigenvalue equation for the modes of the system an analytic continuation passing from the upper half-plane ($ \im z \geq 0 $) than the lower half-plane ($ \im z <0 $) through the natural interval separating the first two branch points $ z = \epsilon_1 = E_0 $ and $ z = \epsilon_2 $ of the eigenvalue equation on the positive real half-axis, even if this fact is implicit in reference \cite{AndrianovPopov}. What would happen if we made an analytic continuation while passing rather through the interval $ [\epsilon_2,+\infty [$ or, in the case where there is a third branch point $ \epsilon_3> \epsilon_2 $, through the intervals $ [\epsilon_2, \epsilon_3] $ and $ [\epsilon_3,+\infty [$ ~? Could we thus discover other continuum collective branches ~? Could there even be a continuum collective branch for $ \mu <0 $, where the pairs tend to become bosonic ~? Finally, we have limited ourselves for the moment to neutral fermion gases, the physical system we had in mind being that of cold atoms, which has the advantage of allowing an adjustment at will of the interaction strength by magnetic Feshbach resonance \cite{Thomas2002, Salomon2003, Grimm2004b, Ketterle2004, Salomon2010, Zwierlein2012}. What about the electron gases in BCS superconductors, i.e.\ idealized superconductors, where the effect of the crystal lattice and its phonons is modeled by an attractive short-range interaction between $ \uparrow $ and $ \downarrow $ ~? We must add to the latter the repulsive Coulomb interaction between electrons, independent of spin and long range, which, as we know, suppresses the acoustic branch \cite{Anderson1958}. What about the continuum collective branch ~? In reference \cite{AndrianovPopov}, the eigenvalue equation on the angular frequency of the modes was modified to take account of the Coulomb interaction; it was then shown, in the weak interaction limit $ E_0 \ll \epsilon_{\rm F} $ where $ \epsilon_{\rm F} $ is the Fermi energy of the gas, that the continuum collective branch survives and that its complex dispersion relation remains the same as in the presence of the only short-range interaction, at least as long as $ q \xi \lesssim 1 $ therefore that the dispersion relation remains approximately quadratic. Can we, while remaining in the weak interaction regime relevant for BCS superconductors ($ E_0/\epsilon_{\rm F} $ is typically of the order of $ 10^{-3} $ or $ 10^{-4} $), extend the study of the branch beyond the quadratic regime by imposing only $ q \xi = O (1) $ when $ E_0/\epsilon_{\rm F} \to 0 $, and see if Coulomb's interaction still has no effect in this case ~? All these haunting questions call for an in-depth complementary study, which is the subject of this article. 

To complete our introduction, let's give the outline of the article. Section \ref{sec:reminders} briefly presents the formalism used: the interaction between neutral fermions is modeled by an on-site contact attraction, after discretization of space into a cubic lattice of spacing $ b $; to describe the evolution of the independent collective variables $ \rho $, $ \Delta $ and $ \Delta^* $ of the system after a weak and bounded in time perturbation of the ground state, we can use the linearized time-dependent BCS equations around their stationary solution; the eigenvalue equation is written as the determinant of a matrix function of $ z $ and $ \qq $, a $ 3 \times 3 $ matrix since there are three independent variables, and of which we recall how to carry out the analytic continuation at $ \im z <0 $ using the spectral densities of the broken pair continuum. In section \ref{sec:lcf}, we focus on the weak coupling limit $ E_0 \lll \epsilon_{\rm F} $, easier to reach in the ground state with BCS superconductors than with cold atoms, and which we treat mathematically by making $ E_0/\epsilon_{\rm F} $ tend to zero at $ q \xi $ fixed. We find that the continuum collective branch is decoupled from the other excitations in the form of a modulus mode (bringing into play oscillations of the only modulus of the order parameter, excluding its phase), for the pure contact interaction (section \ref{subsubsec:dmmicontact}) but also in the presence of the Coulomb interaction (section \ref{subsubsec:avec_coulomb}), with the same eigenvalue equation in both cases. The spectral densities useful for the analytic continuation can be expressed in terms of elliptic integrals (section \ref{subsec:cpf_denspec}). The branch is first calculated numerically, then studied analytically at low and at large $ q \xi $ in section \ref{subsec:exci_mod}. To be complete, we perform the same study on the acoustic branch (phase mode) for a pure contact interaction in section \ref{subsec:son}. The long section \ref{sec:cold atoms} is devoted to the arbitrary interaction regime, appanage of the cold atom gases; the interaction is pure contact, and it is resonant (its scattering length $ a $ in the $s$ wave greatly exceeds its range $ b $ in absolute value), which obliges to take the continuous limit $ b \to 0 $ of our lattice model: the eigenvalue equation is reduced to the determinant of a $2\times 2$ matrix (section \ref{subsec:bzero}). Its analytic continuation is performed in the case $ \mu> 0 $ (where $ k_0 = (2m \mu)^{1/2}/\hbar $) passing through the interval $ [\epsilon_1, \epsilon_2] $, except in section \ref{subsec:autrefen_muneg}. We give in section \ref{subsec:rhosspgen} the position on the positive real semi-axis of the branch points $ \epsilon_1 $, $ \epsilon_2 $, even $ \epsilon_3 $, as a function of $ q $, and the expression of the spectral densities on each interval between the branch points and infinity. We enrich our study of the continuum branch (compared to section \ref{sec:lcf}) by introducing into section \ref{subsec:carac} other observables than its complex dispersion relation $ q \mapsto z_\qq $, namely its spectral weight (residue) on the small phase or modulus deviations from equilibrium of the order parameter, and the coherent superposition of these two channels making it possible to maximize the spectral weight in the experimental signal. We present a complete numerical study of the continuum branch in section \ref{subsec:numcom}, then carry out an analytical study at low $ q $ in section \ref{subsec:faibleq}, specifying this time (at contrary to reference \cite{PRL2019}) the domain of validity of the quadratic approximation, which shows the expected condition $ q \xi \ll 1 $ in the weak interaction regime but the other and non-intuitive condition $ q/k_0 \ll (\mu/\Delta)^{1/2} $ in the regime $ \mu/\Delta \to 0^+$ of vanishing chemical potential. In section \ref{subsec:faibledelta}, devoted to the weak interaction limit $ \Delta/\mu \ll 1 $, we need, to exhaust the problem, to study the branch successively at three wave number scales, at $ q \approx 1/\xi $ i.e.\ $ q \approx k_0 \Delta/\mu $ (section \ref{subsubsec:q_approx_d}), at $ q \approx k_0 $ (section \ref{subsubsec:q_approx_d0}) and at the unexpected scale $ q \approx k_0 (\Delta/\mu)^{2/3} $ governing the position of the minimum of $ q \mapsto \re z_\qq $ (section \ref{subsubsec:q_approx_d23}). In section \ref{subsec:grand_d}, devoted to the inverse limit $ \mu/\Delta \to 0^+$, it suffices to consider successively the scale $ q \approx k_0 $ (section \ref{subsubsec:grand_d_qapproxd0}) and the scale $ q \approx k_0 (\mu/\Delta)^{1/2} $ (section \ref{subsubsec:grand_d_qapproxd12}) for a full description. Finally, in section \ref{subsec:autrefen_muneg}, we seek and sometimes find more exotic continuum modes by performing the analytic continuation for $ \mu> 0 $ through the other windows $ [\epsilon_2, \epsilon_3] $ and $ [\epsilon_3,+\infty [$, and even extending the study to the case $ \mu <0 $ (where $ k_0 = 0 $). We conclude in section \ref{sec:conclusion}.

\section{Formalism reminders: model, linearized time-dependent BCS equations, spectral densities, analytic continuation} 
\label{sec:reminders}

The model and formalism of this article are those of references \cite{theseHadrien, Annalen}, the main lines of which we recall here. Fermions with two internal states $ \uparrow $ and $ \downarrow $ live with periodic boundary conditions in the quantization volume $ [0, L]^3 $ discretized into a cubic lattice of spacing $ b $, in the grand-canonical ensemble of chemical potential $ \mu = \mu_\uparrow = \mu_\downarrow $. The dispersion relation of a fermion alone on the lattice is that $ \kk \mapsto E_\kk = \hbar^2 k^2/2m $ of free space on the first Brillouin zone $ \mathcal{D} = [-\pi/b, \pi/b [^ 3 $ of the lattice, and is extended periodically beyond. Fermions of opposite spins undergo the binary contact interaction $ V (\rr_1, \rr_2) = g_0 \delta_{\rr_1, \rr_2}/b^3 $, where $ \delta $ is a Kronecker delta and the bare coupling constant is adjusted as a function of the lattice spacing to reproduce the desired scattering length $ a $ between fermions in the $s$ wave: $ 1/g_0 = 1/g-\int_{\mathcal{D}} \frac{\dd^3k}{(2 \pi)^3} \frac{1}{2E_\kk} $ where $ g = 4 \pi \hbar^2a/m $ is the effective coupling constant \cite{Houchesdimred}.

Initially, the fermion gas is prepared in the ground state, unpolarized by symmetry between $ \uparrow $ and $ \downarrow $. The state is described in an approximate variational manner by the usual BCS ansatz, coherent state of bound pairs $ \uparrow \downarrow $ of particles $ | \psi_{\rm BCS}^{(0)} \rangle $ minimizing the energy. It has an order parameter $ \Delta $ defined below and a pair-breaking quasiparticle excitation spectrum of the usual form $ \epsilon_\kk = (\xi_\kk^2+\Delta^2)^{1/2} $, where $ \kk $ spans $ \mathcal{D} $ and $ \xi_\kk = E_\kk-\mu+g_0 \rho/2 $ contains the grand-canonical energy shift and Hartree's mean field proportional to the total density $ \rho $. 

The gas is then subjected to an arbitrary external perturbation, except that it conserves the number of particles in $ \uparrow $ and in $ \downarrow $, and that it is sufficiently weak so that one can limit its action, in the BCS variational framework, to creation of a coherent state of pairs of quasiparticles of weak amplitude $ | \beta | \ll 1 $. At times $ t> 0 $ after the perturbation, the gas state vector is therefore written 
\be
\label{eq:ansatz}
|\psi_{\rm BCS}(t)\rangle= \mathcal{N}(t) \exp\left[-\sum_\qq\sum_{\kk\in\mathcal{D}} \beta_{\kk\qq}(t)\, \hat{\gamma}^\dagger_{\qq/2+\kk\uparrow}\hat{\gamma}^\dagger_{\qq/2-\kk\downarrow}\right]|\psi_{\rm BCS}^0\rangle
\ee
where the operators of annihilation $ \hat{\gamma}_{\kk \sigma} $ and of creation $ \hat{\gamma}_{\kk \sigma}^\dagger $ of a BCS quasiparticle of wave vector $ \kk $ and spin $ \sigma $ obey the usual fermionic anticommutation relations. The independent variational parameters are the amplitudes of probability $ \beta_{\kk \qq} (t) $ to have broken a bound pair into two quasiparticles of opposite spins and wave vectors $ \qq/2 \pm \kk $; $ \qq $ therefore represents the total (or center of mass) wave vector of the broken pair, and $ \kk $ its internal (or relative) wave vector; $ \mathcal{N} (t) $ is a normalization factor. 

At order one in the amplitudes $ \beta $, we extract from the ansatz (\ref{eq:ansatz}) the expressions of the order parameter $ \Delta (\rr, t) $ and the total density $ \rho (\rr, t) $ of the gas excited at point $ \rr $ and at time $ t $: 
\bea
\label{eq:deltart}
\Delta(\rr,t)&\equiv& g_0 \langle\hat{\psi}_\downarrow(\rr)\hat{\psi}_\uparrow(\rr)\rangle=-\frac{g_0}{L^3}\sum_{\kk\in\mathcal{D}} U_k V_k +\sum_{\qq} \delta\Delta_\qq(t)\eee^{\ii\qq\cdot\rr}+O(\beta^2)\\
\rho(\rr,t)&\equiv& \sum_{\sigma=\uparrow,\downarrow} \langle \hat{\psi}^\dagger_\sigma(\rr)\hat{\psi}_\sigma(\rr)\rangle=\frac{2}{L^3}\sum_{\kk\in\mathcal{D}} V_k^2 +\sum_{\qq} \delta\rho_\qq(t)\eee^{\ii\qq\cdot\rr}+O(\beta^2)
\label{eq:rhort}
\eea
These are the variables through which the collective modes of the system can show up. In the right-hand side of these expressions appears first the constant contribution of the initial stationary state, function of the amplitudes $ U_k = [\frac{1}{2} (1+\xi_\kk/\epsilon_\kk)]^{1/2} $ and $ V_k = [\frac{1}{2} (1-\xi_\kk/\epsilon_\kk)]^{1/2} $ of the quasiparticle modes on particles and holes. Then come linear terms in $ \beta $ spatially modulated with the wave vectors $ \qq $ of the pairs; their time dependence can be derived from that of $ \beta_{\kk \qq} $, itself deduced from the usual variational calculus. As in reference \cite{theseHadrien}, it is clever to introduce partially decoupling linear plus and minus combinations of $ \beta $ and $ \beta^* $; here we only need their value at $ t = 0^+$: 
\be
\label{eq:valtzero}
y^{\pm}_{\kk\qq}=\beta_{\kk\qq}(0^+)\pm\beta_{\kk(-\qq)}^*(0^+)
\ee
which already gives access to the values of the Fourier amplitudes in equations (\ref{eq:deltart}, \ref{eq:rhort}) just after the excitation: 
\be
\label{eq:vc0p}
\delta\Delta_\qq(0^+)\mp\delta\Delta_{-\qq}^*(0^+) = -\frac{g_0}{L^3} \sum_{\kk\in\mathcal{D}} W_{\kk\qq}^{\pm} y_{\kk\qq}^{\mp}
\ \ \ \mbox{and}\ \ \ \delta\rho_\qq (0^+) = \frac{1}{L^3} \sum_{\kk\in\mathcal{D}} W_{\kk\qq}^0 y_{\kk\qq}^+
\ee
The linearized motion equations of the variables $ \beta \pm \beta^* $ are given in reference \cite{theseHadrien}. The added value here is to take the Laplace transform, to deduce that of the Fourier amplitudes in (\ref{eq:deltart}, \ref{eq:rhort}) then, by inverse Laplace transformation, get the time dependence of the latter: 
\be
\label{eq:cauchysable}
\begin{pmatrix}
\delta\Delta_\qq(t)-\delta\Delta_{-\qq}^*(t) \\
\delta\Delta_\qq(t)+\delta\Delta_{-\qq}^*(t) \\
\delta\rho_\qq(t)
\end{pmatrix}
=\int_{\ii\eta+\infty}^{\ii\eta-\infty} \frac{\dd z}{2\ii\pi}\, \frac{\eee^{-\ii zt/\hbar}}{M(z,\qq)}
\begin{pmatrix}
\Sigma_{W^+y^-}^z(z,\qq)+\Sigma_{W^+y^+}^\epsilon(z,\qq) \\
\Sigma_{W^-y^+}^z(z,\qq)+\Sigma_{W^-y^-}^\epsilon(z,\qq) \\
\Sigma_{W^0y^+}^z(z,\qq)+\Sigma_{W^0y^-}^\epsilon(z,\qq)
\end{pmatrix}
\ee
different values of $ \qq $ being decoupled by momentum conservation. The integral on $ z $ is taken in the complex plane by flying over the real axis ($ \eta> 0 $, even $ \eta \to 0^+$). The integrand contains, in addition to the exponential oscillating at the complex angular frequency $ z/\hbar $, a source vector with three components and the inverse of a $3\times 3$ matrix, both functions of $ z $. After passing to the thermodynamic limit $ L \to+\infty $, we arrive at a compact writing, in particular 
\be
\label{eq:defM}
M(z,\qq) =
\begin{pmatrix}
\breve{\Sigma}_{W^+W^+}^\epsilon(z,\qq) && \Sigma_{W^+W^-}^z(z,\qq) && -g_0\Sigma_{W^+W^0}^z(z,\qq) \\
\Sigma_{W^-W^+}^z(z,\qq) && \breve{\Sigma}_{W^-W^-}^\epsilon(z,\qq) && -g_0\Sigma_{W^-W^0}^\epsilon(z,\qq) \\
\Sigma_{W^0W^+}^z(z,\qq) && \Sigma_{W^0W^-}^\epsilon(z,\qq) && 1-g_0\Sigma_{W^0W^0}^\epsilon(z,\qq)
\end{pmatrix}
\equiv
\begin{pmatrix}
M_{++}(z,\qq) && M_{+-}(z,\qq) && M_{+0}(z,\qq) \\
M_{-+}(z,\qq) && M_{--}(z,\qq) && M_{-0}(z,\qq) \\
M_{0+}(z,\qq) && M_{0-}(z,\qq) && M_{00}(z,\qq)
\end{pmatrix}
\ee
thanks to the short-hand notation of integrals on the internal wave vector $ \kk $ of the pairs: 
\be
\label{eq:defSigma}
\Sigma_{ab}^z(z,\qq)=\int_{\mathcal{D}}\frac{\dd^3k}{(2\pi)^3} \frac{z\, a_{\kk\qq} b_{\kk\qq}}{z^2-\epsilon_{\kk\qq}^2} \ ,
\ \ \
\Sigma_{ab}^\epsilon(z,\qq)=\int_{\mathcal{D}}\frac{\dd^3k}{(2\pi)^3} \frac{\epsilon_{\kk\qq} a_{\kk\qq} b_{\kk\qq}}{z^2-\epsilon_{\kk\qq}^2} \ ,
\ \ \
\breve{\Sigma}_{ab}^\epsilon(z,\qq)=\int_{\mathcal{D}}\frac{\dd^3k}{(2\pi)^3} \left[\frac{\epsilon_{\kk\qq} a_{\kk\qq} b_{\kk\qq}}{z^2-\epsilon_{\kk\qq}^2}
+\frac{1}{2\epsilon_\kk}\right]
\ee
The version with accent $ \breve{\Sigma} $ contains a counter term $ 1/2 \epsilon_\kk $, obtained by eliminating $ g_0 $ in terms of the order parameter ($-1/g_0 = \int_{\mathcal{D}} \frac{\dd^3k}{(2 \pi)^3} \frac{1}{2 \epsilon_\kk} $), which provides ultraviolet convergence for the integral in the continuous limit $ b \to 0 $; versions without accent converge as is when $ b \to 0 $. Each symbol $ \Sigma $ displays under the integral sign the expected energy denominator, containing the energy $ \epsilon_{\kk \qq} $ of the broken pair, and in the numerator coupling amplitudes between quasiparticles involving their modal amplitudes: 
\be
\label{eq:defW}
W^{\pm}_{\kk\qq}=U_{\kk+\qq/2}U_{\kk-\qq/2}\pm V_{\kk+\qq/2}V_{\kk-\qq/2}\ , \ \ \
W^0_{\kk\qq}=U_{\kk+\qq/2}V_{\kk-\qq/2}+U_{\kk-\qq/2}V_{\kk+\qq/2}\ , \ \ \
\epsilon_{\kk\qq} = \epsilon_{\kk+\qq/2}+\epsilon_{\kk-\qq/2}
\ee
In matrix $ M $, they appear by pairs of factors, which have the following more explicit expressions: 
\be
\label{eq:Wexpli}
\begin{matrix}
\displaystyle (W^+)^2=\frac{\epsilon_+\epsilon_-+\xi_+\xi_-+\Delta^2}{2\epsilon_+\epsilon_-} &\quad& & \displaystyle W^+W^-=\frac{\xi_+}{2\epsilon_+}+\frac{\xi_-}{2\epsilon_-} &\quad & \displaystyle W^+W^0=\frac{\Delta}{2\epsilon_+}+\frac{\Delta}{2\epsilon_-} \\
\displaystyle (W^-)^2=\frac{\epsilon_+\epsilon_-+\xi_+\xi_--\Delta^2}{2\epsilon_+\epsilon_-} &\quad& & \displaystyle W^-W^0=\frac{(\xi_++\xi_-)\Delta}{2\epsilon_+\epsilon_-} & \quad & \displaystyle (W^0)^2=\frac{\epsilon_+\epsilon_--\xi_+\xi_-+\Delta^2}{2\epsilon_+\epsilon_-}
\end{matrix}
\ee
where one introduced the short-hand notations $ \epsilon_\pm = \epsilon_{\kk \pm \qq/2} $ and $ \xi_\pm = \xi_{\kk \pm \qq/2} $. The whole (starting from equation (\ref{eq:cauchysable})) constitutes a masterful generalization of equation (12) of reference \cite{PRLGurarie} to a nonzero excitation wave vector $ \qq $ and to an arbitrary infinitesimal non-polarizing perturbation of the initial stationary BCS state. 

\begin{figure} [tb] 
\centerline{\includegraphics [width = 0.8 \textwidth, clip =]{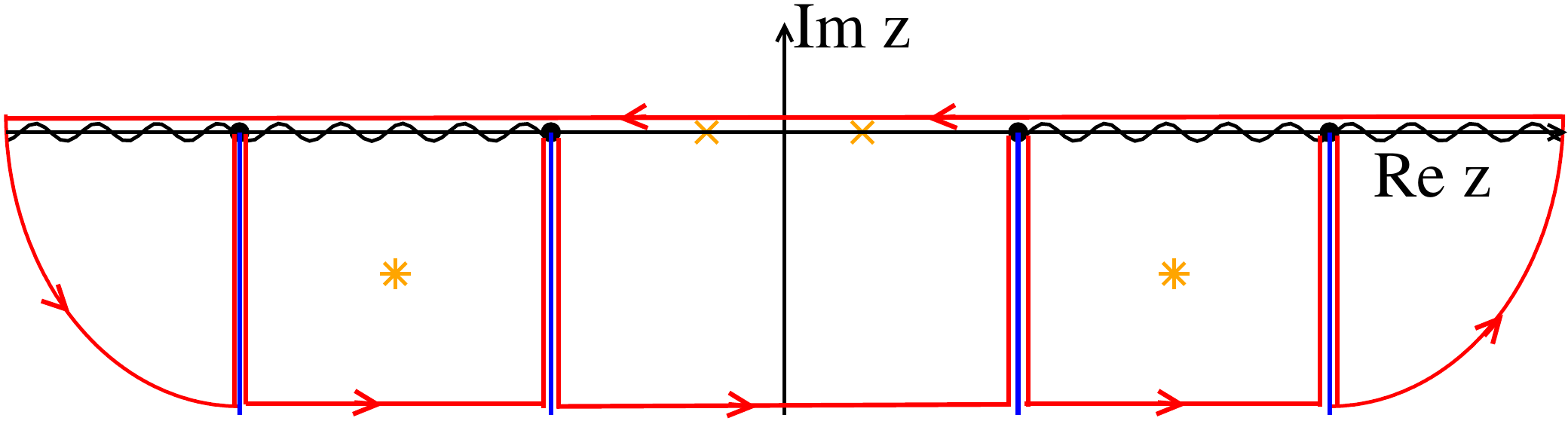}} 
\caption{In the complex plane, integration path to follow and analytic continuations to be performed in order to be able to apply Cauchy's theorem to integral (\ref{eq:cauchysable}). We consider the case where the integrand has two branch points (black discs) on the positive real semi-axis, and two others on the negative real semi-axis taking into account the symmetry $ z \leftrightarrow (-z) $. Its original branch cuts are shown with wavy lines. To fix the ideas, the branch cuts shifted by analytic continuation are arranged vertically in the form of half-lines (solid blue lines). The integration path, which bypasses them, is represented by the red arrow line. Orange stars: poles appearing in the analytic continuation (they are roots of $ \det M_\downarrow (z, \qq) $ of integer degree). Orange cross: poles appearing on the real axis, without the need for analytic continuation (these are the roots of $ \det M (z, \qq) $ of integer degree).} 
\label{fig:contour} 
\end{figure}

Equation (\ref{eq:cauchysable}) strongly evokes the formalism of Green's functions and of the resolvent. It is natural to apply the residue formula to it, that is to say Cauchy's theorem, by closing the integration path with a large half-circle in the lower complex half-plane, oriented counterclockwise and of divergent radius. As usual, the idea must be adapted to circumvent by an appropriate path the branch cuts resulting from vanishing denominators in the symbols $ \Sigma $ of (\ref{eq:defSigma}) when $ z $ or $-z $ falls in the broken pair continuum $ \{\epsilon_{\kk \qq}, \kk \in \mathbb{R}^3 \} $, or rather the {\sl new} branch cuts obtained by deformation and rotation of the original branch cuts {\sl} around their branch point (the new branch cuts are often arranged vertically in the lower half-plane), see supplement A${}_{\rm III} $ of reference \cite{CCTlivre} and our figure \ref{fig:contour}; after bypassing a {\sl new} branch cut, the integrand of (\ref{eq:cauchysable}) must of course be replaced by its analytic continuation from the upper half-plane to the lower half-plane across the moved {\sl original} branch cut.

In the residue formula applied to (\ref{eq:cauchysable}) thus transformed, we associate by definition with each pole $ z_\qq $ of the integrand \footnote{Given the even parity in $ z $ of $ \det M (z, \qq) $, we agree that $ \re z_\qq> 0 $ in the definition of the complex energy of a mode.} a collective mode of eigenenergy $ z_\qq $, and the residue includes the exponential factor $ \exp (-\ii z_\qq t/\hbar) $, only dependent on time. As the matrix elements of $ M $ do not present a pole even after analytic continuation, nor the source vector besides, and taking into account the writing of the inverse matrix $ M^{-1} ={}^t \mathrm{com} \, M/\det M $ in terms of the comatrix and its determinant, any pole can only come from $ \det M $ (or its analytic continuation) vanishing with an integer power law in $ z_\qq $. There are therefore two possible cases: $ (i) $ the pole is located on the real axis, no analytic continuation is required, $ z_\qq $ is actually a real energy $ \hbar \omega_\qq $ and the mode is not damped, $ (ii) $ the pole is located in the lower complex half-plane, it was necessary to carry out an analytic continuation of $ \det M $ to make it appear, $ z_\qq $ is complex ($ \im z_\qq <0 $) and the mode is damped exponentially. To find all the collective modes at wave vector $ \qq $, we must therefore solve the following two equations: 
\be
\label{eq:defmode}
\boxed{
\det M(z_\qq,\qq)=0, \quad \det M_\downarrow(z_\qq,\qq)=0
}
\ee
the vertical arrow in the second equation representing the analytic continuation to $ \im z <0 $. Note that, if the excitation of the gas is poorly chosen, it may be that the source vector in (\ref{eq:cauchysable}) or its analytic continuation vanishes at $ z = z_\qq $, in which case the residue of the integrand is zero at $ z_\qq $ and the mode is not expressed. 

It remains to be seen how to carry out in practice the analytic continuation of the matrix elements of $ M $ to the lower half-plane. We can use \cite{PRL2019} a result appearing in appendix C of \cite{livreNozieres}, for any function $ f (z) $ written as an integral on the real variable $ \epsilon $, with a energy denominator $ z-\epsilon $ and a spectral density $ \rho (\epsilon) $: 
\be
\label{eq:Noz}
f(z)=\int_{\epsilon_a}^{\epsilon_b} \dd\epsilon \frac{\rho(\epsilon)}{z-\epsilon}\quad\forall z\in\mathbb{C}\setminus[\epsilon_a,\epsilon_b] \Longrightarrow f_\downarrow(z)=f(z)-2\ii\pi \rho(z) \ \ \mbox{for}\ \ \im z\leq 0
\ee
the analytic continuation taking place of course through the branch cut $ [\epsilon_a, \epsilon_b] $ of $ f (z) $, which it folds over $]-\infty, \epsilon_a] \cup [\epsilon_b,+\infty [$. For (\ref{eq:Noz}) to apply, the spectral density $ \rho (\epsilon) $ must have an analytic continuation $ \rho (z) $ in the lower half-plane, so that it be already analytical on the interval $ [\epsilon_a, \epsilon_b] $. In practice, we are led to split the function to be continued into a sum of terms of type $ f (z) $, the bounds $ \epsilon_a $, $ \epsilon_b $, etc. corresponding to the successive positions of the singularities of $ \rho (\epsilon) $ on the real axis (points of divergence, discontinuity, kinks). In the case of matrix elements (\ref{eq:defSigma}), this is done as in \cite{PRL2019} using spectral densities \footnote{So that $ \rho_{ss'} $ is the spectral density of $ M_{ss'} $ within the meaning of (\ref{eq:Noz}), a factor $ (2 \pi)^{-3} $ should be included in its definition. We did not do it here, so that the functions $ \rho_{ss'} $ coincide with those of reference \cite{PRL2019}.} 
\be
\label{eq:defdensspec}
\rho_{ss'}(\epsilon,\qq) \equiv \int \dd^3k \frac{1}{2} W_{\kk\qq}^s W_{\kk\qq}^{s'} \delta(\epsilon-\epsilon_{\kk\qq})\quad \forall s,s'\in\{+,-,0\}
\ee
which we will calculate in due course later. 

\section{Weak coupling limit $ \Delta \lll \mu $: the case of BCS superconductors} 
\label{sec:lcf} 

\subsection{Regime considered}

In this section, the contact interaction between opposite spins fermions is very weakly attractive: its scattering length $ a $ is negative and very small in absolute value compared to the average distance between particles, or if we prefer compared to the inverse $ 1/k_{\rm F} $ of the Fermi wave number of the gas, $ k_{\rm F} = (3 \pi^2 \rho)^{1/3} $. We can also suppose that $ | a | $ is weaker than the spacing $ b $ of the lattice model, therefore than the interaction range: the binary interaction is non-resonant and in the Born regime of the scattering theory. We finally have the chain of inequalities $ 0 <-a \ll b \ll 1/k_{\rm F} $.

It is under these conditions that BCS theory is most quantitative. It gives the right equation of state of the gas to order one in $ k_{\rm F} a $ and correctly takes into account Hartree's shift of the position $ k_0 $ of the minimum of the BCS excitation spectrum (see section 3.4.4 of \cite{Varenna06} and reference \cite{Ketterlegap}). In the chain of inequalities, the last ensures that physics is little affected by the discretization of space and makes it possible to replace the integration domain $ \mathcal{D} $ by $ \mathbb{R}^3 $ in (\ref{eq:defSigma}); the penultimate makes it possible to replace the bare coupling constant $ g_0 $ by the effective constant $ g = 4 \pi \hbar^2a/m $ in the gas equation of state $ \mu = \epsilon_{\rm F}+g \rho/2 $, in the BCS excitation spectrum $ \epsilon_{\kk} = [(E_\kk-\epsilon_{\rm F})^2+\Delta^2]^{1/2} $ (hence the position of the minimum $ k_0 = k_{\rm F} $) and in the third column of matrix (\ref{eq:defM}). 

As the order parameter is exponentially small in the interaction strength, $ \Delta/\epsilon_{\rm F} \simeq 8 \eee^{-2} \exp (-\pi/2 k_{\rm F} | a |) $ in BCS theory, we study in this section the continuum collective branch to leading order in $ \Delta $, i.e.\  we make $ \Delta $ mathematically tend to zero after having performed the appropriate scale changes on the complex energy $ z_\qq $ and the wave number $ q $ ensuring the existence of a finite limit. 

\subsection{Decoupling of modulus mode} 
\subsubsection{In the presence of a contact interaction} 
\label{subsubsec:dmmicontact} 

Before going to the mathematical limit $ \Delta \to 0 $, we must look at the continuum collective branch to the correct energy and wave number scales which are naturally the binding energy $ 2 \Delta $ and the inverse $ 1/\xi $ of the size $ \xi \propto \hbar^2 k_0/m \Delta $ \cite{Marini1998} of a bound pair of fermions $ \uparrow \downarrow $, so keep the reduced variables constant 
\be
\label{eq:adim}
\bar{q}=\frac{\hbar^2 k_0 q}{m\Delta} \ \ \ \mbox{and}\ \ \ \bar{z}=z/\Delta
\ee
So, to obtain a nonzero limit in the matrix elements (\ref{eq:defM}) of $ M $, that is to say in integrals (\ref{eq:defSigma}), it is necessary to make in spherical coordinates of polar axis $ \qq $ the change of variable 
\be
\label{eq:Kcouche}
k=k_0 +\frac{m\Delta}{\hbar^2 k_0} \bar{K}
\ee
on the modulus of the internal wave vector of the pairs and make $ \Delta $ tend to zero at $ \bar{K} $ fixed. In other words, integrals (\ref{eq:defSigma}) are dominated here by a thin layer of energy width $ \propto \Delta $ around the Fermi surface of the gas. \footnote{If we make $ \Delta \to 0 $ at fixed $ \kk $, with $ q = O (\Delta) $ and $ z = O (\Delta) $, the integrands of all $ \Sigma $ and $ \breve{\Sigma} $ tend to zero, which justifies this assertion and our procedure.} 
Let us first carry out the passage to the limit in the BCS excitation spectrum:
\be
\label{eq:defxpmepm}
\xi_\pm/\Delta \stackrel{\bar{K},\bar{q}\, \scriptsize\mbox{fixed}}{\underset{\Delta\to 0}{\to}} \bar{K}\pm \frac{1}{2} \bar{q}u \equiv x_\pm(\bar{K},u)\ \ \ \ \mbox{and}
\ \ \ \
\epsilon_\pm/\Delta \stackrel{\bar{K},\bar{q}\,\scriptsize \mbox{fixed}}{\underset{\Delta\to 0}{\to}} (1+x_\pm^2)^{1/2} \equiv e_\pm(\bar{K},u)
\ee
$ u $ being the cosine of the angle between the vectors $ \kk $ and $ \qq $, then in the integrals themselves in order to obtain their zero order approximation in $ \Delta/\mu $, identified by the exponent $ (0) $: 
\bea
\label{eq:sig0++}
\breve{\Sigma}_{W^+W^+}^{\epsilon\,(0)}(\bar{z},\bar{q})
&=& \frac{mk_0}{(2\pi\hbar)^2} \int_\mathbb{R} \dd \bar{K} \int_{-1}^{1} \dd u \left[
\frac{(e_++e_-)(e_+e_-+x_+x_-+1)}{2 e_+e_-[\bar{z}^2-(e_++e_-)^2]} +\frac{1}{2(\bar{K}^2+1)^{1/2}} \right] \not\equiv 0\\
\label{eq:sig0--}
\breve{\Sigma}_{W^-W^-}^{\epsilon\,(0)}(\bar{z},\bar{q})
&=& \frac{mk_0}{(2\pi\hbar)^2} \int_\mathbb{R} \dd \bar{K} \int_{-1}^{1} \dd u \left[
\frac{(e_++e_-)(e_+e_-+x_+x_--1)}{2 e_+e_-[\bar{z}^2-(e_++e_-)^2]} +\frac{1}{2(\bar{K}^2+1)^{1/2}}\right] \not\equiv 0\\
\label{eq:sig0+-}
\Sigma_{W^+W^-}^{z\,(0)}(\bar{z},\bar{q})
&=& \frac{mk_0}{(2\pi\hbar)^2} \int_\mathbb{R} \dd \bar{K} \int_{-1}^{1} \dd u  \,
\frac{\bar{z}(x_+e_-+x_-e_+)}{2 e_+e_-[\bar{z}^2-(e_++e_-)^2]}  \equiv 0  \\
\label{eq:sig0+0}
\Sigma_{W^+W^0}^{z\,(0)}(\bar{z},\bar{q})
&=& \frac{mk_0}{(2\pi\hbar)^2} \int_\mathbb{R} \dd \bar{K} \int_{-1}^{1} \dd u  \,
\frac{\bar{z}(e_++e_-)}{2 e_+e_-[\bar{z}^2-(e_++e_-)^2]} \not\equiv 0 \\
\label{eq:sig0-0}
\Sigma_{W^-W^0}^{\epsilon\,(0)}(\bar{z},\bar{q})
&=& \frac{mk_0}{(2\pi\hbar)^2} \int_\mathbb{R} \dd \bar{K} \int_{-1}^{1} \dd u  \,
\frac{(e_++e_-)(x_++x_-)}{2 e_+e_-[\bar{z}^2-(e_++e_-)^2]}  \equiv 0  \\
\label{eq:sig000}
\Sigma_{W^0W^0}^{\epsilon\,(0)}(\bar{z},\bar{q})
&=& \frac{mk_0}{(2\pi\hbar)^2} \int_\mathbb{R} \dd \bar{K} \int_{-1}^{1} \dd u  \,
\frac{(e_++e_-)(e_+e_--x_+x_-+1)}{2 e_+e_-[\bar{z}^2-(e_++e_-)^2]}  \not\equiv 0
\eea
A simplification occurs: integrals (\ref{eq:sig0+-}) and (\ref{eq:sig0-0}) are identically zero by antisymmetry of their integrand under the exchange $ (K, u) $ to $ (-K,-u) $, which transforms $ x_+$ and $ x_-$ into their opposite but leaves $ e_+$ and $ e_-$ unchanged. As a consequence, the second column of matrix $ M $ in (\ref{eq:defM}) is reduced to its element on the diagonal and the first eigenvalue equation in (\ref{eq:defmode}) is split in two independent equations: 
\be
\label{eq:decouplcontact}
\boxed{
M_{--}^{(0)}(\bar{z}_q,\bar{q})\equiv\breve{\Sigma}^{\epsilon\,(0)}_{W^-W^-}(\bar{z}_q,\bar{q})=0 \ \ \ \mbox{or}\ \ \ \left[1-g\Sigma_{W^0W^0}^{\epsilon\,(0)}(\bar{z}_q,\bar{q})\right]\breve{\Sigma}_{W^+W^+}^{\epsilon\,(0)}(\bar{z}_q,\bar{q})+g \left[\Sigma_{W^+W^0}^{z\,(0)}(\bar{z}_q,\bar{q})\right]^2=0
}
\ee
The first relates only to the variable $-$, that is to say the second component of vector (\ref{eq:cauchysable}) of the collective variables, the linear combination of $ \delta \Delta_\qq $ and $ \delta \Delta^*_{-\qq} $ reconstructing the deviation of the modulus $ | \Delta (\rr, t) | $ of the order parameter from its equilibrium value. It will give birth, after analytic continuation and explicit calculation in section \ref{subsec:exci_mod}, to the continuum mode, which is therefore, in the weak coupling limit and at the energy scale $ \Delta $, a {\sl modulus} mode. This result was known for a zero-range interaction between $ \uparrow $ and $ \downarrow $ ($ b \to 0 $) \cite{AndrianovPopov}; we see that it survives here at a nonzero range, even much greater than $ | a | $. The second equation in (\ref{eq:decouplcontact}) gives rise to the acoustic mode. It couples the variable $+$ (weak phase deviation of the order parameter from its equilibrium value) and the variable $ 0 $ (weak deviation of the gas density); it is only if we make $ g $ tend to zero, as in section \ref{subsec:son}, that the variable $+$ decouples from the density and that the acoustic mode becomes a pure {\sl phase} mode as in \cite{AndrianovPopov}. 

\subsubsection{In the presence also of the Coulomb interaction} 
\label{subsubsec:avec_coulomb} 

In a BCS superconductor, the paired fermions are not neutral particles, but electrons of charge $-e $. We would therefore have to complete our model Hamiltonian, by adding to the short-range interaction $ b $ between $ \uparrow $ and $ \downarrow $ the long-range, spin-independent Coulomb interaction potential, 
\be
V(\rr)=\frac{e^2}{4\pi\epsilon_0 r} = \frac{\tilde{e}^2}{r}
\ee
therefore acting both between same-spin and opposite-spin particles.  It would not be an easy task. Fortunately, the work has already been done in reference \cite{AndrianovPopov}, in the limit where the range $ b $ tends to zero at fixed scattering length $ a $ of the short-range interaction. The eigenvalue equation on the energy $ z $ of the modes of wave vector $ \qq $ takes the form of a determinant $ \det \hat{K} = 0 $, where $ \hat{K} $ is a $3\times 3$ matrix which we do not write, but whose coefficients are expressed in terms of functions $ A (\qq, \pm z) $, $ B (\qq, z) $, $ C (\qq, z) $ and $ D (\qq, \pm z) $. \footnote{Reference \cite{AndrianovPopov} wrongly states, after its equation (3.9), that $ D (z, \qq) $ is an odd function of the quadrivector $ (z, \qq) $; this is true only in the weak coupling limit at the scales of (\ref{eq:adim}).} 
Remarkably, we succeeded, after an integration at zero temperature on the Matsubara frequency not carried out in \cite{AndrianovPopov}, to connect these functions of the Coulomb problem to the functions (\ref{eq:defSigma}) of the non Coulomb problem as follows: 
\be
\label{eq:corAPnous}
\begin{array}{lll}
\displaystyle
A(\qq,\ii z)=\frac{1}{2}\left[\breve{\Sigma}^\epsilon_{W^+W^+}(z,\qq)+\breve{\Sigma}^\epsilon_{W^-W^-}(z,\qq)\right]-\Sigma_{W^+W^-}^z(z,\qq)
& \quad & \displaystyle C(\qq,\ii z) = \frac{q^2}{4\pi} - 2\tilde{e}^2\Sigma^\epsilon_{W^0W^0}(z,\qq) \\
&& \\
\displaystyle
B(\qq,\ii z)=\frac{1}{2}\left[\breve{\Sigma}^\epsilon_{W^-W^-}(z,\qq)-\breve{\Sigma}^\epsilon_{W^+W^+}(z,\qq)\right] &\quad &
D(\qq,\ii z)=\ii\tilde{e}\left[\Sigma^z_{W^+W^0}(z,\qq)-\Sigma^\epsilon_{W^-W^0}(z,\qq)\right]
\end{array}
\ee
In addition, in the weak coupling limit, the eigenvalue equation decouples in the same way as in the non-Coulombian case (\ref{eq:decouplcontact}): 
\be
\label{eq:decouplcoul}
\boxed{
M_{--}^{(0)}(\bar{z}_q,\bar{q})=0 \ \ \ \mbox{or}\ \ \ \left[\frac{q^2}{8\pi\tilde{e}^2}-\Sigma_{W^0W^0}^{\epsilon\,(0)}(\bar{z}_q,\bar{q})\right]\breve{\Sigma}_{W^+W^+}^{\epsilon\,(0)}(\bar{z}_q,\bar{q})+\left[\Sigma_{W^+W^0}^{z\,(0)}(\bar{z}_q,\bar{q})\right]^2=0
}
\ee
The first equation remains exactly the same: at the energy and wave number scales of (\ref{eq:adim}), the continuum mode is therefore absolutely not affected by the Coulomb interaction to zeroth order in $ \Delta $. The second equation is deduced from that of (\ref{eq:decouplcontact}) by the following substitution on the effective coupling constant: 
\be
\label{eq:subs}
g \to 2\tilde{V}(\qq) = \frac{8\pi \tilde{e}^2}{q^2}
\ee
where $ \tilde{V} (\qq) $ is the Fourier transform of $ V (\rr) $. The interpretation is simple: $ g $ represents, in fact, the Fourier transform of the interaction potential at wave vector $ \qq $, which goes unnoticed for a zero-range interaction (the Fourier transform of a Dirac distribution is constant), but is obvious for the Coulomb interaction; the factor $ 2 $ takes into account the fact that $ V (\rr) $ couples each spin state to the other two (unlike contact interaction); finally, $ g $ disappears completely in the substitution (\ref{eq:subs}), instead of being simply supplemented by the term $ 2 \tilde{V} (\qq) $, because the formalism of \cite{AndrianovPopov} is written in the zero range limit $ b/| a | \to 0 $, where the bare coupling constant $ g_0 $ appears instead of $ g $ as in the third column of (\ref{eq:defM}) and tends to zero.

Result (\ref{eq:decouplcoul}) was already in \cite{AndrianovPopov}, but only up to $ \bar{q} \lesssim $ 1. Let us add that, to give it a precise mathematical meaning, the term $ q^2/8 \pi \tilde{e}^2 $ must remain constant when $ \Delta \to 0 $. It is therefore not enough to fix $ \bar{q} $ and $ \bar{z} $ as in (\ref{eq:adim}), it is also necessary to fix the ratio $ \hbar \omega_0/\Delta $ between the plasma angular frequency $ \omega_0 $ of the Coulomb gas in its normal phase (see equation (3.8) of reference \cite{AndrianovPopov}) and its order parameter $ \Delta $ in the ground state. 
We have 
\be
\frac{q^2}{8\pi\tilde{e}^2}= \frac{2\bar{q}^2}{3} \frac{\Delta^2}{(\hbar\omega_0)^2} \frac{mk_{\rm F}}{(2\pi\hbar)^2}
\quad\mbox{with}\quad \frac{(\hbar\omega_0)^2}{\Delta^2} \equiv \frac{4\pi\hbar^2\tilde{e}^2\rho/m}{\Delta^2}
\ee

\subsection{Analytical expression of spectral densities} 
\label{subsec:cpf_denspec} 

\subsubsection{The spectral density $ \rho_{--}^{(0)} (\bar{z}, \bar{q}) $} 
\label{subsubsec:rho-0} 

To determine the dispersion relation of the modulus mode, that is to say the continuum collective branch, we must calculate the function $ M_{--}^{(0)} (\bar{z}, \bar{q}) $ appearing in (\ref{eq:decouplcontact}, \ref{eq:decouplcoul}) and find its roots after analytic continuation to the lower half-plane. We need for that, see the procedure (\ref{eq:Noz}), to know the associated spectral density (\ref{eq:defdensspec}), with $ s = s' =-$, in the weak coupling limit. As in section \ref{subsubsec:dmmicontact}, we make 
$ \Delta $ tend to zero in $ \rho_{--} (\epsilon, \qq) $ by fixing the reduced energy $ \bar{\epsilon} = \epsilon/\Delta $ and the reduced wave numbers $ \bar{q} $ and $ \bar{K} $ of equations (\ref{eq:adim}, \ref{eq:Kcouche}), and obtain the finite 
limit
\be
\label{eq:rho--0}
\rho_{--}^{(0)}(\bar{\epsilon},\bar{q}) \equiv \frac{\pi m k_0}{\hbar^2} \int_\mathbb{R} \dd \bar{K} \int_{-1}^{1} \dd u \frac{e_+e_-+x_+x_--1}{2 e_+e_-}
\delta(e_++e_--\bar{\epsilon})
\ee
Thanks to Dirac's distribution $ \delta $ fixing the total energy in the integrand, it is possible to integrate explicitly at least once by solving the equation $ e_++e_-= \bar{\epsilon} $, for example on the variable $ \bar{K} $, as it is done in \ref{subapp:lcf}: 
\be
\label{eq:firho--0}
\rho_{--}^{(0)}(\bar{\epsilon},\bar{q}) = \frac{2\pi m k_0}{\hbar^2} \int_0^1 \dd u \,\Theta(\bar{\epsilon}^2-(4+u^2\bar{q}^2)) \left(\frac{4+u^2\bar{q}^2-\bar{\epsilon}^2}{u^2\bar{q}^2-\bar{\epsilon}^2}\right)^{1/2}
\ee
where $ \Theta $ is the Heaviside function. 
To carry out the remaining angular integral, we must distinguish three cases according to the value of $ \bar{\epsilon} $: $ (i) $ the argument of $ \Theta $ is negative over the whole integration interval, $ (ii) $ it changes sign in the interior of the interval, and $ (iii) $ it is positive over the entire interval. At the boundary energies between these cases, the spectral density has a kink. The three corresponding analytical expressions are different, and are identified by exponents in Roman numerals, see table \ref{tab:rholcf}. 

\begin{table} [htb] 
\[
\begin{array}{|l|l|}
\hline
\bar{\epsilon} < 2 & \rho_{--}^{(0)[\mathrm{I}]}(\bar{\epsilon},\bar{q})=0 \\
                   & \rho_{++}^{(0)[\mathrm{I}]}(\bar{\epsilon},\bar{q})=0 \\
\hline
\!\!2<\bar{\epsilon}<\!(4+\bar{q}^2)^{1/2}\!\! & \displaystyle\rho_{--}^{(0)[\mathrm{II}]}(\bar{\epsilon},\bar{q})\equiv \frac{4\pi mk_0}{\hbar^2} \left[\frac{\bar{\epsilon}}{2\bar{q}} E\left(\frac{(\bar{\epsilon}^2-4)^{1/2}}{\bar{\epsilon}}\right) - \frac{2}{\bar{q}\bar{\epsilon}}K\left(\frac{(\bar{\epsilon}^2-4)^{1/2}}{\bar{\epsilon}}\right)\right] \\
 & \displaystyle\rho_{++}^{(0)[\mathrm{II}]}(\bar{\epsilon},\bar{q})\equiv\frac{4\pi mk_0}{\hbar^2} \left[\frac{\bar{\epsilon}}{2\bar{q}} E\left(\frac{(\bar{\epsilon}^2-4)^{1/2}}{\bar{\epsilon}}\right)\right] \\
\hline
(4+\bar{q}^2)^{1/2} < \bar{\epsilon}
& \displaystyle\rho_{--}^{(0)[\mathrm{III}]}(\bar{\epsilon},\bar{q})\equiv \frac{4\pi mk_0}{\hbar^2} \left[\frac{\bar{\epsilon}}{2\bar{q}} E\left(\asin \frac{\bar{q}}{(\bar{\epsilon}^2-4)^{1/2}},\frac{(\bar{\epsilon}^2-4)^{1/2}}{\bar{\epsilon}}\right) - \frac{2}{\bar{q}\bar{\epsilon}}F\left(\asin \frac{\bar{q}}{(\bar{\epsilon}^2-4)^{1/2}},\frac{(\bar{\epsilon}^2-4)^{1/2}}{\bar{\epsilon}}\right)\right]\!\!\\
& \displaystyle\rho_{++}^{(0)[\mathrm{III}]}(\bar{\epsilon},\bar{q})\equiv \frac{4\pi mk_0}{\hbar^2} \left[\frac{\bar{\epsilon}}{2\bar{q}} E\left(\asin \frac{\bar{q}}{(\bar{\epsilon}^2-4)^{1/2}},\frac{(\bar{\epsilon}^2-4)^{1/2}}{\bar{\epsilon}}\right) \right]\\
\hline
\end{array}
\]
\caption{Spectral densities $--$ and $++$ in the weak coupling limit. We took the limit $ \Delta/\mu \to 0 $ at fixed reduced energy $ \bar{\epsilon} = \epsilon/\Delta $ and reduced wave number $ \bar{q} = \hbar^2 k_0 q/m \Delta $, $ k_0 $ being the position of the minimum $ \Delta $ of the BCS dispersion relation. On each energy interval considered, these are regular functions, that can be analytically continued to complex energies. $ E (\phi, k) $ and $ F (\phi, k) $ are the elliptical integrals of second and first kind, in the definition \S 8.111 of reference \cite{GR}; $ E (k) \equiv E (\pi/2, k) $ and $ K (k) \equiv F (\pi/2, k) $ are their full forms.} 
\label{tab:rholcf} 
\end{table} 

Let us take advantage of our knowledge of the spectral density to express the function $ M_{--}^{(0)} (\bar{z}, \bar{q}) $ in the convenient form of a single integral on energy. The presence of the counter-term in integral (\ref{eq:sig0--}) ensuring its ultraviolet convergence poses a small technical difficulty. We get around it by introducing an arbitrarily large cut-off on $ \bar{K} $, which allows us to separate the counter-term and to express the rest as an integral with an energy cut-off; we subtract from the whole the quantity $ M_{--}^{(0)} (2,0) $, transformed beforehand in the same way, then we make the cut-off tend to infinity. Now $ M_{--}^{(0)} (2,0) = 0 $, taking into account the exact relation (\ref{eq:zeroexact}) to come. There remains therefore 
\be
\label{eq:M--0}
M_{--}^{(0)}(\bar{z},\bar{q})= (2\pi)^{-3} \int_2^{+\infty} \dd\bar{\epsilon} \left[\rho_{--}^{(0)}(\bar{\epsilon},\bar{q})
\left(\frac{1}{\bar{z}-\bar{\epsilon}}-\frac{1}{\bar{z}+\bar{\epsilon}}\right) -\rho_{--}^{(0)}(\bar{\epsilon},0)
\left(\frac{1}{2-\bar{\epsilon}}-\frac{1}{2+\bar{\epsilon}}\right)
\right]
\ee
where the spectral density at zero wave number, taken for example from table \ref{tab:rholcf} by making $ \bar{q} $ tend to zero over the interval III, is simply equal to 
\be
\label{eq:rho--00}
\rho_{--}^{(0)}(\bar{\epsilon},0) = \frac{2\pi m k_0}{\hbar^2} \Theta(\bar{\epsilon}-2) \frac{(\bar{\epsilon}^2-4)^{1/2}}{\bar{\epsilon}}
\ee

\subsubsection{Spectral density $ \rho_{++}^{(0)} (\bar{z}, \bar{q}) $} 
\label{subsubsec:rho++0} 

To determine the dispersion relation of the phase mode of the gas of neutral particles, that is to say the acoustic branch after taking the limit $ g \to 0 $ in the second equation of (\ref{eq:decouplcontact}), we need calculate the function $ M_{++}^{(0)} (\bar{z}, \bar{q}) \equiv \breve{\Sigma}_{W^+W^+}^{\epsilon \, (0)} (\bar{z}, \bar{q}) $. No analytic continuation is required, but knowledge of the associated spectral density $ \rho_{++}^{(0)} (\bar{\epsilon}, \bar{q}) $ leads us to a single integral on energy, rather than the double integral (\ref{eq:sig0++}). We therefore proceed as in section \ref{subsubsec:rho-0} to obtain 
\be
\label{eq:M++0}
M_{++}^{(0)}(\bar{z},\bar{q})= (2\pi)^{-3} \int_2^{+\infty} \dd\bar{\epsilon} \left[\rho_{++}^{(0)}(\bar{\epsilon},\bar{q})
\left(\frac{1}{\bar{z}-\bar{\epsilon}}-\frac{1}{\bar{z}+\bar{\epsilon}}\right) -\rho_{--}^{(0)}(\bar{\epsilon},0)
\left(\frac{1}{2-\bar{\epsilon}}-\frac{1}{2+\bar{\epsilon}}\right)
\right]
\ee
The analytical expression of the spectral density $++$ for arbitrary $ \bar{q} $ appears in table \ref{tab:rholcf}, that $--$ for $ \bar{q} = 0 $ appears in (\ref{eq:rho--00}), and we have as in (\ref{eq:M--0}) used the intermediate point $ (\bar{z}, \bar{q}) = (2,0) $ where $ M_{--}^{(0)} $ vanishes, to make the counter-term of integral (\ref{eq:sig0++}) disappear in the difference $ M_{++}^{(0)} (\bar{z}, \bar{q})-M_{--}^{(0)} (2.0) $. 

\subsection{Study of the modulus excitation branch} 
\label{subsec:exci_mod} 

The first equation of (\ref{eq:decouplcontact}) and (\ref{eq:decouplcoul}) has no solution on the real axis. We therefore carry out the analytic continuation to the lower complex half-plane, passing here between its branch points $ \bar{z} = 2 $ and $ \bar{z} = (4+\bar{q}^2)^{1/2} $, that is to say using, in the sense of table \ref{tab:rholcf}, the type II expression of the spectral density $--$ in the Nozi\`eres term from (\ref{eq:Noz}). So we have to look for solutions $ \bar{z}_{\bar{q}}^{(0)} $ of $ <0 $ imaginary part of the equation 
\be
\label{eq:defzqbar0}
{M_{--}^{(0)}}_{\downarrow[\mathrm{II}]}(\bar{z}^{(0)}_{\bar{q}},\bar{q})=0
\ee
with 
\be
\label{eq:m--0e}
{M_{--}^{(0)}}_{\downarrow[\mathrm{II}]}(\bar{z},\bar{q})= M_{--}^{(0)}(\bar{z},\bar{q})-(2\pi)^{-3} 2\ii\pi \rho_{--}^{(0)[\mathrm{II}]}(\bar{z},\bar{q})
\ee

\subsubsection{Numerical results} 
\label{subsubsec:resnumlcf} 

\begin{figure} [tb] 
\centerline{\includegraphics [width = 0.33 \textwidth, clip =]{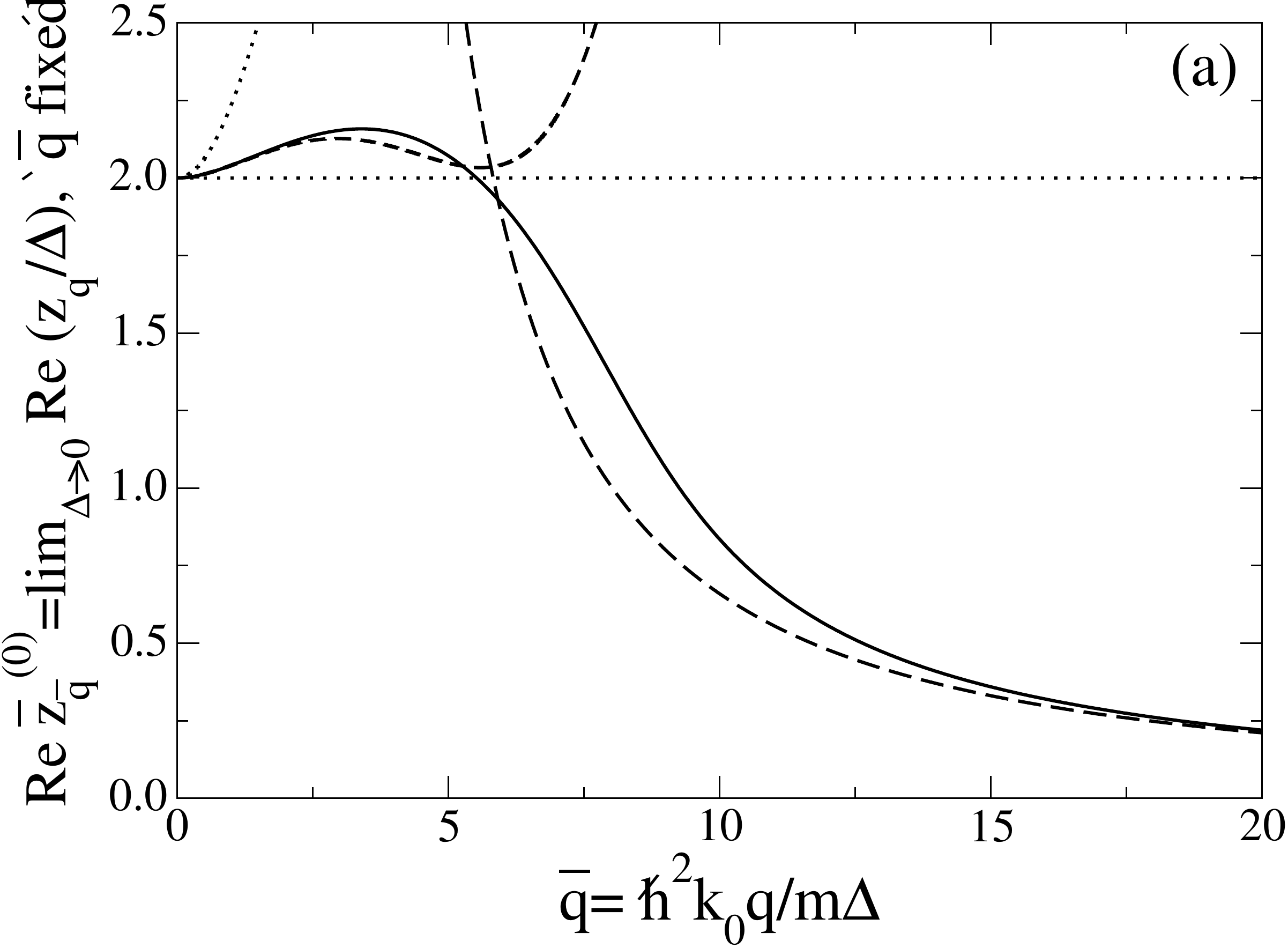} \hspace{3mm} \includegraphics [width = 0.33 \textwidth, clip =]{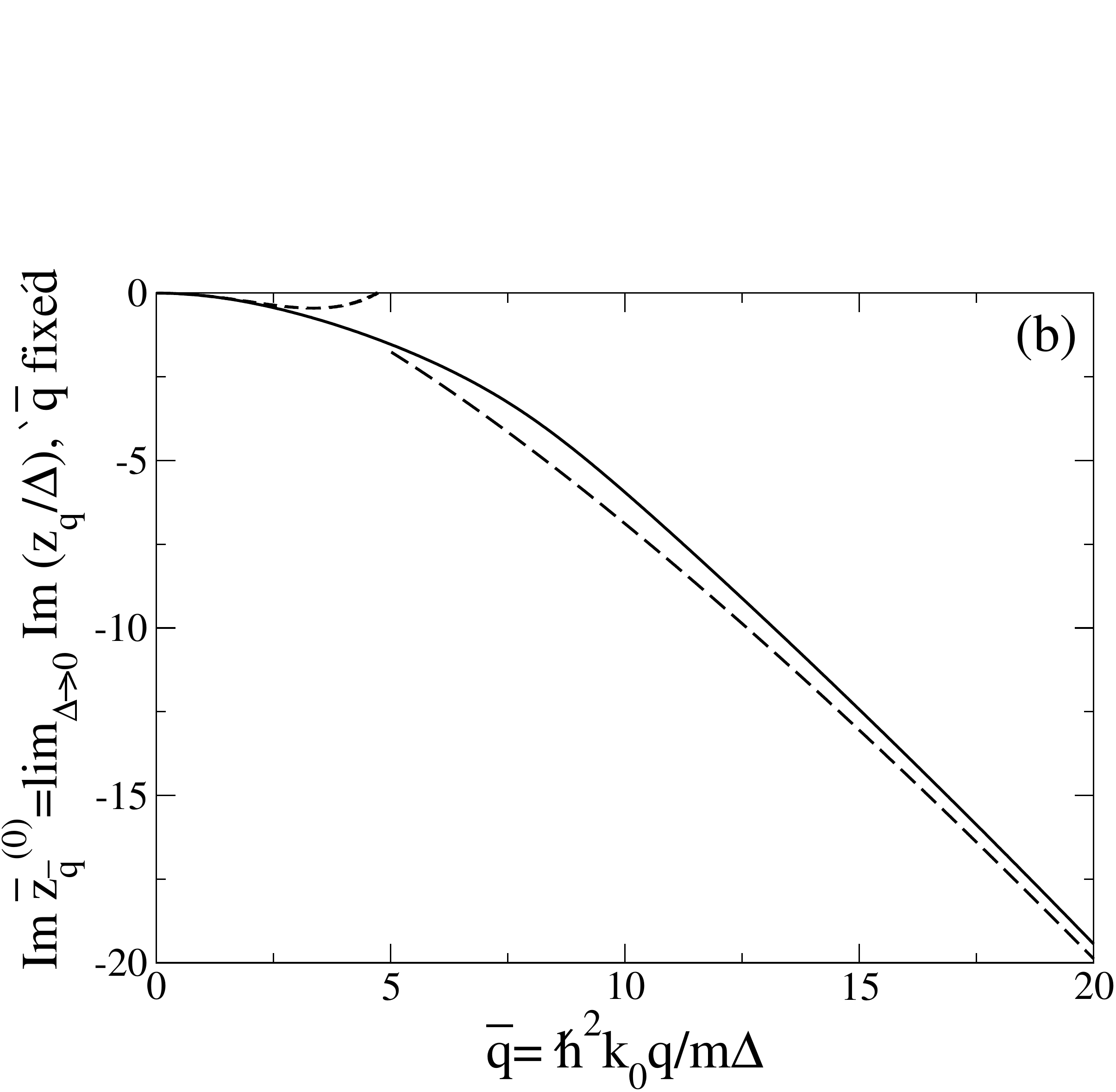} \hspace{3mm} \includegraphics [width = 0.33 \textwidth, clip =]{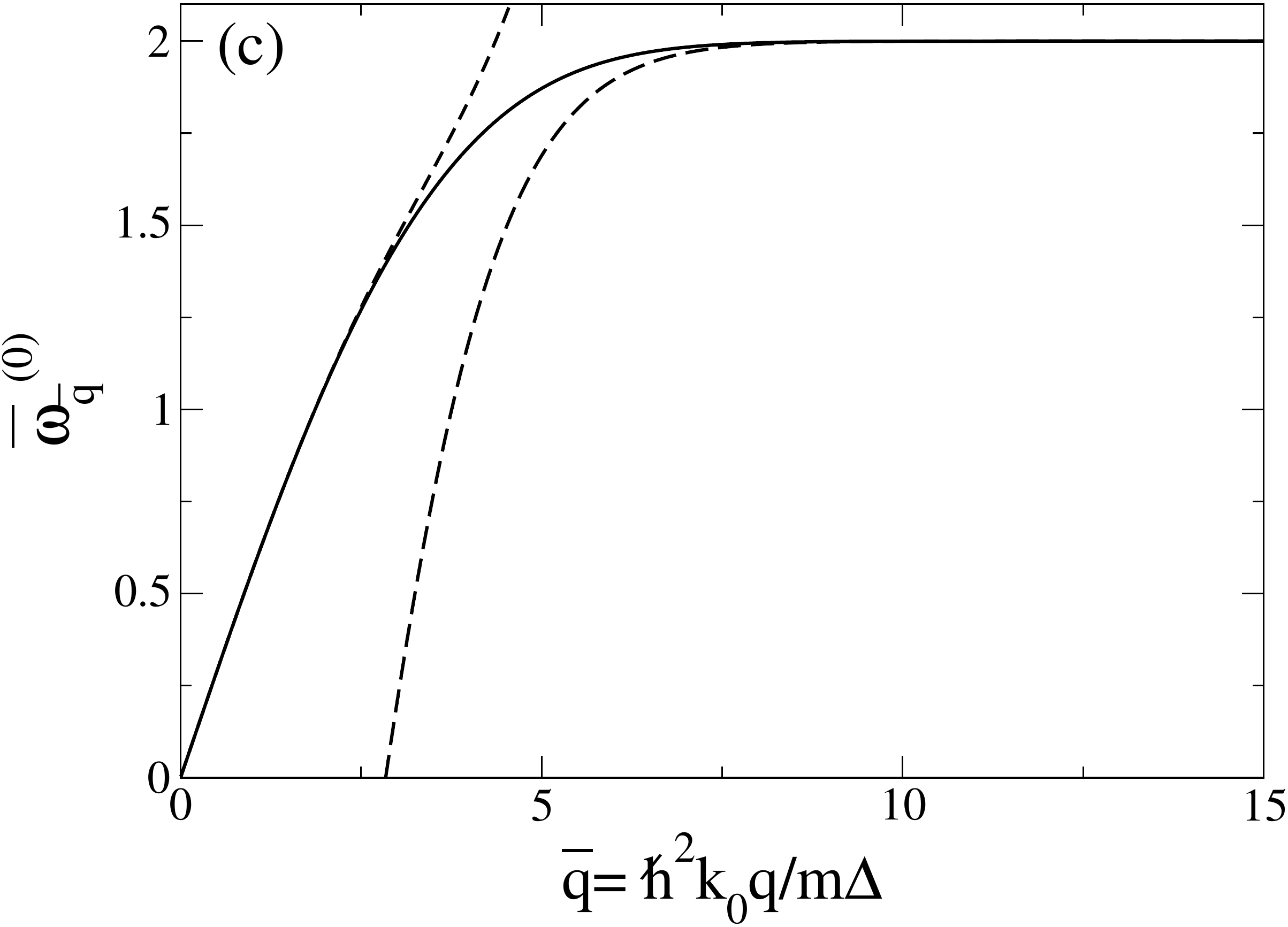}} 
\caption{In the weak coupling limit $ \Delta/\mu \to 0 $, real part (a) and imaginary part (b) of the complex eigenenergy $ z_\qq $ of the continuum collective branch, i.e.\  of the modulus mode, of an unpolarized gas of spin $1/2$ fermions, as functions of the wave number $ q $. The rescaling of $ q $ and $ z_\qq $ into $ \bar{q} $ and $ \bar{z}_{q} $ according to equation (\ref{eq:adim}) ensures that $ \bar{z}_{q} $ has a finite limit $ \bar{z}_{\bar{q}}^{(0)} $ when $ \Delta/\mu \to 0 $ at fixed $ \bar{q} $. Solid line: numerical calculation of the root of function (\ref{eq:m--0e}) in the lower complex half-plane. Short dashed line: Taylor expansion (\ref{eq:dl4}) of order 4 at low $ \bar{q} $. Long dashed line: asymptotic expansion (\ref{eq:asympt}) with large $ \bar{q} $, limited to leading order $ \bar{q} $ for the imaginary part and $ 1/\bar{q} $ for the real part. Dotted lines in (a): lower $ \bar{\epsilon}_1^{(0)} (\bar{q}) = 2 $ and upper $ \bar{\epsilon}_2^{(0)} (\bar{q}) = (4+\bar{q}^2)^{1/2} $ bounds of the energy interval which we have passed to analytically continue the equation on $ \bar{z}_{\bar{q}}^{(0)} $ in (\ref{eq:m--0e}); it is also the window for direct observability of the mode in the gas response functions \cite{PRL2019}. These results are valid at zero temperature for a contact interaction in the $ s $ wave, as in neutral particle superfluids, to which the Coulomb interaction can be added, as in the purely electronic BCS superconductors. We give in (c) the acoustic branch, that is to say the phase mode dispersion relation, in a superfluid of neutral fermions, obtained by numerically solving the second equation in (\ref{eq:decouplcontact}) at order 0 in the coupling constant $ g $, i.e.\ the equation $ M_{++}^{(0)} (\bar{\omega}^{(0)}_{\bar{q}}, \bar{q}) = $ 0 of unknown $ \bar{\omega}^{(0)}_{\bar{q}} \in [0,2 [$; no analytic continuation is required. The short and long dashed lines correspond to the low and large $ \bar{q} $ approximations in (\ref{eq:devson}). Here, the position of the minimum of the BCS dispersion relation $ \epsilon_\kk $ is $ k_0 = k_{\rm F} $, the Fermi wave number of the gas. 
} 
\label{fig:cpf} 
\end{figure}

The numerical solution of equation (\ref{eq:defzqbar0}) is easy, if we use a certain formal calculation software, in which numerical integration, evaluation of elliptical integrals and the search for roots of a function of the complex variable were preprogrammed. The result is presented in solid line in figure \ref{fig:cpf}a for the real part and in figure \ref{fig:cpf}b for the imaginary part of the continuum collective branch. As a function of the reduced wave number $ \bar{q} $, the imaginary part is monotonically decreasing and seems to tend asymptotically to $-\infty $. The real part has a more interesting structure: starting from $ 2^+$ (which corresponds to the expected unreduced eigenenergy $ 2 \Delta^+$), \footnote{We write $ 2^+$ because the continuum collective mode disappears at $ q = 0 $, see \cite{PRL2019} and references cited.} it goes through a maximum $ \re \bar{z} \simeq 2.158 $ at $ \bar{q} \simeq 3.403 $ before decreasing and, it seems, tending to zero at infinity. 

We also went in search of a type III continuum branch, analytically continuing the eigenenergy equation from the upper half-plane to the lower half-plane via the energy interval $ [(4+\bar{q}^2)^{1/2},+\infty [$. To do this, we replace $ \rho_{--}^{(0) [\mathrm{II}]} $ with $ \rho_{--}^{(0) [\mathrm{III}]} $ in (\ref{eq:defzqbar0}, \ref{eq:m--0e}), where the notations are those of table \ref{tab:rholcf}. However, we did not find a solution. It seems indeed that $ \im M_{--\downarrow [\mathrm{III}]}^{(0)} (\bar{z}, \bar{q}) <0 $ all over the interior of the fourth quadrant of the complex plane ($ \re \bar{z}> 0 $ and $ \im \bar{z} <0 $), for all $ \bar{q} $. \footnote{We checked it analytically at the edges of the domain, at infinity ($ | \bar{z} | \to+\infty $ at fixed phase) and for $ \bar{q} \to+\infty $ at fixed $ \bar{z} $ or at fixed $ \bar{z}/\bar{q} $.} 

\subsubsection{Analytical results at low $ \bar{q} $} 
\label{subsubsec:raafqb} 

To analytically determine the behavior of the continuum collective branch with low reduced wave number, we parametrize its eigenenergy as in reference \cite{PRL2019}, i.e.\  we set 
\be
\bar{z}=2+\frac{\zeta}{4}\bar{q}^2
\ee
in function (\ref{eq:m--0e}) then we make $\bar{q}$ tend to zero with a fixed coefficient $ \zeta $ in the lower complex half-plane. From the integral form (\ref{eq:M--0}) and from the expression of the spectral density $--$ in table \ref{tab:rholcf}, we get \footnote{In (\ref{eq:M--0}), we first change the variable $ \bar{\epsilon} = 2 \ch \Omega $. We then distinguish the intervals $ \Omega \in [0, A \bar{q}/2] $ and $ \Omega \in [A \bar{q}/2,+\infty [$, where $ A \gg $ 1 is arbitrary. On the first interval, we change the variable $ \Omega = x \bar{q}/2 $ then we expand the integrand in powers of $ \bar{q} $ at fixed  $ x $, by separating the cases $ x <1 $ and $ x> 1 $. On the second interval, we expand the integrand in powers of $ \bar{q} $ at fixed $ \Omega $. In the final result, we collect the terms order by order in $ \bar{q} $ and we make $ A $ tend to $+\infty $ in the coefficients. It remains to calculate some integrals of the form $ \int \dd x \sqrt{x^2-1} x/(x^2-\zeta)^n $ or $ \int \dd x \asin (1/x) x/(x^2-\zeta)^n $, $ n \in \{0,1,2 \} $. For example, $ \int_1^{+\infty} \dd x \asin (1/x) x/(x^2-\zeta) =-(\pi/4) \ln (1-1/\zeta)-(\ii \pi/2) \asin (1/\sqrt{\zeta}) $ ($ \im \zeta <0 $) knowing that $ \asin z =-\ii \ln (\sqrt{1-z^2}+\ii z), \forall z \in \mathbb{C} \setminus (]-\infty,-1 [\cup] 1,+\infty [$).} 
\begin{multline}
\label{eq:m--0efq}
{M_{--}^{(0)}}_{\downarrow[\mathrm{II}]}(\bar{z},\bar{q})\stackrel{\zeta\scriptsize\,\mbox{fixed},\, \im\zeta<0}{\underset{\bar{q}\to 0}{=}}
\frac{4\pi m k_0}{(2\pi)^3\hbar^2} \left\{\ii\frac{\pi}{8}\bar{q}\left(\sqrt{\zeta-1}+\zeta\asin\frac{1}{\sqrt{\zeta}}-\pi\zeta\right)
+\bar{q}^2\left(\frac{\zeta}{4}-\frac{1}{12}\right) \right. \\
\left. +\frac{\ii\pi\bar{q}^3}{256}\left[\left(\pi-\asin\frac{1}{\sqrt{\zeta}}\right)\zeta^2-(5\zeta-2)\sqrt{\zeta-1}\right]\right\}
+ O(\bar{q}^4)
\end{multline}
where the arrow reminds us that the analytic continuation has been made. It remains to replace $ \zeta $ by a Taylor expansion in powers of $ \bar{q} $, whose successive coefficients are determined by canceling the right-hand side of (\ref{eq:m--0efq}) order by order. The result, 
\be
\label{eq:dl4}
\boxed{
\bar{z}_{\bar{q}}^{(0)}\underset{\bar{q}\to 0}{=} 2+\frac{1}{4}\bar{q}^2 \left[\zeta_0+\zeta_1\bar{q}+\zeta_2\bar{q}^2+O(\bar{q}^3)\right] \ \ \ \mbox{with}\ \ \
\left\{
\begin{array}{ll}
\zeta_0=\sqrt{\zeta_0-1}/(\pi-\asin\frac{1}{\sqrt{\zeta_0}}) &\!\!\!\!\!\!\simeq \phantom{+}0. 236\, 883 - 0. 295\, 632\, \ii  \\
\zeta_1=2\ii\zeta_0(1-3\zeta_0)/(3\pi\sqrt{\zeta_0-1})&\!\!\!\!\!\!\simeq -0. 081\, 630 - 0. 014\, 462\, \ii \\
\zeta_2=\frac{\zeta_0^2(1-3\zeta_0)(\zeta_0-11/9)}{\pi^2(\zeta_0-1)^2} +\frac{\zeta_0(1-2\zeta_0)}{16} &\!\!\!\!\!\!\simeq \phantom{+}0. 007\, 158 + 0. 016\, 367\, \ii
\end{array}
\right.
}
\ee
shown in short dashed lines in figures \ref{fig:cpf}a and \ref{fig:cpf}b, qualitatively accounts for the existence of a maximum on the real part of the branch. The value of $ \zeta_0 $ was already included in references \cite{AndrianovPopov} and \cite{PRL2019}. It is interesting to note that expansion (\ref{eq:dl4}) has a nonzero cubic term, unlike the acoustic branch, see reference \cite{PRAConcav} and our equation (\ref{eq:devson}) to come from which it is obtained. 

\subsubsection{Analytical results for a large $ \bar{q} $} 
\label{subsubsec:raagqb} 

To analytically determine the behavior of the continuum collective branch for a large reduced wave number, we postulate a quasi-linear divergence of $ \bar{z}_{\bar{q}}^{(0)} $ (in fact only of its imaginary part, see below). We therefore set 
\be
\bar{z} = \mathcal{Z}\bar{q}
\ee
in function (\ref{eq:m--0e}) and make $ \bar{q} $ tend to infinity with a fixed coefficient $ \mathcal{Z} $ in the lower half-plane. From the integral form (\ref{eq:M--0}) and from the expression of the spectral density $--$ in table \ref{tab:rholcf}, we derive the asymptotic expansion \footnote{\label{note:aidint} We start by assuming that $ \im \mathcal{Z}> 0 $. In the integral expression (\ref{eq:M--0}), we change the variable $ \bar{\epsilon} = \bar{q} x $ then we expand the integrand in powers of $ 1/\bar{q} $ at fixed $ x $, distinguishing the cases $ x <1 $ and $ x> 1 $. In the first case, $ \rho_{--}^{(0)} (\bar{q} x, \bar{q}) = \frac{2 \pi mk_0}{\hbar^2} [x-\frac{1}{\bar{q}^2x} \ln (4 \eee \bar{q}^2x^2)+o (1/\bar{q}^2)] $; in the second, $ \rho_{--}^{(0)} (\bar{q} x, \bar{q}) = \frac{2 \pi mk_0}{\hbar^2} [1+\frac{1}{\bar{q}^2x} \ln \frac{x-1}{x+1}+o (1/\bar{q}^2)] $; these expansions are taken from form (\ref{eq:firho--0}), considering when $ x <1 $ the intervals $ u \in [0, xA/\bar{q}^2] $ and $ u \in [xA/\bar{q}^2, (x^2-4/\bar{q}^2)^{1/2}] $ where $ A \to+\infty $ after expansion in $ 1/\bar{q} $. We come across the integrals $ I = \int_0^1 \dd x \, \frac{\ln x}{xz} = g_2 (1/z) $ and $ J = \int_1^{+\infty} \dd x \, \ln [(x-1)/(x+1)]/(xz) = g_2 ((z+1)/(z-1))-g_2 (1) $ (we made in $ J $ the change of variable $ X = (x-1)/(x+1) \in [0,1] $ then we performed a partial fraction decomposition). The final result is analytically continued from $ \im \mathcal{Z}> 0 $ to $ \im \mathcal{Z} <0 $ through $ [0,1] $ by the substitution $ g_2 (1/\mathcal{Z}) \to g_2 (1/\mathcal{Z})+2 \ii \pi \ln \mathcal{Z} $, while $ g_2 (-1/\mathcal{Z}) $, $ g_2 ((\mathcal{Z}-1)/(\mathcal{Z}+1)) $ and $ g_2 ((\mathcal{Z}+1)/(\mathcal{Z}-1)) $ remain unchanged since $ g_2 (z) $ has $ [1,+\infty [$ as branch cut; similarly, the cleverly introduced functions $ \ln (1 \pm \mathcal{Z}) $ remain unchanged.} 
\be
\label{eq:m--0gq}
{M_{--}^{(0)}}_{\downarrow[\mathrm{II}]}(\bar{z},\bar{q})\stackrel{\mathcal{Z}\scriptsize\,\mbox{fixed},\, \im \mathcal{Z}<0}{\underset{\bar{q}\to +\infty}{=}}
\frac{4\pi m k_0}{(2\pi)^3\hbar^2} \left[\mathcal{M}_0(\mathcal{Z},\bar{q}) + \frac{1}{\bar{q}^2}\mathcal{M}_2(\mathcal{Z},\bar{q}) + o\left(\frac{1}{\bar{q}^2}\right)\right]
\ee
with coefficients that remarkably, depend slowly (logarithmically) on $ \bar{q} $: 
\bea
\mathcal{M}_0(\mathcal{Z},\bar{q}) &\!\!\!\!\!=\!\!\!\!\!& \ln\left(\frac{2\bar{q}}{\eee}\right) -\frac{\ii\pi \mathcal{Z}}{2} + \frac{1-\mathcal{Z}}{2}\ln\frac{1-\mathcal{Z}}{2} + \frac{1+\mathcal{Z}}{2}\ln\frac{1+\mathcal{Z}}{2} \\
\mathcal{M}_2(\mathcal{Z},\bar{q}) &\!\!\!\!\!=\!\!\!\!\!& \frac{1}{2\mathcal{Z}} \left[g_2\left(\frac{\mathcal{Z}-1}{\mathcal{Z}+1}\right)-g_2\left(\frac{\mathcal{Z}+1}{\mathcal{Z}-1}\right)-\ln\left(4\eee \bar{q}^2\right) \left(\ln\frac{1+\mathcal{Z}}{2}\!-\!\ln\frac{1-\mathcal{Z}}{2}-\!\ii\pi\right)+\!2g_2\left(\frac{1}{\mathcal{Z}}\right)-\!2g_2\left(-\frac{1}{\mathcal{Z}}\right)+\!4\ii\pi \ln \mathcal{Z}\right] \nonumber \\
&\!\!\!\!\!\!\!\!\!\!&
\eea
where $ g_\alpha (z) $ is a Bose function or polylogarithm. 
From the cancellation of the right-hand side of (\ref{eq:m--0gq}) order by order in $ \bar{q} $, we therefore deduce an asymptotic expansion 
\be
\label{eq:asympt}
\boxed{
\bar{z}_{\bar{q}}^{(0)}\underset{\bar{q}\to+\infty}{=} \mathcal{Z}_0 \bar{q} +\frac{\mathcal{Z}_2}{\bar{q}}+o\left(\frac{1}{\bar{q}}\right)
}
\ee
whose coefficients also depend logarithmically on $ \bar{q} $. As one would expect from figures \ref{fig:cpf}a and \ref{fig:cpf}b, the coefficient of the dominant term is pure imaginary: 
\be
\label{eq:Y0}
\mathcal{Z}_0=-\ii Y_0 \ \ \ \mbox{with}\ \ \ \ln\left(\bar{q}\sqrt{1+Y_0^2}/\eee\right)=\left(\frac{\pi}{2}+\atan Y_0\right)Y_0
\ee
We don't know how to solve the transcendental equation on the imaginary part $ Y_0> 0 $ other than numerically. \footnote{By committing a relative error $ O (1/Y_0^3) $ on $ Y_0 $, we can replace the equation by $ \pi Y_0 = \ln (\bar{q} Y_0) $ whose solution is $ Y_0 = (-1/\pi) W_{-1} (-\pi/\bar{q}) $ where $ W_{-1} $ is the secondary branch of the Lambert function $ W $.} The corresponding prediction is shown in long dashed lines in figure \ref{fig:cpf}b. In the sub-leading term, the real part is the most physically interesting part, because it provides an equivalent of the real part of the continuum collective branch. Luckily, it is also the simplest one to write, since only the imaginary part of $ \mathcal{Z} \mapsto \mathcal{M}_2 (\mathcal{Z}, \bar{q}) $ takes a simple form on the pure imaginary axis: 
\be
\im \mathcal{M}_2(-\ii Y,\bar{q})=\frac{\pi^2}{Y}\quad \forall Y>0 \ \ \ \mbox{then}\ \ \ \re \mathcal{Z}_2 = \frac{\pi^2}{\left(\frac{\pi}{2}+\atan Y_0\right)Y_0}
\ee
The result, tending to zero as $ 1/\bar{q} $ up to a logarithmic factor, is shown in long dashed lines in figure \ref{fig:cpf}a. 

\subsection{Acoustic branch for a contact interaction in the limit $ g \to 0 $} 
\label{subsec:son} 

As we know, our neutral fermion gas has an acoustic excitation branch $ q \mapsto \hbar \omega_\qq $, purely real in the time-dependent linearized BCS formalism used here, and extensively studied in reference \cite{CKS}. We consider it useful, however, to give some additional results in the weak coupling limit, where it can be rescaled as the continuum collective branch: 
\be
\frac{\hbar\omega_{\qq}}{\Delta}\stackrel{\bar{q}\scriptsize\,\mbox{fixed}}{\underset{\check{\Delta}\to 0}{\to}}
\bar{\omega}_{\bar{q}}^{(0)}
\ee
and where it solves the second equation of (\ref{eq:decouplcontact}). To simplify, let us neglect in the latter the small corrections proportional to the effective coupling constant $ g $. Then $ \bar{\omega}_{\bar{q}}^{(0)} $ is a root of $ \bar{z} \mapsto M_{++}^{(0)} (\bar{z}, \bar{q}) $, without any need for analytic continuation (unlike the case at nonzero temperature, where the branch cut on the eigenenergy equation reaches zero energy, and where up to four complex acoustic branches can be found \cite{JLTP, sontempnn}). 
By numerical solution, we plot the acoustic branch in figure \ref{fig:cpf}c: after a linear start in the reduced wave number, it saturates quickly (faster than a power law) to the reduced value $ 2 $, that is, to the energy $ 2 \Delta $ of the lower edge of the broken pair continuum. On the same figure are shown with a dashed line the analytical predictions 
\be
\label{eq:devson}
\boxed{
\bar{\omega}_{\bar{q}}^{(0)} \underset{\bar{q}\to 0}{=} \frac{\bar{q}}{\sqrt{3}} \left[1-\frac{\bar{q}^2}{45} +\frac{17\bar{q}^4}{28350}+O(\bar{q}^6)\right]
\quad\mbox{and}\quad
\bar{\omega}_{\bar{q}}^{(0)}-2\underset{\bar{q}\to +\infty}{\sim} -16\eee^{-2} \exp[-(2\bar{q}/\pi)\ln(\bar{q}/\eee)]
}
\ee
The one at low $ \bar{q} $ is taken from reference \cite{PRAConcav}. The one at large $ \bar{q} $ is original: we deduce it from the integral form (\ref{eq:M++0}) of $ M_{++}^{(0)} $. \footnote{To do this, we had to overcome a non-trivial integral, 
$ \int_0^1 \dd x \frac{E (x)-E (0)-x^2 [E (1)-E (0) ]}{x (1-x^2)^{3/2}} = \int_0^1 \dd u \frac{E (u ')-E (1)}{u^2}+\int_0^1 \dd u \frac{E (u ')-E (0)}{1-u^2} = 1+\frac{\pi}{2} (\ln 2-1) $ 
where we made the change of variable $ x = u '$, with the notation $ u' = (1-u^2)^{1/2} $ as in \S 8.111 of reference \cite{GR}.} 

\section{Arbitrary interaction regime: the case of cold atoms in the BEC-BCS crossover} 
\label{sec:cold atoms}

In a cold atom gas, opposite spins fermions are subject to short-range van der Waals interactions, of scattering length $ a $ virtually adjustable between $-\infty $ and $+\infty $ by Feshbach resonance. In the case $ a <0 $, however, the gases are always prepared in the strong interaction regime $ 1 \lesssim k_{\rm F} | a | $, which alone ensures an accessible reduced critical temperature for superfluidity $ T_c/T_{\rm F} $ (we hardly know how to cool these gases in a controlled manner below $ T = 0.05 T_{\rm F} $, where $ T_{\rm F} $ is the Fermi temperature  \cite{Zwierlein2012, KetterleRecord}). This contrasts with the weak interaction limit $ k_{\rm F} | a | \ll 1 $ in section \ref{sec:lcf}. The ground state of the gas remains qualitatively the same, that is to say a condensate of $ \uparrow \downarrow $ pairs bound by the Cooper mechanism in the presence of Fermi seas, but BCS theory becomes less quantitative. In addition, under pain of having heavy three-body losses, the experiments do not leave the dilute regime $ k_{\rm F} b \ll 1 $, where $ b $ is the interaction range \footnote{We limit here, as in most experiments, to a broad magnetic Feshbach resonance, where the true range $ b $ and the effective range $ r_e $ are of the order of the van der Waals length $ \ell = (m C_6/\hbar^2)^{1/4} $, unlike narrow resonances for which $-r_e \gg b \approx \ell $.}; this condition, joined to $ 1 \lesssim k_{\rm F} | a | $, now imposes that $ | a | \gg b $: the interaction becomes resonant in the $s$ wave. In the case $ a> 0 $, there exists for $ b \lesssim a $ a dimer $ \uparrow \downarrow $ of characteristic diameter $ a $ in free space and the bound pairs can be formed independently of the existence of Fermi seas; at sufficiently low density, the average distance between fermions is $ \gg a $ and the ground state of the gas tends to a Bose-Einstein condensate (BEC) of dimers. Even in this case, the experiments are limited to the resonant regime $ a \gg b $, otherwise the dimer gas would have a short lifetime under the effect of three-body losses. \footnote{The elastic collision rate would become lower than the inelastic rate, which would compromise evaporative cooling \cite{DPGSCS}.} In practice, experimental studies focus on the crossover between the two limiting cases BCS $ k_{\rm F} a \to 0^-$ and BEC $ k_{\rm F} a \to 0^+$, in the crossing interval $-1 \lesssim 1/k_{\rm F} a \lesssim 1 $ made non trivial by the interaction strength and made universal by the condition everywhere satisfied $ | a | \gg b $ \cite{Leggett1980, Engelbrecht, livreZwerger}.

In our model, the interaction range is represented by the spacing $ b $ of the cubic lattice. The resonant scattering condition $ | a | \gg b $ (knowing that we always have $ k_{\rm F} b \ll 1 $) therefore allows us to take the continuous limit $ b \to 0 $, which brings some welcome simplifications (section \ref{subsec:bzero}). The resulting spectral densities can be calculated analytically (section \ref{subsec:rhosspgen}). After having presented observables of the continuum mode complementary to its dispersion relation (section \ref{subsec:carac}), we carry out a numerical study (section \ref{subsec:numcom}), then a detailed analytical study by successively considering the limit of low wave numbers (section \ref{subsec:faibleq}), that of a weak order parameter $ \Delta/\mu \ll 1 $ (section \ref{subsec:faibledelta}) and that of a weak and positive chemical potential $ \Delta/\mu \gg 1 $ (section \ref{subsec:grand_d}). All these results on the continuum collective branch are obtained for $ \mu> 0 $, where the BCS excitation spectrum $ \epsilon_\kk = [(E_\kk-\mu)^2+\Delta^2]^{1/2} $ reaches its minimum at the nonzero wave number $ k_0 = (2m \mu)^{1/2}/\hbar $, and by analytic continuation of the eigenenergy equation through the natural window delimited by its first two branch points $ \epsilon_1 (q) $ and $ \epsilon_2 (q) $, which imposes $ 0 <q <2 k_0 $ \cite{PRL2019}. We generalize however the study in section \ref{subsec:autrefen_muneg} by carrying out an analytic continuation through the other windows (of energy higher than $ \epsilon_2 (q) $) and by considering the case $ \mu <0 $ where the BCS spectrum is minimal at $ k = $ 0. 

\subsection{Continuous or zero-range limit $ b \to 0 $} 
\label{subsec:bzero} 

Let us make the spacing $ b $ of the lattice model tend to zero at fixed scattering length $ a $. We can first replace the first Brillouin zone $ \mathcal{D} $ by $ \mathbb{R}^3 $ in the integrals $ \Sigma $ of (\ref{eq:defSigma}); they were built to have a finite value in this limit. We then notice, on its expression given at the beginning of section \ref{sec:reminders}, that the on-site coupling constant $ g_0 $ tends to zero. We can therefore replace $ g_0 $ by zero in the third column of matrix $ M $  in equation (\ref{eq:defM}), which greatly simplifies the writing of the determinant of $ M $ in the eigenenergy equation (\ref{eq:defmode}): 
\be
\label{eq:detMbzero}
\det M(z,\qq) = M_{++}(z,\qq) M_{--}(z,\qq)-[M_{+-}(z,\qq)]^2
\ee
where $ M_{ss} $ are the diagonal elements and $ M_{+-} = M_{-+} $ the non-diagonal elements of the upper left $ 2 \times 2 $ block of $ M $. \footnote{Equation (\ref{eq:detMbzero}) is already in \cite{CKS}. It is also equivalent, given the identities (\ref{eq:corAPnous}), to the zero temperature limit of equation (2.6) of \cite{AndrianovPopov}, except that this reference approximates the chemical potential $ \mu $ by Fermi energy $ \epsilon_{\rm F} $ in the BCS dispersion relation.} For the same reason, the inverse $ M^{-1} $ appearing in time dependence (\ref{eq:cauchysable}) of collective variables is simplified: small phase and modulus deviations from equilibrium of the order parameter only depend on the first two components of the source vector of (\ref{eq:cauchysable}), which allows, without taking up too much space, to give the inspiring writing displaying the determinant of $ M $ in the denominator: 
\be
\label{eq:signaltbzero}
\begin{pmatrix}
\delta\Delta_\qq(t)-\delta\Delta_{-\qq}^*(t) \\
\delta\Delta_\qq(t)+\delta\Delta_{-\qq}^*(t)
\end{pmatrix}
=\int_{\ii\eta+\infty}^{\ii\eta-\infty} \frac{\dd z}{2\ii\pi}\, \frac{\eee^{-\ii zt/\hbar}}{\det M(z,\qq)}
\begin{pmatrix}
M_{--}(z,\qq) & -M_{+-}(z,\qq) \\
-M_{+-}(z,\qq) & M_{++}(z,\qq)
\end{pmatrix}
\begin{pmatrix}
\Sigma_{W^+y^-}^z(z,\qq)+\Sigma_{W^+y^+}^\epsilon(z,\qq) \\
\Sigma_{W^-y^+}^z(z,\qq)+\Sigma_{W^-y^-}^\epsilon(z,\qq)
\end{pmatrix}
\ee
On the other hand, the small density deviation $ \delta \rho_\qq (t) $ depends on the three components of the source vector and involves other matrix elements of $ M $ than those of (\ref{eq:detMbzero}); for simplicity, we will therefore not study it here, although it is more directly measurable than the deviations of the order parameter. 

For the numerical and analytical studies to come, it is advisable to express the useful matrix elements of $ M $ in terms of the corresponding spectral densities (\ref{eq:defdensspec}), these being anyway required by the Nozi\`eres analytic continuation (\ref{eq:Noz}). It is straightforwardly done in the case of $ M_{+-} $: it is enough to insert in the integrand (\ref{eq:defSigma}) of $ \Sigma_{W^+W^-}^z $ the integral of $ \delta (\epsilon-\epsilon_{\kk \qq}) $ on all energies $ \epsilon \geq 2 \Delta $, which is obviously equal to unity, then exchange the integration on the wave vector $ \kk $ and on $ \epsilon $ to reveal the spectral density $ \rho_{+-} $ as in (\ref{eq:defdensspec}). We get 
\be
\label{eq:M+-foncrho}
M_{+-}(z,\qq)=(2\pi)^{-3} \int_{2\Delta}^{+\infty} \dd\epsilon\, \rho_{+-}(\epsilon,\qq) \left(\frac{1}{z-\epsilon}+\frac{1}{z+\epsilon}\right)
\ee
The case of the diagonal elements $ M_{ss} $, $ s = \pm $, is made more delicate by the necessary presence of the counter term $ 1/2 \epsilon_\kk $; before applying the same procedure as for $ M_{+-} $, we therefore introduce an arbitrarily large cut-off $ A $ on the energy $ \epsilon_{\kk \qq} $ of a broken pair, using a factor $ \Theta (A-\epsilon_{\kk \qq}) $ ($ \Theta $ is the Heaviside function), then we separate the counter-term from the rest. In the rest, we use a minus-plus trick, taking advantage of the fact that the spectral densities $ \rho_{ss} (\epsilon, \qq) $ are equivalent at high energy to $ \rho_\infty (\epsilon) = \pi (m/\hbar^2)^{3/2} \epsilon^{1/2} $, proportional to the density of states of a free particle of mass $ m/2 $. \footnote{Asymptotically, the introduction of the factor $ \Theta (A-\epsilon_{\kk \qq}) $ amounts to cutting the wave number $ k $ or the kinetic energy $ E_\kk $ since $ \epsilon_{\kk \qq} = 2E_\kk+O (1) $ when $ k \to+\infty $. Since the weights $ W_{\kk \qq}^{\pm} $ tend to $ 1 $ in this limit, with a difference $ O (1/k^4) $, we deduce that $ \rho_{ss} (\epsilon, \qq) \sim \rho_\infty (\epsilon) $ as advertised. In the integral on $ \epsilon $ still cut-off at $ A $ by the factor $ \Theta (A-\epsilon) $, we subtract from the integrand its equivalent $ (-2/\epsilon) \rho_\infty (\epsilon) $, which allows us to take the limit $ A \to+\infty $; the integral of the remaining bit $ \int_{2 \Delta}^{A} \dd \epsilon \, (-2/\epsilon) \rho_\infty (\epsilon) $ is easy to calculate. In the integral on $ \kk $ of the counter-term, still cut-off by $ \Theta (A-\epsilon_{\kk \qq}) $, we subtract from $ 1/2 \epsilon_\kk $ its equivalent $ 1/2E_\kk $, which makes it possible to take the limit $ A \to+\infty $; the integral of the remaining bit $ \int \dd^3 \kk \, \Theta (A-\epsilon_{\kk \qq})/2E_\kk $ is easily calculated in the limit of large $ A $ since we can replace $ \epsilon_{\kk \qq} $ with $ 2 E_\kk $.} It finally comes: 
\be
\label{eq:Mssgen}
M_{ss}(z,\qq)=\frac{C(\Delta)}{(2\pi)^3} +(2\pi)^{-3} \int_{2\Delta}^{+\infty} \dd\epsilon \left[\rho_{ss}(\epsilon,\qq)\left(\frac{1}{z-\epsilon}-\frac{1}{z+\epsilon}\right)+\frac{2}{\epsilon} \rho_\infty(\epsilon)\right],
\quad \forall s\in\{+,-\}
\ee
In the regularization scoria 
\be
C(\Delta)=4\rho_\infty(2\Delta) + \int_{\mathbb{R}^3}\frac{\dd^3k}{2}\left(\frac{1}{\epsilon_\kk}-\frac{2m}{\hbar^2 k^2}\right) = 2\pi\left(\frac{2m}{\hbar^2}\right)^{3/2}\Delta^{1/2}\left[1+I_1(\mu/\Delta)\right]
\ee
appears an integral on $ \kk $ already calculated analytically in \cite{Marini1998}, see the function $ I_1 (\mu/\Delta) $ of this reference, but of which we give here, by hyperbolic parametrization, the more compact expression $ I_1 (\sh \tau) = \ch \tau \left (2 \eee^\tau \right)^{1/2} K \left (\ii \eee^\tau \right)-\left (2 \eee^{-\tau} \right)^{1/2} E (\ii \eee^\tau) $, where $ E $ and $ K $ are the complete elliptical integrals of second and first kind as in \S 8.112 of reference \cite{GR}. To be complete, let us note that a more beautiful form, without regularization scoria but with a more elaborate counter-term, is easily deduced from identities \footnote{We establish them by direct substitution in integral forms (\ref{eq:defSigma}): taking into account (\ref{eq:Wexpli}) and $ \epsilon_\kk^2 = \xi_\kk^2+\Delta^2 $, the integrand is identically zero.} 
\be
\label{eq:zeroexact}
M_{++}(0,\mathbf{0})=M_{--}(2\Delta,\mathbf{0})=0
\ee
Just subtract from (\ref{eq:Mssgen}) with $ s =+$ its value for $ (z = 0, \qq = \mathbf{0}) $, which is zero, and from (\ref{eq:Mssgen}) with $ s =-$ its value for $ (z = 2 \Delta, \qq = \mathbf{0}) $, also zero, to obtain 
\bea
\label{eq:m++fintastu}
M_{++}(z,\qq) &=& (2\pi)^{-3} \int_{2\Delta}^{+\infty} \dd\epsilon \left[\rho_{++}(\epsilon,\qq)\left(\frac{1}{z-\epsilon}-\frac{1}{z+\epsilon}\right)
+\frac{2}{\epsilon}\rho_{++}(\epsilon,\mathbf{0})\right] \\
\label{eq:m--fintastu}
M_{--}(z,\qq) &=& (2\pi)^{-3} \int_{2\Delta}^{+\infty} \dd\epsilon \left[\rho_{--}(\epsilon,\qq)\left(\frac{1}{z-\epsilon}-\frac{1}{z+\epsilon}\right)
-\rho_{--}(\epsilon,\mathbf{0})\left(\frac{1}{2\Delta-\epsilon}-\frac{1}{2\Delta+\epsilon}\right)\right]
\eea
The spectral densities have a simple analytic expression at zero vector wave, \footnote{\label{note:rhoqz} We limit ourselves for brevity to the case $ \mu> 0 $ and introduce the intermediates $ \rho_{\pm} (\epsilon) = (\pi/2) (2m/\hbar^2)^{3/2} [\mu \pm (\epsilon^2-4 \Delta^2)^{1/2}/2]^{1/2} $, with $ \rho_+(\epsilon) $ of support $ [2 \Delta,+\infty [$ and $ \rho_-(\epsilon) $ of support $ [2 \Delta, 2 (\Delta^2+\mu^2)^{1/2}] $ (any function is by definition zero outside its support). Then $ \rho_{--} (\epsilon, \mathbf{0}) = [\rho_+(\epsilon)+\rho_-(\epsilon)] (\epsilon^2-4 \Delta^2)^{1/2}/\epsilon $, $ \rho_{++} (\epsilon, \mathbf{0}) = [\rho_+(\epsilon)+\rho_-(\epsilon)] \epsilon/(\epsilon^2-4 \Delta^2)^{1/2} $ and $ \rho_{+-} (\epsilon, \mathbf{0}) = \rho_+(\epsilon)-\rho_-(\epsilon) $.} but much less simple at nonzero $ \qq $ as it will appear in the next section \ref{subsec:rhosspgen}. 

\subsection{Analytical computation of spectral densities} 
\label{subsec:rhosspgen} 

\subsubsection{Non-analyticity points $ \epsilon_1 (q) $, $ \epsilon_2 (q) $ and $ \epsilon_3 (q) $} 
\label{subsubsec:lpdnae1e2e3} 

To perform the analytic continuation of the matrix elements $ M_{ss'} (z, \qq) $ as in (\ref{eq:Noz}), we must first find the points of non-analyticity in $ \epsilon $ of the corresponding spectral densities $ \rho_{ss'} (\epsilon, \qq) $ on the real axis. This work was carried out qualitatively in reference \cite{PRL2019} and the physical origin of these points was explained. We therefore give here a purely quantitative discussion. 

\begin{figure} [tb] 
\centerline{\includegraphics [width = 0.33 \textwidth, clip =]{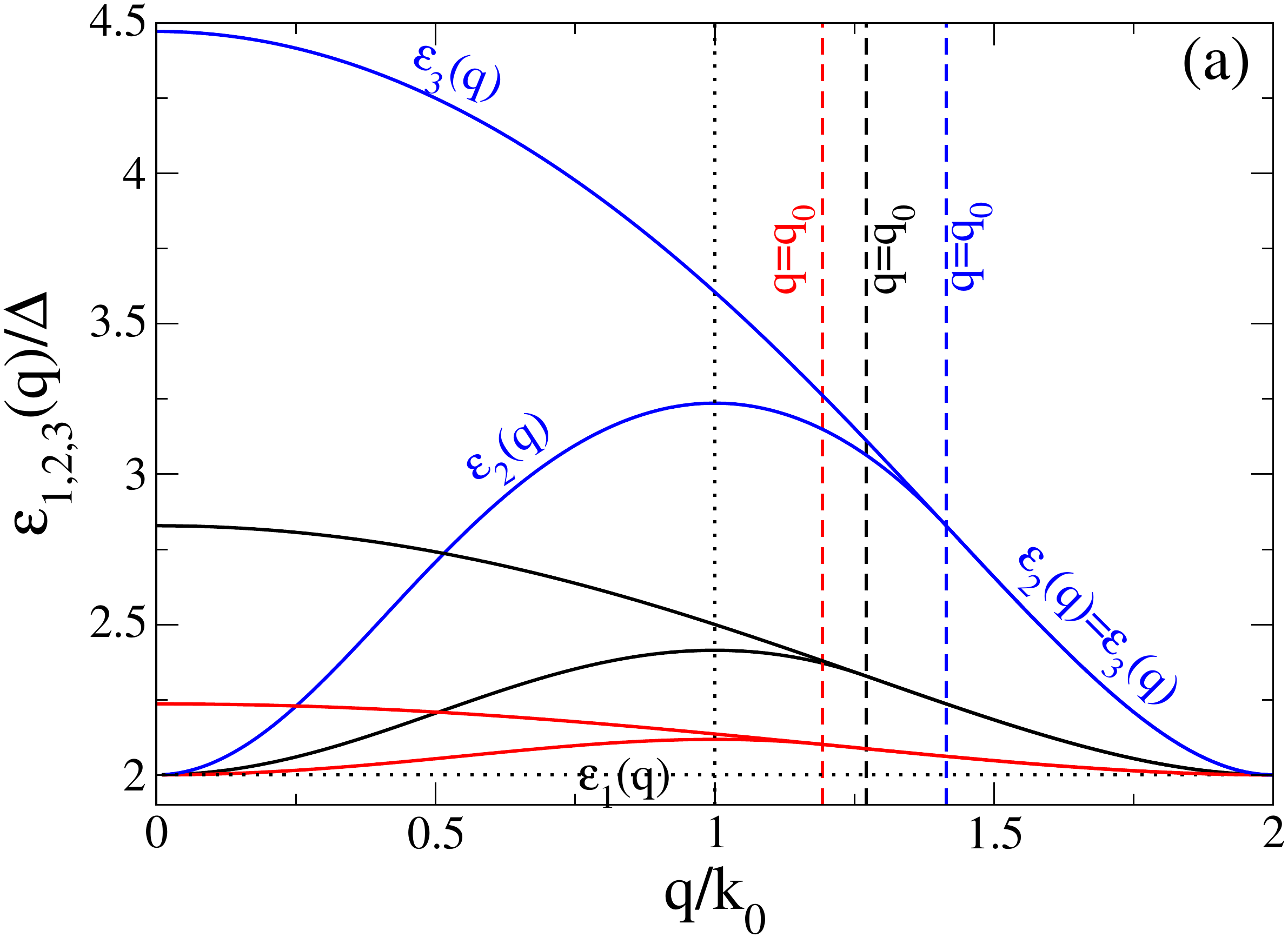} \hspace{6mm} \includegraphics [width = 0.33 \textwidth, clip =]{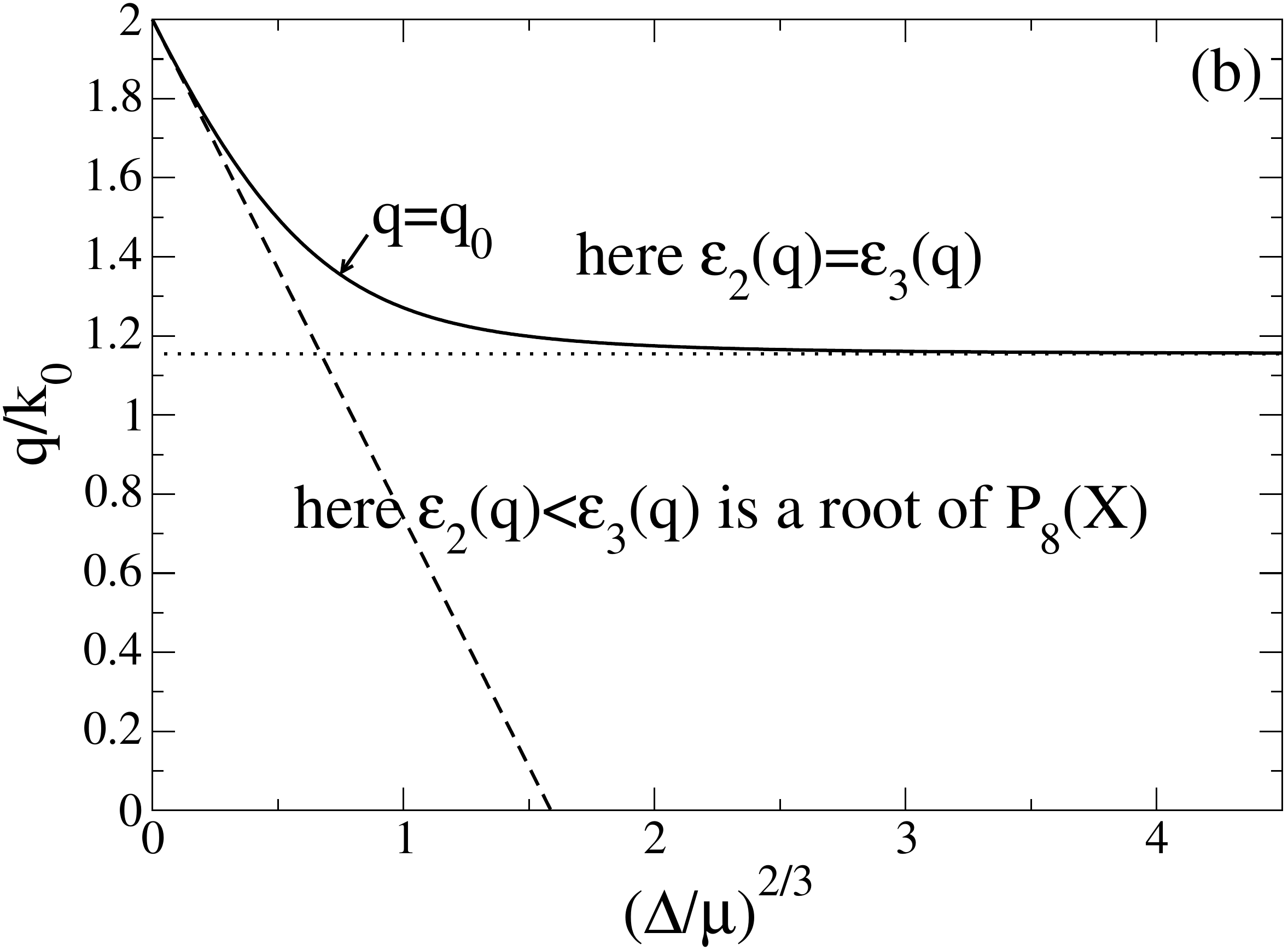} \hspace{6mm} \includegraphics [width = 0.33 \textwidth, clip =]{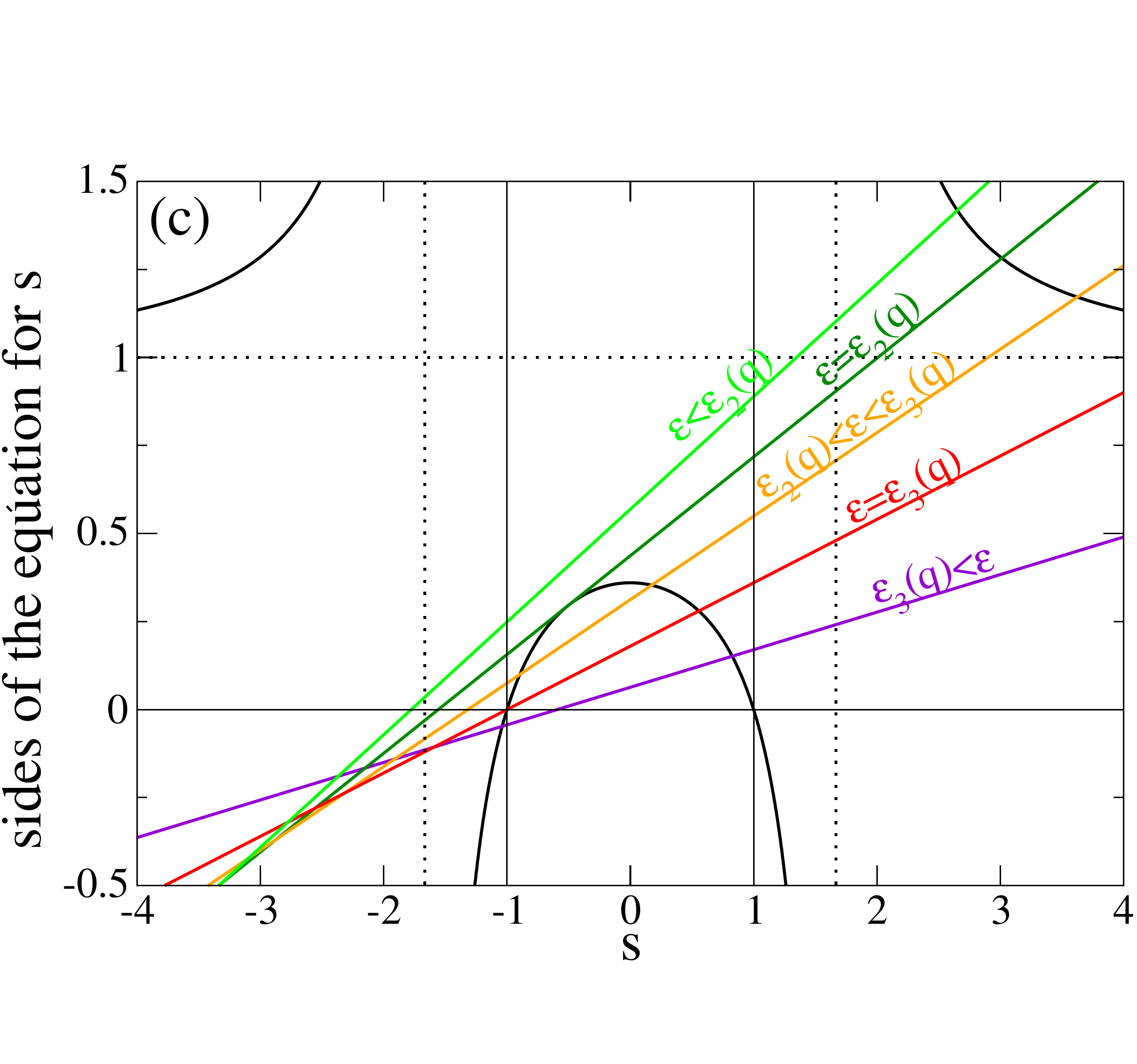}} 
\caption{For a gas of fermions of zero-range interaction (case of cold atoms) and chemical potential $ \mu > 0 $, we show in (a) the energies $ \epsilon_n (q) $, $ n \in \{1,2,3 \} $, points of non-analyticity in the spectral densities $ \epsilon \mapsto \rho_{ss'} (\epsilon, \qq) $ on the real axis as functions of the wave number $ q $ in units of $ k_0 = (2m \mu)^{1/2}/\hbar $, for different values of the order parameter $ \Delta/\mu = 1 $ (black), $ \Delta/\mu = 2 $ (red), $ \Delta/\mu = 1/2 $ (blue). The horizontal black dotted line showing $ \epsilon_1 (q) = 2 \Delta $ is common to the three coupling values since the energies are expressed in units of $ \Delta $. For a given value of $ \Delta/\mu $, the lower solid line shows $ \epsilon_2 (q) $ and the upper solid line shows $ \epsilon_3 (q) $, the two curves joining at a certain critical wave number $ q = q_0 (\Delta) $ (vertical dashed line), given by (\ref{eq:valq0}), beyond which they coincide by convention ($ \epsilon_2 (q) = \epsilon_3 (q) $). The two energies $ \epsilon_{1,2} (q) $ separate at $ q = 0 $ and the three energies $ \epsilon_n (q) $ join at $ q = 2 k_0 $, whatever $ \Delta $. Similarly, the curve $ \epsilon_2 (q) $ reaches its maximum invariably at $ q = k_0 $ (black vertical dotted line), to the left of $ q = q_0 $. We show in (b) the critical wave number $ q_0 $ as a function of $ \Delta $ with a solid black line. This one therefore separates the band $ \mathbb{R}^+\times [0,2k_0] $ in the plane $ (\Delta, q) $ into a higher domain where $ \epsilon_2 (q) \equiv \epsilon_3 (q) $ is given by equation (\ref{eq:om1om3}), and a lower domain where $ \epsilon_2 (q)/\mu <\epsilon_3 (q)/\mu $ is the largest real root of polynomial $ P_8 (X) $ of equation (\ref{eq:P8}). At the border, $ \epsilon_2 (q) $ undergoes a second order transition (continuity of $ \epsilon_2 (q) $ and its first derivative but discontinuity of its second derivative $ \dd^2 \epsilon_2 (q)/\dd q^2 $, see footnote \ref{note:tangenter}). Dotted and dotted lines: limiting behaviors (\ref{eq:devq0}) of $ q_0 $. We carry out in (c) a graphic discussion of equation (\ref{eq:surs}) on $ s $ by representing the right-hand side (solid black line, vertical and horizontal asymptotes in dotted lines) and the left-hand side (oblique straight lines in color) as functions of the unknown, for $ q = 2k_0/3 $, $ \epsilon = 5 \Delta/2 $ and different values of $ \Delta/\mu $ [$ 2/3 $ (light green), $ 0.7607084 $ (dark green), $ 9/10 $ (orange), $ 32/27 $ (red), $ 2 $ (purple), from top to bottom at $ s = 0 $]: in the interval $ [-1,1] $, delimited by thin continuous vertical lines, there is then successively $ 0 $ root (case $ \epsilon <\epsilon_2 (q) $), $ 1 $ double root (case $ \epsilon = \epsilon_2 (q) <\epsilon_3 (q) $), $ 2 $ roots of absolute value $ <1 $ (case $ \epsilon_2 (q) <\epsilon <\epsilon_3 (q) $), $ 2 $ roots of which one is equal to $-1 $ (case $ \epsilon = \epsilon_3 (q) $), and finally $ 1 $ root (case $ \epsilon> \epsilon_3 (q) $). In the spectral densities, it is necessary to integrate on $ s \in [-1,1] $ where the black curve is below the oblique line plotted at the considered energy (see discussion around (\ref{eq:surs})).} 
\label{fig:discugraph} 
\end{figure}

Let's start with the richest case, $ \mu> 0 $ and $ 0 <q <2 k_0 $, with $ k_0 = (2m \mu)^{1/2}/\hbar $ here, which has two or three points of non-analyticity. The extreme points are given by 
\be
\label{eq:om1om3}
\epsilon_1(q)=2\Delta \ \ \ \mbox{and}\ \ \ \epsilon_3(q)=2\epsilon_{\qq/2}=2\left[\left(\frac{\hbar^2q^2}{8m}-\mu\right)^2+\Delta^2\right]^{1/2}
\ee
where $ \epsilon_1 (q) $ is simply the lower edge of the broken pair continuum \cite{CKS, vcLandau}; their expression appeared in \cite{PRL2019}. Note that $ \epsilon_3 (q) $ merges with $ \epsilon_1 (q) $ at $ q = 2 k_0 $, which is the upper edge of the considered $q$ interval. Let us now introduce the wave number $ q_0 $, function of the interaction strength and always located in the interval $ [0,2k_0] $: 
\be
\label{eq:valq0}
\frac{1}{4}\check{q}_0^2 = 1 +\check{\Delta}^{2/3}\left[\left(1+\check{\Delta}^2\right)^{1/2}-1\right]^{1/3} -\check{\Delta}^{2/3}\left[\left(1+\check{\Delta}^2\right)^{1/2}+1\right]^{1/3}
\ee
where we thought it best, for simplicity, to express $ q_0 $ in units of $ k_0 $ and the order parameter $ \Delta $ (here width of the BCS gap) in units of $ \mu $. This rescaling, identified by a Czech accent, amounts to setting $ \hbar = 2m = k_0 = 1 $ and can be applied to any wave number $ q $ and any energy $ \epsilon $, as follows: 
\be
\label{eq:varadim}
\check{q}=\frac{q}{k_0} \ \mbox{and}\ \ \check{\epsilon}=\frac{\epsilon}{\mu} \ \ \ \mbox{with}\ \ \ k_0=(2m\mu)^{1/2}/\hbar\ \ \ \mbox{here}
\ee
Then, for all $ q \in] 0, q_0 [$, there is a midpoint of non-analyticity $ \epsilon_2 (q) $, therefore strictly between $ \epsilon_1 (q) $ and $ \epsilon_3 (q) $, given after rescaling by the largest real root of the polynomial of degree eight: 
\be
\label{eq:P8}
P_8(X)\equiv [X^2-\check{q}^2(2+\check{q})^2]^2 [X^2-\check{q}^2(2-\check{q})^2]^2- 4\check{\Delta}^2 [X^2-\check{q}^2(4-\check{q}^2)]\left\{[X^2-\check{q}^2(4-\check{q}^2)]^2-36\check{q}^4X^2\right\}-432\check{\Delta}^4 \check{q}^4 X^2
\ee
Expression (\ref{eq:valq0}) of $ \check{q}_0 $ and that (\ref{eq:P8}) of the polynomial $ P_8 (X) $ will be justified in section \ref{subsubsec:exdesp}. 
In $ q = q_0 $, this midpoint disappears by merging with $ \epsilon_3 (q) $. For $ q \in] q_0,2 k_0 [$, there remains only the two points (\ref{eq:om1om3}), but we will however set $ \epsilon_2 (q) = \epsilon_3 (q) $ in the following for simplicity. \footnote{\label{note:tangenter} The function \og largest real root of $ P_8 (X) $ \fg \, naturally continues to exist for $ q> q_0 $; it tangentially meets the function $ \epsilon_3 (q) $ at $ q = q_0 $ (same first but not second derivative) and becomes smaller than $ \epsilon_3 (q) $ beyond this point. Note that $ \epsilon_3 (q) $ has an inflection point in $ q = q_0 $, $ \dd^2 \epsilon_3 (q = q_0)/\dd q^2 = 0 $, as predicted by the method of reference \cite{PRL2019}: the first function $ k \mapsto \epsilon_{k+q/2}+\epsilon_{| kq/2 |} $ introduced in footnote \ref{note:methodeprl} to come has a maximum at $ k = 0 $ for $ q <q_0 $ and a minimum for $ q> q_0 $.} 
This discussion of the points of non-analyticity is illustrated and summarized in figures \ref{fig:discugraph}a and \ref{fig:discugraph}b. We complete it with some analytical results. First, as seen in figure \ref{fig:discugraph}b, $ \check{q}_0 $ has the following limiting behaviors at weak coupling $ \Delta/\mu \to 0^+$ and for a vanishing $ \mu/\Delta \to 0^+$ chemical potential: 
\be
\label{eq:devq0}
\check{q}_0\underset{\check{\Delta}\to 0}{=}2-2^{1/3}{\check{\Delta}}^{2/3}+O(\check{\Delta}^{4/3}) \quad\mbox{and}\quad\check{q}_0\underset{\check{\Delta}\to+\infty}{=}\frac{2}{\sqrt{3}}+O(\check{\Delta}^{-2})
\ee
and remains always greater than $ 2/\sqrt{3} $ therefore than 1. Then, as we see in figure \ref{fig:discugraph}a, $ \epsilon_2 (q) $ exhibits, as a function of $ q $, an absolute maximum at $ q = k_0 $, around which we have the Taylor expansion: 
\be
\check{\epsilon}_2(\check{q}) \underset{\check{q}\to 1}{=}\check{\Delta}+\left(1+\check{\Delta}^2\right)^{1/2}-\frac{(\check{q}-1)^2}{(1+\check{\Delta}^2)^{1/2}-\check{\Delta}/2} + O(\check{q}-1)^3 
\ee
In $ q = 0 $, $ \epsilon_2 (q) $ quadratically merges the point of non-analyticity $ \epsilon_1 (q) $, as already written by reference \cite{PRL2019}: 
\be
\label{eq:eps2fq}
\check{\epsilon}_2(\check{q})\underset{\check{q}\to 0}{=}2\check{\Delta}+\check{q}^2/\check{\Delta}+O(\check{q}^4)
\ee
Finally, in the limits of large and of weak $ \Delta/\mu $ taken at fixed $ q/k_0 $, one has simple expressions 
\be
\label{eq:eps2gdfd}
\check{\epsilon}_2(\check{q}) \stackrel{\check{q}<\check{q}_0(\infty)=2/\sqrt{3}}{\underset{\check{\Delta}\to+\infty}{=}} 2\check{\Delta} +\frac{\check{q}^2(2-\check{q}^2)}{2\check{\Delta}} + O(1/\check{\Delta}^3)
\quad\mbox{and}\quad
\check{\epsilon}_2(\check{q}) \stackrel{\check{q}\,\scriptsize\mbox{fixed}\,\in]0,2[}{\underset{\check{\Delta}\to 0^+}{=}}
\check{q}(2-\check{q}) +\check{\Delta} \check{q}^{1/2}(2-\check{q})^{1/2} +\check{\Delta}^2 \frac{1+\check{q}^2}{4\check{q}} + O(\check{\Delta}^3)
\ee
Let us specify the nature of the singularities of the spectral densities at energies $ \epsilon_i $. $ \epsilon_2 (q) $ and $ \epsilon_3 (q) $ are kinks for the three spectral densities $ \rho_{++} $, $ \rho_{--} $ and $ \rho_{+-} $. $ \epsilon_1 (q) $ is a point of discontinuity for $ \rho_{++} $ and a kink for $ \rho_{--} $, see (\ref{eq:devrho++rho--seuil}), but is nothing for the spectral density $ \rho_{+-} $ which remains zero at positive energies up to $ \epsilon = \epsilon_2 (q) $. 

Let us end with the poorest cases: when $ \mu> 0 $ but $ q> 2 k_0 $, or when $ \mu <0 $ (the BCS dispersion relation $ \epsilon_\kk $ is then minimal at $ k = 0 $) but $ q $ arbitrary, there remains only one point of non-analyticity of the spectral densities on the real axis, namely $ \epsilon = \epsilon_3 (q) $ \cite{PRL2019}, which is none other than the lower edge of the broken pair continuum at wave number $ q $ \cite{CKS, vcLandau}. \footnote{\label{note:methodeprl} These results are obtained with the method of reference \cite{PRL2019}, which connects the points of non-analyticity in energy of the spectral densities with the absolute or relative extrema of the functions $ k \mapsto \max_u \epsilon_{\kk+\qq/2}+\epsilon_{\kk-\qq/2} = \epsilon_{k+q/2}+\epsilon_{| kq/2 |} $ and $ k \mapsto \min_u \epsilon_{\kk+\qq/2}+\epsilon_{\kk-\qq/2} = 2 \epsilon_{(k^2+q^2/4)^{1/2}} $ where $ u = \cos \widehat{(\kk, \qq)} $. In the two poor cases considered, these functions are strictly increasing over $ \mathbb{R}^+$ because $-\mu+\hbar^2q^2/8m> 0 $. 
} 

\subsubsection{Expressions of spectral densities} 
\label{subsubsec:exdesp} 

In the continuous limit of the lattice model, the spectral densities (\ref{eq:defdensspec}) can be expressed analytically in terms of elliptical integrals. This computational breakthrough will greatly simplify the numerical and analytical studies that follow; in particular, it makes it possible to immediately carry out the analytic continuation of the spectral densities to complex energies $ z $ as required by the procedure (\ref{eq:Noz}), since the properties of elliptic integrals in the complex plane are known. \footnote{\label{note:ldcie} The elliptical integrals $ E (\phi, k) $ and $ F (\phi, k) $ have as functions of $ k^2 $, in their principal value, the branch cut $ k^2 \sin^2 \phi \in [1,+\infty [$, that of $ (1-k^2 \sin^2 \phi)^{1/2} $.} We give here the expressions of the spectral densities in the most interesting case $ \mu> 0 $ and $ 0 <q <2 k_0 $, see table \ref{tab:rhogen}. 
To make the writing compact, 
we have rescaled the energy $ \epsilon $ and the wave number $ q $ as in (\ref{eq:varadim}), the spectral densities as follows: 
\be
\label{eq:rhoadim}
\rho_{ss'}(\epsilon,\qq)=\frac{2mk_0}{\hbar^2}\,\check{\rho}_{ss'}(\check{\epsilon},\check{q})
\ee
then we have introduced for the energies $ \epsilon \geq 2 \Delta $ the hyperbolic changes of variable ($ \Omega \geq 0 $), 
\be
\label{eq:chgtvar}
\ch\Omega=\frac{\check{\epsilon}}{2\check{\Delta}}\, ,\quad \sh\Omega=\frac{(\check{\epsilon}^2-4\check{\Delta}^2)^{1/2}}{2\check{\Delta}}\quad\mbox{and}\quad
\thf\Omega=\frac{(\check{\epsilon}^2-4\check{\Delta}^2)^{1/2}}{\check{\epsilon}}
\ee
and defined the following functions of an angle $ \psi \in [0, \pi] $, 
\bea
\label{eq:fmm}
\!\!\!\!\!\!f_{--}(\psi)&\!\!\!\!\!\equiv\!\!\!\!\!\!\!&\int_0^\psi\!\!\!\frac{\dd\alpha\cos^2\!\alpha\sh^2\!\Omega}{(1\!+\!\sh^2\!\Omega\sin^2\!\alpha)^{3/2}}= E(\psi,\ii\sh\Omega)+\frac{\sh^2\!\Omega\sin\psi\cos\psi}{(1\!+\!\sh^2\!\Omega\sin^2\!\psi)^{1/2}}-F(\psi,\ii\sh\Omega)=
\ch\Omega\, E(u,\thf\Omega)-\frac{F(u,\thf\Omega)}{\ch\Omega} \\
\label{eq:fpp}
\!\!\!\!\!\!f_{++}(\psi)&\!\!\!\!\!\equiv\!\!\!\!\!\!\!&\int_0^\psi\!\!\!\frac{\dd\alpha\ch^2\Omega}{(1\!+\!\sh^2\!\Omega\sin^2\!\alpha)^{3/2}}=E(\psi,\ii\sh\Omega) +\frac{\sh^2\!\Omega\sin\psi\cos\psi}{(1\!+\!\sh^2\!\Omega\sin^2\!\psi)^{1/2}}=\ch\Omega\, E(u,\thf\Omega) \\
\label{eq:fpm}
\!\!\!\!\!\!f_{+-}(\psi)&\!\!\!\!\!\equiv\!\!\!\!\!\!\!&\int_0^\psi\!\!\!\frac{\dd\alpha\cos\alpha\sh\Omega\ch\Omega}{(1\!+\!\sh^2\!\Omega\sin^2\!\alpha)^{3/2}} = \frac{\sh\Omega\ch\Omega\sin\psi}
{(1\!+\!\sh^2\!\Omega\sin^2\!\psi)^{1/2}}
\eea
where $ E (\phi, k) $ and $ F (\phi, k) $ are the elliptic integrals of second and first kind in the definition \S8.111 of \cite{GR}, we used \S8.127 of \cite{GR} and we set 
\be
u=\acos\frac{\cos\psi}{(1+\sh^2\Omega\sin^2\psi)^{1/2}}
\ee
The calculations leading to these results are exposed in \ref{subapp:gen}, but let's give the main lines here. In the integral form (\ref{eq:defdensspec}) of the spectral densities, written in spherical coordinates of axis $ \qq $, one carries out immediately the integration on the azimuthal angle thanks to the rotational invariance of axis $ \qq $ then, with more effort, on the polar angle thanks to the Dirac distribution fixing the energy of the broken pair in the integrand. It remains to integrate on the wave number $ k $. However, the polar angles where the argument of the Dirac distribution vanishes must take physical values, which imposes constraints on $ k $. First, they must be real, which restricts $ k $ to a certain segment which is cleverly 
parametrized as follows: 
\be
\check{k}^2+\frac{\check{q}^2}{4}-1\equiv \frac{1}{2} \left(\check{\epsilon}^2-4\check{\Delta}^2\right)^{1/2}s \quad \mbox{where} \quad s\in [-1,1]
\ee
The integral on $ k $ therefore boils down to an integral on $ s $. Then they must be in the range $ [0, \pi] $. To implement this constraint, we must solve the following third degree equation on $ s $, which expresses the fact that the polar angles reach the edges $ 0 $ and $ \pi $ of the allowed interval: 
\be
\label{eq:surs}
\boxed{
\frac{4\check{q}^2}{\check{\epsilon}^2} \left[\frac{1}{2}\left(\check{\epsilon}^2-4\check{\Delta}^2\right)^{1/2}s+1-\frac{\check{q}^2}{4}\right]=\frac{1-s^2}{\frac{\check{\epsilon}^2}{\check{\epsilon}^2-4\check{\Delta}^2}-s^2}
}
\ee
The polar angles are between $ 0 $ and $ \pi $ when the right-hand side of (\ref{eq:surs}) is less than the left-hand side. The graphical discussion in figure \ref{fig:discugraph}c shows that, for $ q $ small enough, three cases arise: $ (i) $ at fairly low energy, equation (\ref{eq:surs}) has no root in the interval $ [-1,1] $, and we can therefore integrate on $ s $ in all $ [-1,1] $; $ (ii) $ at intermediate energies, equation (\ref{eq:surs}) has two roots $ s_1 $ and $ s_2 $ ($ s_1 <s_2 $) in $ [-1,1] $, and we must integrate on $ s $ in $ [-1, s_1] \cup [s_2,1] $; $ (iii) $ at fairly high energy, the root $ s_1 $ goes below $-1 $ and equation (\ref{eq:surs}) has $ s_2 $ as the only root in $ [-1,1] $, so that we must integrate on $ s $ in $ [s_2,1] $. These topology changes of the integration domain on $ s $ are at the origin of the points of non-analyticity of the spectral densities. The point $ \epsilon = \epsilon_2 (q) $ in section \ref{subsubsec:lpdnae1e2e3} corresponds to the transition between the cases $ (i) $ and $ (ii) $: $ s_1 = s_2 $ is double root and the discriminant of equation (\ref{eq:surs}) vanishes, which leads to $ P_8 (\check{\epsilon}_2 (q)) = 0 $, where $ P_8 (X) $ is the polynomial (\ref{eq:P8}). \footnote{\label{note:Lagrange} It remains to show that $ \check{\epsilon}_2 (\check{q}) $ is the largest real root of $ P_8 (X) $. To this end, let us work on the polynomial $ P_4 (X) $ of degree 4 with real coefficients such that $ P_8 (X) = P_4 (X^2) $. In Lagrange's method, the roots of $ P_4 (X) $ are expressed as functions of those of the associated degree 3 polynomial $ R (X) $ called resolvent cubic. We have the following result: if $ R (X) $ (with real coefficients) has only one real root, this one is positive and $ P_4 (X) $ has two real roots $ x_1 $ and $ x_2 $ and two complex conjugate roots $ z_0 $ and $ z_0^* $. Now, $ R (X) $ actually has only one real root since its Cardan discriminant $ \delta = (512^2/27) \check{\Delta}^4 (1+\check{\Delta}^2) \check{q}^{12} [27 \check{\Delta}^4+72 \check{\Delta}^2 \check{q}^4+16 \check{q}^6 (4-\check{q}^2)]^3 $ is $> 0 $ for $ 0 <\check{q} <2 $. To show that $ x_1 $ and $ x_2 $ are of the same sign, we use the Vi\`ete relation $ x_1 x_2 | z_0 |^2 = P_4 (0) = \check{q}^6 (4-\check{q}^2)^3 [4 \check{\Delta}^2+\check{q}^2 (4-\check{q}^2)]> 0 $. Finally, the roots $ x_1 $ and $ x_2 $ are $> 0 $ since one of them is none other than $ \check{\epsilon}^2_2 (q) $; to show that this is the larger of the two, we verify that this is true for $ \check{\Delta} \to+\infty $ ($ \check{\epsilon}^2_2 (q) $ is the only one root of $ P_4 (X) $ which diverges) then we use a continuity argument up to $ \Delta = 0^+$ (the resultant of $ P_4 (X) $ and of its derivative $ P_4'(X) $, proportional to $ \delta $, never vanishes so $ P_4 (X) $ has no double root, and $ x_1 $ and $ x_2 $ cannot cross). This demonstration does not apply for $ \check{\Delta} = 0 $, where $ z_0 $ has a real limit, $ x_1 $ and $ x_2 $ merge and $ \check{\epsilon}^2_2 (q) $ is this time the smallest root of $ P_4 (X) $.} The point $ \epsilon = \epsilon_3 (q) $ in section \ref{subsubsec:lpdnae1e2e3} corresponds to the transition between cases $ (ii) $ and $ (iii) $: we then have $ s_1 =-1 $, as we can verify by direct substitution of $ s $ by $-1 $ in (\ref{eq:surs}). When $ q $ is too large, the scenario is simplified: when we increase the energy starting from the case $ (i) $, a double root $ s_1 = s_2 $ certainly appears in (\ref{eq:surs}), but this root is $ <-1 $, it is {\sl not} in the interval $ [-1,1] $ and we can continue to integrate on $ s \in [-1 , 1] $; the energy continuing to increase, $ s_2 $ reaches $-1 $ then exceeds it while $ s_1 $ remains $ <-1 $: we jumped directly from the case $ (i) $ to the case $ (iii) $. The value $ q_0 $ of $ q $ separating the two scenarios $ (i) \to (ii) \to (iii) $ and $ (i) \to (iii) $ is therefore such that the double root $ s_1 = s_2 $ of (\ref{eq:surs}) appears exactly in $-1 $; by writing that $ s =-1 $ is the root of (\ref{eq:surs}) and its derivative with respect to $ s $, we come across a cubic equation on $ q^2 $, equivalent to the condition $ \dd^2 \epsilon_3 (q)/\dd q^2 = 0 $ of footnote \ref{note:tangenter}, and of which (\ref{eq:valq0}) is actually the real solution. 

To be complete, we give, when $ \epsilon> \epsilon_2 (q) $, the expression of the two smallest real roots $ s_1 $ and $ s_2 $ of equation (\ref{eq:surs}) in terms of the Cardan angle $ \gamma $ spanning $ [0, \pi] $: 
\begin{multline}
\label{eq:vals1s2}
s_{1,2}=\frac{Y_q}{6\check{q}^2(\check{\epsilon}^2-4\check{\Delta}^2)^{1/2}} - \frac{(Y_q^2+12\check{q}^4\check{\epsilon}^2)^{1/2}}{3\check{q}^2(\check{\epsilon}^2-4\check{\Delta}^2)^{1/2}} \cos\frac{\pi\mp\gamma}{3} \ \ \mbox{where}\ \ Y_q=\check{\epsilon}^2-\check{q}^2(4-\check{q}^2)\\ \ \ \mbox{and}\ \ 
\gamma=\acos\frac{Y_q(Y_q^2-36\check{\epsilon}^2\check{q}^4)+216\,\check{\epsilon}^2\check{q}^4\check{\Delta}^2}{(Y_q^2+12\check{q}^4\check{\epsilon}^2)^{3/2}}
\end{multline}
Note also that the change of variable $ s = \cos \alpha $, where $ \alpha \in [0, \pi] $, was carried out, as we see on the integration variable in the definition of the functions $ f_{ss'} (\psi) $ and on the value of their argument in table \ref{tab:rhogen}. Finally, we can verify that, in the weak coupling limit $ \Delta/\mu \to 0^+$ at $ \bar{q} = 2 \check{q}/\check{\Delta} $ and $ \bar{\epsilon} = \check{\epsilon}/\check{\Delta} $ fixed as prescribed by (\ref{eq:adim}), tables \ref{tab:rholcf} and \ref{tab:rhogen} agree; funny remark, there is in fact a perfect agreement on the spectral densities $ \rho_{ss}^{[\mathrm{II}]} $ even {\sl before} taking the limit (only the upper bound of the interval II is different for finite $ \Delta/\mu $). The calculations leading to table \ref{tab:rhogen} easily extend to any $ q $ and to any sign of $ \mu $; thus, for $ \mu> 0 $ but $ q> 2k_0 $, the spectral densities are identically zero for $ \epsilon <\epsilon_3 (q) $ and of the form III otherwise, as it appears from the end of section \ref{subsubsec:lpdnae1e2e3}. 

\begin{table} [htb] 
\[
\begin{array}{|l|l|}
\hline
\check{\epsilon} < 2\check{\Delta} & \check{\rho}_{--}^{[\mathrm{I}]}(\check{\epsilon},\check{q})=0 \\
                                   & \check{\rho}_{++}^{[\mathrm{I}]}(\check{\epsilon},\check{q})=0 \\
                                   & \check{\rho}_{+-}^{[\mathrm{I}]}(\check{\epsilon},\check{q})=0 \\
\hline
2\check{\Delta}<\check{\epsilon}<\check{\epsilon}_2(\check{q}) \ & \displaystyle\check{\rho}_{--}^{[\mathrm{II}]}(\check{\epsilon},\check{q})\equiv \pi \left[\frac{\check{\epsilon}}{2\check{q}} E\left(\thf\Omega\right) - \frac{2\check{\Delta}^2}{\check{q}\check{\epsilon}}K\left(\thf\Omega\right)\right]=\frac{\pi\check{\Delta}}{\check{q}} \left[E\left(\ii\sh\Omega\right)-K\left(\ii\sh\Omega\right)\right]\\
 & \displaystyle\check{\rho}_{++}^{[\mathrm{II}]}(\check{\epsilon},\check{q})\equiv \pi\left[\frac{\check{\epsilon}}{2\check{q}} E\left(\thf\Omega\right)\right]=\frac{\pi\check{\Delta}}{\check{q}}E(\ii\sh\Omega) \\
      & \check{\rho}_{+-}^{[\mathrm{II}]}(\check{\epsilon},\check{q})\equiv 0 \\
\hline
\check{\epsilon}_2(\check{q})<\check{\epsilon}<\check{\epsilon}_3(\check{q}) 
& \displaystyle\check{\rho}_{--}^{[\mathrm{III}]}(\check{\epsilon},\check{q})\equiv  
\frac{\pi\check{\Delta}}{2\check{q}}\left[f_{--}\left(\frac{\pi}{2}-\asin s_2\right)+f_{--}\left(\frac{\pi}{2}+\asin s_1\right)\right]
\\
& \displaystyle\check{\rho}_{++}^{[\mathrm{III}]}(\check{\epsilon},\check{q})\equiv 
\frac{\pi\check{\Delta}}{2\check{q}}\left[f_{++}\left(\frac{\pi}{2}-\asin s_2\right)+f_{++}\left(\frac{\pi}{2}+\asin s_1\right)\right]
\\
& \displaystyle\check{\rho}_{+-}^{[\mathrm{III}]}(\check{\epsilon},\check{q})\equiv 
\frac{\pi\check{\Delta}}{2\check{q}}\left[f_{+-}\left(\frac{\pi}{2}-\asin s_2\right)-f_{+-}\left(\frac{\pi}{2}+\asin s_1\right)\right] \\
\hline
\check{\epsilon}_3(\check{q})<\check{\epsilon} 
& \displaystyle\check{\rho}_{--}^{[\mathrm{IV}]}(\check{\epsilon},\check{q})\equiv \frac{\pi\check{\Delta}}{2\check{q}} \left[f_{--}\left(\frac{\pi}{2}-\asin s_2\right)\right] \\
& \displaystyle \check{\rho}_{++}^{[\mathrm{IV}]}(\check{\epsilon},\check{q})\equiv \frac{\pi\check{\Delta}}{2\check{q}} \left[f_{++}\left(\frac{\pi}{2}-\asin s_2\right)\right]\\
& \displaystyle \check{\rho}_{+-}^{[\mathrm{IV}]}(\check{\epsilon},\check{q})\equiv \frac{\pi\check{\Delta}}{2\check{q}} \left[f_{+-}\left(\frac{\pi}{2}-\asin s_2\right)\right]\\
\hline
\end{array}
\]

\caption{Spectral densities $--$, $++$ and $+-$ of (\ref{eq:defdensspec}) in the continuous limit $ b \to 0 $ of our lattice model, as functions of energy $ \epsilon> 0 $. The interaction strength is arbitrary but we have limited ourselves to the richest case $ \mu> 0 $ and $ 0 <q <2 k_0 $, where $ k_0 = (2 m \mu)^{1/2}/\hbar $ is the wave number minimizing the BCS dispersion relation $ \epsilon_\kk $. On each interval between the zero energy, the non-analyticity points $ \epsilon_i $ given in section \ref{subsubsec:lpdnae1e2e3} and the infinite energy, these are smooth functions of $ \epsilon $ with different analytic continuations to complex energies, hence their numbering in Roman numerals. $ E (k) $ and $ K (k) $ are the complete elliptic integrals of second and first kind (see \S 8.112 of \cite{GR}); $ \Omega \in] 0,+\infty [$ is a convenient hyperbolic parametrization (\ref{eq:chgtvar}) of the energy when it is greater than $ 2 \Delta $; in the fourth and fifth row of the table, the two forms given are equivalent since $ E (\thf \Omega) = E (\ii \sh \Omega)/\ch \Omega $ and $ K (\thf \Omega) = \ch \Omega \, K (\ii \sh \Omega) $, including for complex $ \Omega $ as long as $ \re \ch \Omega> 0 $; the functions $ f_{--} $, $ f_{++} $ and $ f_{+-} $ are those of equations (\ref{eq:fmm}), (\ref{eq:fpp}) and (\ref{eq:fpm}); $ s_1 $ and $ s_2 $ are the real roots of (\ref{eq:surs}) less than 1 sorted in ascending order, when they exist, and are given by (\ref{eq:vals1s2}). When $ q> q_0 $, where $ q_0 $ is given by (\ref{eq:valq0}), we actually have $ \epsilon_2 (q) = \epsilon_3 (q) $ (see section \ref{subsubsec:lpdnae1e2e3}). The energies and wave numbers are rescaled as in (\ref{eq:varadim}) and the spectral densities as in (\ref{eq:rhoadim}).} 
\label{tab:rhogen} 
\end{table} 

\subsubsection{Behavior of spectral densities at high energy and at the edge of the continuum} 

Let us give two very simple applications of the expressions of spectral densities from table \ref{tab:rhogen}. The first is a high energy asymptotic expansion, 
\be
\begin{pmatrix}
\check{\rho}_{++}(\check{\epsilon},\check{q}) \\
\check{\rho}_{--}(\check{\epsilon},\check{q}) \\
\check{\rho}_{+-}(\check{\epsilon},\check{q})
\end{pmatrix}
\underset{\check{\epsilon}\to+\infty}{=}
\frac{\pi}{8^{1/2}} \left[\left(\check{\epsilon}^2+4\check{\Delta}^2\right)^{1/2}+2-\frac{\check{q}^2}{2}\right]^{1/2}
\left(
\begin{array}{rl}
 &[1 + (4\check{\Delta}^2/\check{\epsilon}^3) (1-5\check{q}^2/12) +O(\check{\epsilon}^{-4})]\\
(\check{\epsilon}^2-4\check{\Delta}^2)\check{\epsilon}^{-2} &[1 + (4\check{\Delta}^2/\check{\epsilon}^3) (1-13\check{q}^2/12) +O(\check{\epsilon}^{-4})] \\
(\check{\epsilon}^2-4\check{\Delta}^2)^{1/2}\check{\epsilon}^{-1} &[1 + (4\check{\Delta}^2/\check{\epsilon}^3) (1-3\check{q}^2/4) +O(\check{\epsilon}^{-4})]
\end{array}
\right)
\ee
useful above all for controlling and reducing energy truncation errors in the numerical calculations of section \ref{subsec:numcom}. At the order of this expansion, however quite high, we have the impression that $ \rho_{+-} $ is the geometric mean of $ \rho_{++} $ and $ \rho_{--} $, which is not true on any neighborhood of infinity [see the integrals in (\ref{eq:fmm}), (\ref{eq:fpp}) and (\ref{eq:fpm})] unless $ q = 0 $ (see footnote \ref{note:rhoqz}). The second application is a Taylor expansion at the lower edge $ 2 \Delta $ of the broken pair continuum (here $ 0 <\check{q} <2 $): 
\be
\label{eq:devrho++rho--seuil}
\check{\rho}_{++}(\check{\epsilon},\check{q}) \underset{\check{\epsilon}\to2\check{\Delta}^+}{=} \frac{\pi^2}{2\check{q}} \left[\check{\Delta} +\frac{1}{4}(\check{\epsilon}-2\check{\Delta})+
O(\check{\epsilon}-2\check{\Delta})^2\right]
\quad\mbox{and}\quad
\check{\rho}_{--}(\check{\epsilon},\check{q}) \underset{\check{\epsilon}\to2\check{\Delta}^+}{=} \frac{\pi^2}{4\check{q}}\left[(\check{\epsilon}-2\check{\Delta})-\frac{1}{8\check{\Delta}}(\check{\epsilon}-2\check{\Delta})^2+O(\check{\epsilon}-2\check{\Delta})^3\right]
\ee
We then see, by means of (\ref{eq:Mssgen}), that $ \check{M}_{++} (\check{\epsilon}+\ii 0^+, \check{q}) $ exhibits on the real axis at $ \check{\epsilon} = 2 \check{\Delta} $ a logarithmic singularity in its real part and a discontinuity in its imaginary part: 
\be
\check{M}_{++}(\check{\epsilon}+\ii 0^+,\check{q}) \underset{\check{\epsilon}\to2\check{\Delta}}{=} (2\pi)^{-3}\frac{\pi^2\check{\Delta}}{2\check{q}}[\ln|\check{\epsilon}-2\check{\Delta}|-\ii\pi\,\Theta(\check{\epsilon}-2\check{\Delta})][1+O(\check{\epsilon}-2\check{\Delta})]+ \mbox{smooth function of}\,\check{\epsilon}
\ee
which explains the sharp peak (with vertical tangent) in the intensity of the modulus-modulus response function at the angular frequency $ 2 \Delta/\hbar $, observed but not interpreted in reference \cite{PRL2019}, and which has nothing to do with the Lorentzian peak of the continuum mode (see section \ref{subsec:carac}). In $ \check{M}_{--} (\check{\epsilon}+\ii 0^+, \check{q}) $, this edge singularity is reduced by a factor $ \check{\epsilon}-2 \check{\Delta} $. As for $ \check{M}_{+-} (\check{\epsilon}, \check{q}) $, it is of course a smooth function of $ \check{\epsilon} $ around the edge since the spectral density $ \rho_{+-} $ is identically zero on a neighborhood of the edge. The matrix elements are here rescaled as in (\ref{eq:adimbis}). 

\subsection{Characterizing the continuum collective branch: dispersion relation, spectral weights, optimal excitation and observation channels} 
\label{subsec:carac} 

From an experimental point of view, the presence of a complex energy $ z_\qq $ in the broken pair continuum results in a visible Lorentzian peak in the intensity of the gas frequency response function introduced in \cite{PRL2019}, or if one prefers in the modulus squared of the Fourier transform of the time response (\ref{eq:cauchysable}), of the form $ \omega \mapsto | Z/(\hbar \omega-z_\qq) |^2 $ where $ \omega $ is the angular frequency, at least if $ (i) $ the central energy $ \re z_\qq $ of the Lorentzian is in the interval $ [\epsilon_a, \epsilon_b] $ between the points of non-analyticity through which the analytic continuation is carried out ($ \epsilon_a/\hbar $ and $ \epsilon_b/\hbar $ thus delimit an interval of observability in angular frequency) and $ (ii) $ half-energy-width $ | \im z_\qq | $ of the Lorentzian is sufficiently weak that one can easily separate it from the broad contribution of the continuum \cite{PRL2019}. We thus see that the mode is characterized by three real quantities, its angular frequency $ \re z_\qq/\hbar $, its damping rate $ | \im z_\qq |/\hbar $ and its spectral weight, proportional to $ | Z |^2 $, where $ Z $ is the residue of the pole $ z_\qq $ in the analytic continuation of the experimental response function. 

To characterize the continuum mode, we must first calculate its complex energy $ z_\qq $, by solving equation (\ref{eq:detMbzero}) after analytic continuation as follows: 
\be
\label{eq:surzqgen}
0=\det M_\downarrow(z_\qq,\qq)\equiv {M_{++}}_\downarrow(z_\qq,\qq){M_{--}}_\downarrow(z_\qq,\qq)-[{M_{+-}}_\downarrow(z_\qq,\qq)]^2
\ \mbox{with}\ {M_{ss'}}_\downarrow(z,\qq)= M_{ss'}(z,\qq) - (2\pi)^{-3} 2\ii\pi \rho_{ss'}^{[\mathrm{II}]}(z,\qq)
\ee
We indeed perform especially the study for $ \mu> 0 $ and $ 0 <q <2 k_0 $, and for an analytic continuation through the type II interval, that is $ 2 \Delta <\epsilon <\epsilon_2 (q) $, see table \ref{tab:rhogen}; the other cases are treated in section \ref{subsec:autrefen_muneg}. 
Then we calculate the residue $ Z $. For this, we start from the temporal expression (\ref{eq:signaltbzero}) of small deviations of the order parameter and, in the integrand, we expand in the vicinity of $ z = z_\qq $ the inverse of matrix $ M $ analytically continued and here restricted to its upper left $ 2 \times 2 $ block, which brings up a matrix residue $ \mathcal{M} $:
\be
\label{eq:residmat}
M_\downarrow(z_\qq,\qq)^{-1} \underset{z\to z_\qq}{=} \frac{\mathcal{M}}{z-z_\qq} + O(1) \quad\mbox{with}\quad
\mathcal{M}=\frac{1}{\left(\frac{\dd}{\dd z} \det M_\downarrow\right)(z_\qq,\qq)}
\begin{pmatrix}
{M_{--}}_\downarrow(z_\qq,\qq) & -{M_{+-}}_\downarrow(z_\qq,\qq) \\
-{M_{+-}}_\downarrow(z_\qq,\qq) & {M_{++}}_\downarrow(z_\qq,\qq)
\end{pmatrix}
\ee
If the excitation only perturbs the phase of the order parameter ($+$ channel, first coordinate) and we measure the response on the phase of the order parameter, the relevant residue is $ Z_+= \langle+| \mathcal{M} |+\rangle = M_{--\downarrow} (z_\qq, \qq)/(\frac{\dd}{\dd z} \det M_\downarrow) (z_\qq, \qq) $. Conversely, if we initially excite and finally measure only the modulus of the order parameter ($-$ channel, second coordinate), the relevant residue is $ Z_-= \langle-| \mathcal{M} |-\rangle = M_{++\downarrow} (z_\qq, \qq)/(\frac{\dd}{\dd z} \det M_\downarrow) (z_\qq, \qq) $. In the general case, the residue $ Z $ in the gas response is a linear combination of the matrix elements of $ \mathcal{M} $, with coefficients depending on the excitation applied to the gas and the measured observable. By applying the residue theorem to (\ref{eq:signaltbzero}) as in figure \ref{fig:contour}, we find the contribution of the continuum mode to the time signal: 
\be
\label{eq:contribpolesignal}
\mbox{signal}(t)|_{\scriptsize\mbox{pole}\, z_\qq}= Z \, \eee^{-\ii z_\qq t/\hbar} \quad\mbox{with}\quad Z=\langle\chi_{\rm obs}| \mathcal{M} |\chi_{\rm exc}\rangle
\ee
where the amplitudes $ \langle \chi_{\rm obs} |+\rangle $ and $ \langle \chi_{\rm obs} |-\rangle $ are those of the observable measured in the phase and modulus channels, and the amplitudes $ \langle+| \chi_{\rm exc} \rangle $ and $ \langle-| \chi_{\rm exc} \rangle $ are those of the excitation applied in these channels. Note that $ | \chi_{\rm exc} \rangle $ has no simple expression, but involves the source vector in (\ref{eq:signaltbzero}) analytically continued to $ z = z_\qq $, itself a function of the applied excitation through $ y^{\pm}_{\kk \qq} $ therefore of the initial perturbation of the BCS state vector (\ref{eq:ansatz}). 
However, since matrix $ \mathcal{M} $ is by construction of zero determinant, it is generally of rank one; after some thought, we manage to put it in a particularly simple dyadic form, in terms of vectors $ \hat{\chi} $ normalized to unity, after having factored out a reduced real global residue $ Z_{\rm opt} $: 
\be
\label{eq:voiesoptimales}
\mathcal{M}=Z_{\rm opt}|\hat{\chi}_{\rm obs}^{\rm opt}\rangle \langle\hat{\chi}_{\rm exc}^{\rm opt}|
\quad\ \mbox{with}\quad
\left\{
\begin{array}{l}
|\hat{\chi}_{\rm exc}^{\rm opt}\rangle\equiv Z_{\rm opt}^{-1/2}\binom{(Z_+^{1/2})^*}{-\varepsilon (Z_-^{1/2})^*}\propto
\binom{|Z_+|^{1/2}}{\eee^{-\ii\theta}|Z_-|^{1/2}}\\
\\
|\hat{\chi}_{\rm obs}^{\rm opt}\rangle\equiv Z_{\rm opt}^{-1/2}\binom{Z_+^{1/2}}{-\varepsilon Z_-^{1/2}} \propto
\binom{|Z_+|^{1/2}}{\eee^{\ii\theta}|Z_-|^{1/2}}
\end{array}
\right.
\quad \mbox{and}\quad \varepsilon \in\{+,-\}
\ee
The reduced global residue is simply the sum of the moduli of the particular residues $ Z_+$ and $ Z_-$. As we can see, the vectors of the dyad are expressed in terms of these moduli and of an angle $ \theta $, relative phase of the modulus-phase coupling amplitude and of a diagonal coupling amplitude in matrix $ M $ analytically continued in $ z_\qq $: 
\be
\label{eq:zopt_thetagen}
Z_{\rm opt}=|Z_+|+|Z_-| \quad\mbox{and}\quad \theta=\arg[-{M_{+-}}_\downarrow\!(z_\qq,\qq)/{M_{--}}_\downarrow\!(z_\qq,\qq)]
\ee
We understand that $ \arg (Z_-/ Z_+) = 2 \theta $ since $ Z_+/ Z_-= M_{--\downarrow} (z_\qq, \qq)/M_{++\downarrow} (z_\qq, \qq) $. By comparing (\ref{eq:voiesoptimales}) and (\ref{eq:contribpolesignal}), we see that $ \langle \hat{\chi}_{\rm obs}^{\rm opt} | $ gives the optimal observable, that is to say the relative linear combination of the phase and modulus channels to be observed to maximize the residue $ | Z | $ therefore the weight of the continuum mode in the experimental signal; similarly, $ | \hat{\chi}_{\rm exc}^{\rm opt} \rangle $ gives the relative amplitudes of the excitation which must be produced on the phase and the modulus of the order parameter to maximize the weight of the continuum mode in the signal, for a fixed observable. If these optimizations are made jointly on the excitation and the observation, one achieves the maximum modal weight accessible for fixed norms of $ \langle \chi_{\rm obs} | $ and $ | \chi_{\rm exc} \rangle $, of residue $ Z_{\rm opt} || \chi_{\rm obs} || \, || \chi_{\rm exc} || $. 
To compute the denominator of $ \mathcal{M} $ in (\ref{eq:residmat}), we need the spectral density derivatives with respect to $ z $, which can be written simply (see \S8.123 of \cite{GR}): 
\be
\frac{\dd}{\dd\check{z}} \rho_{++}^{[\mathrm{II}]}(\check{z},\check{q}) = \frac{\check{z}}{\check{z}^2-4\check{\Delta}^2} \check{\rho}_{--}^{[\mathrm{II}]}(\check{z},\check{q})
\quad\mbox{and}\quad \frac{\dd}{\dd\check{z}}\rho_{--}^{[\mathrm{II}]}(\check{z},\check{q}) = \frac{1}{\check{z}} \check{\rho}_{++}^{[\mathrm{II}]}(\check{z},\check{q})
\ee
where $ \check{z} = z/\mu $ as in (\ref{eq:varadim}). The derivative of the matrix elements of $ M $ is obtained by differentiation of the forms (\ref{eq:M+-foncrho}, \ref{eq:Mssgen}) under the integral sign. 

\subsection{Numerical results on the continuum collective branch} 
\label{subsec:numcom} 

\begin{figure} [tb] 
\centerline{\includegraphics [width = 0.3 \textwidth, clip =]{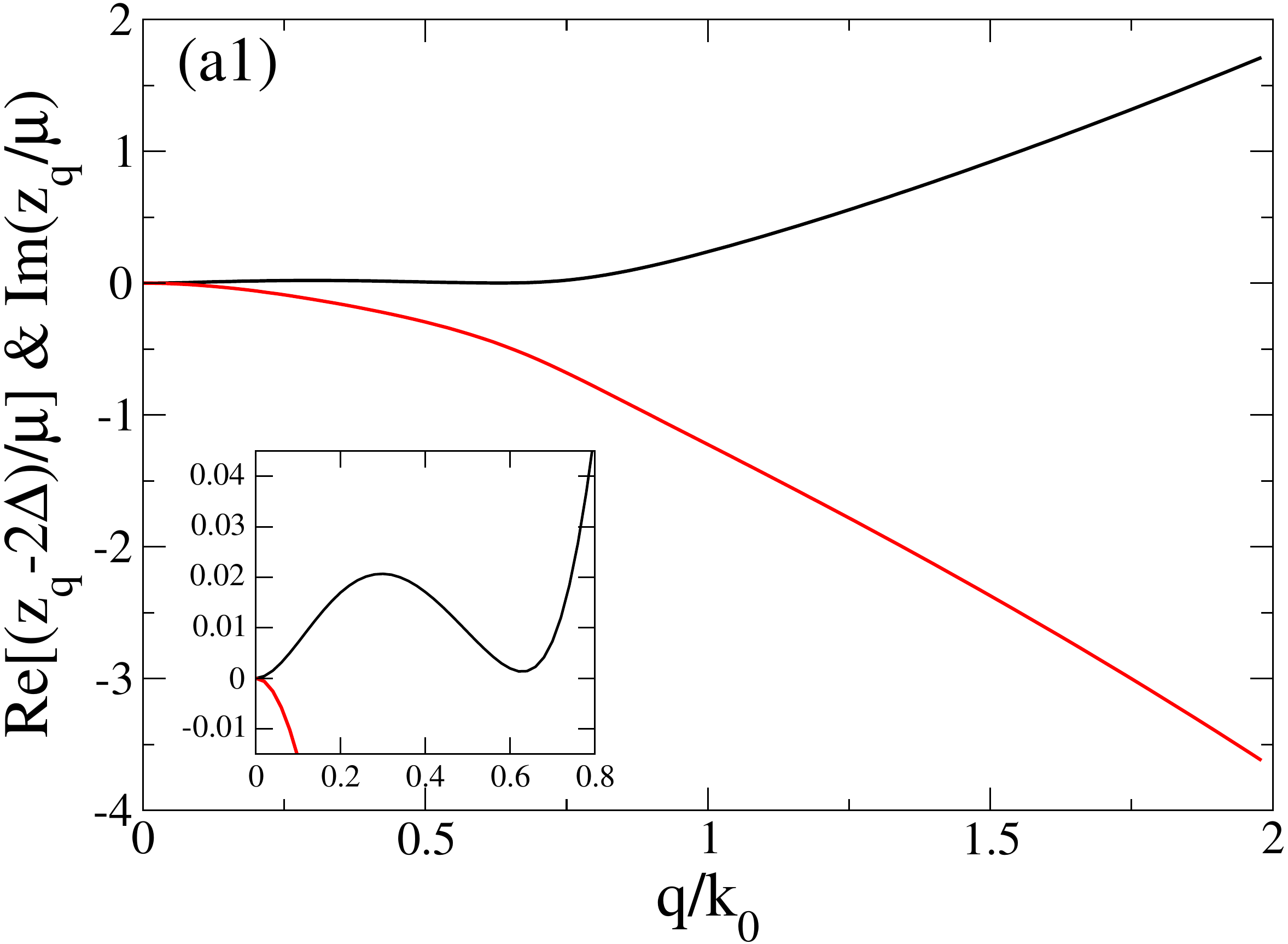} \hspace{5mm} \includegraphics [width = 0.3 \textwidth, clip =]{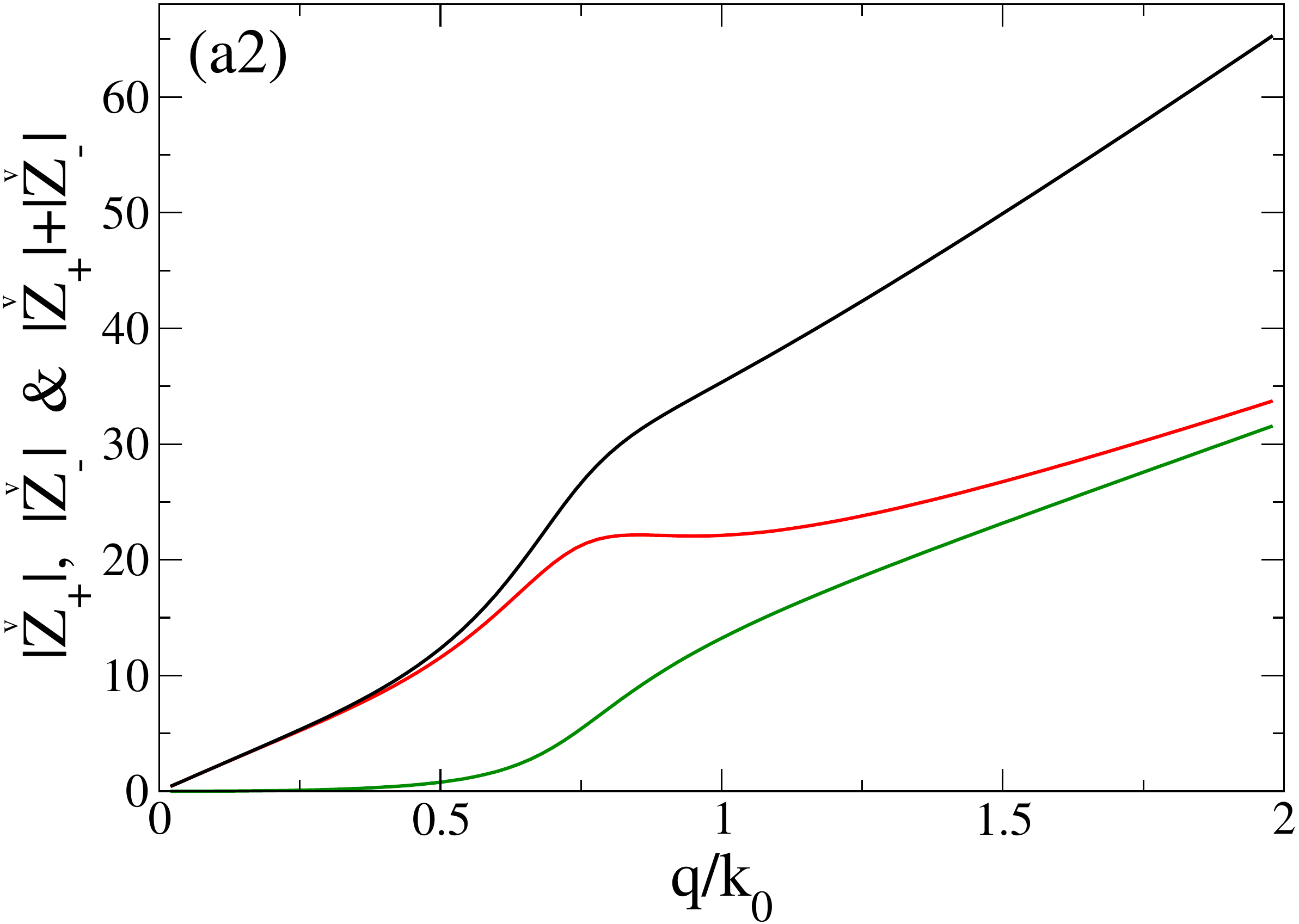} \hspace{5mm} \includegraphics [width = 0.3 \textwidth, clip =]{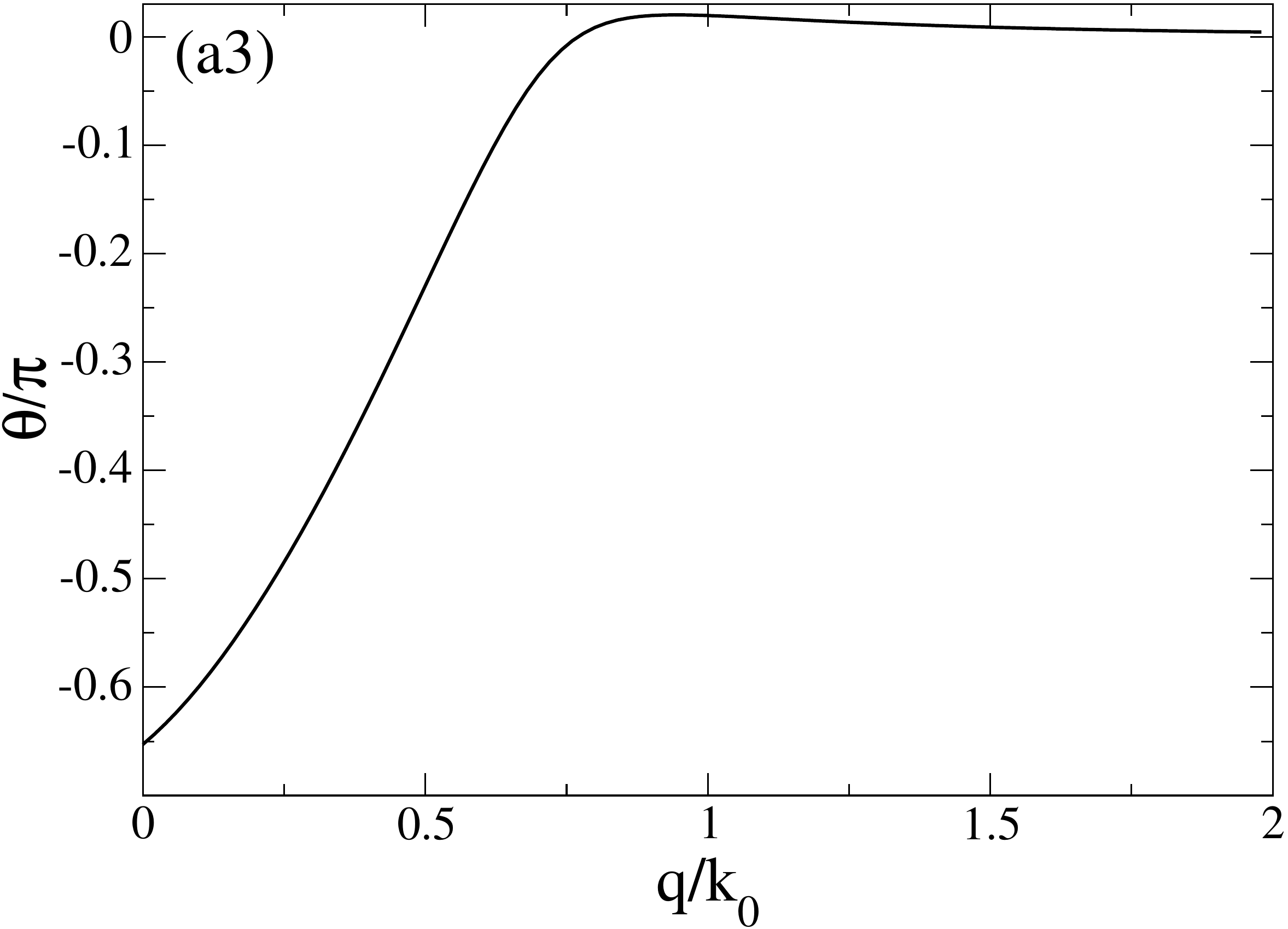}} 
\centerline{\includegraphics [width = 0.3 \textwidth, clip =]{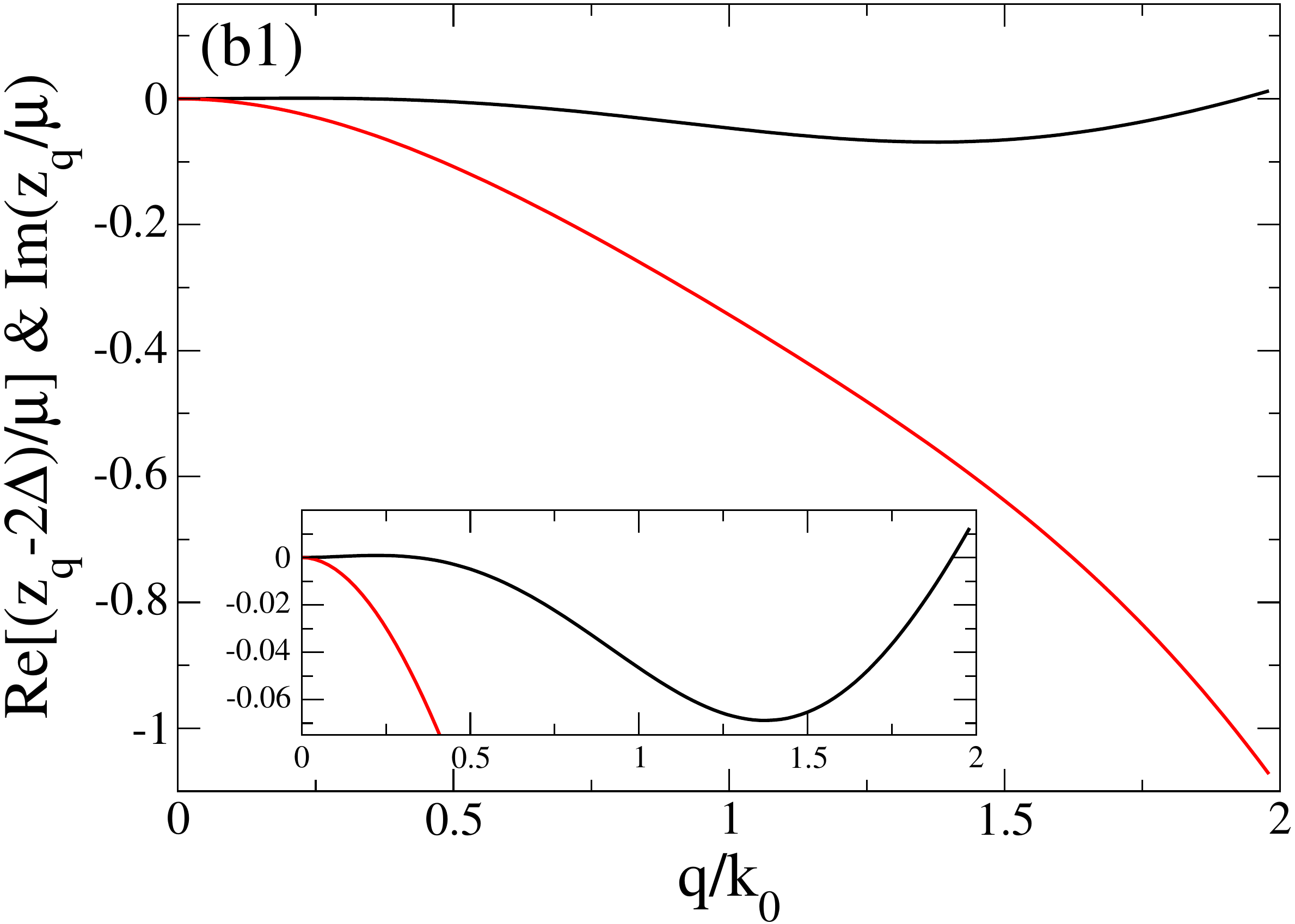} \hspace{5mm} \includegraphics [width = 0.3 \textwidth, clip =]{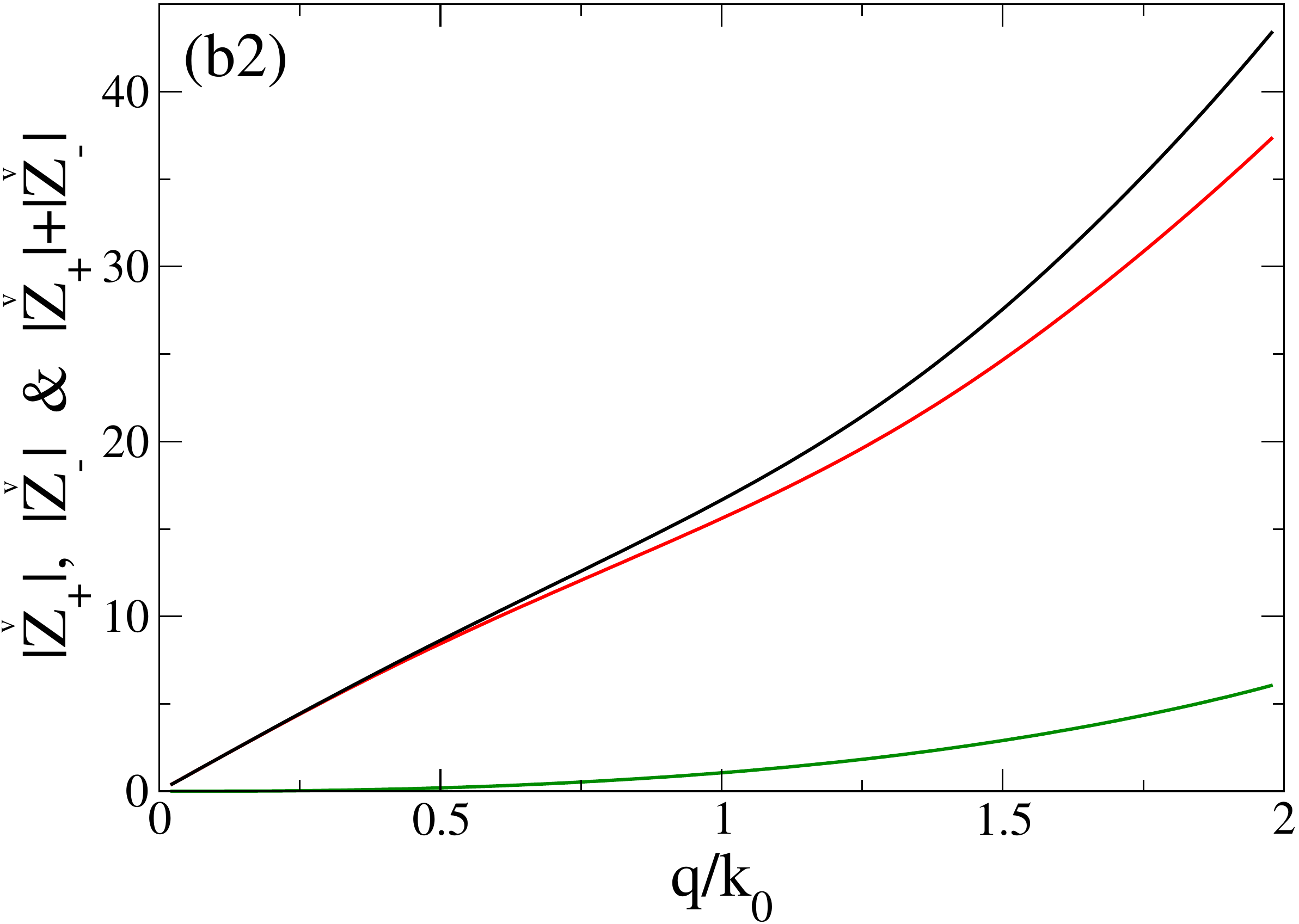} \hspace{5mm} \includegraphics [width = 0.3 \textwidth, clip =]{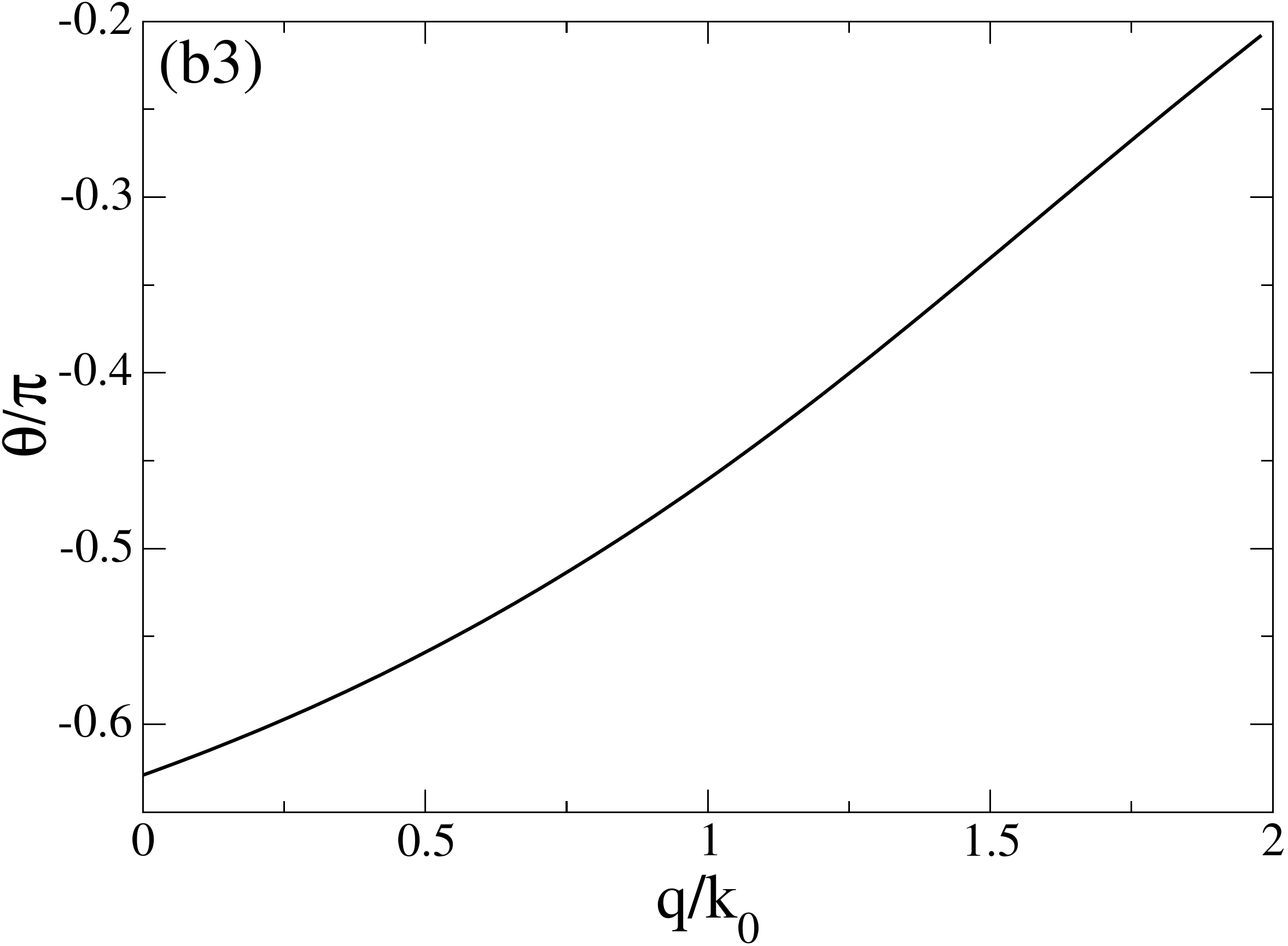}} 
\centerline{\includegraphics [width = 0.3 \textwidth, clip =]{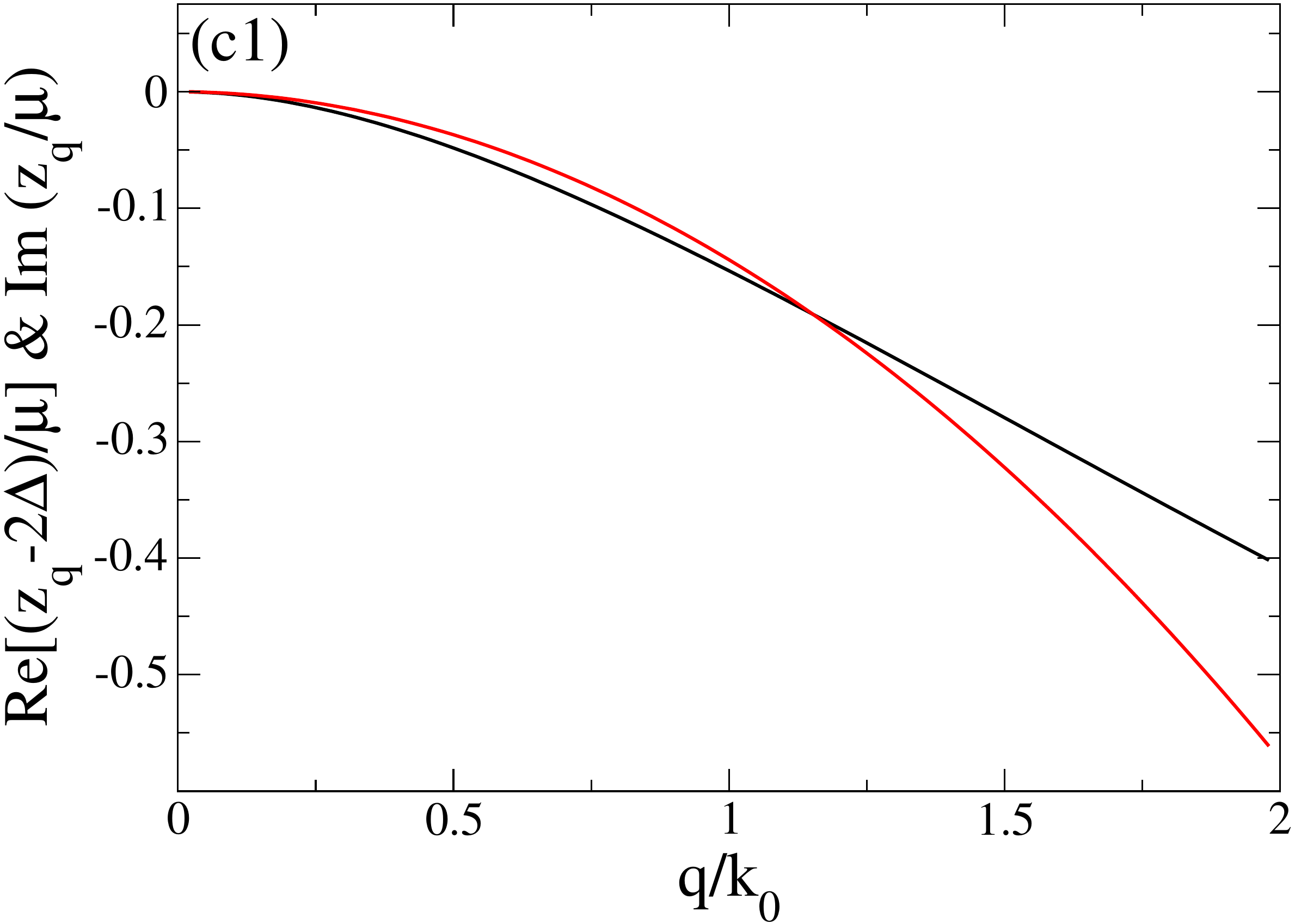} \hspace{5mm} \includegraphics [width = 0.3 \textwidth, clip =]{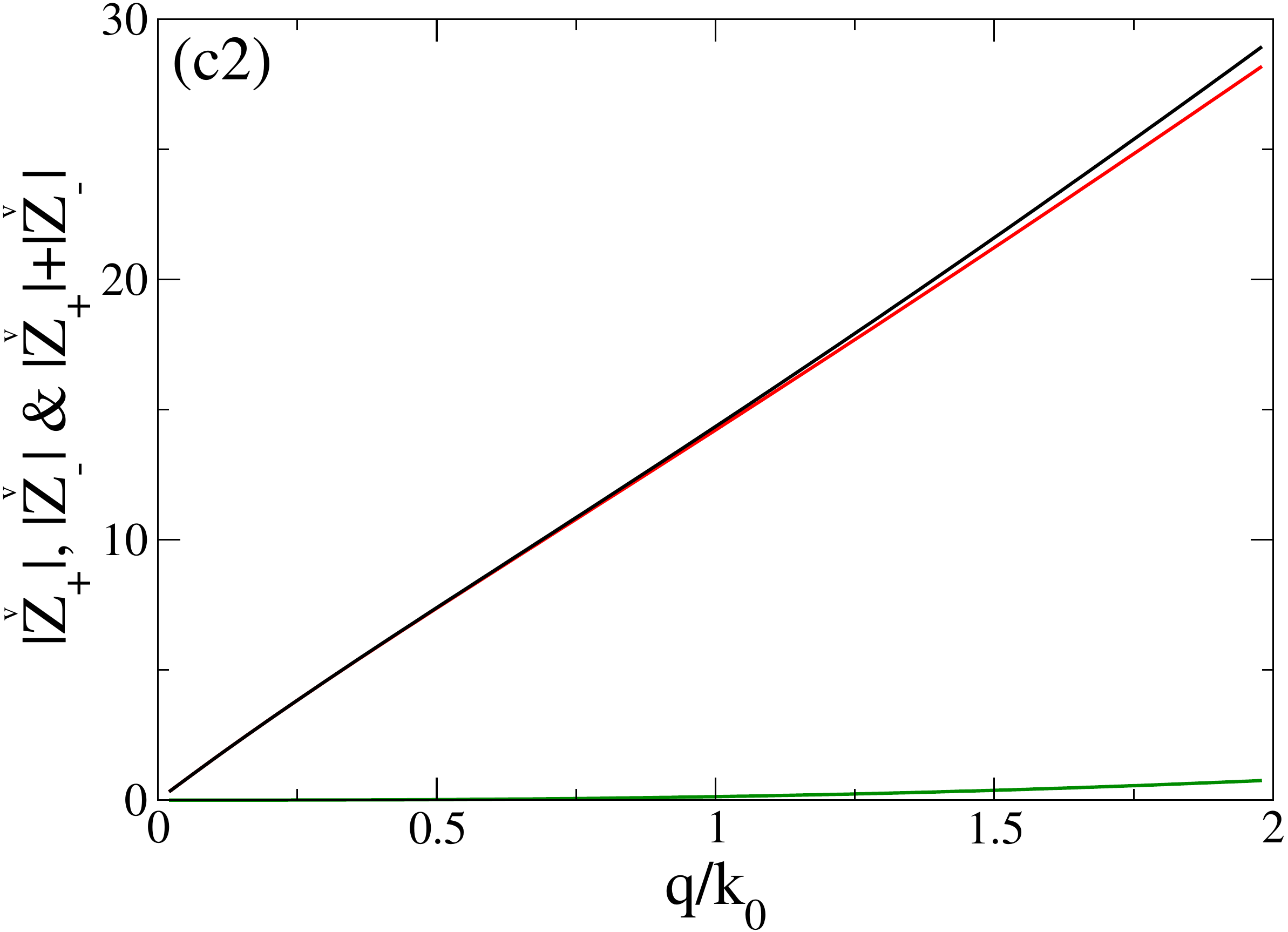} \hspace{5mm} \includegraphics [width = 0.3 \textwidth, clip =]{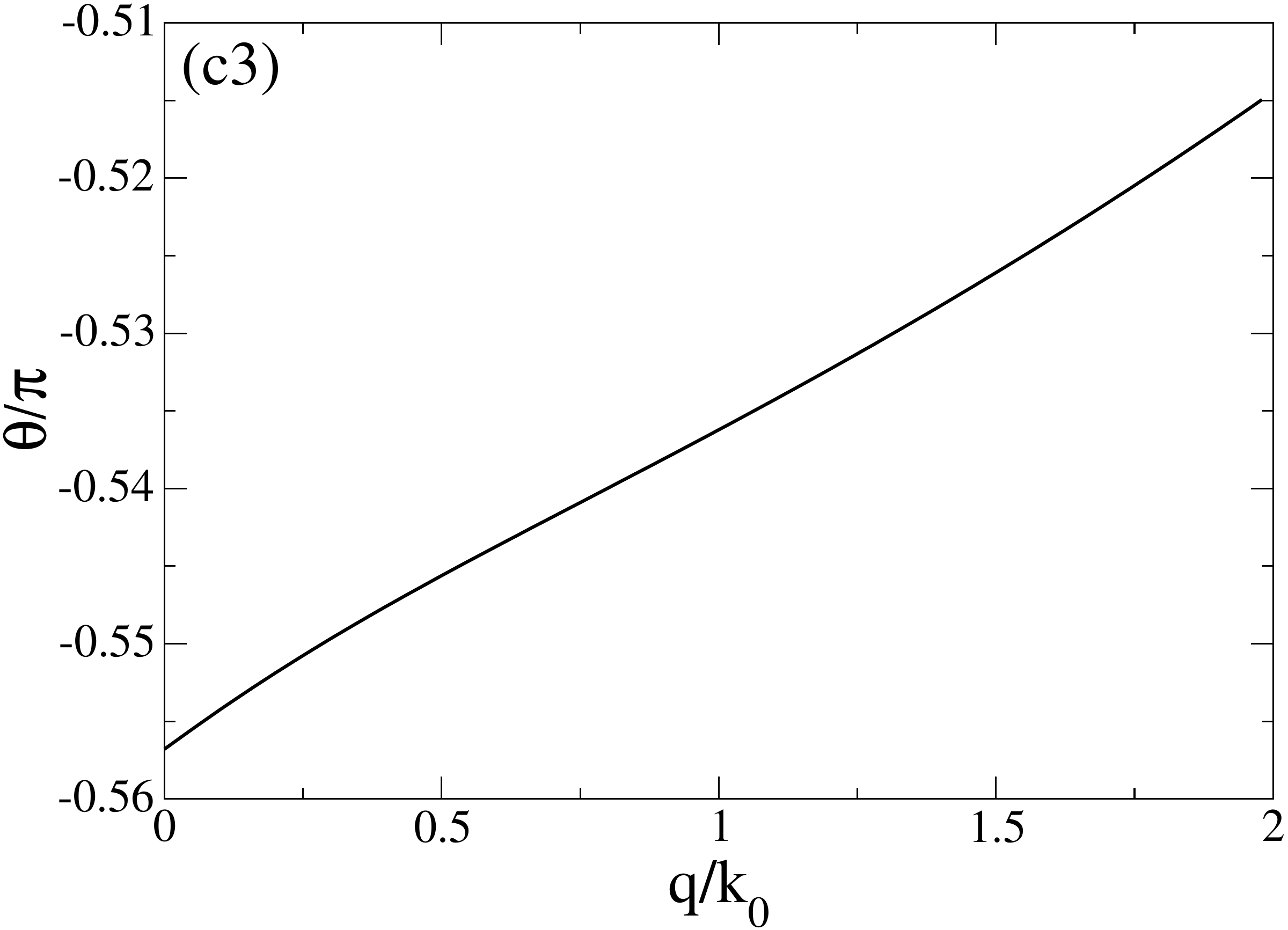}} 
\caption{In a gas of fermions with zero range interactions (case of cold atoms) and chemical potential $ \mu> 0 $, in column 1: complex dispersion relation $ q \mapsto z_\qq $ of the continuum mode, obtained by numerical solution of equation (\ref{eq:surzqgen}); the analytic continuation is of type II, therefore $ q $ spans $] 0, 2k_0 [$ where $ k_0 = (2m \mu)^{1/2}/\hbar $. Solid black line: real part shifted from its limit $ 2 \Delta $ at the origin; solid red line: imaginary part. The order parameter is equal to on row (a) $ \Delta/\mu = 1/5 $, on row  (b) $ \Delta/\mu = 1 $ and on row (c) $ \Delta/\mu = $ 10. In column 2: residues $ | Z_+| $ (green) and $ | Z_-| $ (red) of the mode in the phase (index $+$) and modulus (index $-$) channels of deviations of the order parameter from equilibrium, and accessible reduced maximum residue $ | Z_+|+| Z_-| $ (black), as functions of the wave number $ q $ in units of $ k_0 $, after rescaling as in equation (\ref{eq:adimbis}). In column 3: angle $ \theta $ appearing in the optimal excitation and observation channels (\ref{eq:voiesoptimales}) allowing to reach the reduced maximum residue $ | Z_+|+| Z_-| $. The insets in (a1) and (b1) are enlargements.} 
\label{fig:resnum} 
\end{figure}

We solve numerically the eigenenergy equation (\ref{eq:surzqgen}) on the continuum mode, by evaluating the integral forms (\ref{eq:M+-foncrho}, \ref{eq:Mssgen}) and the elliptic functions in the spectral densities of table \ref{tab:rhogen} as in section \ref{subsubsec:resnumlcf}. Here, $ \mu> 0 $ and the analytic continuation is done through the interval II, $ 2 \Delta <\epsilon <\epsilon_2 (q) $, so that the wave number $ q $ varies between opening point $ q = 0 $ and closing point $ q = 2 k_0 $ of the interval. The dispersion relation, the residues in the phase and modulus channels, the optimal reduced residue $ Z_{\rm opt} $ and the angle $ \theta $ defining the optimal observation channel thus obtained (see section \ref{subsec:carac}) are shown in figure \ref{fig:resnum}, for a weak coupling $ \Delta/\mu = 1/5 $, a strong coupling $ \Delta/\mu = 1 $ and a coupling $ \Delta/\mu = 10 $ close to a zero chemical potential.

In all cases, the imaginary part of $ z_\qq $ is a decreasing negative function of the wave number $ q $. The real part has a richer behavior in weak coupling: it passes by a relative maximum $> 2 \Delta $ then by an absolute minimum $ <2 \Delta $ at weak $ q $, before starting to grow approximately linearly at large $ q $. In strong coupling, the rich structure is maintained but is stretched in wave number (the absolute minimum is located at $ q \approx k_0 $) and the relative maximum is hardly noticeable. Near the zero of $ \mu $, the real part in turn becomes a decreasing function of $ q $, and is everywhere less than $ 2 \Delta $, which does not really bring the mode into the gap $ [0,2 \Delta] $ since its complex energy $ z_\qq $ remains separated by the part of the branch cut folded over $]-\infty, 2 \Delta] $ in the prescription (\ref{eq:Noz}); in particular, we do not expect a Lorentzian peak in the response functions of the system at the angular frequency $ \omega = \re z_\qq/\hbar $ because it is outside the observability window (see section \ref{subsec:carac} and reference \cite{PRL2019}). The situation is the same for the part of the branch with $ \re z_\qq> \epsilon_2 (q) $, which is separated from the real physical axis by two superimposed branch cuts, the one existing and the one coming from the folding back on $ [\epsilon_2 (q),+\infty [$ of part of the branch cut $ [2 \Delta, \epsilon_2 (q)] $. Paradoxically, we notice that the largest real and imaginary parts of $ \check{z}_q-2 \check{\Delta} $ in absolute value over $] 0,2 k_0 [$ are reached in the weak coupling regime rather than in the $ \Delta/\mu \geq $ 1 regime. 

The residue $ Z_-$ in the $-$ channel of the modulus deviations of the order parameter strongly dominates that $ Z_+$ of the phase deviations, as we see in the second column of figure \ref{fig:resnum}, except in weak coupling for $ q \approx k_0 $ where $ | Z_+| \approx | Z_-| $; the dominance of $ | Z_-| $ is particularly clear at low $ q $, where $ | Z_+| $ tends to zero much faster than the linear law followed by $ | Z_-| $. From the point of view of its spectral weight, the continuum mode therefore remains here essentially a modulus mode, as it was in the weak coupling limit at $ q/k_0 = O (\Delta/\mu) $, even if the phase-modulus coupling $ M_{+-} $ in matrix $ M $ cannot absolutely be neglected in the calculation of its eigenenergy $ z_\qq $, on the contrary of equation (\ref{eq:defzqbar0}). As for the relative phase $ \theta $ between the two channels maximizing the spectral weight, in the third column of figure \ref{fig:resnum}, it grows vaguely linearly in $ q $ over the entire interval $] 0, 2k_0 [$, except in weak coupling where it reaches a plateau close to zero. 

The analytical studies of sections \ref{subsec:faibleq}, \ref{subsec:faibledelta} and \ref{subsec:grand_d} provide a physical and quantitative explanation of these observations. 

\subsection{The continuum collective branch at low $ q $} 
\label{subsec:faibleq} 

As in the pioneer reference \cite{AndrianovPopov}, first of all it is necessary to study analytically the continuum mode at low wave number. Here we largely resume the results of \cite{PRL2019} at infinitesimal $q$ (section \ref{subsubsec:ldqesc}) by refining them (section \ref{subsubsec:rdvdlaq} specifies the domain of applicability in wave number) and extending them to other quantities than $ z_\qq $ and $ Z_-$ (section \ref{subsubsec:aoafq}). 

\subsubsection{The quadratic start and its coefficient} 
\label{subsubsec:ldqesc} 

As shown by reference \cite{AndrianovPopov} in the weak coupling limit $ \Delta/\mu \to 0 $ and reference \cite{PRL2019} for any coupling $ \Delta/\mu> 0 $, the continuum collective branch (for an analytic continuation of type II) has a quadratic start at low wave number $ q $: 
\be
\label{eq:quadzq}
\check{z}_q \underset{\check{q}\to 0}{=} 2\check{\Delta} + \zeta \frac{\check{q}^2}{\check{\Delta}} + O(\check{q}^3)
\ee
with a complexe coefficient $ \zeta $ solution in the lower complex half-plane of the transcendental equation \footnote{By setting $ S = 2 (\pi-\asin \frac{1}{\sqrt{\zeta}}) $, we also have $ \zeta = 2/(1-\cos S) $ with $ S $ solution of $ S (S+\sin S)+256 \pi^2 \, [\check{M}_{+-} (2 \check{\Delta}, 0)]^2 (1-\cos S) = 0 $ (knowing that $ [(1+\cos S)/(1-\cos S)]^{1/2} =-\sin S/(1-\cos S) $), which in the weak coupling limit looks like equation (2.16) of \cite{AndrianovPopov}.} 
\be
\label{eq:transzeta}
\boxed{
\left[\pi-\asin\frac{1}{\sqrt{\zeta}}\right]\left[\left(\pi-\asin\frac{1}{\sqrt{\zeta}}\right)\zeta-\sqrt{\zeta-1}\right]
+128\pi^2\, [\check{M}_{+-}(2\check{\Delta},0)]^2=0
}
\ee
The non-diagonal element of $ M $, that is to say the phase-modulus coupling, is written here at zero order in $ q $ therefore at energy $ 2 \Delta $ and at zero wave number, in a dimensionless form which we will apply in the following to all the matrix elements and by ricochet to the residues: 
\be
\label{eq:adimbis}
M_{ss'}(z,\qq)=\frac{2mk_0}{\hbar^2} \check{M}_{ss'}(\check{z},\check{q}) \quad\mbox{and}\quad Z_{\pm}=\frac{\hbar^4 k_0}{4m^2} \check{Z}_{\pm}
\ee
By setting $ 1/\check{\Delta} = \sh \tau $, we take from \cite{PRL2019} the explicit form 
\be
\label{eq:valM+-}
\check{M}_{+-}(2\check{\Delta},0)= -\frac{(\eee^{2\tau}-1)^{1/2}}{(2\pi)^2} \left[\re\Pi(\eee^\tau,\ii\eee^\tau)-\Pi(-\eee^\tau,\ii\eee^\tau)+\frac{K(\ii\eee^\tau)}{\sh\tau}\right] \, <0 \quad\forall \Delta/\mu>0
\ee
We also reproduce the low and large $ \Delta/\mu $ expansions of \cite{PRL2019}, by pushing the calculation one order further at large $ \Delta/\mu $: \footnote{In (\ref{eq:zetapetitgranddelta}), we used, among other things, the non-trivial fact that $ \re \Pi (\eee^\tau, \ii \eee^\tau) \underset{\tau \to 0^+}{\to} K (\ii)-E (\ii)/2 $; compared to \cite{PRL2019}, we preferred to show a $ \Gamma $ function rather than elliptical integrals, using identities $ K (\ii) = (8/\pi)^{1/2} [\Gamma (5/4)]^2 $ and $ E (\ii)-K (\ii) = [\Gamma (3/4)]^2/(2 \pi)^{1/2} $.} 
\be
\label{eq:zetapetitgranddelta}
\zeta\underset{\check{\Delta}\to 0^+}{=}\zeta_0 -\frac{2\zeta_0^2}{\zeta_0-1}\left(\frac{\check{\Delta}}{\pi}\right)^2 \ln^2\frac{\check{\Delta}}{8\eee} +
O(\check{\Delta}^3 \ln^\alpha\check{\Delta})\quad\mbox{and}\quad 
\zeta\underset{\check{\Delta}\to +\infty}{=}-\frac{128}{\pi^5}[\Gamma(5/4)]^4 \check{\Delta} -\frac{24\ii \sqrt{2}}{\pi^{7/2}} [\Gamma(5/4)]^2\check{\Delta}^{1/2} + \frac{5}{2\pi^2} +\frac{4}{\pi^3} + O(\check{\Delta}^{-1/2})
\ee
The value $ \zeta_0 $ is that (\ref{eq:dl4}) of the weak coupling limit, and it should be noted at large $ \Delta/\mu $ that $ \zeta $ is real at the dominant order. Reference \cite{PRL2019} also gives an equivalent of $ Z_-$, see our equation (\ref{eq:equivzmp}). It is not useful here to specify the value of the exponent $ \alpha> 0 $ in (\ref{eq:zetapetitgranddelta}). 

\subsubsection{Validity regime of the quadratic approximation} 
\label{subsubsec:rdvdlaq} 

Reference \cite{PRL2019} does not specify up to what value of $ q $ we can be satisfied with the quadratic approximation (\ref{eq:quadzq}) on $ z_\qq $. Let us therefore make a critical review of the various approximations leading to it. Essentially, \cite{PRL2019} sets $ \check{k} = 1+\check{q} K $ in the integral definition (\ref{eq:defSigma}) of the matrix elements of $ M $ and, in spherical coordinates of axis $ \qq $, made $ \check{q} $ tend to zero under the integral sign at fixed $ K $. In the denominator of the integrand, this supposes a quadratization of the BCS dispersion relation $ \epsilon_\kk $ around its minimum (reached in $ \check{k} = 1 $), which actually concatenates the {\sl two} following quadratic approximations, 
\be
\label{eq:epskquad}
\check{\epsilon}_k =[(\check{k}^2-1)^2+\check{\Delta}^2]^{1/2}\simeq[4(\check{k}-1)^2+\check{\Delta}^2]^{1/2}\simeq \check{\Delta}+\frac{2(\check{k}-1)^2}{\check{\Delta}}
\ee
whose global validity condition is 
\be
\label{eq:condquad}
|\check{k}-1| \ll \min(1,\check{\Delta})
\ee
We end up {\sl in fine} with equivalents ($ u $ is the cosine of the polar angle): \footnote{The second integral of (\ref{eq:integfq}) appears in equation (2.13) of reference \cite{AndrianovPopov} up to the trivial change of variable $ K = K'/ 2 $.} 
\be
\label{eq:integfq}
\check{M}_{++}(\check{z},\check{q}) \underset{\check{q}\to 0}{\sim} -\frac{\pi\check{\Delta}}{4\check{q}(2\pi)^3} \int_{-1}^{1}\dd u
\int_\mathbb{R} \frac{\dd K}{K^2+(u^2-\zeta)/4}\quad\mbox{and}\quad\check{M}_{--}(\check{z},\check{q}) \underset{\check{q}\to 0}{\sim}
-\frac{\pi\check{q}}{4\check{\Delta}(2\pi)^3} \int_{-1}^{1} \dd u \int_\mathbb{R} \frac{\dd K\,(\zeta-u^2)}{K^2+(u^2-\zeta)/4}
\ee
whose explicit integration and analytic continuation from $ \im \zeta> 0 $ to $ \im \zeta <0 $ through $ [0,1] $ (this is the interval to choose given the behavior of $ \check{\epsilon}_2 (\check{q}) $ at low $ \check{q} $ in section \ref{subsubsec:lpdnae1e2e3}) gives \cite{PRL2019} 
\be
\label{eq:mequivfq}
\check{M}_{++\downarrow}(\check{z}_q,\check{q})\underset{\check{q}\to 0}{\sim} -\frac{\ii\pi^2\check{\Delta}}{(2\pi)^3\check{q}} \left(\pi-\asin\frac{1}{\sqrt{\zeta}}\right)\quad\mbox{and}\quad \check{M}_{--\downarrow}(\check{z}_q,\check{q})\underset{\check{q}\to 0}{\sim} -\frac{\ii\pi^2\check{q}}{(2\pi)^32\check{\Delta}} \left[\left(\pi-\asin\frac{1}{\sqrt{\zeta}}\right)\zeta-\sqrt{\zeta-1}\right]
\ee
and reproduces (\ref{eq:transzeta}). In reality, integrating with respect to $ K $ over all $ \mathbb{R} $ only makes sense if the integrals converge within the validity domain of the quadratic approximation (\ref{eq:epskquad}), so if $ \check{k}-1 = \check{q} K $ satisfies the condition (\ref{eq:condquad}). Since the width in $ K $ space in (\ref{eq:integfq}) is of the order of $ | \zeta |^{1/2} $ if $ 1 \lesssim | \zeta | $ and of the order of unity otherwise, we end up with the constraint 
\be
\label{eq:limvalapproxquad}
\check{q}\ll\frac{\min(1,\check{\Delta})}{(1+|\zeta|)^{1/2}}
\approx
\left\{
\begin{array}{lcl}
\check{\Delta} &\mbox{if}& \check{\Delta}\to 0 \\
\check{\Delta}^{-1/2} &\mbox{if}& \check{\Delta}\to +\infty
\end{array}
\right.
\ee
In the weak coupling limit, it reproduces the expected inequality $ \check{q} \ll \check{\Delta} $ \cite{AndrianovPopov} i.e.\  $ q \xi \ll 1 $ where $ \xi $ is the size of a bound pair. In the vicinity of the chemical potential zero, it shows the unexpected wave number scale $ \check{q} \approx \check{\Delta}^{-1/2} $, of which we do not have any physical interpretation but which will emerge in section \ref{subsec:grand_d}. 

\subsubsection{Other observables at low $ q $} 
\label{subsubsec:aoafq} 

From equivalents (\ref{eq:mequivfq}) on the analytically continued matrix elements of $ M $, we get the equivalents of the residues in the phase channel $+$ and the modulus channel $-$, and of the relative phase  between the two channels maximizing the global residue: 
\bea
\label{eq:equivzmp}
\check{Z}_- &\!\!\!\underset{\check{q}\to 0}{\sim}\!\!\!& \frac{2\ii\check{q}}{\pi^2} \frac{(2\pi)^3(\pi-\asin\frac{1}{\sqrt{\zeta}})}{(\pi-\asin\frac{1}{\sqrt{\zeta}})^2+\frac{(\pi-\asin\frac{1}{\sqrt{\zeta}})\zeta-\sqrt{\zeta-1}}{2\zeta\sqrt{\zeta-1}}} \quad\mbox{and}\quad
\check{Z}_+ \underset{\check{q}\to 0}{\sim}\frac{\ii\check{q}^3}{\pi^2\check{\Delta}^2}\frac{(2\pi)^3\left[\left(\pi-\asin\frac{1}{\sqrt{\zeta}}\right)\zeta-\sqrt{\zeta-1}\right]}{(\pi-\asin\frac{1}{\sqrt{\zeta}})^2+\frac{(\pi-\asin\frac{1}{\sqrt{\zeta}})\zeta-\sqrt{\zeta-1}}{2\zeta\sqrt{\zeta-1}}}\\
\label{eq:equivtheta}
\theta &\!\!\!\underset{\check{q}\to 0}{\sim}\!\!\!& \arg\left[(-\ii)\left(\pi-\asin\frac{1}{\sqrt{\zeta}}\right)\right]
\eea
It was enough for that to return to the definitions of these quantities in section \ref{subsec:carac} and take the derivative of (\ref{eq:mequivfq}) with respect to $ \zeta $. The equivalent of $ Z_-$ was already in \cite{PRL2019}. That of $ Z_+$, which vanishes cubically rather than linearly, explains the overwhelming domination of the $-$ channel at low $ q $ in the second column of Figure \ref{fig:resnum}. Finally, relation (\ref{eq:equivtheta}) gives the starting point of the curves in the third column of figure \ref{fig:resnum}. 

\subsection{The continuum collective branch for $ \Delta/\mu \ll 1 $} 
\label{subsec:faibledelta} 

In the weak coupling limit, we study the continuum collective branch (by type II analytic continuation) at  wave number scales $ q/k_0 \approx \Delta/\mu $ (section \ref{subsubsec:q_approx_d}), $ q \approx k_0 $ (section \ref{subsubsec:q_approx_d0}) and $ q/k_0 \approx (\Delta/\mu)^{2/3} $ (section \ref{subsubsec:q_approx_d23}). The first scale is natural, it corresponds to $ 1/\xi $ where $ \xi $ is the size of a $ \uparrow \downarrow $ bound pair; the spectral study having already been made in section \ref{sec:lcf}, we will content ourselves with adding some results on the residues and on the angle $ \theta $ of section \ref{subsec:carac}. The second scale was ignored in section \ref{sec:lcf} but it remains natural: it corresponds to the width $ 2 k_0 $ of the existence interval of the branch in $ q $ space. The third scale results from an in-depth study: it is the missing link making it possible to connect the large wave number limit of the first scale and the low wave number limit of the second; it also contains the minimum of the real part of the dispersion relation. 

\subsubsection{At the wave number scale $ \check{q} \approx \check{\Delta} $} 
\label{subsubsec:q_approx_d} 

\begin{figure} [tb] 
\centerline{\includegraphics [width = 0.45 \textwidth, clip =]{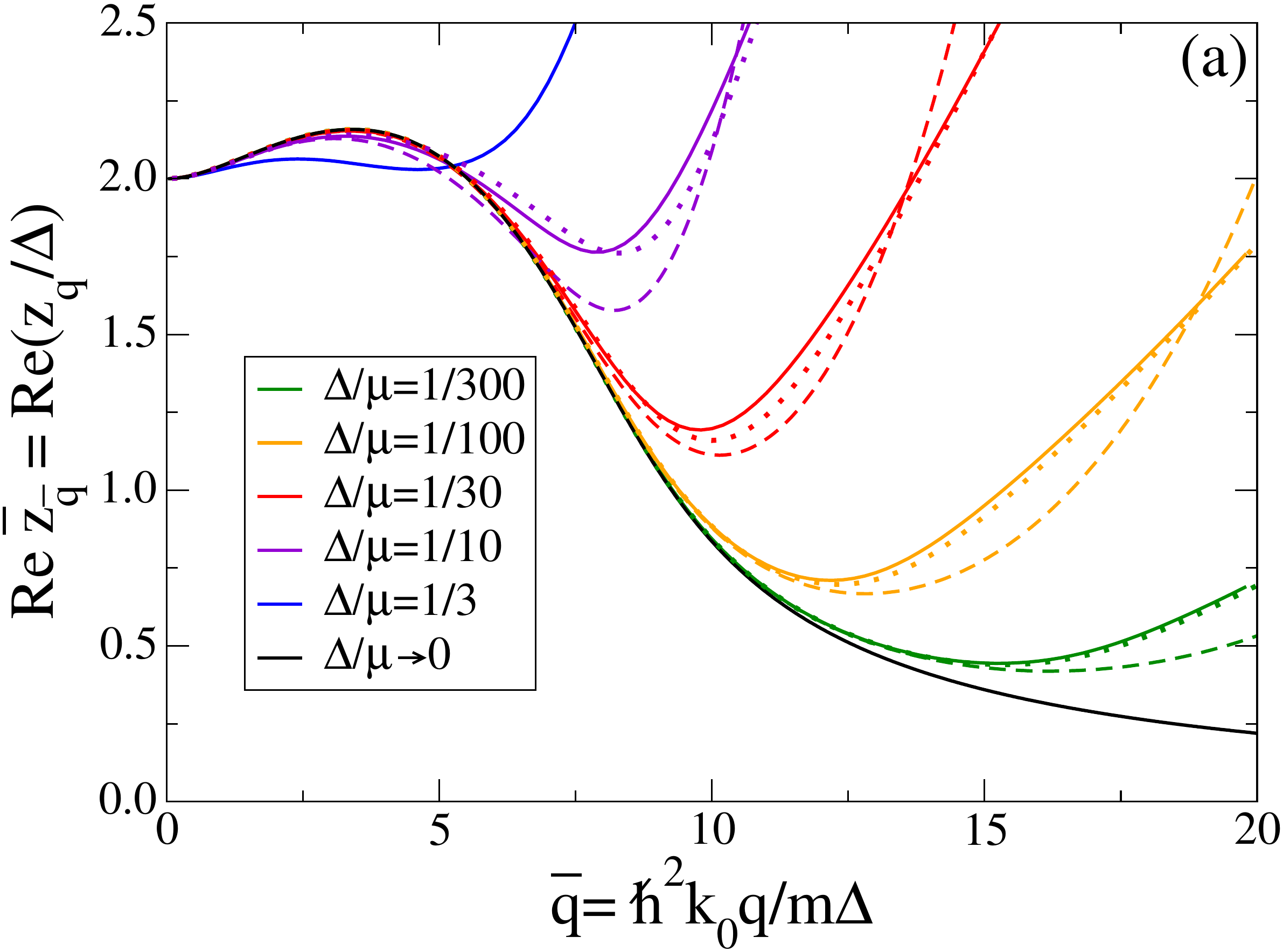} \hspace{5mm} \includegraphics [width = 0.45 \textwidth, clip =]{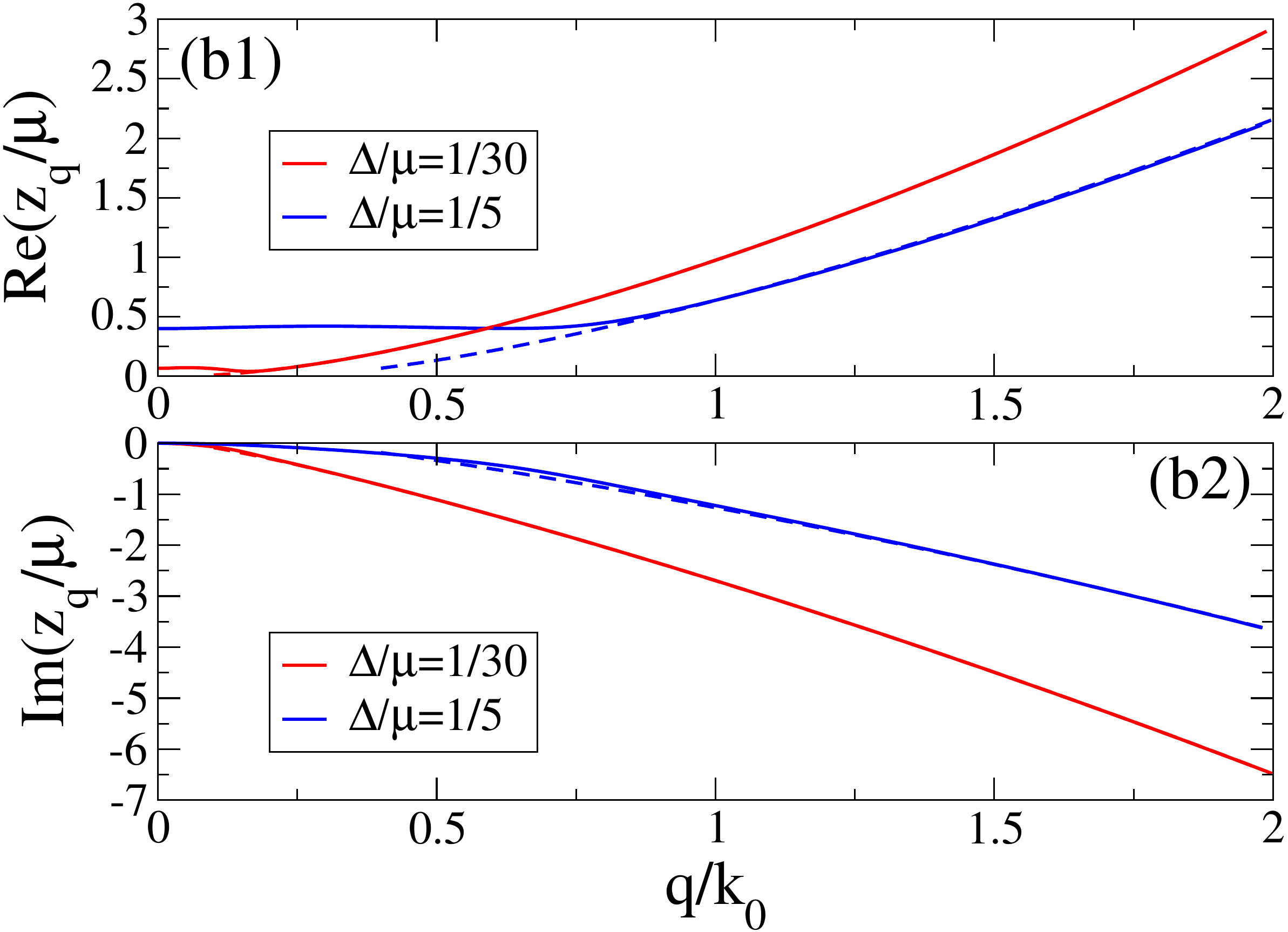}} 
\caption{In the limit $ \Delta/\mu \ll 1 $, convergence of the continuum collective branch $ q \mapsto z_\qq $ (a) at fixed $ \bar{q} = \hbar^2k_0q/m \Delta $ to the limiting law $ \bar{q} \mapsto \bar{z}_{\bar{q}}^{(0)} $, where $ \bar{z}_{\bar{q}}^{(0)} $ is solution of equation (\ref{eq:defzqbar0}), and (b) at fixed $ \check{q} = q/k_0 $ to the leading order law $ \check{q} \mapsto \check{z}_{\check{q}}^{(0)} $ defined by equation (\ref{eq:defzqch0}). In (a), restricted for simplicity to the real part of $ z_\qq $, in solid line: exact dispersion relation, obtained by numerical solution of equation (\ref{eq:surzqgen}) (in color), or equation (\ref{eq:defzqbar0}) for $ \Delta/\mu \to 0 $ (black); dashed: dispersion relation to second order in $ \Delta $ (\ref{eq:zqordredelta2}); roughly dotted: prediction $ \bar{z}_q^{\rm trin} $ (\ref{eq:zqtrin}) of the second degree equation (\ref{eq:trinome}) on $ z_\qq $. In (b1) for the real part and (b2) for the imaginary part: in solid lines, exact dispersion relation; in dashed line: leading order law (\ref{eq:defzqch0}) (for $ \Delta/\mu = 1/30 $, the dashed line is present but it is masked by the solid line). The values of $ \Delta/\mu $ used are specified in the figure.} 
\label{fig:convfd} 
\end{figure}

As in section \ref{subsubsec:dmmicontact}, we make $ \Delta/\mu $ tend to zero at fixed $ \bar{q} = 2 \check{q}/\check{\Delta} $, after rescaling of energies by $ \Delta $, $ \bar{z} = z/\Delta $. At zero order in $ \Delta/\mu $, we find the same limiting dispersion relation $ \bar{q} \mapsto \bar{z}^{(0)}_{\bar{q}} $, solution of (\ref{eq:defzqbar0}) and the study of which has already been done in section \ref{subsec:exci_mod}. However, let us give some additional results. Numerically, we show in figure \ref{fig:convfd} the real part of the dispersion relation, with the rescaling $ \bar{q} $ and $ \bar{z} $ of the variables, for weak and decreasing values of $ \Delta/\mu $, in order to illustrate the convergence towards the limiting law and to reveal the presence for increasing $ \bar{q} $ of a decreasing minimum, totally absent from this limiting law and whose elucidation is postponed to section \ref{subsubsec:q_approx_d23}. Analytically, we complete section \ref{subsec:exci_mod} by giving an equivalent of the residue of the continuum mode, within the meaning of section \ref{subsec:carac}, in the modulus channel: 
\be
\label{eq:defzb-zero}
\check{Z}_-\stackrel{\bar{q}\,\scriptsize\mbox{fixed}}{\underset{\check{\Delta}\to 0}{\sim}}\check{\Delta}\,\bar{Z}_-^{(0)}\quad\mbox{with}\quad
\bar{Z}_-^{(0)}\equiv\frac{1}{\frac{\dd}{\dd\bar{z}}\check{M}_{--}^{(0)}\!\downarrow\!\!(\bar{z}_{\bar{q}}^{(0)},\bar{q})}
\ee
From sections \ref{subsubsec:raafqb} and \ref{subsubsec:raagqb}, we derive its behavior at low reduced wave number 
\be
\label{eq:dlZ4}
\boxed{
\bar{Z}_{-}^{(0)}\underset{\bar{q}\to 0}{=} (2\pi)^2 \left[Z_1 \bar{q} + Z_2 \bar{q}^2 + Z_3 \bar{q}^3 +O(\bar{q}^4) \right] \ \ \ \mbox{with}\ \ \
\left\{
\begin{array}{ll}
Z_1=2\ii\zeta_0/(\pi\sqrt{\zeta_0-1}) &\!\!\!\!\!\!\simeq -0. 125\, 634 + 0. 235\, 132\, \ii  \\
Z_2=2\zeta_0^2(3\zeta_0-5)/[3\pi^2(\zeta_0-1)^2] &\!\!\!\!\!\!\simeq \phantom{+}0. 043\, 448 + 0. 046\, 167\, \ii \\
Z_3=\frac{-2\ii\zeta_0^2}{\pi^3\sqrt{\zeta_0-1}} \left[1+\frac{3\pi^2}{32}+\frac{8/9}{(\zeta_0-1)^3}\right]&\!\!\!\!\!\!\simeq \phantom{+}0. 008\, 812 - 0. 017\, 157\, \ii
\end{array}
\right.
}
\ee
and at large reduced wave number (it is not useful to specify the value of the exponent $ \alpha $): 
\be
\label{eq:domZ-}
\boxed{
\im \bar{Z}_-^{(0)} \underset{\bar{q}\to+\infty}{=} \frac{(2\pi)^2\bar{q}}{\pi/2+\atan Y_0} + O(\ln^\alpha\bar{q}/\bar{q}) \quad\mbox{and}\quad
\re \bar{Z}_-^{(0)} \underset{\bar{q}\to+\infty}{=} \frac{(2\pi^2)^2}{\bar{q}} \frac{[\pi/2+\atan Y_0+Y_0/(1+Y_0^2)]}{Y_0^2(\pi/2+\atan Y_0)^3}
+o(1/\bar{q})
}
\ee
where the real number $ Y_0 $ is the solution of (\ref{eq:Y0}). 
Finally, we give to leading order the residue in the phase channel and the relative phase $ \theta $ between the two channels maximizing the total residue: 
\be
\label{eq:lcfz+theta}
\check{Z}_+^{(3)}= \check{\Delta}\, \bar{Z}_-^{(0)} \left[\frac{\check{M}_{+-}^{(1)}(\bar{z}_{\bar{q}}^{(0)},\bar{q})}{\check{M}_{++\downarrow}^{(0)}(\bar{z}_{\bar{q}}^{(0)},\bar{q})}\right]^2
\quad\mbox{and}\quad
\theta^{(0)}=\arg\left[-\frac{\check{M}_{++\downarrow}^{(0)}(\bar{z}_{\bar{q}}^{(0)},\bar{q})}{\check{M}_{+-}^{(1)}(\bar{z}_{\bar{q}}^{(0)},\bar{q})}\right]
\ee
The expressions of the matrix elements are given at zero order in $ \Delta/\mu $ by (\ref{eq:M--0}, \ref{eq:M++0}) (before analytic continuation and without rescaling (\ref{eq:adimbis})), and to order one for the phase-modulus coupling by equation (\ref{eq:M+-1}) to come. We thus see that the residue in the phase channel is, relative to the modulus channel, reduced by a factor $ \check{\Delta}^2 \ll 1 $. Figure \ref{fig:surprise} shows all of these quantities for a fixed, very small value of $ \Delta/\mu $. 

\subsubsection{At the wave number scale $ \check{q} \approx \check{\Delta}^0 $} 
\label{subsubsec:q_approx_d0} 

Whether we are in weak coupling or not, most of the continuum collective branch (for an analytic continuation of type II) lives with a wave number  $ q $ of order $ k_0 $. We therefore study it here by moving to the limit $ \check{\Delta} \to 0 $ at fixed $ \check{q} \in] 0,2 [$; given the numerical results of figure \ref{fig:resnum}, we expect a reduced eigenenergy $ \check{z}_q $ of order $ 0 $ in $ \check{\Delta} $. Starting from section \ref{subsubsec:lpdnae1e2e3}, we note in this limit that $ \check{q}_0 \to2> \check{q} $, so that the spectral densities include {\sl a priori} three distinct points of non-analyticity on the positive real semi-axis, whose expressions are very simple at zero order in $ \check{\Delta} $: 
\be
\check{\epsilon}_1(\check{q})\to 0,\quad\check{\epsilon}_2(\check{q})\to\check{\epsilon}^{(0)}_2(\check{q})=\check{q}(2-\check{q}) \quad \mbox{and}\quad \check{\epsilon}_3(\check{q})\to\check{\epsilon}^{(0)}_3(\check{q})=2(1-\check{q}^2/4)
\ee
However, the computation of the spectral densities of table \ref{tab:rhogen} when $ \check{\Delta} \to 0 $ at fixed energy $ \check{\epsilon} $ brings up a {\sl new} kink $ \check{\epsilon}_{0} (\check{q}) = \check{q} (2+\check{q}) $, located, depending on the position of $ \check{q} $  with respect to $ 2/3 $, between $ \check{\epsilon}_2 (\check{q}) $ and $ \check{\epsilon}_3 (\check{q}) $ or beyond $ \check{\epsilon}_3 (\check{q}) $: 
\be
\label{eq:rho0zfix}
\begin{array}{|c|c|c|c|c|}
\hline
0<\check{q}<2/3 & 0<\check{\epsilon}<\check{\epsilon}_2^{(0)}(\check{q}) & \check{\epsilon}_2^{(0)}(\check{q})<\check{\epsilon}<\check{q}(2+\check{q}) & \check{q}(2+\check{q})<\check{\epsilon}<\check{\epsilon}_3^{(0)}(\check{q}) & \check{\epsilon}_3^{(0)}(\check{q})<\check{\epsilon} \\
\hline
\check{\rho}_{ss}^{(0)}(\check{\epsilon},\check{q})= &\frac{\pi\check{\epsilon}}{2\check{q}} & \frac{\pi\check{\epsilon}}{4\check{q}}+\frac{\pi}{\sqrt{8}} (\check{\epsilon}_3^{(0)}-\check{\epsilon})^{1/2} & \frac{\pi}{\sqrt{8}}[(\check{\epsilon}+\check{\epsilon}_3^{(0)})^{1/2}+(\check{\epsilon}_3^{(0)}-\check{\epsilon})^{1/2}]& \frac{\pi}{\sqrt{8}}(\check{\epsilon}+\check{\epsilon}_3^{(0)})^{1/2} \\
\check{\rho}_{+-}^{(0)}(\check{\epsilon},\check{q})= & 0 & \frac{\pi\check{\epsilon}}{4\check{q}}-\frac{\pi}{\sqrt{8}} (\check{\epsilon}_3^{(0)}-\check{\epsilon})^{1/2} & \frac{\pi}{\sqrt{8}}[(\check{\epsilon}+\check{\epsilon}_3^{(0)})^{1/2}-(\check{\epsilon}_3^{(0)}-\check{\epsilon})^{1/2}]& \frac{\pi}{\sqrt{8}}(\check{\epsilon}+\check{\epsilon}_3^{(0)})^{1/2} \\
\hline
2/3<\check{q}<2 & 0<\check{\epsilon}<\check{\epsilon}_2^{(0)}(\check{q}) & \check{\epsilon}_2^{(0)}(\check{q})<\check{\epsilon}<\check{\epsilon}_3^{(0)}(\check{q}) & \check{\epsilon}_3^{(0)}(\check{q}) < \check{\epsilon} < \check{q}(2+\check{q}) & \check{q}(2+\check{q}) < \check{\epsilon} \\
\hline
\check{\rho}_{ss}^{(0)}(\check{\epsilon},\check{q})= & \frac{\pi\check{\epsilon}}{2\check{q}} & \frac{\pi\check{\epsilon}}{4\check{q}}+\frac{\pi}{\sqrt{8}}(\check{\epsilon}_3^{(0)}-\check{\epsilon})^{1/2}& \frac{\pi\check{\epsilon}}{4\check{q}} & \frac{\pi}{\sqrt{8}}(\check{\epsilon}+\check{\epsilon}_3^{(0)})^{1/2} \\
\check{\rho}_{+-}^{(0)}(\check{\epsilon},\check{q})= & 0 & \frac{\pi\check{\epsilon}}{4\check{q}}-\frac{\pi}{\sqrt{8}}(\check{\epsilon}_3^{(0)}-\check{\epsilon})^{1/2} & \frac{\pi\check{\epsilon}}{4\check{q}} & \frac{\pi}{\sqrt{8}}(\check{\epsilon}+\check{\epsilon}_3^{(0)})^{1/2} \\
\hline
\end{array}
\ee
This additional kink remains simple to interpret: as we can see on (\ref{eq:P8}), the polynomial $ P_8 (X) $ has for $ \check{\Delta} = 0 $ two positive real roots, one is $ \check{\epsilon}_2^{(0)} (\check{q}) $ as it should be, and the other is precisely $ \check{\epsilon}_{0} (\check{q}) $; the fact that the first root is always less than the second, while $ \check{\epsilon}_2 (\check{q}) $ is the largest real root of $ P_8 (X) $ at nonzero $ \check{\Delta} $, is only a paradox that the last sentence of footnote \ref{note:Lagrange} explains perfectly. Then let's take the limit $ \check{\Delta} \to 0 $ at fixed $ \check{z} $ in expression (\ref{eq:M+-foncrho}, \ref{eq:Mssgen}) of the matrix elements of $ M $, using (\ref{eq:rho0zfix}) and the limiting behavior $ C (\Delta) = 2 \pi [\ln (8/\check{\Delta})-2]+o (1) $ of the regularization scoria. We come across energy integrals that are fairly simple but a little long to calculate; after analytic continuation of type II as in (\ref{eq:surzqgen}), we finally find for $ \im \check{z} <0 $ to zero order in $ \check{\Delta} $: 
\be
\check{M}_\downarrow(\check{z},\check{q})\stackrel{\check{z}\scriptsize\,\mbox{and}\,\check{q}\,\mbox{fixed}}{\underset{\check{\Delta}\to 0}{=}}\check{M}^{(0)}_\downarrow(\check{z},\check{q})+o(1)
\quad\mbox{with}\quad\check{M}^{(0)}_\downarrow(\check{z},\check{q})=
\begin{pmatrix}
\alpha(\check{z}) & \beta(\check{z}) \\
\beta(\check{z})  & \alpha(\check{z})
\end{pmatrix}
\ee
The eigenenergy equation therefore separates into $ \alpha+\beta = 0 $ and $ \alpha-\beta = 0 $; as we can verify, the first of the two equations contains the continuum mode, and it is therefore the only one that we will write explicitly here: 
\begin{multline}
(2\pi)^2[\alpha(\check{z})+\beta(\check{z})]=\ln\left(\frac{8}{\check{\Delta}\eee}\right) +\frac{\check{z}}{4\check{q}}[\ln(\check{z}+2\check{q}-\check{q}^2)-\ln(\check{z}-2\check{q}-\check{q}^2)-2\ii\pi] \\
-\left(\frac{\check{z}+\check{\epsilon}_3^{(0)}(\check{q})}{2}\right)^{1/2} \left\{\argth\frac{[2(\check{z}+\check{\epsilon}_3^{(0)}(\check{q}))]^{1/2}}{\check{q}+2}+\argth\frac{2-\check{q}}{[2(\check{z}+\check{\epsilon}_3^{(0)}(\check{q}))]^{1/2}}\right\}
\end{multline}
As we can see, there remains in the equation a logarithmic dependence with the rescaled order parameter $ \check{\Delta} $. The eigenenergy thus has a well defined zero order in $ \check{\Delta} $, but not a finite limit. We write to summarize: 
\be
\label{eq:defzqch0}
\boxed{
\check{z}_q\stackrel{\scriptsize\check{q}\,\mbox{fixed}}{\underset{\check{\Delta}\to 0}{=}}\check{z}^{(0)}_{\check{q}}+o(1)
\quad\mbox{with}\quad
\alpha(\check{z}_{\check{q}}^{(0)})+\beta(\check{z}_{\check{q}}^{(0)})=0
}
\ee
By numerical solution, we obtain the dashed curves in figures \ref{fig:convfd}b1 and \ref{fig:convfd}b2, in remarkable agreement with the exact result except in the neighborhood of $ \check{q} = 0 $, although the values of $ \check{\Delta} $ chosen are not extremely small. We also obtain analytically the behavior of the mode eigenenergy to order zero at low wave number: 
\be
\label{eq:devfqdtv0}
\check{z}^{(0)}_{\check{q}}\underset{\check{q}\to 0}{=}\check{\zeta}_0 \check{q}+ \check{\zeta}_1\check{q}^2+ o(\check{q}^2)
\quad\mbox{with}\quad
\left\{
\begin{array}{ll}
\check{\zeta}_0 = -2\ii Y_0|_{\bar{q}=2\check{q}/\check{\Delta}} = 2\mathcal{Z}_0|_{\bar{q}=2\check{q}/\check{\Delta}} \\
\\
\check{\zeta}_1 = \displaystyle\left.-\frac{Y_0\left[\ln\frac{\check{\Delta}}{8}+\left(\frac{\pi}{2}+\atan Y_0\right)Y_0\right]}{\frac{\pi}{2}+\atan Y_0}\right|_{\bar{q}=2\check{q}/\check{\Delta}}
\end{array}
\right.
\ee
where $ Y_0 $ is solution of the transcendental equation (\ref{eq:Y0}). 
It is understood that the limits $ \check{\Delta} \to 0 $ and $ \check{q} \to 0 $ do not commute. One can wonder however, in the limit $ \check{\Delta} \to 0 $, if the regimes $ \bar{q} = 2 \check{q}/\check{\Delta} \gg 1 $ and $ \check{q} \ll 1 $ are connected. By comparing the coefficients of the asymptotic expansion (\ref{eq:asympt}) of section \ref{sec:lcf} and of the Taylor expansion (\ref{eq:devfqdtv0}), we find that the imaginary parts perfectly match to dominant order $ q $ (we find in (\ref{eq:devfqdtv0}) the coefficient $ \mathcal{Z}_0 $ of (\ref{eq:asympt})), but the real parts do not match at all, even in terms of their power law ($ 1/\bar{q} $ and $ \check{q}^2 $ have different exponents $-1 $ and $+2 $ in $ q $); this very interesting fact will be exploited and explained in section \ref{subsubsec:q_approx_d23}. We also obtain the asymptotic expansion of eigenenergy in the logarithm of the order parameter: 
\be
\label{eq:devlndqf}
\check{z}^{(0)}_{\check{q}} \stackrel{\check{q}\in]0,2[\,\scriptsize\mbox{fixed}}{\underset{\ln\frac{1}{\check{\Delta}}\to+\infty}{=}} -\frac{2\ii\check{q}}{\pi} \ln\frac{8}{\check{\Delta}\eee^2}
+\frac{\check{q}^{3/2}\eee^{-\ii\pi/4}}{\pi^{1/2}}\left(\ln\frac{8}{\check{\Delta}\eee^2}\right)^{1/2}+\frac{\check{q}^2}{4}
+\frac{\eee^{\ii\pi/4}}{2}(\pi\check{q})^{1/2}\left(1-\frac{3\check{q}^2}{16}\right)\left(\ln\frac{8}{\check{\Delta}\eee^2}\right)^{-1/2}
-\frac{4}{3} \left(\ln\frac{8}{\check{\Delta}\eee^2}\right)^{-1} + O\left(\ln\frac{1}{\check{\Delta}}\right)^{-3/2}
\ee
The logarithm is never very large in a cold atom gas, the practical interest of (\ref{eq:devlndqf}) is quite small, although we went to an unreasonably high order. We get from it however, by removing all rescaling, a very suggestive equivalent in the mathematical limit of weak interaction \footnote{We used the BCS result $ \check{\Delta} \sim 8 \eee^{-2} \eee^{-\pi/2k_{\rm F} | a |} $ when $ k_{\rm F} a \to 0^-$ and replaced $ k_0 $ by its limit $ k_{\rm F} $.} 
\be
\boxed{
z_q\stackrel{q\in]0,2k_{\rm F}[\scriptsize\mbox{fixed}}{\underset{k_{\rm F} a\to 0^-}{\sim}} -\frac{\ii\hbar^2 q}{2m|a|}
}
\ee
where, remember, $ a $ is the scattering length and $ k_{\rm F} $ the Fermi wave number. Finally, to be complete, let us give the expression of the other observables of section \ref{subsec:carac} at zero order in $ \check{\Delta} $ at fixed $ \check{q} $, namely the residues in the phase channel $+$ and in the modulus channel $-$, as well as the relative phase in the superposition of these two channels maximizing the total residue: 
\be
\label{eq:zpzm00ettheta00}
\check{Z}_{\pm}^{(0)}=\frac{1/2}{\alpha'(\check{z}^{(0)}_{\check{q}})+\beta'(\check{z}^{(0)}_{\check{q}})}=
\frac{(4\pi)^2 \check{q}\,[\check{z}^{(0)}_{\check{q}}+\check{\epsilon}_3^{(0)}(\check{q})]}{[\check{z}^{(0)}_{\check{q}}+2\check{\epsilon}_3^{(0)}(\check{q})][\ln(\check{z}^{(0)}_{\check{q}}+2\check{q}-\check{q}^2)-\ln(\check{z}^{(0)}_{\check{q}}-2\check{q}-\check{q}^2)-2\ii\pi]-4\check{q}\ln\frac{8}{\check{\Delta}\eee}}\ \mbox{and}\ \theta^{(0)}=\arg 1 =0
\ee
These expressions result directly from the fact that the matrix residue in (\ref{eq:residmat}) is written after rescaling
\be
\check{\mathcal{M}}=\frac{1/2}{\alpha'(\check{z}_{\check{q}}^{(0)})+\beta'(\check{z}_{\check{q}}^{(0)})}
\begin{pmatrix}
 1 &  1 \\
 1 &  1
\end{pmatrix}
\ee
which shows that the continuum mode is, in this limit, a linear superposition with equal amplitudes of small phase and modulus deviations of the order parameter. Results (\ref{eq:defzqch0}) and (\ref{eq:zpzm00ettheta00}) are shown in green solid line in figure \ref{fig:surprise} for a very small value of $ \Delta/\mu $, which makes them very close to the exact result for a not too small reduced wave number $ \check{q} $. 

\subsubsection{In the vicinity of the eigenfrequency minimum:  the unexpected wave number scale $ \check{q} \approx \check{\Delta}^{2/3} $} 
\label{subsubsec:q_approx_d23}

Our motivation to go beyond sections \ref{subsubsec:q_approx_d} and \ref{subsubsec:q_approx_d0} was to understand the origin of the absolute minimum that we can see very well in the real part of the continuum collective branch in figure \ref{fig:convfd}a, in weak coupling regime $ \check{\Delta} \ll 1 $. In particular, this minimum does not appear at all in the limiting dispersion relation $ \bar{q} \mapsto \bar{z}_{\bar{q}}^{(0)} $ obtained for $ \check{\Delta} \to 0 $ at fixed $ \bar{q} = 2 \check{q}/\check{\Delta} $ in section \ref{subsec:exci_mod} (see the black curve in figure \ref{fig:convfd}a), to our dissatisfaction. This led us to the discovery of the non-trivial wave number scale $ \check{q} \approx \check{\Delta}^{2/3} $, as we expose it here. 

The first idea to account for the minimum at $ \check{q} = \check{q}_{\rm min}>0 $ in figure \ref{fig:convfd}a is to go to next order in the small parameter $ \check{\Delta} $. At fixed $ \bar{q} $ and $ \bar{z} = z/\Delta $, we therefore continue the expansion of the matrix elements of $ \check{M} $, to order one for the phase-modulus coupling and to order two for the modulus-modulus coupling: 
\be
\label{eq:M--02M+-1}
\check{M}_{--}(\bar{z},\bar{q}) \stackrel{\scriptsize\bar{z}\,\mbox{and}\,\bar{q}\,\mbox{fixed}}{\underset{\check{\Delta}\to 0}{=}}\check{M}_{--}^{(0)}(\bar{z},\bar{q})+\check{M}_{--}^{(2)}(\bar{z},\bar{q})+o(\check{\Delta}^2) \quad\mbox{and}\quad
\check{M}_{+-}(\bar{z},\bar{q}) \stackrel{\scriptsize\bar{z}\,\mbox{and}\,\bar{q}\,\mbox{fixed}}{\underset{\check{\Delta}\to 0}{=}} \check{M}_{+-}^{(1)}(\bar{z},\bar{q})+o(\check{\Delta})
\ee
The zero order appears, in non-dimensionless integral form, in equation (\ref{eq:M--0}). The rest of the calculation, which is a bit long, is based on integral representations (\ref{eq:M+-foncrho}, \ref{eq:m--fintastu}). Let's give the result, explicit: \footnote{To calculate $ \check{M}_{+-}^{(1)} $, we split in rescaled (\ref{eq:M+-foncrho}) the integration domain into two sub-intervals $ I_<= [\check{\epsilon}_2 (\check{q}), \eta] $ and $ I_> = [\eta,+\infty [$ where $ \eta \ll $ 1 is fixed. On the first one, we expand the integrand in powers of $ \check{\Delta} $ at fixed $ \bar{\epsilon} = \check{\epsilon}/\check{\Delta} $. On the second one, we expand at fixed $ \check{\epsilon} $. To do this, you must expand the roots $ s_1 $ and $ s_2 $ of (\ref{eq:surs}). On $ I_<$, $ s_{1,2} \simeq \mp s_0-A \check{\Delta} $ with $ s_0^2 = \bar{\epsilon}^2 (\bar{\epsilon}^2-4-\bar{q}^2)/[(\bar{\epsilon}^2-4) (\bar{\epsilon}^2-\bar{q}^2)] $ and $ A = \bar{\epsilon}^2 \bar{q}^2/[(\bar{\epsilon}^2-4)^{1/2} (\bar{\epsilon}^2-\bar{q}^2)^2] $ so that $ \check{\rho}_{+-} \sim \frac{\pi}{4} \check{\Delta} \bar{\epsilon} (\bar{\epsilon}^2-4-\bar{q}^2)^{1/2}/(\bar{\epsilon}^2-\bar{q}^2)^{1/2} $. On $ I_> $, $ s_{1,2} \simeq \mp 1-\bar{q}^2 \check{\Delta}^4 (\check{\epsilon} \mp 2)/\check{\epsilon}^4 $ hence $ \check{\rho}_{+-} \sim \frac{\pi}{2} [(1+\check{\epsilon}/2)^{1/2}-\Theta (2-\check{\epsilon}) (1-\check{\epsilon}/2)^{1/2}] $. Integration on energy brings up formally divergent terms of the form $ \check{\Delta} \ln \eta $, which exactly cancel between the upper and lower parts, and we obtain (\ref{eq:M+-1}). The calculation of $ \check{M}_{--}^{(2)} $ proceeds in the same way, starting from the clever form (\ref{eq:m--fintastu}), but it is heavier: $ (i) $ you must add the subinterval $ [\check{\epsilon}_1, \check{\epsilon}_2 (\check{q})] $, on which there is however no deviation from the spectral density $ \rho^{(0) [\mathrm{II}]}_{--} $ of table \ref{tab:rholcf}, $ (ii) $ on $ I_<$, you have to expand $ s_{1,2} $ up to order $ \check{\Delta}^2 $, to get $ \check{\rho}_{--}-\check{\rho}^{(0) [\mathrm{III}]}_{--} \sim-\frac{\pi}{32} \check{\Delta}^2 (\bar{\epsilon}^2-4-\bar{q}^2)^{1/2} [\bar{\epsilon}^6-(4+\bar{q}^2) \bar{\epsilon}^4-\bar{q}^2 (\bar{q}^2-16) \bar{\epsilon}^2+\bar{q}^6]/(\bar{\epsilon}^2-\bar{q}^2)^{5/2} $, $ (iii) $ on $ I_> $, you have to expand $ s_{1,2} $ up to order $ \check{\Delta}^6 $ to get $ \check{\rho}_{--} = \frac{\pi}{2} [(1+\check{\epsilon}/2)^{1/2}+\Theta (2-\check{\epsilon}) (1-\check{\epsilon}/2)^{1/2}]-\frac{\pi \check{\Delta}^2}{32 \sqrt{2} \check{\epsilon}^2} [\frac{\bar{q}^2 \check{\epsilon}^2+48 \check{\epsilon}+64}{(2+\check{\epsilon})^{1/2}}+\Theta (2-\check{\epsilon}) \frac{\bar{q}^2 \check{\epsilon}^2-48 \check{\epsilon}+64}{(2-\check{\epsilon})^{1/2}} ]+O (\check{\Delta}^4) $.} 
\bea
\label{eq:M+-1}
&& \!\!\!\!\!\!\!\!\!\!\!\!\!\!\!\!\!\!\!\!\!\check{M}_{+-}^{(1)}(\bar{z},\bar{q}) = \frac{\pi\check{\Delta}\bar{z}}{2(2\pi)^3}\left[\ln\frac{\check{\Delta}}{8\eee}+\phi(\bar{z},\bar{q})\right]
\quad\mbox{with}\quad
\phi(\bar{z},\bar{q})\equiv\frac{(4+\bar{q}^2-\bar{z}^2)^{1/2}}{(\bar{z}^2-\bar{q}^2)^{1/2}} \acos \frac{(4+\bar{q}^2-\bar{z}^2)^{1/2}}{2}  \\
&& \!\!\!\!\!\!\!\!\!\!\!\!\!\!\!\!\!\!\!\!\!\check{M}_{--}^{(2)}(\bar{z},\bar{q}) =  -\frac{\pi\check{\Delta}^2}{16(2\pi)^3} \left[(\bar{q}^2\!+\!\bar{z}^2\!-\!4)\ln\frac{\check{\Delta}\eee^{1/2}}{8} \!-\!\frac{\bar{q}^2(\bar{q}^4\!+\!(4\!-\!\bar{z}^2)\bar{q}^2\!+\!8\bar{z}^2)}{(\bar{z}^2\!-\!\bar{q}^2)^2} \!+\!\frac{\bar{z}^6\!-\!(4\!+\!\bar{q}^2)\bar{z}^4\!+\!\bar{q}^2(16\!-\!\bar{q}^2)\bar{z}^2\!+\!\bar{q}^6}{(\bar{z}^2\!-\!\bar{q}^2)^2}\phi(\bar{z},\bar{q})\right] 
\eea
Note the appearance of logarithmic terms in $ \check{\Delta} $, and the fact that the function $ \phi (\bar{z}, \bar{q}) $, introduced for simplicity, has, in terms of the variable $ \bar{z} $, no branch cut on the interval $ [\bar{\epsilon}_1, \bar{\epsilon}^{(0)}_2] = [2, (4+\bar{q}^2)^{1/2}] $, so the first order correction to $ M_{+-} $ and the second order correction to $ M_{--} $ are analytic and their continuation of type II to the lower complex half-plane is all done, $ \check{M}_{+-\downarrow}^{(1)} (\bar{z}, \bar{q}) = \check{M}_{+-}^{(1)} (\bar{z}, \bar{q}) $ and $ \check{M}_{--\downarrow}^{(2)} (\bar{z}, \bar{q}) = \check{M}_{--}^{(2)} (\bar{z}, \bar{q}) $. For the phase-phase coupling, it will not be useful to go beyond the zero order, already given in the integral form (\ref{eq:M++0}). Although the phase-modulus coupling is first order in $ \check{\Delta} $, it appears squared in the eigenenergy equation (\ref{eq:surzqgen}); the first eigenenergy correction is therefore second order: 
\be
\label{eq:zqordredelta2}
\bar{z}_q \stackrel{\scriptsize\bar{q}\,\mbox{fixed}}{\underset{\check{\Delta}\to 0}{=}} \bar{z}_{\bar{q}}^{(0)}+\bar{z}_{\bar{q}}^{(2)}+o(\check{\Delta}^2)
\quad\mbox{with}\quad \left[\frac{\bar{z}_{\bar{q}}^{(2)}}{\bar{Z}_-^{(0)}}+\check{M}_{--}^{(2)}(\bar{z}_{\bar{q}}^{(0)})\right] \check{M}_{++\downarrow}^{(0)}(\bar{z}_{\bar{q}}^{(0)},\bar{q})
 =[\check{M}_{+-}^{(1)}(\bar{z}_{\bar{q}}^{(0)},\bar{q})]^2
\ee
To obtain this expression, it was enough to replace in (\ref{eq:surzqgen}) $ M_{+-} $ and $ M_{++} $ by their dominant behavior taken at unperturbed energy (after analytic continuation of type II for $ M_{++} $); in the second order term of $ M_{--} $, see (\ref{eq:M--02M+-1}), we could do the same, but in the zero order contribution (after analytic continuation of type II), we had to do a first order Taylor expansion: 
\be
\label{eq:devTaylor}
\check{M}_{--\downarrow}^{(0)}(\bar{z}_q,\bar{q})=\check{M}_{--\downarrow}^{(0)}(\bar{z}^{(0)}_{\bar{q}},\bar{q})+ (\bar{z}_q-\bar{z}^{(0)}_{\bar{q}}) \frac{\dd}{\dd\bar{z}}
\check{M}_{--\downarrow}^{(0)}(\bar{z}_{\bar{q}}^{(0)},\bar{q})+\ldots
\ee
then use relations (\ref{eq:defzqbar0}, \ref{eq:defzb-zero}). In other words, the first correction to the universal dispersion relation $ \bar{q} \mapsto \bar{z}_{\bar{q}}^{(0)} $ of the weak coupling limit results from two physical effects,  {\sl a priori} of the same importance: $ (i) $ the effect of the phase-modulus coupling, completely ignored in section \ref{subsec:exci_mod}, and $ (ii) $ a correction to the equation in the modulus channel, that is to say to the modulus-modulus coupling, also omitted. 

Let us analyze result (\ref{eq:zqordredelta2}) in simple limiting cases. {\sl At low $ \bar{q} $,} the dominant contribution to the energy correction is quadratic in wave number and comes from phase-modulus coupling, that coming from $ M_{--}^{(2)} $ being only cubic. We thus find 
\be
\bar{z}_{\bar{q}}^{(2)} \underset{\bar{q}\to 0}{\sim} \bar{z}_{\bar{q}}^{(2)}|_{M_{+-}^{(1)}} \underset{\bar{q}\to 0}{\sim} -\frac{\check{\Delta}^2\bar{q}^2}{2\pi^2} \frac{\zeta_0^2}{\zeta_0-1}
\left(\ln\frac{\check{\Delta}}{8\eee}\right)^2
\ee
which reproduces the term of order $ \check{\Delta}^2 $ of the quadratic-start coefficient of the continuum collective branch, see (\ref{eq:zetapetitgranddelta}). {\sl At large $ \bar{q} $,} the calculation is simple when we have explicit expressions, and we get the dominant order in $ \bar{q} $: 
\be
\label{eq:domM+-M--}
\check{M}_{+-}^{(1)}(\bar{z}_{\bar{q}}^{(0)}\!,\bar{q})\!\!\!\underset{\bar{q}\gg 1}{\simeq}\!\!\!-\frac{\ii\pi\check{\Delta}Y_0\bar{q}}{2(2\pi)^3} \ln\frac{\check{\Delta}\bar{q}(1+Y_0^2)^{1/2}}{8\eee}\ \mbox{and}\ \check{M}_{--}^{(2)}(\bar{z}_{\bar{q}}^{(0)}\!,\bar{q})\!\!\!\underset{\bar{q}\gg 1}{\simeq}\!\!\!-\frac{\pi\check{\Delta}^2\bar{q}^2(1-Y_0^2)}{16(2\pi)^3}\left\{\pi\ln\check{\Delta}+\ln\left[\frac{\bar{q}(1+Y_0^2)^{1/2}\eee^{1/2}}{8}\right]-(1\!-\!Y_0^4)^{-1}\right\} 
\ee
where $ Y_0 $ is given by (\ref{eq:Y0}). The calculation is more laborious on integral forms: at $ \bar{z} \approx \bar{q} \approx \bar{z}_{\bar{q}}^{(0)} $, the zero order matrix elements $ \check{M}_{++}^{(0)} $ and $ \check{M}_{--}^{(0)} $ behave in the same way to dominant order in $ \bar{q} $, \footnote{After analytic continuation, they are both equivalent up to a constant factor to $ \mathcal{M}_0 (\mathcal{Z}, \bar{q}) $, as in (\ref{eq:m--0gq}).} therefore at this order both vanish when $ \bar{z} = \bar{z}_{\bar{q}}^{(0)} $: we work a lot for a zero result... It is therefore clever to work directly on the difference between (\ref{eq:M++0}) and (\ref{eq:M--0}), by setting $ \check{D} (\bar{z}, \bar{q}) \equiv \check{M}_{++}^{(0)} (\bar{z}, \bar{q})-\check{M}_{--}^{(0)} (\bar{z}, \bar{q}) $. So, after analytic continuation and always to leading order in $ \bar{q} $, \footnote{We come across integrals like those $ I $ and $ J $ of footnote \ref{note:aidint}, that we treat in the same way.} we obtain in terms of the Bose function $ g_2 (z) $: 
\be
\label{eq:domM++}
\check{D}_\downarrow(\bar{z}_{\bar{q}}^{(0)}\!,\bar{q})\underset{\bar{q}\gg 1}{\simeq}\frac{8\pi}{(2\pi)^3Y_0\bar{q}^2}\left\{\pi\ln(-2\ii\bar{q}Y_0)-\ln\left[\bar{q}(1+Y_0^2)^{1/2}\right]\atan\left(\frac{1}{Y_0}\right)-\im g_2\left(\frac{1}{\ii Y_0}\right)+\frac{1}{2}\im g_2\left(\frac{2}{\ii Y_0+1}\right)\right\}
\ee
The behavior of the residue $ \bar{Z}_-^{(0)} $ at large wave number is in (\ref{eq:domZ-}). We deduce from all this that again, the phase-modulus coupling brings the dominant correction of order $ \check{\Delta}^2 $ to the continuum collective branch and we can keep 
\be
\label{eq:d2q5}
\bar{z}_{\bar{q}}^{(2)}\underset{\bar{q}\gg1}{\simeq} \bar{Z}_-^{(0)}\frac{[\check{M}_{+-}^{(1)}(\bar{z},\bar{q})]^2}{\check{D}_\downarrow(\bar{z}_{\bar{q}}^{(0)}\!,\bar{q})}
\approx \check{\Delta}^2\bar{q}^5
\ee
 {\sl At intermediate wave numbers,} $ \bar{q} \approx 1 $, it turns out that the phase-modulus coupling contribution to $ \re \bar{z}^{(2)}_{\bar{q}} $ in (\ref{eq:zqordredelta2}) vanishes, the point where it vanishes depending little on $ \check{\Delta} $ and remaining close to $ \bar{q} = 5 $ for the parameters of figure \ref{fig:convfd}a. The contribution of $ M_{--}^{(2)} $ to this observable is certainly no longer negligible compared to that of $ M_{+-}^{(1)} $ but it remains small in absolute value, and we can ignore it in practice; this explains why the different solid lines (including the black limiting curve) seem to intersect at $ \bar{q} \simeq 5 $ in the figure. 

As we can judge in figure \ref{fig:convfd}a, by comparing the dashed curves to those in solid line, the calculation (\ref{eq:zqordredelta2}) of $ \bar{z}_q $ to second order in $ \check{\Delta} $ explains the existence of a minimum on $ \re z_q $, but in a more qualitative than quantitative way. In particular, it fails to properly describe the rise in the curve beyond the minimum, the agreement with the numerical results in this area not even improving at very weak coupling. The problem comes from the approximation $ \check{M}_{++\downarrow}^{(0)} (\bar{z}_q, \bar{q}) \approx \check{M}_{++\downarrow}^{(0)} (\bar{z}^{(0)}_{\bar{q}}, \bar{q}) $ made in equation (\ref{eq:zqordredelta2}). Indeed, as we said, $ \check{M}_{++}^{(0)} $ and $ \check{M}_{--}^{(0)} $ evaluated in $ \bar{z} \approx \bar{q} $ are equivalent and of order $ \approx \bar{q}^0 $ at large $ \bar{q} $; as $ \check{M}_{--\downarrow}^{(0)} $ vanishes by construction in $ \bar{z}^{(0)}_{\bar{q}} $, $ \check{M}_{++\downarrow}^{(0)} $ is abnormally weak at this point, that is to say of order $ \bar{q}^{-2} $ as we read it on (\ref{eq:domM++}), rather than of order $ \bar{q}^0 $ as you might expect. In this case, we must write $ \check{M}_{++\downarrow} = \check{D}_\downarrow+\check{M}_{--\downarrow} $, approximate the small contribution $ \check{D}_\downarrow $ by its value in $ \bar{z}^{(0)}_{\bar{q}} $ but make on the remaining bit $ M_{--\downarrow} $ the  {\sl same} Taylor expansion as in equation (\ref{eq:devTaylor}). We can also omit the correction $ M_{--}^{(2)} $, which we showed in the previous paragraph to be very small at any $ \bar{q} $ and even sub-leading at large $ \bar{q} $. We finally replace the first degree equation (\ref{eq:zqordredelta2}) with a second degree equation (trinomial) on $ \bar{z}_q $, hence the notation \og trin \fg \, in exponent: \footnote{By anticipating the law $ \bar{q}_{\rm min} \approx \bar{\Delta}^{-1/3} $, we check around the minimum that the imaginary parts of the two terms in square brackets in (\ref{eq:trinome}) are effectively of the same order of magnitude $ \approx 1/\bar{q}_{\rm min}^2 $; we have to consider their imaginary part here because we want to access the real part of $ (\bar{z}_q^{\rm trin}-\bar{z}^{(0)}_{\bar{q}}) $ in prefactor of the square brackets; however the residue in the denominator of this prefactor is purely imaginary to leading order in $ \bar{q} $ and the right-hand side of (\ref{eq:trinome}) is real.} 
\be
\label{eq:trinome}
\frac{(\bar{z}_q^{\rm trin}-\bar{z}_{\bar{q}}^{(0)})}{\bar{Z}_-^{(0)}}\left[\frac{\bar{z}_q^{\rm trin}-\bar{z}_{\bar{q}}^{(0)}}{\bar{Z}_{-}^{(0)}}+\check{D}_\downarrow(\bar{z}_{\bar{q}}^{(0)},\bar{q})\right] =[\check{M}_{+-}^{(1)}(\bar{z}_{\bar{q}}^{(0)},\bar{q})]^2
\ee
The correct solution of the equation is written 
\be
\label{eq:zqtrin}
\boxed{
\frac{\bar{z}_q^{\rm trin}-\bar{z}_{\bar{q}}^{(0)}}{\bar{Z}_-^{(0)}} = -\frac{1}{2} \check{D}_\downarrow(\bar{z}_{\bar{q}}^{(0)},\bar{q})+\eta
\left[\frac{1}{4} \check{D}_\downarrow(\bar{z}_{\bar{q}}^{(0)},\bar{q})^2 + \check{M}_{+-}^{(1)}(\bar{z}_{\bar{q}}^{(0)},\bar{q})^2\right]^{1/2}
}
\ee
where the sign $ \eta $ is that of $ \re \check{D}_\downarrow (\bar{z}_{\bar{q}}^{(0)}, \bar{q}) $, positive for $ \bar{q} $ large enough, negative for $ \bar{q} $ small enough. The corresponding prediction, shown in dotted lines in figure \ref{fig:convfd}a, is in good agreement with the exact results, an agreement all the better since $ \check{\Delta} $ is weaker, even in the rise beyond the minimum. When we move away from the minimum in the direction of increasing $ \bar{q} $, the contribution of $ \check{D}_\downarrow $ in (\ref{eq:zqtrin}) becomes more and more negligible and the rise in the curve takes the simple form \footnote{There is a $-$ sign before the term $ \check{M}_{+-}^{(1)} $ in (\ref{eq:remontee}) because, for $ 1 \ll \bar{q} \ll 1/\check{\Delta} $, its imaginary part is positive, its square is negative and $ \im (\check{D}_\downarrow^2) <0 $.} 
\be
\label{eq:remontee}
\frac{\bar{z}_q^{\rm trin}-\bar{z}_{\bar{q}}^{(0)}}{\bar{Z}_-^{(0)}} \underset{\bar{q}\gg\bar{q}_{\rm min}}= -\check{M}_{+-}^{(1)}(\bar{z}_{\bar{q}}^{(0)},\bar{q})-\frac{1}{2} \check{D}_\downarrow(\bar{z}_{\bar{q}}^{(0)},\bar{q})+\ldots
\ee
When we move away from the minimum in the direction of decreasing $ \bar{q} $, the contribution of $ \check{D}_\downarrow $ becomes dominant and (\ref{eq:trinome}) is reduced to expansion (\ref{eq:zqordredelta2}) (up to the omission of $ \check{M}_{--}^{(2)} $), which is therefore only retrospectively applicable at $ \bar{q} \ll \bar{q}_{\rm min} $. To be complete, we give the expression taken from (\ref{eq:trinome}) of the residues in the phase and modulus channels, \footnote{We obtain the derivative of the analytically continued determinant of $ M $ by taking the derivative of the left-hand side of (\ref{eq:trinome}) with respect to $ \bar{z}^{\rm trin}_q $; the right-hand side is constant and gives zero contribution.} as well as the relative phase in the superposition of these two channels maximizing the global residue: 
\be
\label{eq:zpmtrinthetatrin}
\theta^{\rm trin} = \arg\left(\frac{-\bar{Z}_-^{(0)} \check{M}_{+-}^{(1)}(\bar{z}_{\bar{q}}^{(0)},\bar{q})}{\bar{z}_q^{\rm trin}-\bar{z}_{\bar{q}}^{(0)}}\right) \quad\mbox{and}\quad
\bar{Z}^{\rm trin}_{\pm} =  \frac{1}{2} \bar{Z}_-^{(0)}\mp\eta \frac{\bar{Z}_-^{(0)} \check{D}_\downarrow(\bar{z}_{\bar{q}}^{(0)},\bar{q})/4}
{\left[\frac{1}{4} \check{D}_\downarrow(\bar{z}_{\bar{q}}^{(0)},\bar{q})^2 + \check{M}_{+-}^{(1)}(\bar{z}_{\bar{q}}^{(0)},\bar{q})^2\right]^{1/2}} 
\ee

\begin{figure} [tb] 
\centerline{\includegraphics [width = 0.45 \textwidth, clip =]{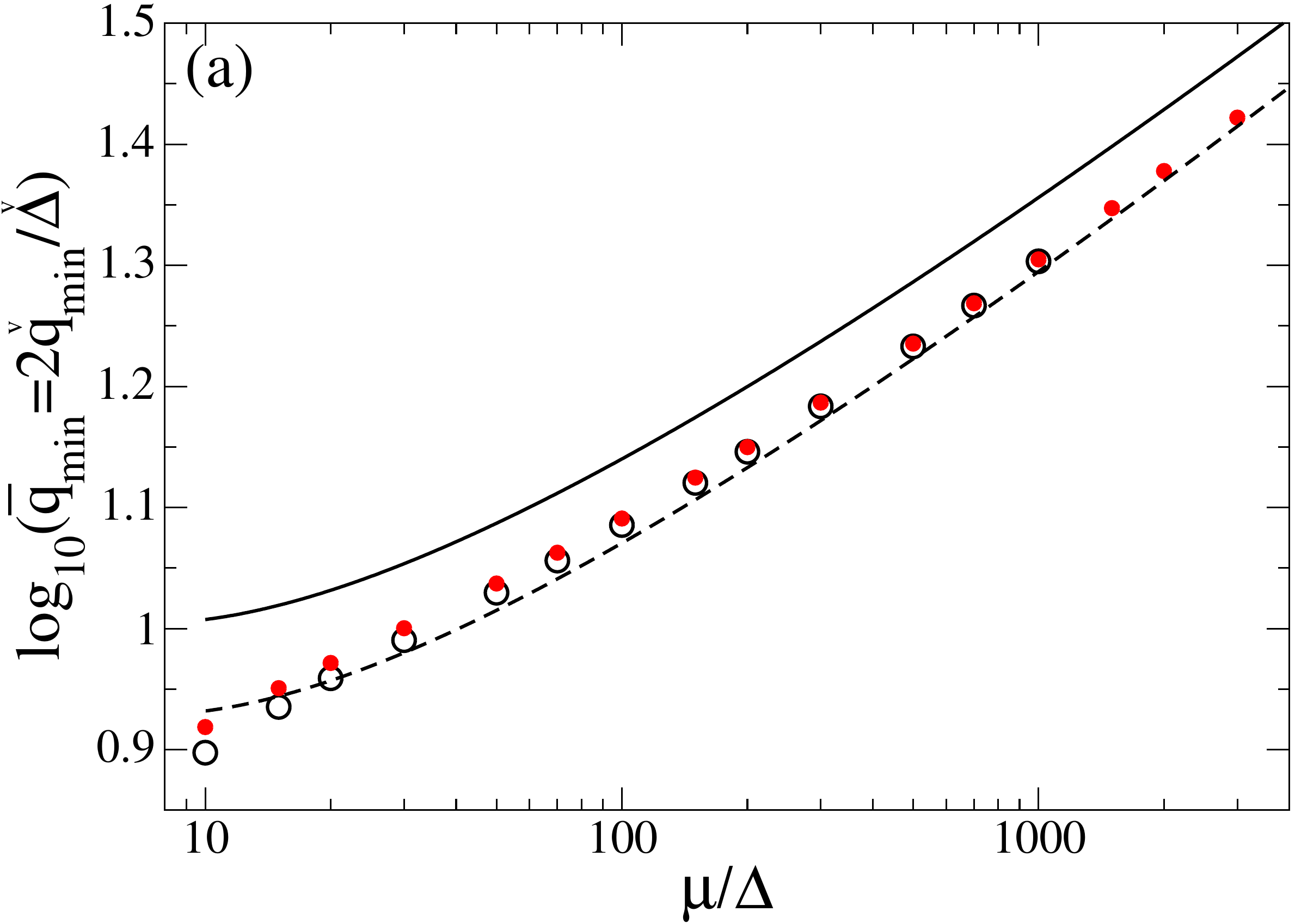} \hspace{5mm} \includegraphics [width = 0.45 \textwidth, clip =]{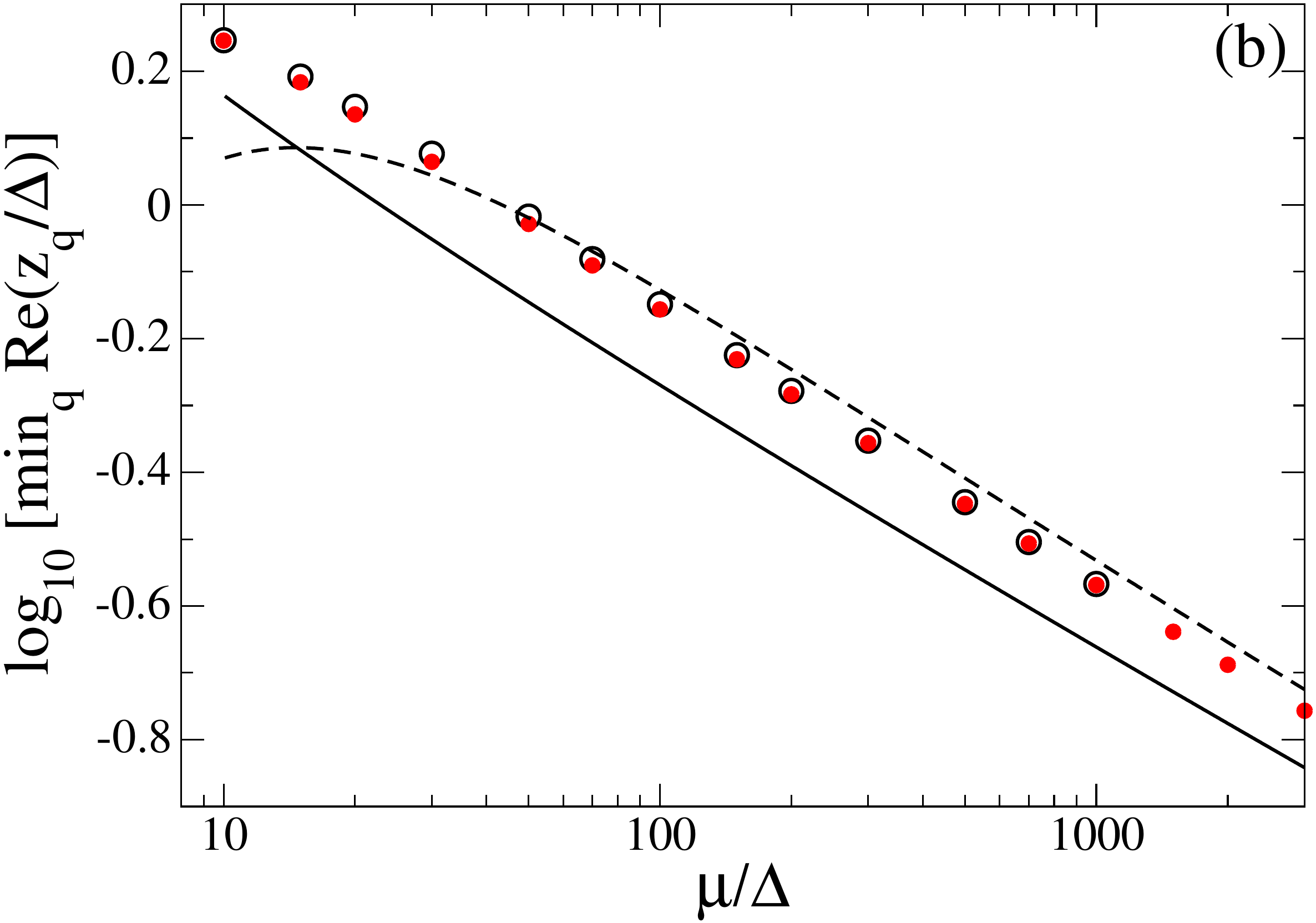}} 
\caption{For interactions of zero range (case of cold atoms) in weak coupling regime, position $ q_{\rm min} $ (a) and value $ \re z_{q_{\rm min}} $ (b) of the minimum of the real part of the continuum collective branch $ q \mapsto \re z_\qq $ for an analytic continuation of type II, as functions of the inverse of the order parameter. 
Black circles: exact values from a numerical solution of equation (\ref{eq:surzqgen}). Red disks: predictions of the quadratic equation (\ref{eq:zqtrin}) on $ z_\qq $. Solid black line: asymptotic equivalents (\ref{eq:asympqmin_et_rezqmin}). Black dashed line: Pad\'e approximants (\ref{eq:padeqmin}) and (\ref{eq:paderezqmin}). These last two elements belong to the unexpected wave number scale $ \check{q} \approx \check{\Delta}^{2/3} $.} 
\label{fig:qmin} 
\end{figure}

Let us now proceed to the in-depth study of the real part of the continuum collective branch in the vicinity of its minimum at $ q = q_{\rm min}> 0 $ in the limit $ \check{\Delta} \to 0 $. For this, we must first determine the scale law of $ q_{\rm min} $. A first way to do this is to say that $ \bar{z}^{(0)}_{\bar{q}} \approx \bar{z}^{(2)}_{\bar{q}} $ at $ q = q_{\rm min} $, which, taking into account (\ref{eq:asympt}) and (\ref{eq:d2q5}), leads to $ \bar{q}_{\rm min} \approx \check{\Delta}^{-1/3} $ so to $ \check{q}_{\rm min} \approx \check{\Delta}^{2/3} $, which is the unexpected wave number scale announced in the title of this section \ref{subsubsec:q_approx_d23}. A second way is to require that $ \check{D}_\downarrow \approx \check{M}^{(1)}_{+-} $ in the expression between square brackets in (\ref{eq:zqtrin}), which, given (\ref{eq:domM+-M--}) and (\ref{eq:domM++}), leads to the same result. We deduce that $ \re \bar{z}_{q_{\rm min}} \approx \check{\Delta}^{1/3} $ or $ \re \check{z}_{q_{\rm min}} \approx \check{\Delta}^{4/3} $. The rescaling of the dispersion relation adapted to the new scale will be indicated by a hat. So we first put $ \check{q} = \check{\Delta}^{2/3} \hat{q} $, that is to say $ \bar{q} = 2 \check{\Delta}^{-1/3} \hat{q} $. Then we obtain the expression of the real part of the branch thus rescaled to leading order in the limit $ \check{\Delta} \to 0 $ at fixed $ \hat{q} $, by working a little on (\ref{eq:trinome}): \footnote{An important simplification comes from the fact that to leading order in $ \bar{q} $, $ \re \bar{z}_{\bar{q}}^{(0)} = \frac{1}{2} \re [\bar{Z}_-^{(0)} \check{D}_\downarrow (\bar{z}_q^{(0)}, \bar{q})] $.} 
\be
\label{eq:simplif}
\re\check{z}_q \stackrel{\hat{q}\, \scriptsize\mbox{fixed}}{\underset{\check{\Delta}\to 0}{=}}\check{\Delta}^{4/3}\left[\re\hat{z}_{\hat{q}}^{(0)}+o(1)\right]\quad\mbox{with}\quad \re\hat{z}_{\hat{q}}^{(0)}= -\frac{(2\pi)^2\bar{q}\check{\Delta}^{-1/3}}{\frac{\pi}{2}+\atan Y_0} \im \left[\frac{1}{4} \check{D}_\downarrow^{\rm asymp}(\bar{z}_{\bar{q}}^{(0)},\bar{q})^2 + \check{M}_{+-}^{(1),{\rm asymp}}(\bar{z}_{\bar{q}}^{(0)},\bar{q})^2\right]^{1/2}
\ee
where $ \check{D}_\downarrow (\bar{z}_{\bar{q}}^{(0)}, \bar{q}) $ and $ \check{M}_{+-}^{(1)} (\bar{z}_{\bar{q}}^{(0)}) $ have been replaced by their dominant order (\ref{eq:domM++}) and (\ref{eq:domM+-M--}) at large $ \bar{q} $, as it was also necessary to do for $ \bar{Z}_-^{(0)} $ using (\ref{eq:domZ-}), and as indicated by the exponent \og asymp \fg. The wave number dependence of result (\ref{eq:simplif}) unfortunately remains rather obscure, since $ Y_0 $ in (\ref{eq:Y0}) has no simple dependence on $ \bar{q} $. \footnote{On the other hand, we know how to express $ \bar{q} $ as a function of $ Y_0 $, and (\ref{eq:simplif}) has a complicated explicit form in terms of $ Y_0 $.} 
To progress, let's make the clever change of variable 
\be
\label{eq:defkappa}
\hat{q}\equiv\frac{\pi\kappa}{|\ln\varepsilon\,|^{2/3}}\quad\mbox{with}\quad  \varepsilon\equiv \check{\Delta}^{2/3}
\ee
then make $ \varepsilon $ tend to zero at fixed $ \kappa $ in (\ref{eq:simplif}). The dominant order in $ | \ln \varepsilon | $ is simple to calculate, \footnote{Let's also give $ \im \check{z}_q \stackrel{\scriptsize \kappa \, \mbox{fixed}}{\underset{\varepsilon \to 0}{\sim}}-\varepsilon \, | \ln \varepsilon \, |^{1/3} \kappa $, taken from (\ref{eq:trinome}) and in agreement with (\ref{eq:asympt}, \ref{eq:Y0}) since at this order $ \im \bar{z}_q^{\rm trin} $ is dominated by $ \im \bar{z}_{\bar{q}}^{(0)} $. This means that to leading order, $ \hat{z}_{\hat{q}} $ is pure imaginary of order $ \varepsilon^{-1} $. The relative error in these expressions, including (\ref{eq:equivrerznouv}), is $O((\ln | \ln \varepsilon \, |)^\alpha/| \ln \varepsilon \, |) $, where the value of the exponent $ \alpha $ does not matter.} 
\be
\label{eq:equivrerznouv}
\re\hat{z}^{(0)}_{\hat{q}} \stackrel{\scriptsize \kappa\,\mbox{fixed}}{\underset{\varepsilon\to 0}{\sim}}
\left\{
\begin{array}{ll}
\displaystyle\frac{2\pi}{|\ln\varepsilon\,|^{1/3}}\,\frac{1}{(4\kappa^2-\kappa^8)^{1/2}} & \mbox{if}\quad\kappa < 2^{1/3} \\
\\
\displaystyle|\ln\varepsilon\,|^{2/3}\frac{(\kappa^6-4)^{1/2}}{2\kappa} &\mbox{if}\quad\kappa>2^{1/3}
\end{array}
\right.
\ee
and reaches its minimum at $ \kappa = 1 $. After returning to the original dimensionless variables (\ref{eq:varadim}), it leads to the very beautiful equivalents 
\be
\label{eq:asympqmin_et_rezqmin}
\boxed{
\check{q}_{\rm min} \underset{\varepsilon\to 0}{\sim} \frac{\pi\varepsilon}{|\ln\varepsilon\,|^{2/3}} \quad\mbox{and}\quad
\re\check{z}_{q_{\rm min}} \underset{\varepsilon\to 0}{\sim} \frac{2\pi\varepsilon^2}{3^{1/2}|\ln\varepsilon\,|^{1/3}}
}
\ee
shown in solid lines in figure \ref{fig:qmin}, and which are already quite close to the exact values (symbols). We pushed the computation to the sub-sub-leading order in (\ref{eq:simplif}), the small expansion parameter being $ 1/| \ln \varepsilon \, | $: \footnote{Let's set $ x = | \ln \varepsilon \, |^{-1/3} \to 0 $ and $ \eta = \ln (\kappa/x) $. From the expansion $ \pi Y_0 = \frac{1}{2x^3}+\eta+2 \eta x^3+O (\eta^2 x^6) $ we get for $ \kappa <2^{1/3} $: $ \re \hat{z}^{(0)}_{\hat{q}} = \frac{2 \pi x}{\kappa (4-\kappa^6)^{1/2}} \left [1+\frac{x^3}{4-\kappa^6} (b_0+b_6 \kappa^6)+\frac{x^6}{(4-\kappa^6)^2} (c_0+c_6 \kappa^6+c_{12} \kappa^{12})+O (x^9 \eta^3) \right] $ with $ b_0 = 8 (1-\eta) $, $ b_6 = 1+\ln 2+3 \eta $, $ c_0 = 64 (1-3 \eta+\eta^2) $, $ c_6 = 34-2 \pi^2+4 \ln 2-6 \ln^22+4 (25+7 \ln 2) \eta-46 \eta^2 $ and $ c_{12} = 1+8 \ln 2+3 \ln^22-4 (1+\ln 2) \eta+9 \eta^2 $.} 
\bea
\label{eq:padeqmin}
\check{q}_{\rm min} &\underset{\varepsilon\to 0}{=}& \frac{\pi\varepsilon}{|\ln\varepsilon\,|^{2/3}}
\left[\frac{c_1+(c_1^2-c_2)|\ln\varepsilon\,|^{-1}}{c_1-c_2|\ln\varepsilon\,|^{-1}}+O(w^\alpha|\ln\varepsilon\,|^{-3})\right] \\
\label{eq:paderezqmin}
\re\check{z}_{q_{\rm min}} &\underset{\varepsilon\to 0}{=}& \frac{2\pi\varepsilon^2}{3^{1/2}|\ln\varepsilon\,|^{1/3}}
\left[\frac{1}{1-d_1|\ln\varepsilon\,|^{-1}+(d_1^2-d_2)|\ln\varepsilon\,|^{-2}}+O(w^\alpha|\ln\varepsilon\,|^{-3})\right]
\eea
where we introduced $ w = \ln (2 \, | \ln \varepsilon \, |^{1/3}) $, coefficients $ c_n $ and $ d_n $ functions of $ w $, 
\be
\begin{array}{lcl}
c_1=-\frac{19}{24} -\frac{1}{3}w &\quad & c_2=\frac{959}{1152} +\frac{5\pi^2}{18}+2\ln 2 +4\ln^2 2+\left(\frac{5}{12}-4\ln 2\right)w+\frac{8}{9} w^2 \\
d_1=3+2\ln 2-\frac{5}{3}w &\quad & d_2=\frac{1223}{144}-\frac{2\pi^2}{9}+12\ln 2 -\left(\frac{115}{9}+\frac{10}{3}\ln 2\right)w+\frac{23}{9}w^2
\end{array}
\ee
and we leave the exponent $ \alpha $ undetermined in the error terms. Of course, these coefficients are those of the expansion in powers of $ | \ln \varepsilon \, |^{-1} $ of the expressions in square brackets in (\ref{eq:padeqmin}, \ref{eq:paderezqmin}), but we preferred to write the result in the Pad\'e form (1,1) or (0,2) because this greatly improves the speed of convergence with $ | \ln \varepsilon \, |^{-1} $: dashed lines in figure \ref{fig:qmin} already agree very well with numerical results for modest values of $ \mu/\Delta $. 

\begin{figure} [tb] 
\centerline{\includegraphics [width = 0.3 \textwidth, clip =]{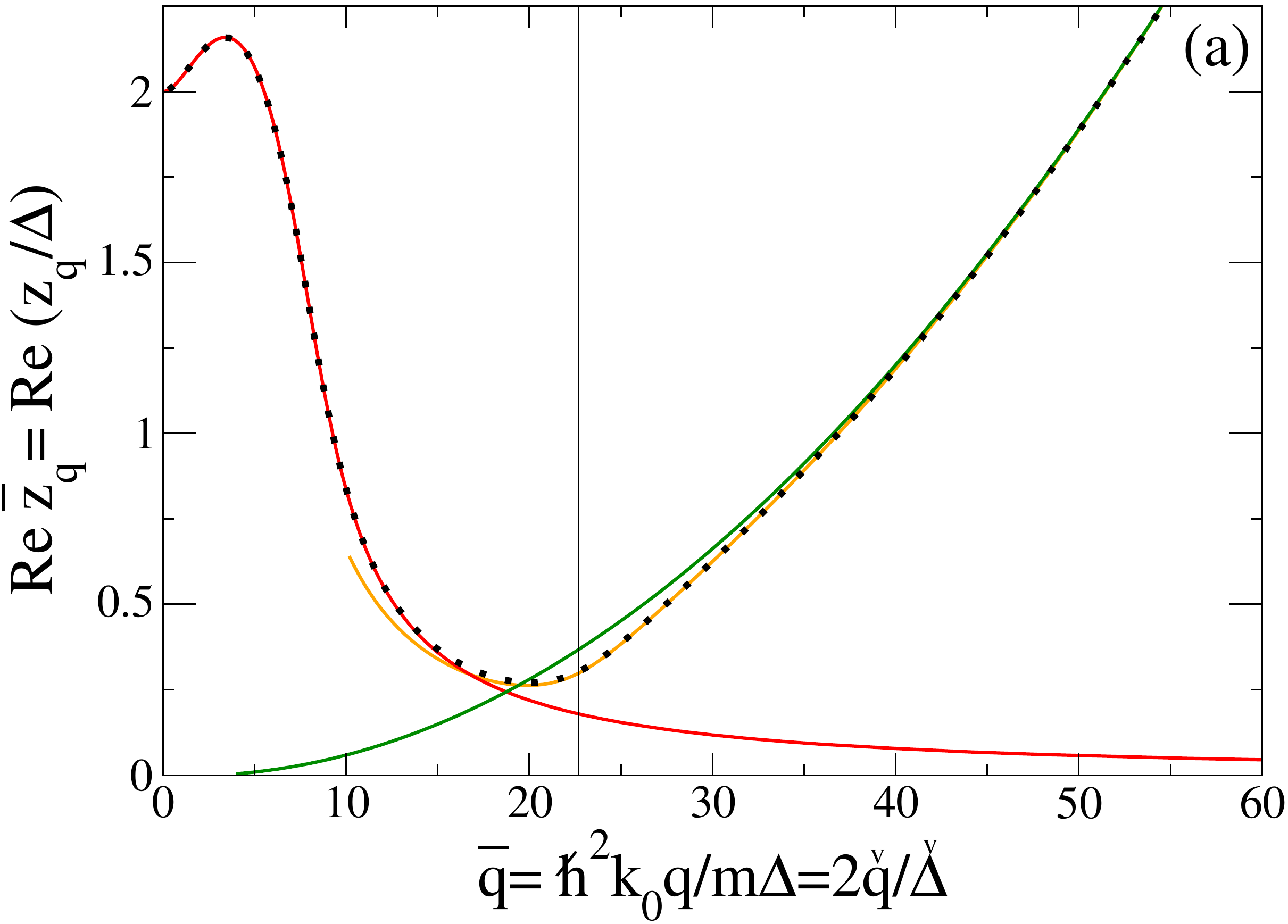} \hspace{5mm} \includegraphics [width = 0.3 \textwidth, clip =]{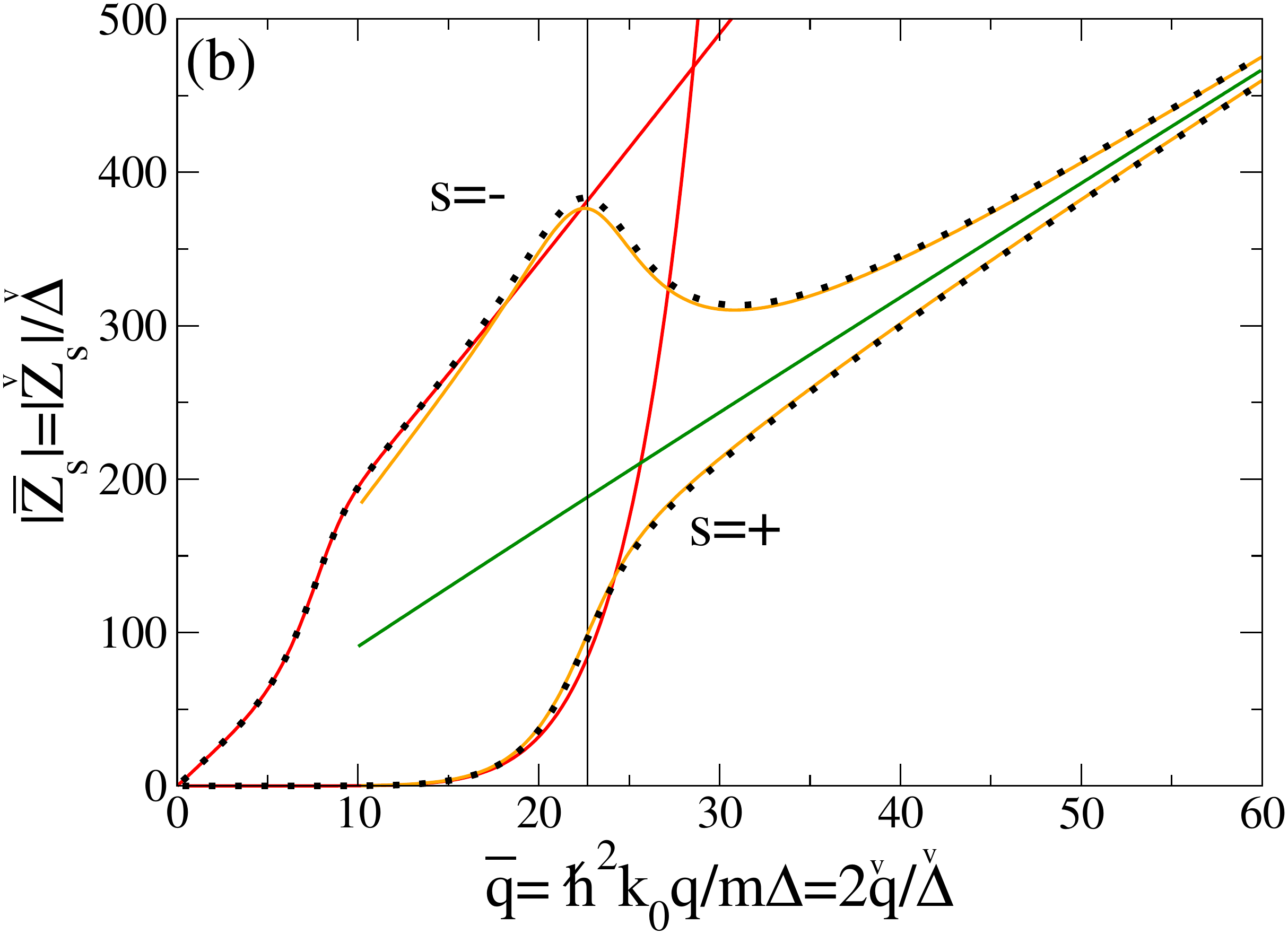} \hspace{5mm} \includegraphics [width = 0.3 \textwidth, clip =]{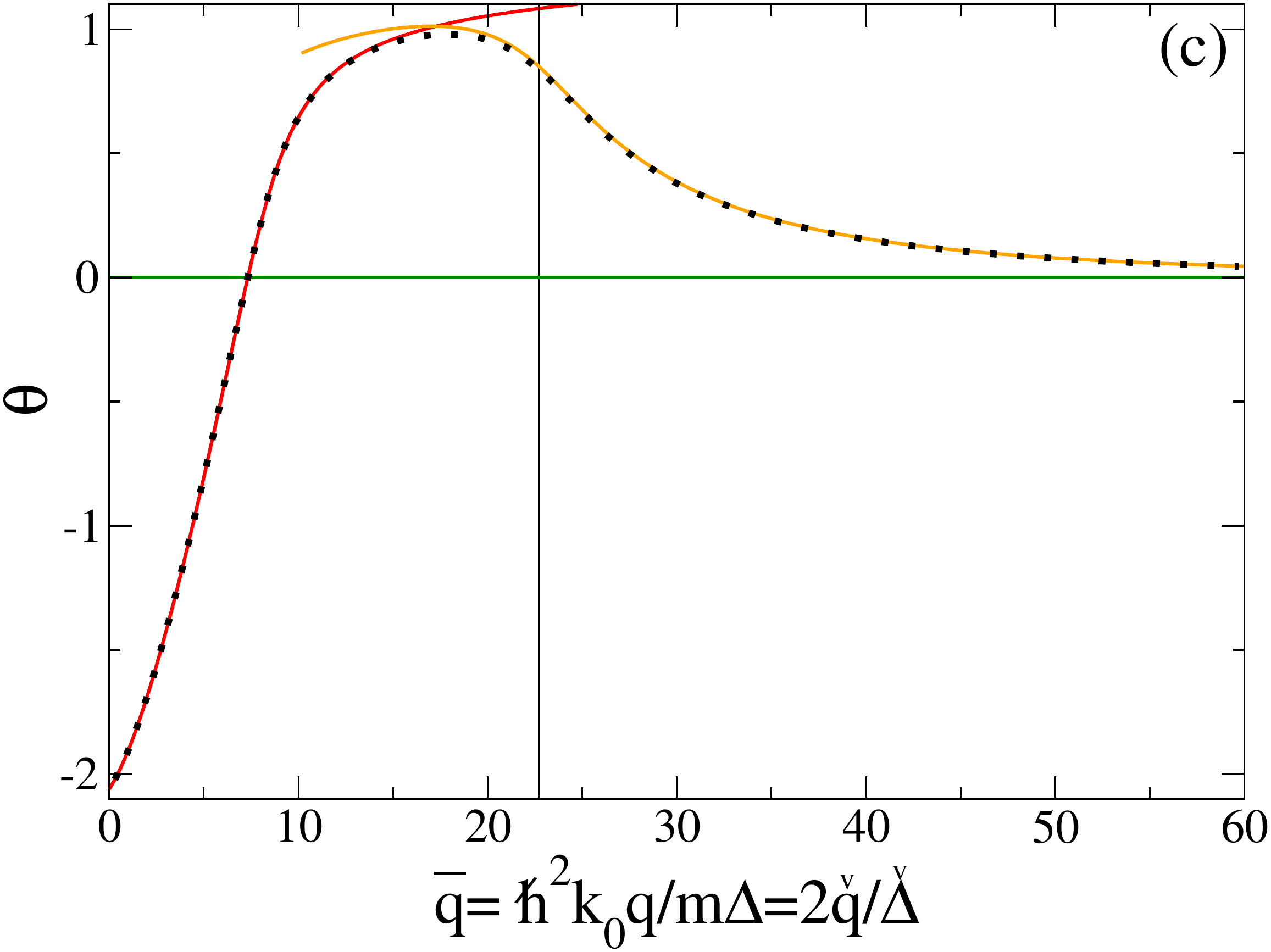}} 
\caption{For a gas of fermions with zero-range interaction in the limit $ \check{\Delta} = \Delta/\mu \to 0 $, illustration of the unexpected wave number scale $ \check{q} \approx \check{\Delta}^{2/3} $ and its connection with the scales $ \check{q} \approx \check{\Delta} $ and $ \check{q} \approx \check{\Delta}^0 $ on (a) the real part of the dispersion relation of the continuum mode of type II, (b) the spectral weights of the mode in the phase channel $+$ and modulus channel $-$ of the deviations of the order parameter from equilibrium and (c) the relative phase $ \theta $ between the two channels maximizing the mode spectral weight according to equation (\ref{eq:voiesoptimales}). 
Big black dotted line: numerical solution of equation (\ref{eq:surzqgen}) and its insertion in the definition of $ Z_\pm $ below (\ref{eq:residmat}) and in (\ref{eq:zopt_thetagen}). 
Solid red line: dominant order in $ \check{\Delta} $ at the scale $ \check{q} \approx \check{\Delta} $ ($ \check{\Delta} \to 0 $ at fixed $ \bar{q} = 2 \check{q}/\check{\Delta} $), namely the numerical solution of equation (\ref{eq:defzqbar0}) and its insertion into expressions (\ref{eq:defzb-zero}) and (\ref{eq:lcfz+theta}). Solid green line: dominant order in $ \check{\Delta} $ at the scale $ \check{q} \approx \check{\Delta}^0 $ ($ \check{\Delta} \to 0 $ at fixed $ \check{q} = q/k_0 $), namely expressions (\ref{eq:defzqch0}) and (\ref{eq:zpzm00ettheta00}). Solid orange line: dominant order in $ \check{\Delta} $ at the scale $ \check{q} \approx \check{\Delta}^{2/3} $ ($ \check{\Delta} \to 0 $ at $ \kappa $ fixed in the change of variable (\ref{eq:defkappa})), namely expressions (\ref{eq:simplif}) and (\ref{eq:zpmtrinthetatrin}) in which we wrote the coefficients to leading order in $ \check{\Delta} $ as in equations (\ref{eq:domZ-}), (\ref{eq:domM+-M--}) and (\ref{eq:domM++}). 
Thin vertical line: wave number at the center of the scale $ \check{q} \approx \check{\Delta}^{2/3} $, i.e.\ value of $ \bar{q} $ corresponding to $ \kappa = 1 $. 
We took $ \check{\Delta} = 1/$ 1000.} 
\label{fig:surprise} 
\end{figure}

Finally, it must be verified that, in the limit $ \check{\Delta} \to 0 $, no wavelength scale of the continuum collective branch has been forgotten between the new $ \check{q} \approx \check{\Delta}^{2/3} $ and the old $ \check{q} \approx \check{\Delta}^0 $. \footnote{We have already seen how (\ref{eq:trinome}) is connected to the scale $ \check{q} \approx \check{\Delta} $. Indeed, if $ \check{q} \approx \check{\Delta} $, we have $ \check{q} \ll \check{q}_{\rm min} \approx \check{\Delta}^{2/3} $. However, as it is said below (\ref{eq:remontee}), equation (\ref{eq:trinome}) for $ \check{q} \ll \check{q}_{\rm min} $ is essentially reduced to (\ref{eq:zqordredelta2}), therefore gives the right behavior of $ \bar{z}_q $ for $ \check{q} \approx \check{\Delta} $.} Analytically, we get from (\ref{eq:remontee}), limited to the first term in its right-hand side, that $ \re \hat{z}^{(0)}_{\hat{q}} \sim \check{\zeta}_1 \hat{q}^2 $ at large wave numbers compared to the new scale, $ \check{q} \gg \check{\Delta}^{2/3} $ in other words $ \hat{q} \gg 1 $, where $ \check{\zeta}_1 $ is exactly the same coefficient as in expansion (\ref{eq:devfqdtv0}) of $ \re \check{z}^{(0)}_{\check{q}} $ at low wave numbers compared to the old scale, $ \check{q} \ll \check{\Delta}^0 $: the connection is perfect. Numerically, we produced the synthetic figure \ref{fig:surprise}, fairly mathematical given the very small value of $ \check{\Delta} $ chosen, but which shows how the prediction (\ref{eq:trinome}) on $ \bar{z}_q $ and its derived predictions (\ref{eq:zpmtrinthetatrin}) on the residues and the optimal phase-modulus relative phase interpolate beautifully between the predictions at the scale $ \check{q} \approx \check{\Delta} $ in section \ref{subsubsec:q_approx_d} and those at the scale $ \check{q} \approx \check{\Delta}^0 $ in section \ref{subsubsec:q_approx_d0}, all agreeing very closely to the numerical results in their validity regime. 

\subsection{The continuum collective branch for $ \Delta/\mu \gg 1 $} 
\label{subsec:grand_d} 

In the vicinity of the chemical potential zero, on the positive side, we find that the continuum collective branch (for an analytic continuation of type II) presents two scales of variation with wave number, that $ \check{q} \approx \check{\Delta}^0 $ fixed by the width of the branch existence interval (section \ref{subsubsec:grand_d_qapproxd0}) and that $ \check{q} \approx \check{\Delta}^{-1/2} $ required by the quadratic approximation (\ref{eq:quadzq}) of the branch (section \ref{subsubsec:grand_d_qapproxd12}). These two regimes are perfectly connected as we will see, there is no other wave number scale to consider. 

\subsubsection{At the wave number scale $ \check{q} \approx \check{\Delta}^0 $} 
\label{subsubsec:grand_d_qapproxd0} 

We make here $ \check{\Delta} $ tend $+\infty $ with fixed reduced wave number $ \check{q} $, by postulating a finite-limit difference between the branch and the edge of the broken pair continuum: 
\be
\check{z}_q-2\check{\Delta}\equiv \check{Z}_{\check{q}} \stackrel{\check{q}\,\scriptsize\mbox{fixed}}{\underset{\check{\Delta}\to+\infty}{\approx}} \check{\Delta}^0
\ee
as suggested by the numerical results in figure \ref{fig:resnum} and the quadratic approximation (\ref{eq:quadzq}, \ref{eq:zetapetitgranddelta}) written in the limit (\ref{eq:limvalapproxquad}) of its domain of validity. To obtain an equivalent of the matrix elements of $ \check{M} $, we return to their integral formulation (\ref{eq:defM}, \ref{eq:defSigma}) in the space of relative wave vectors $ \kk $ of the broken pair, with the integration domain $ \mathcal{D} $ replaced by $ \mathbb{R}^3 $ of course. We can identify {\sl a priori} three natural energy scales $ \epsilon_{\kk \pm \qq/2} \approx \epsilon_{\kk} $ in these integrals, so three wave number scales $ \check{k} $: 
\be
\label{eq:3ech}
\mbox{(a)}~: \check{\epsilon}_k-\check{\Delta} \approx \check{\Delta}^{-1} \Longleftrightarrow \check{k}\approx \check{\Delta}^0 \quad;\quad
\mbox{(b)}~: \check{\epsilon}_k-\check{\Delta} \approx \check{\Delta}^0 \Longleftrightarrow \check{k}\approx \check{\Delta}^{1/4} \quad;\quad
\mbox{(c)}~: \check{\epsilon}_k-\check{\Delta} \approx \check{\Delta} \Longleftrightarrow \check{k}\approx \check{\Delta}^{1/2}
\ee
and three corresponding changes of variable 
\be
\label{eq:cvabc}
\check{k}=\check{\Delta}^0\check{K}_{\rm a}=\check{\Delta}^{1/4}\check{K}_{\rm b}=\check{\Delta}^{1/2}\check{K}_{\rm c}
\ee
To obtain the dominant contribution to the integrals of the first (a), the second (b) or the third scale (c), you must make $ \check{\Delta} $ tend to $+\infty $ for fixed values of $ \check{K}_{\rm a} $, $ \check{K}_{\rm b} $ or $ \check{K}_{\rm c} $ in the integrand. In $ \check{M}_{ss} $, scale (b) dominates and we fix $ \check{K}_{\rm b} $; in the phase-modulus coupling, scale (c) wins and we fix $ \check{K}_{\rm c} $. We thus obtain the equivalents 
\bea
\check{M}_{++}(\check{z},\check{q}) &\stackrel{\check{Z}=\check{z}-2\check{\Delta}\scriptsize\,\mbox{fixed}}{\underset{\check{\Delta}\to +\infty}{\sim}}& \check{\Delta}^{3/4} \int_0^{+\infty} \dd\check{K}_{\rm b} \frac{(2\pi)^{-2}\check{K}_{\rm b}^2}{\check{Z}-\check{K}_{\rm b}^4}=-\frac{(-\check{Z})^{-1/4}\check{\Delta}^{3/4}}{8\pi\sqrt{2}} \\
\check{M}_{--}(\check{z},\check{q})&\stackrel{\check{Z}=\check{z}-2\check{\Delta}\scriptsize\,\mbox{fixed}}{\underset{\check{\Delta}\to +\infty}{\sim}}&\check{\Delta}^{-1/4} \int_0^{+\infty}\dd\check{K}_{\rm b} \frac{(2\pi)^{-2}\check{Z}\check{K}_{\rm b}^2}{\check{Z}-\check{K}_{\rm b}^4}
=\frac{(-\check{Z})^{3/4}\check{\Delta}^{-1/4}}{8\pi\sqrt{2}} \\
\check{M}_{+-}(\check{z},\check{q})&\stackrel{\check{Z}=\check{z}-2\check{\Delta}\scriptsize\,\mbox{fixed}}{\underset{\check{\Delta}\to +\infty}{\sim}}& \check{\Delta}^{1/2} \int_0^{+\infty} \dd\check{K}_{\rm c} \frac{-(2\pi)^{-2}}{(1+\check{K}_{\rm c}^4)^{1/2}}
=-\frac{[\Gamma(5/4)]^2\check{\Delta}^{1/2}}{\pi^{5/2}}
\eea
where the explicit value of the integrals is given for $ \im \check{Z} <0 $, in a form adapted to the case to come, $ \im \check{Z} \to 0^-$ at fixed $ \re \check{Z} <0 $. In order to use the procedure of Nozi\`eres (\ref{eq:Noz}), we expand directly the nonzero spectral densities of type II in table \ref{tab:rhogen}, 
\be
\check{\rho}^{[\mathrm{II}]}_{++}(\check{z},\check{q})\stackrel{\check{Z}=\check{z}-2\check{\Delta}\scriptsize\,\mbox{fixed}}{\underset{\check{\Delta}\to +\infty}{\sim}}\frac{\pi^2\check{\Delta}}{2\check{q}}\left[1+\frac{\check{Z}}{4\check{\Delta}}+O\left(\frac{1}{\check{\Delta}^2}\right)\right]
\quad\mbox{and}\quad \check{\rho}^{[\mathrm{II}]}_{--}(\check{z},\check{q})\stackrel{\check{Z}=\check{z}-2\check{\Delta}\scriptsize\,\mbox{fixed}}{\underset{\check{\Delta}\to +\infty}{\sim}} \frac{\pi^2\check{Z}}{4\check{q}}\left[1-\frac{\check{Z}}{8\check{\Delta}}+O\left(\frac{1}{\check{\Delta}^2}\right)\right]
\ee
We note then that the matrix elements $ \check{M}_{ss} $ are dominated by the Nozi\`eres term performing the analytic continuation, larger by a factor $ \check{\Delta}^{1/4} $. This is disturbing and suggests a continuum mode very far from the experimentally accessible energy axis $ \im z = 0^+$ therefore without much physical sense. The solution to leading order of the eigenenergy equation gives 

\be
\label{eq:Zqdom}
\check{Z}_{\check{q}}\stackrel{\check{q}\,\scriptsize\mbox{fixed}}{\underset{\check{\Delta}\to+\infty}{\sim}} -\frac{128}{\pi^5}
[\Gamma(5/4)]^4\check{q}^2
\ee
The real part of the energy of the mode is thus effectively outside the interval $ [2 \Delta, \epsilon_2 (q)] $ of analytic continuation, which is also the interval of observability in the response functions of the system, within the meaning of section \ref{subsec:carac} (the energy $ \epsilon_2 (q) $ is given here by (\ref{eq:eps2gdfd})). It should be remembered, the mode is not really in the physical gap $ [0,2 \Delta] $ even if its real part is mathematically there; it is separated from it by the end of the original branch cut folded over $]-\infty, 2 \Delta] $ by the procedure (\ref{eq:Noz}) \cite{PRL2019}.

Remarkably, the negative real result (\ref{eq:Zqdom}) coincides to leading order with the quadratic approximation in $ \check{q} $ of the branch, see equations (\ref{eq:quadzq}) and (\ref{eq:zetapetitgranddelta}): the quadratic approximation is valid everywhere, and the infinitesimal wave number scale for $ \check{q} $ predicted by its validity condition (\ref{eq:limvalapproxquad}) does not appear. The situation changes on the next orders, the two expected wavelength scales appear and the condition (\ref{eq:limvalapproxquad}) takes on its full meaning. The rather technical calculations are given in \ref{subapp:gdd0}. They are now multi-scales: we must include in the same matrix element $ \check{M}_{ss'} $ scales (b) and (c), as well as the intermediate scale, denoted by (bc), deduced by geometric mean and requiring the change of variable $ \check{k} = \check{\Delta}^{3/8} \check{K}_{\rm bc} $. On the other hand, it is not necessary to treat apart scale (a) at the maximal order considered here (relative error $ O (1/\check{\Delta}) $ on $ \check{M}_{ss'} $ and {\sl in fine} on $ \check{Z}_{\check{q}} $). Here is the result: 
\be
\label{eq:Zqdevloin}
\boxed{
\check{z}_q-2\check{\Delta}\stackrel{\check{q}\scriptsize\,\mbox{fixed}}{\underset{\check{\Delta}\to+\infty}{=}} a_0+a_1 \check{\Delta}^{-1/4}+a_2\check{\Delta}^{-2/4}+a_3\check{\Delta}^{-3/4}+O(\check{\Delta}^{-1})
}
\ee
with 
\bea
a_0 &=& -\frac{128}{\pi^5}[\Gamma(5/4)]^4\check{q}^2 \\
a_1 &=& 16\cdot 2^{1/4}(4-3\ii\sqrt{2})\check{q}^{5/2}[\Gamma(5/4)]^3/\pi^{19/4} \\
a_2 &=& [9\sqrt{2}\pi+16\ii(18+9\pi-\pi^3)]\check{q}^3[\Gamma(5/4)]^2/(3\pi^{11/2}) \\
a_3 &=& \left\{-96(2\sqrt{2}+3\ii)\pi^3+[2\sqrt{2}(112\pi^3-2256-303\pi)+\ii(312\pi^3-5760-585\pi)]\check{q}^2\right\}\frac{\check{q}^{3/2}\Gamma(5/4)}{48\cdot 2^{3/4}\pi^{21/4}}
\eea
If we tried to use (\ref{eq:Zqdevloin}) for $ \check{q} \to 0 $, we would find the real dominant term as we said, but not the pure imaginary sub-leading term of the coefficient of the quadratic start in (\ref{eq:zetapetitgranddelta}), since $ a_2 $ above tends to zero cubically in wave number. On the imaginary part of the dispersion relation, which varies to leading order as $ \check{q}^{5/2} $, the quadratic-start range is reduced to zero at large $ \check{\Delta} $, as predicted by (\ref{eq:limvalapproxquad}), and we cannot exchange the limits $ \check{q} \to 0 $ and $ \check{\Delta} \to+\infty $. To be complete, let us give an expansion of the residues in the phase and modulus channels, and of the relative phase between these two channels corresponding to the maximum global residue: 
\be
|\check{Z}_-|\stackrel{\check{q}\,\scriptsize\mbox{fixed}}{\underset{\check{\Delta}\to+\infty}{=}}16\check{q} \left[1-\frac{(2\pi)^{1/4}\check{q}^{1/2}}{2\Gamma(1/4)\check{\Delta}^{1/4}}+O(\check{\Delta}^{-1/2})\right], \quad |\check{Z}_+|\stackrel{\check{q}\,\scriptsize\mbox{fixed}}{\underset{\check{\Delta}\to+\infty}{\sim}} \frac{4[\Gamma(1/4)]^4\check{q}^3}{\pi^5\check{\Delta}} \quad \mbox{and}\quad \theta+\frac{\pi}{2} \stackrel{\check{q}\,\scriptsize\mbox{fixed}}{\underset{\check{\Delta}\to+\infty}{\sim}} -\frac{(\pi/2)^{1/4}\check{q}^{1/2}}{\Gamma(1/4)\check{\Delta}^{1/4}}
\ee

\subsubsection{At the wave number scale $ \check{q} \approx \check{\Delta}^{-1/2} $} 
\label{subsubsec:grand_d_qapproxd12} 

To study the continuum branch of type II with reduced wave numbers $ \check{q} \approx \check{\Delta}^{-1/2} $ near the chemical potential zero, we set $ \check{q} = \check{Q} \check{\Delta}^{-1/2} $ and let $ \check{\Delta} $ tend to infinity at fixed $ \check{Q} $. According to (\ref{eq:limvalapproxquad}), the wave number scale considered is that of the validity limit of the quadratic approximation (\ref{eq:quadzq}), which we can therefore use to find the energy scaling law. According to (\ref{eq:zetapetitgranddelta}), the coefficient of $ \check{q}^2 $ in (\ref{eq:quadzq}) is of the order $ \check{\Delta}^0 $, so that $ \check{z}_q-2 \check{\Delta} \approx \check{\Delta}^{-1} $ and you have to set 
\be
\label{eq:defzc}
\check{z}-2\check{\Delta}\equiv \frac{\check{\zeta}}{\check{\Delta}}
\ee
It remains to take the limit $ \check{\Delta} \to+\infty $ in the matrix elements of $ \check{M} (\check{z}, \check{q}) $ at fixed $ \check{\zeta} $ and $ \check{Q} $. We proceed in the same way as in section \ref{subsubsec:grand_d_qapproxd0}, by distinguishing the three wave number scales (\ref{eq:3ech}) in the integral on the relative wave vector $ \kk $ of a broken pair. We find that the diagonal elements $ \check{M}_{ss} $ are dominated by wave number scale (a), while the phase-modulus coupling $ \check{M}_{+-} $ is dominated by scale (c): 
\bea
\label{eq:M++gdechdemi}
\check{M}_{++}(\check{z},\check{q})&\stackrel{\scriptsize\check{\zeta}\,\mbox{and}\,\check{Q}\,\mbox{fixed}}{\underset{\check{\Delta}\to+\infty}{\sim}}&
\check{\Delta}^{+1} \int_0^{+\infty}\dd\check{K}_{\rm a} \frac{(2\pi)^{-2}\check{K}_{\rm a}^2}{\check{\zeta}-(\check{K}_{\rm a}^2-1)^2} 
=\frac{-\check{\Delta}}{16\pi(-\check{\zeta})^{1/2}}\left[\left(1+\ii(-\check{\zeta})^{1/2}\right)^{1/2}+\left(1-\ii({-\check{\zeta}})^{1/2}\right)^{1/2}\right]
\\
\label{eq:M--gdechdemi}
\check{M}_{--}(\check{z},\check{q})&\stackrel{\scriptsize\check{\zeta}\,\mbox{and}\,\check{Q}\,\mbox{fixed}}{\underset{\check{\Delta}\to+\infty}{\sim}}&
\check{\Delta}^{-1}\int_0^{+\infty}\dd\check{K}_{\rm a} \frac{(2\pi)^{-2}\check{\zeta}\check{K}_{\rm a}^2}{\check{\zeta}-(\check{K}_{\rm a}^2-1)^2}
=\frac{(-\check{\zeta})^{1/2}}{16\pi\check{\Delta}} \left[\left(1+\ii(-\check{\zeta})^{1/2}\right)^{1/2}+\left(1-\ii(-\check{\zeta})^{1/2}\right)^{1/2}\right]
\\
\label{eq:M+-gdechdemi}
\check{M}_{+-}(\check{z},\check{q})&\stackrel{\scriptsize\check{\zeta}\,\mbox{and}\,\check{Q}\,\mbox{fixed}}{\underset{\check{\Delta}\to+\infty}{\sim}}&
\check{\Delta}^{1/2} \int_0^{+\infty}\dd\check{K}_{\rm c} \frac{-(2\pi)^{-2}}{(1+\check{K}_{\rm c}^4)^{1/2}}
=-\frac{[\Gamma(5/4)]^2}{\pi^{5/2}}\check{\Delta}^{1/2}
\eea
where the explicit form of the integrals is suited to $ \check{\zeta} $ close to a negative real number. The expansion of the spectral densities useful for the analytical continuation through the interval II is carried out directly from table \ref{tab:rhogen}: 
\be
\check{\rho}_{++}^{\mathrm{II}}(\check{z},\check{q})
\stackrel{\scriptsize\check{\zeta}\,\mbox{and}\,\check{Q}\,\mbox{fixed}}{\underset{\check{\Delta}\to+\infty}{\sim}}
\frac{\pi^2\check{\Delta}^{3/2}}{2\check{Q}}+O(\check{\Delta}^{-1/2})\quad\mbox{and}\quad
\check{\rho}_{--}^{\mathrm{II}}(\check{z},\check{q})
\stackrel{\scriptsize\check{\zeta}\,\mbox{and}\,\check{Q}\,\mbox{fixed}}{\underset{\check{\Delta}\to+\infty}{\sim}}
\frac{\pi^2\check{\zeta}\check{\Delta}^{-1/2}}{4\check{Q}}+O(\check{\Delta}^{-5/2})
\ee
As in section \ref{subsubsec:grand_d_qapproxd0}, the diagonal matrix elements $ \check{M}_{ss} $ are much smaller than the spectral densities, this time by a factor $ \approx \check{\Delta}^{1/2} $, and do not contribute to leading order to the analytically continued values $ \check{M}_{ss \downarrow} $. By solving the eigenenergy equation (\ref{eq:surzqgen}), we find the equivalent 
\be
\label{eq:zetaQequiv}
\check{\zeta}_Q\sim -\frac{128}{\pi^5} [\Gamma(5/4)]^4 \check{Q}^2\quad\mbox{that is}\quad \check{z}_q-2\check{\Delta}\stackrel{\check{Q}\scriptsize\,\mbox{fixed}}{\underset{\check{\Delta}\to+\infty}{\sim}} -\frac{128}{\pi^5} [\Gamma(5/4)]^4\check{q}^2
\ee
In first approximation, the dispersion relation is real and coincides with its quadratic approximation, even for $ \check{Q} \gg $ 1. To put an end to this paradox, we must calculate the first correction, of relative order $ \check{\Delta}^{-1/2} $. For the diagonal elements $ \check{M}_{ss \downarrow} $, equations (\ref{eq:M++gdechdemi}, \ref{eq:M--gdechdemi}) are sufficient. For non diagonal elements, in which the spectral density $ \rho_{+-}^{[\mathrm{II}]} $ is identically zero, it remains to determine the sub-leading order $ \check{\Delta}^0 $ of $ \check{M}_{+-} $. This time, we must add the contribution of scale (a) to that of scale (c). The scale (b) contributes indirectly, that is, only in the form of a cut-off $ \Lambda \check{\Delta}^{1/4} $ on the wave number $ \check{k} $, where $ \Lambda $ is a positive arbitrary constant. It is a high (ultraviolet) cut-off for scale (a), without which the integral on $ \check{K}_{\rm a} $ would diverge at infinity. It is also a low (infrared) cut-off for scale (c), without which the integral on $ \check{K}_{\rm c} $ of the sub-leading term would diverge in zero. In the limit $ \check{\Delta} \to+\infty $, scale (b) thus combines the two sub-leading pieces in a $ \Lambda $-independent way and we find the correction sought to (\ref{eq:M+-gdechdemi}), as detailed in \ref{subapp:gddmundemi}: 
\be
\label{eq:sdomm+-}
\check{M}_{+-}(\check{z},\check{q})+\frac{[\Gamma(5/4)]^2}{\pi^{5/2}}\check{\Delta}^{1/2} \stackrel{\scriptsize\check{\zeta}\,\mbox{and}\,\check{Q}\,\mbox{fixed}}{\underset{\check{\Delta}\to+\infty}{\to}}
(2\pi)^{-2} \int_0^{+\infty} \dd\check{K}\left[\frac{\check{K}^2(\check{K}^2-1)}{\check{\zeta}-(\check{K}^2-1)^2}+1\right]=
\frac{1}{16\ii\pi}\left[\left(1+\ii(-\check{\zeta})^{1/2}\right)^{1/2}-\left(1-\ii(-\check{\zeta})^{1/2}\right)^{1/2}\right]
\ee
We end up with the non-quadratic dispersion relation 
\be
\label{eq:zetaQdev}
\boxed{
\check{\zeta}_Q\equiv(\check{z}_q-2\check{\Delta})\check{\Delta}
\stackrel{\check{Q}\scriptsize\,\mbox{fixed}}{\underset{\check{\Delta}\to+\infty}{=}}-C\check{Q}^2
\left\{1+\check{\Delta}^{-1/2}\left[\frac{\ii\sqrt{2}}{\pi\sqrt{C}}(r-r^*)+\frac{3\ii/2}{\pi\sqrt{C}}(r+r^*)\right]+O(\check{\Delta}^{-1})\right\}
}
\ee
with the notations $ C = \frac{128}{\pi^5} [\Gamma (5/4)]^4 $ and $ r = (1+\ii C^{1/2} \check{Q})^{1/2} $. At low $ \check{Q} $, we can approximate $ r \simeq 1 $; we find the first two terms of expansion (\ref{eq:zetapetitgranddelta}) of the branch quadratic start coefficient, i.e.\ the dominant order for the real part and for the imaginary part. At large $ \check{Q} $, we can approximate $ r \simeq \exp (\ii \pi/4) C^{1/4} \check{Q}^{1/2}+\exp (-\ii \pi/4) C^{-1/4} \check{Q}^{-1/2}/2 $, which gives rise to contributions $ \check{Q}^2 $, $ \check{Q}^{5/2} $ and $ \check{Q}^{3/2} $; we then find the low $ \check{q} $ limit of expansion (\ref{eq:Zqdevloin}). \footnote{Let's replace $ \check{q} $ with $ \check{\Delta}^{-1/2} \check{Q} $ in (\ref{eq:Zqdevloin}) and expand to the relative order $ \check{\Delta}^{-1/2} $. Term $ a_0 $ returns the $ \check{Q}^2 $ contribution, term $ a_1 $ returns the $ \check{Q}^{5/2} $ contribution, and the $ \check{q}^{3/2} $ bit of term $ a_3 $ returns the $ \check{Q}^{3/2} $ contribution. Term $ a_2 $ and the $ \check{q}^{7/2} $ bit of $ a_3 $ contribute to the relative orders $ \check{\Delta}^{-1} $ and $ \check{\Delta}^{-3/2} $, out of range of (\ref{eq:zetaQdev}).} The connection is perfect, and there is, in the continuum branch of type II near zero chemical potential, no other wavelength scale than $ \check{q} \approx \check{\Delta}^{-1/2} $ and $ \check{q} \approx \check{\Delta}^0 $. To be complete, let us expand the residues in the phase and modulus channels, and the relative phase between these two channels leading to the maximum global residue: 
\be
|\check{Z}_-|\stackrel{\check{Q}\,\scriptsize\mbox{fixed}}{\underset{\check{\Delta}\to+\infty}{=}}\frac{16\check{Q}}{\check{\Delta}^{1/2}}\left[1-\frac{\sqrt{2}\check{Q}\re r}{2\pi\check{\Delta}^{1/2}\sqrt{1+C\check{Q}^2}}+O(\check{\Delta}^{-1})\right], \quad
|\check{Z}_+|\stackrel{\check{Q}\,\scriptsize\mbox{fixed}}{\underset{\check{\Delta}\to+\infty}{\sim}}\frac{8C\check{Q}^3}{\check{\Delta}^{5/2}}\quad\mbox{and}\quad \theta+\frac{\pi}{2}\stackrel{\check{Q}\,\scriptsize\mbox{fixed}}{\underset{\check{\Delta}\to+\infty}{\sim}}-\check{\Delta}^{-1/2} \frac{\re r}{\pi\sqrt{C}}
\ee

\subsection{Analytical extension for $ \mu> 0 $ through the windows $ \epsilon > \epsilon_2 (q) $ and case $ \mu <0 $} 
\label{subsec:autrefen_muneg} 

We briefly examine in this section the continuum collective excitation branch(es) obtained by analytic continuation of the eigenenergy equation through energy intervals other than $ [\epsilon_1 (q), \epsilon_2 (q)] $. Recall that the branch points $ \epsilon_i (q) $ of the equation and their domain of existence with wave number $ q $ were the subject of section \ref{subsubsec:lpdnae1e2e3}. 

\subsubsection{On the BCS side: $ \mu> 0 $} 
\label{subsubsec:autresmupos} 

Let's start with the case of a positive chemical potential. For wave numbers $ q <q_0 $, where $ q_0 $ is given by (\ref{eq:valq0}), the other possible extension intervals are $ [\epsilon_2 (q), \epsilon_3 (q)] $ and $ [\epsilon_3 (q),+\infty [$. In the procedure (\ref{eq:Noz}), it is then necessary to use respectively form III and form IV of the spectral densities $ \rho_{ss'} (\epsilon, \qq) $, within the meaning of table \ref{tab:rhogen}. 

The essential difference with the type II form is that the $ \rho_{ss'} (\epsilon, \qq) $ depend on the real energy $ \epsilon $ through the first two roots $ s_1 $ and $ s_2 $, or the second root $ s_2 $ of the cubic equation (\ref{eq:surs}) on $ s $, this equation being parametrically dependent on $ \epsilon $. It is now necessary to analytically continue the spectral densities therefore the roots $ s_i $ to complex values $ z $ of the energy in the fourth quadrant $ \re z> 0 $ and $ \im z <0 $ to which our study is limited. Using Cardan formulas, we find that the roots $ s_i $ have, as functions of $ z $, branch cuts where the Cardan discriminant is real negative, and a branch point where it vanishes. \footnote{This phenomenon is general for a polynomial $ P (X, z) $ whose coefficients depend analytically on $ z $. Consider an analytic mapping of a root $ s (z) $ along a path in the complex plane. Taking the derivative of the definitional equation $ P (s (z), z) = 0 $ with respect to $ z $, we find that $ \dd s/\dd z =-\partial_zP (s, z)/\partial_XP (s, z) $, which remains well defined as long as $ \partial_X P (s, z) \neq 0 $ therefore as long as $ s (z) $ is not a multiple root of $ P (X, z) $.} We get rid of the branch cuts by numerically performing a continuous mapping of the three roots $ s_i (z) $ of the cubic equation along a path $ \mathcal{C} $ connecting $ z $ to an origin point $ \epsilon_0 $ fixed in the analytic continuation interval $ [\epsilon_2 (q), \epsilon_3 (q)] $ (for example, its middle) or $ [\epsilon_3 (q),+\infty [$ (by example $ 2 \epsilon_3 (q) $). On the other hand, we cannot eliminate the branch points of $ s_i (z) $. Their positions in the complex plane (other than the expected values $ \pm 2 \Delta $) are given by the roots of the polynomial $ P_8 (X) $ of degree eight, see (\ref{eq:P8}), which is related to the discriminant of the cubic equation, as we have already pointed out in section \ref{subsubsec:exdesp}. Fortunately, it follows from footnote \ref{note:Lagrange} that $ P_8 (X) $ has within the fourth quadrant only one complex root $ \alpha $. It remains to be specified on which side of $ \alpha $ the continuous mapping of $ s_i (z) $ takes place: at fixed starting point $ \epsilon_0 $ and arrival point $ z $, go to one side $ \alpha $ or the other has the effect of exchanging the roots $ s_2 (z) $ and $ s_3 (z) $, as we have verified numerically, which changes the spectral densities. In the present study, we have chosen never to go under the point $ \alpha $, that is to say go around above and to the left if $ \re z <\re \alpha $ and $ \im z <\im \alpha $; the corresponding branch cut on the $ s_i (z) $, and therefore on the analytically continued spectral densities $ \rho_{ss'} (z, \qq) $, \footnote{The roots $ s_i (z) $ are then inserted in the arguments of the incomplete elliptic functions $ E $ and $ F $, see (\ref{eq:fmm}) and (\ref{eq:fpp}) for example. These elliptical functions have branch cuts described in footnote \ref{note:ldcie}, which can always be moved by continuous deformation, and irremovable branch points. Let us show by the absurd that the arguments of $ E $ and $ F $ never reach these branch points, during the journey of $ s_i (z) $ in the complex plane: if this were the case, we would have $ 1+(1-s^2) \sh^2 \Omega = 0 $, with $ \sh^2 \Omega = (z^2-4 \Delta^2)/(4 \Delta^2) $ and $ s $ solution of the equation deduced from (\ref{eq:surs}) by replacing $ \epsilon $ by $ z $; elementary manipulations in (\ref{eq:surs}) thus modified bring down the impossible condition to satisfy $ 1/\sh^2 \Omega = 0 $. To numerically calculate the functions $ E $ and $ F $, we return to their integral definition, see \S 8.111 of \cite{GR}, and if necessary move the branch cut of the square root in the integrand by continuous mapping. We can alternatively work directly on the functions (\ref{eq:fmm}, \ref{eq:fpp}), by writing for example $ f_{--} (\frac{\pi}{2}-\asin s_2) = \int_{s_2}^{1} \frac{\dd X}{\sqrt{1-X^2}} \frac{X^2 \sh^2 \Omega}{[1+(1-X^2) \sh^2 \Omega]^{3/2}} $ thanks to the change of variable $ X = \cos \alpha $.} is the vertical half-line starting from $ \alpha $ downwards. 

\begin{figure} [tbp] 
\centerline{\includegraphics [width = 0.3 \textwidth, clip =]{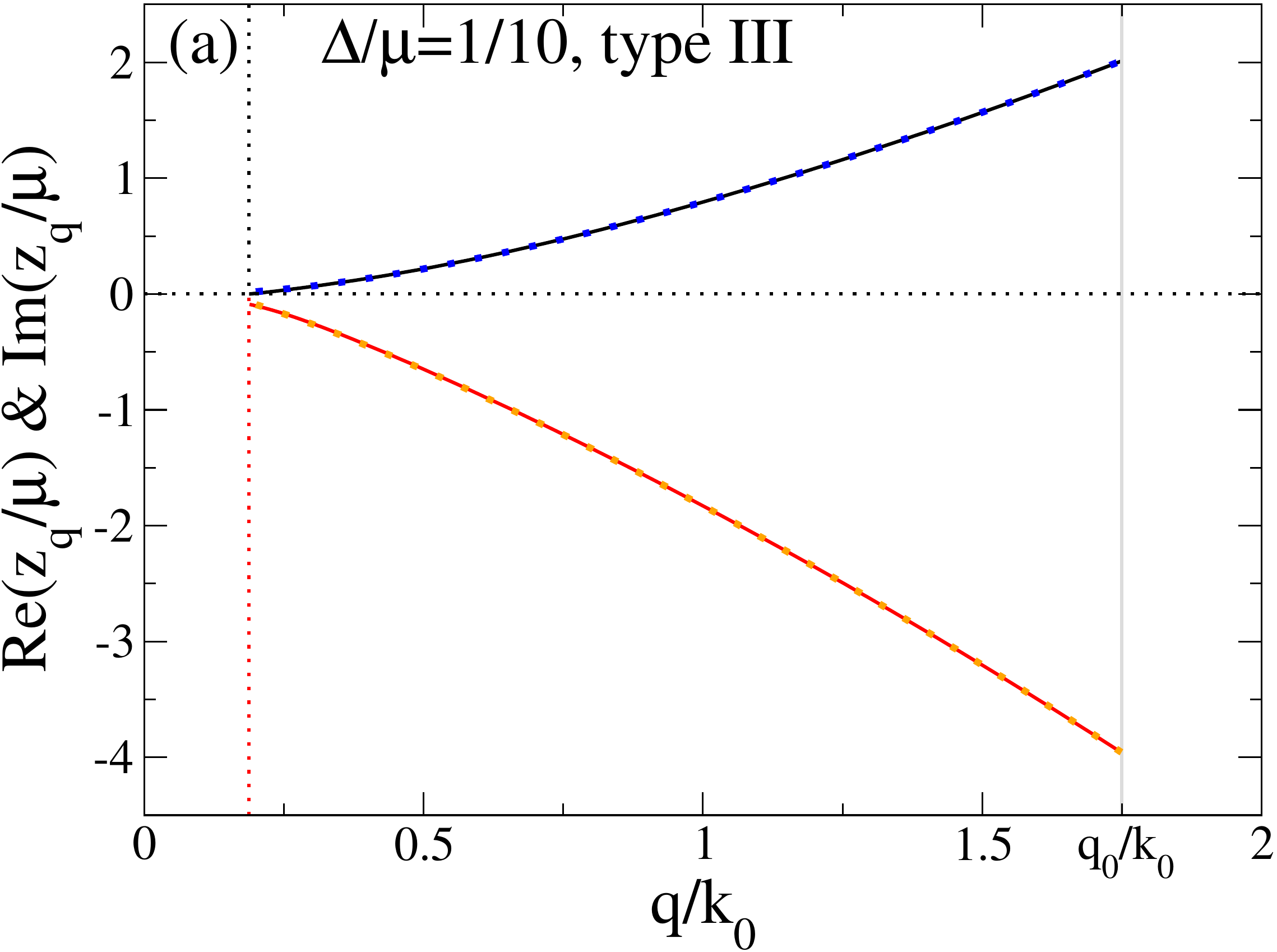} \hspace{5mm} \includegraphics [width = 0.3 \textwidth, clip =]{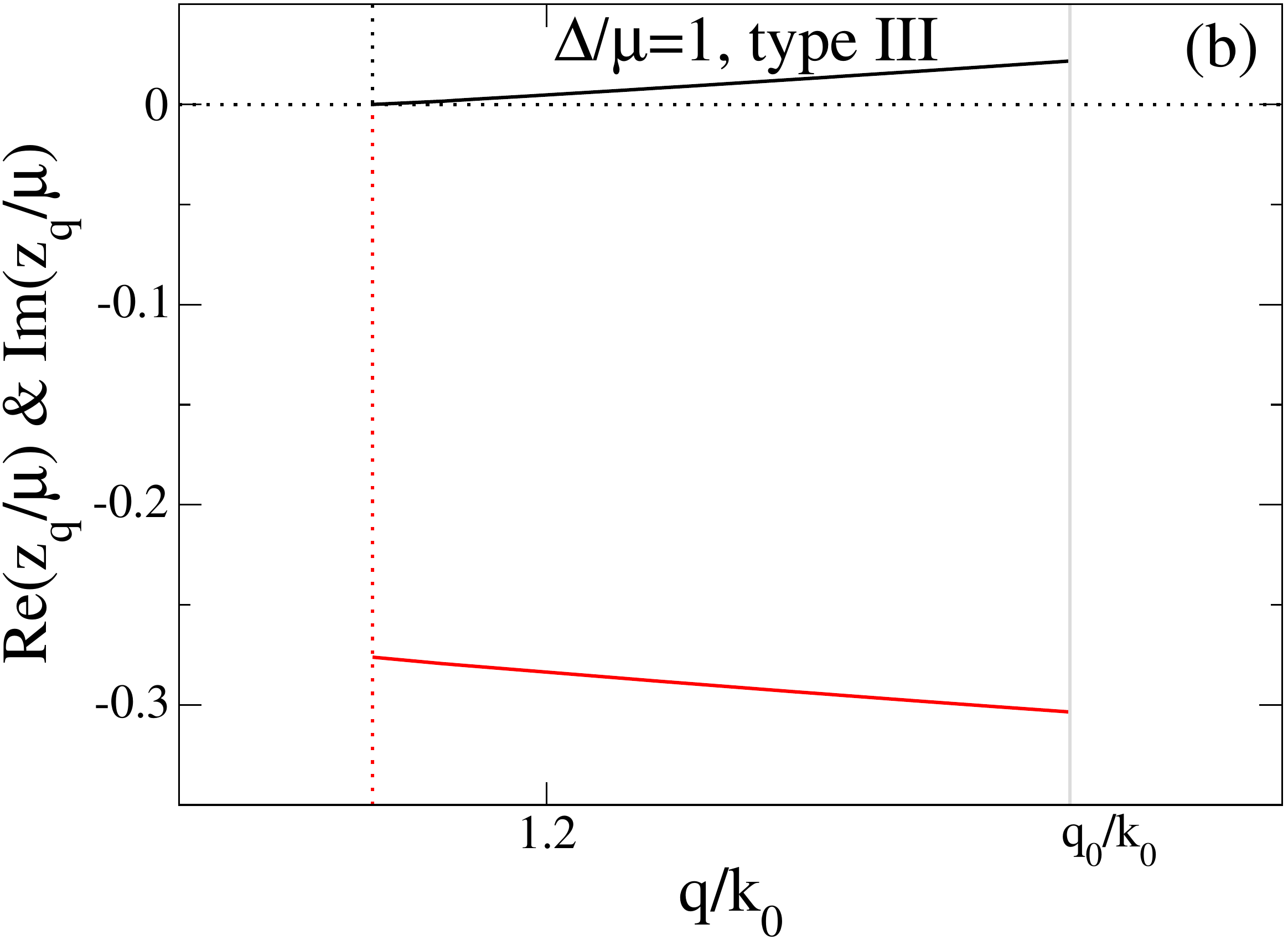} \hspace{5mm} \includegraphics [width = 0.3 \textwidth, clip =]{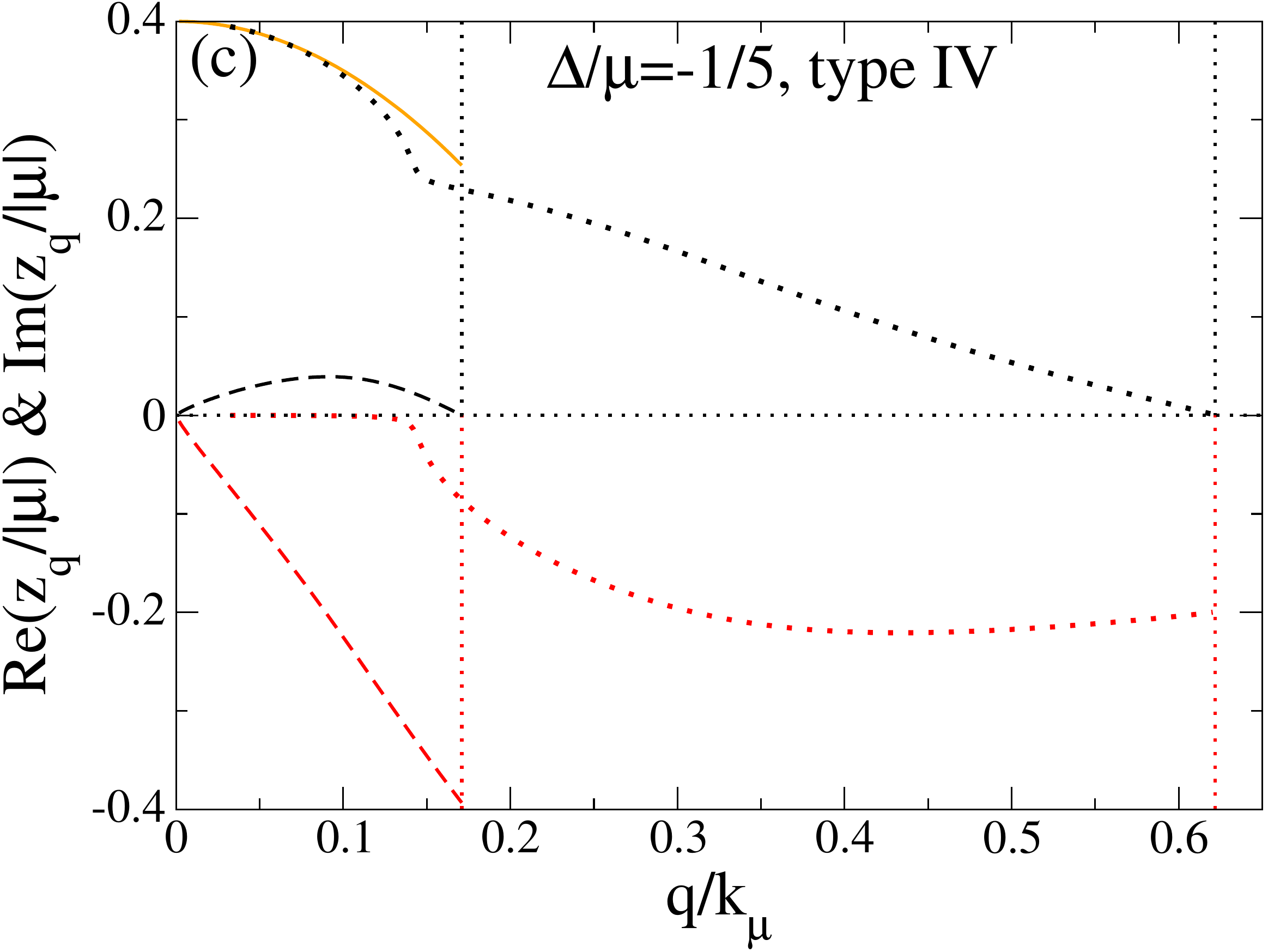}} 
\centerline{\includegraphics [width = 0.3 \textwidth, clip =]{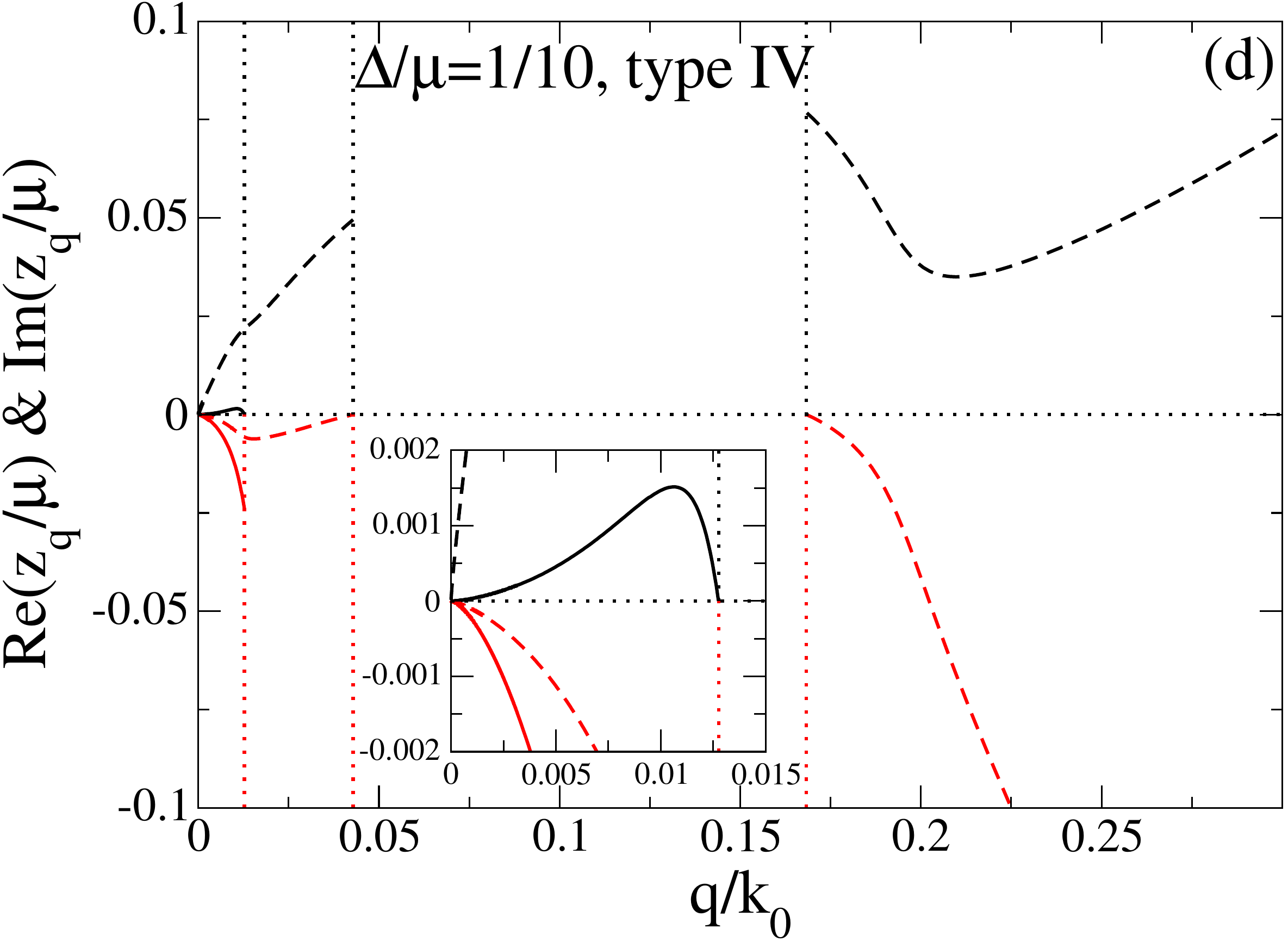} \hspace{5mm} \includegraphics [width = 0.3 \textwidth, clip =]{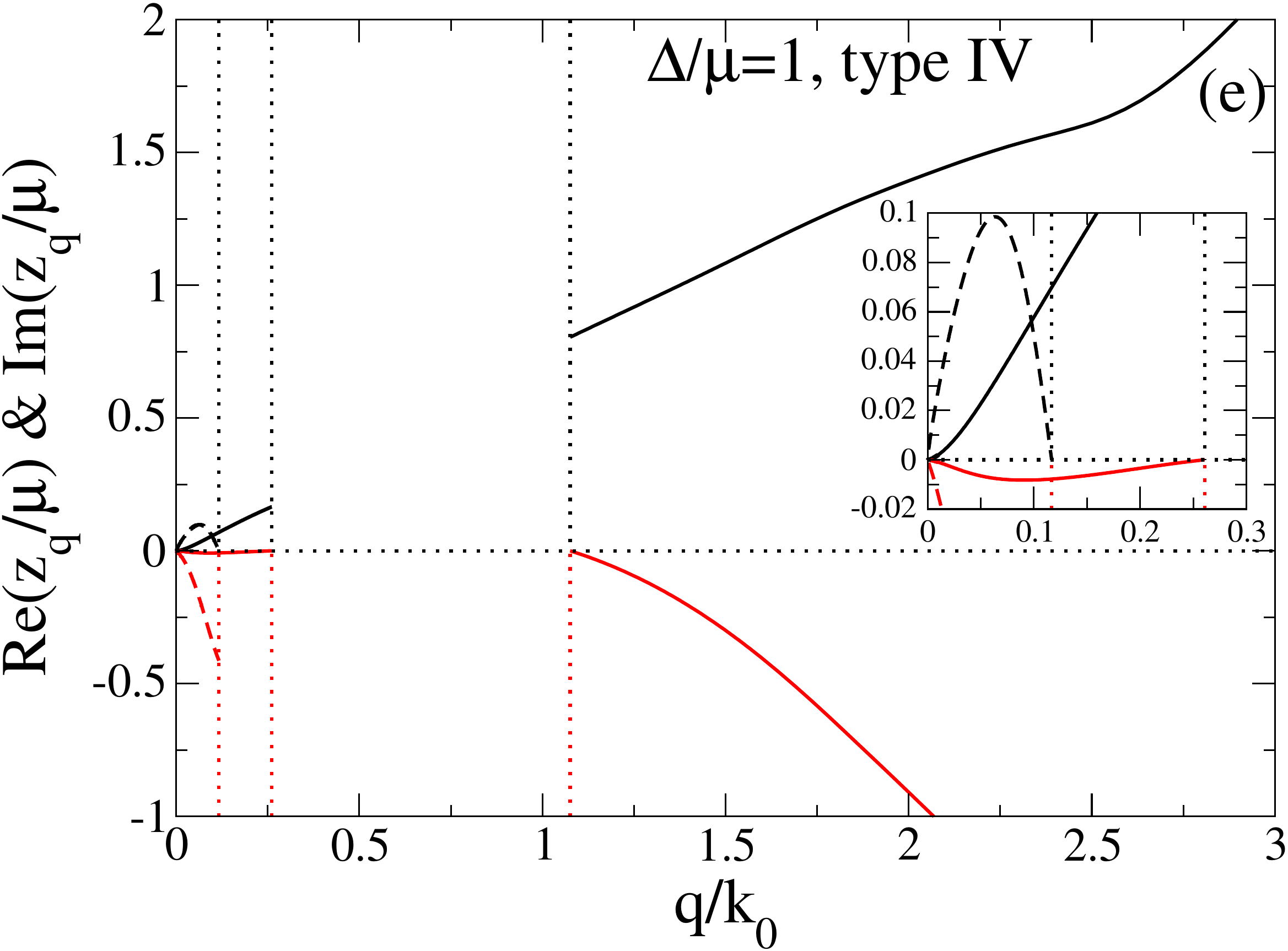} \hspace{5mm} \includegraphics [width = 0.3 \textwidth, clip =]{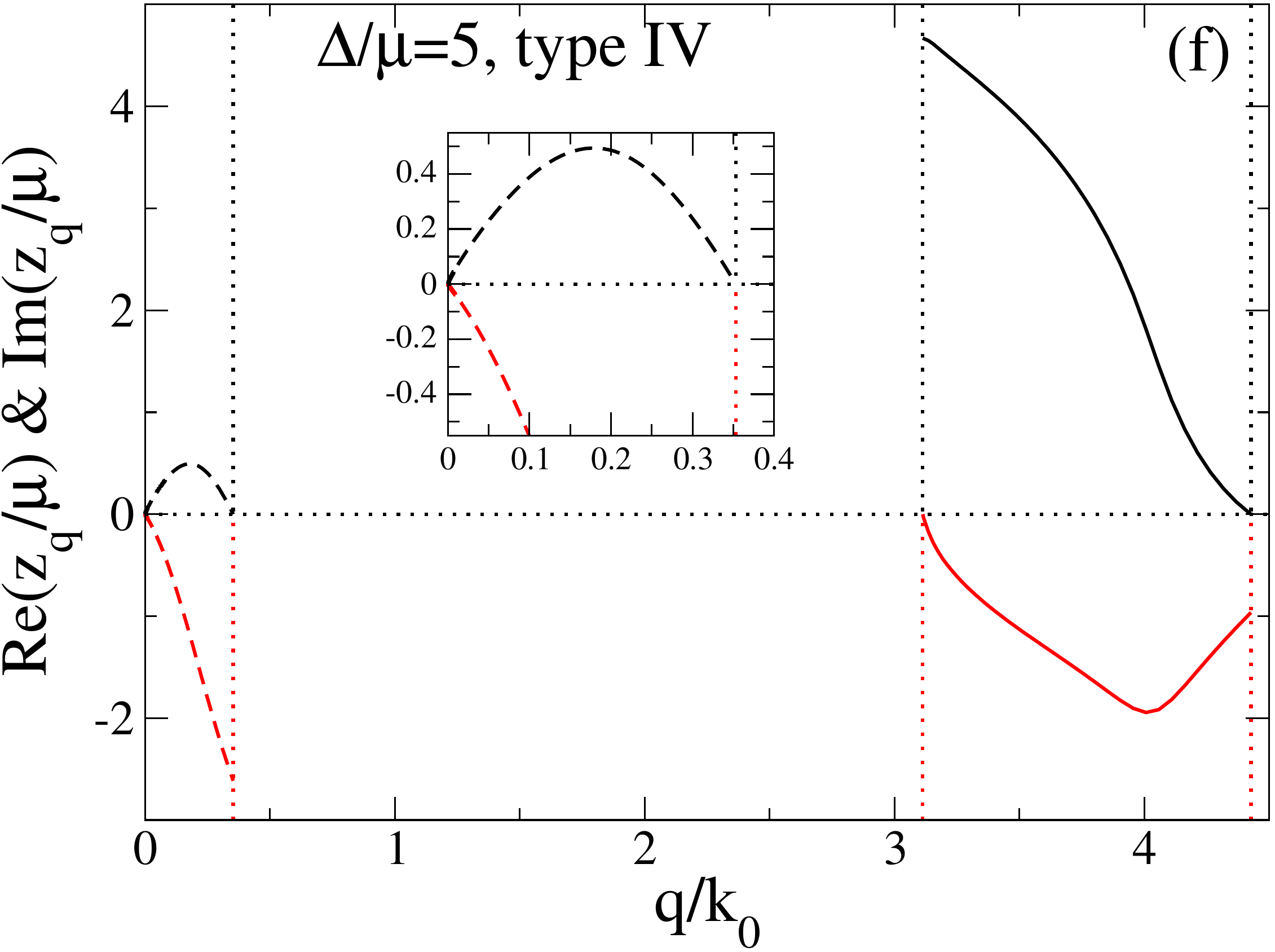}} 
\centerline{\includegraphics [width = 0.3 \textwidth, clip =]{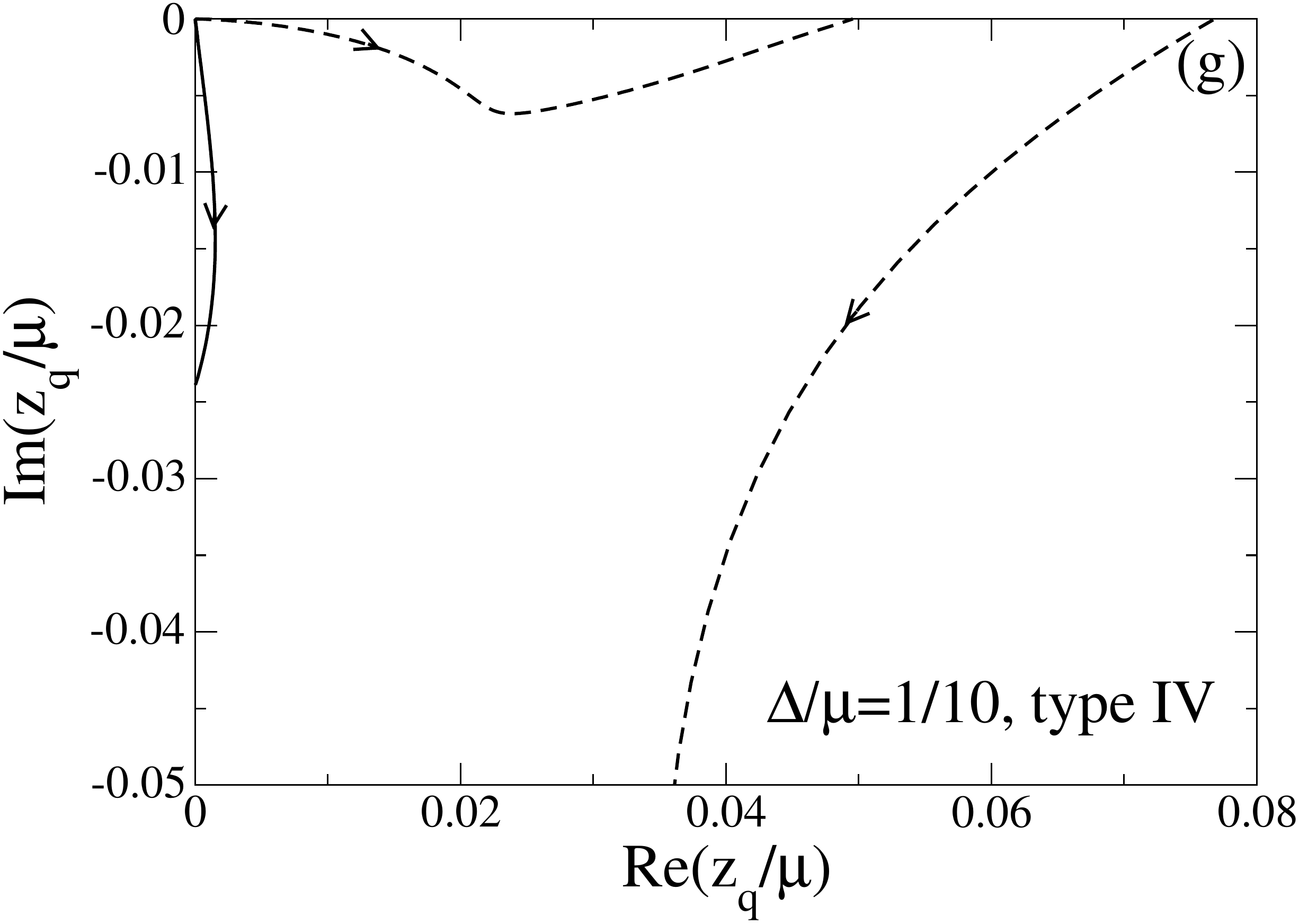} \hspace{5mm} \includegraphics [width = 0.3 \textwidth, clip =]{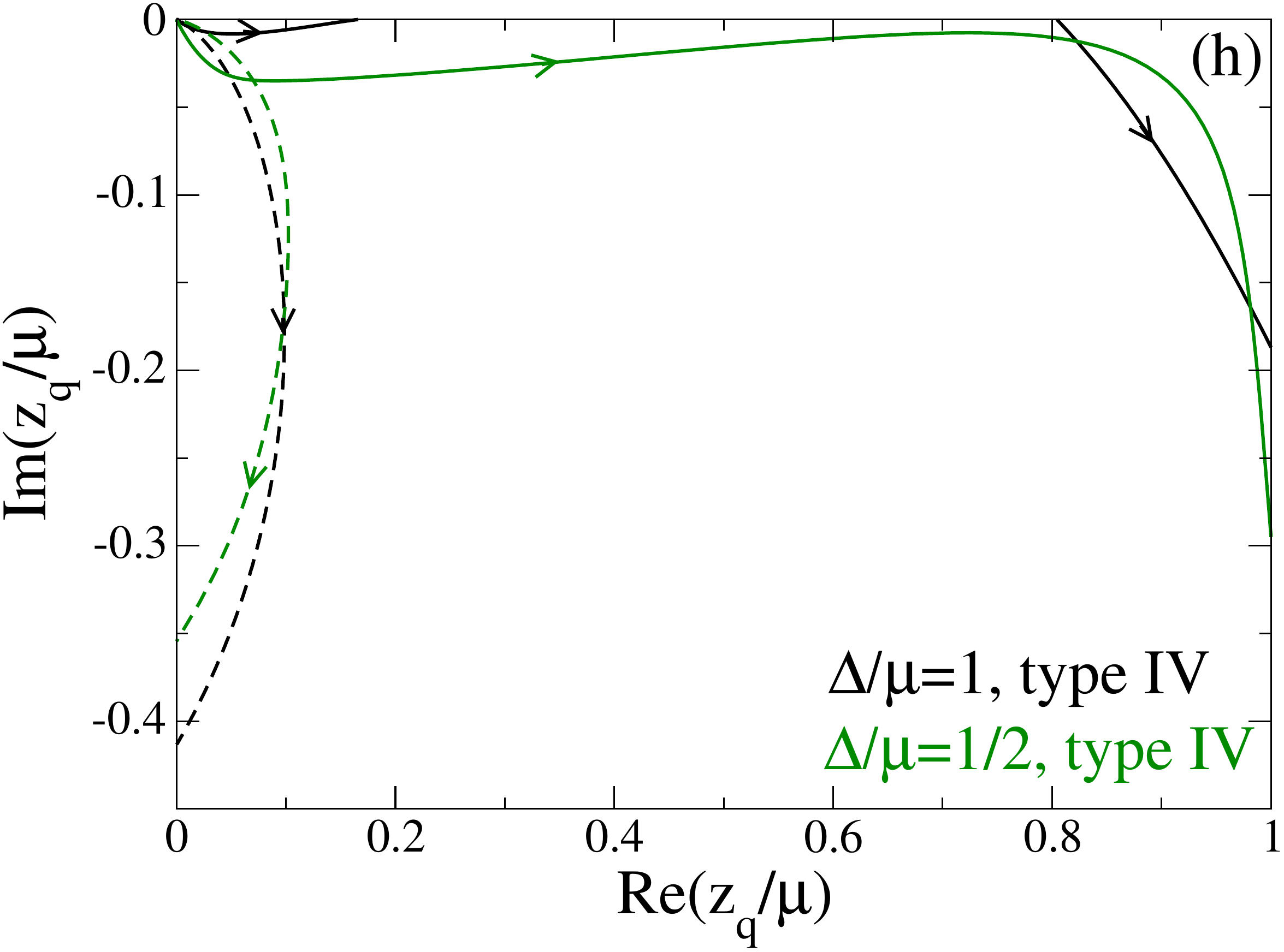} \hspace{5mm} \includegraphics [width = 0.3 \textwidth, clip =]{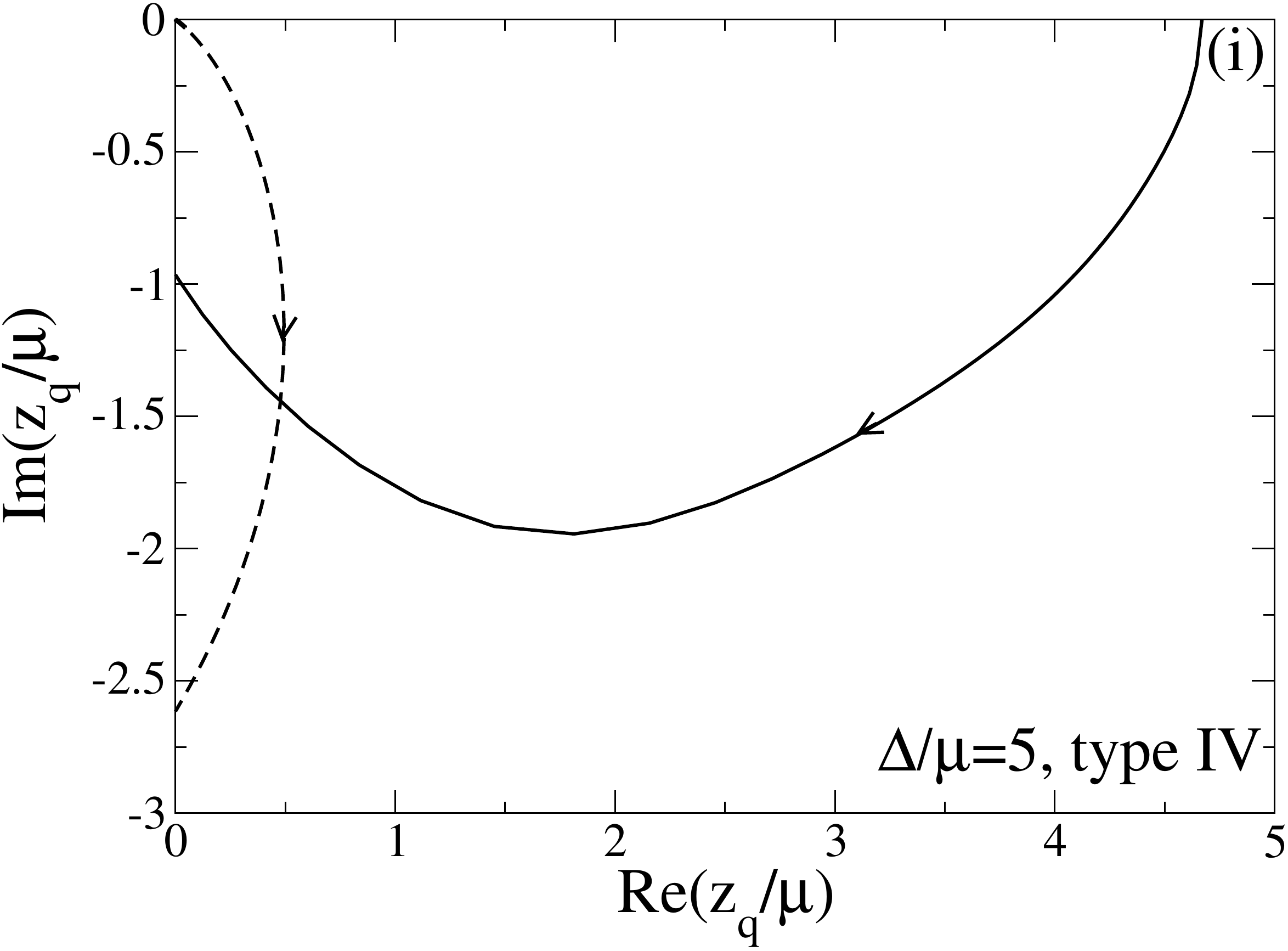}} 
\caption{In a gas of fermions with zero range interactions (case of cold atoms), continuum collective branches $ q \mapsto z_q $ obtained by carrying out the analytic continuation of the eigenenergy equations through other intervals than the one $ [\epsilon_1 (q), \epsilon_2 (q)] $ of type II used so far. Solid black line or black dashed line: real part; solid red line or red dashed line: imaginary part. The interruptions of the branches, due to exits from the fourth quadrant (the only one explored here) (the real or imaginary part of $ z_q $ vanishes) are identified by vertical dotted lines. For $ \mu> 0 $, numerical results for branches of type III (extension through the interval $ [\epsilon_2 (q), \epsilon_3 (q)] $) in (a) and (b), and of type IV (interval $ [\epsilon_3 (q),+\infty [$) in (d), (e) and (f). For $ \mu <0 $, numerical results for type IV branches in (c) (the interval $ [\epsilon_3 (q),+\infty [$ is the only possible). Here $ k_0 = (2m \mu)^{1/2}/\hbar $, $ k_\mu = (2m | \mu |)^{1/2}/\hbar $, $ \Delta/\mu = $ 1/10 in (a) and (d), $ \Delta/\mu = $ 1 in (b) and (e), $ \Delta/\mu = $ 5 in (f) (no type III branch for this value) and $ \Delta/\mu =-1/5 $ in (c). The interval $ [\epsilon_2 (q), \epsilon_3 (q)] $ closes and the type III branch ceases to exist at $ q = q_0 (\Delta) $ given by (\ref{eq:valq0}) and marked with a solid vertical gray line. At weak coupling, in (a), the type III branch almost merges with that previously studied of type II and is very well reproduced by the solution of (\ref{eq:defzqch0}) (large dark blue or orange dots). Insets: enlargements; the values at $ \check{q} <0.001 $ are taken from (\ref{eq:zetas2}) after numerical solution of (\ref{eq:surs2}) rewritten as an equation on $ s_2 (z) $. Trajectories of type IV branches in the complex plane for $ \Delta/\mu = 1/10 $ in (g), $ \Delta/\mu = 1 $ (black) and $ \Delta/\mu = 1/2 $ (green) in (h), $ \Delta/\mu = 5 $ in (i); the arrows indicate the direction of the increasing $ q $. Everywhere for type IV  branches: solid line (dotted line), hypoacoustic branch (\ref{eq:hypoacou}) [hyperacoustic (\ref{eq:hyperacou})] and its possible emanation at large $ q $ after excursion outside the fourth quadrant. In (c): the dotted branch is of another nature, see its analytical behavior (\ref{eq:quadra}) at low $ \check{q} $ (solid orange line). 
\label{fig:autres} 
} 
\end{figure} 

The results of our numerical study are shown in figure \ref{fig:autres}, for three values of the interaction strength, $ \Delta/\mu = 1/10$ in the weak coupling regime, $ \Delta/\mu = 1 $ in the strong coupling regime and $ \Delta/\mu = 5 $ close to a zero chemical potential. We find a branch $ q \mapsto z_q $ of type III, at least as long as $ \Delta/\mu $ is not too large (it is absent from the fourth quadrant for $ \Delta/\mu = 5 $), see figures \ref{fig:autres}a and \ref{fig:autres}b. Its wavelength domain of existence is an interval $ [q_1, q_0] $; the upper bound has the maximum accessible value $ q_0 $, but the lower bound is positive, because the real part of $ z_q $ vanishes (and $ z_q $ passes virtually in the third quadrant) before $ q $ reaches zero. In the weak coupling regime, we can as in section \ref{subsubsec:q_approx_d0} make $ \check{\Delta} $ tend to zero at fixed $ \check{z} $ and reduce the eigenenergy equation to $ \check{M}^{(0)}_{++\downarrow} (\check{z}, \check{q})+\check{M}^{(0)}_{+-\downarrow} (\check{z}, \check{q}) = 0 $ as in (\ref{eq:defzqch0}). For $ \check{q}> 2/3 $, it follows directly from equation (\ref{eq:rho0zfix}) that 
\be
\label{eq:rhoIIIrhoII}
\check{\rho}_{++}^{(0)[\mathrm{III}]}(\check{z},\check{q})+\check{\rho}_{+-}^{(0)[\mathrm{III}]}(\check{z},\check{q})=
\frac{\pi\check{z}}{2\check{q}} = \check{\rho}_{++}^{(0)[\mathrm{II}]}(\check{z},\check{q})+\check{\rho}_{+-}^{(0)[\mathrm{II}]}(\check{z},\check{q})
\ee
so that the Nozi\`eres term takes exactly the same form for the extension intervals $ [\epsilon_1, \epsilon_2] $ and $ [\epsilon_2, \epsilon_3] $. For $ \check{q} <2/3 $, the extension interval $ [\epsilon_2, \epsilon_3] $ is cut in half by $ \check{q} (2+\check{q}) $ but, as we pass above the point $ \alpha $ in the continuous mapping of the roots, it is the third column of (\ref{eq:rho0zfix}) which counts and relation (\ref{eq:rhoIIIrhoII}) remains true. Indeed, the bound $ \check{q} (2+\check{q}) $ is none other than the limit of $ \alpha $ at low $ \check{\Delta} $ (see the sentence below (\ref{eq:rho0zfix}) and footnote \ref{note:Lagrange}). We deduce that the type II and type III branches coincide in the limit $ \check{\Delta} \to 0 $ at fixed $ \check{q} $, which numerical calculation confirms masterfully, see figure \ref{fig:autres}a. On the other hand, we find one or two branches of type IV depending on the case, see figures \ref{fig:autres}d, \ref{fig:autres}e, \ref{fig:autres}f. Their wave number domain is a union of two intervals, according to the vanishing of the real part or the imaginary part of $ z_q $. The upper interval is compact for $ \Delta/\mu $ quite large, but extends to infinity otherwise. Remarkably, the lower interval begins at $ q = 0 $, where $ z_q $ tends to zero; this behavior differs from that of the continuum branch of type II studied in the other sections and rather evokes the acoustic branch.

Let us now show analytically that the collective branches of type IV do not have an acoustic start $ z_q \approx q $ but a hyperacoustic start $ z_q \approx q^{4/5} $ (the group or phase velocity diverges in $ q = 0 $) or hypoacoustic $ z_q \approx q^{3/2} $ (the velocity tends to zero). For this, let's build an approximate form of the equation on $ z_q $ at low energy and low wave number, first by replacing the matrix elements of $ M $ (not yet analytically continued) by their dominant order in $ q $ at fixed $ \zeta \equiv \check{z}/\check{q} $, \footnote{We give $ A = \int \dd^3 \check{k} \frac{\check{\Delta}^2}{2 \check{\epsilon}_k^3} $, $ B = \int \frac{\dd^3 \check{k}}{2 \check{\epsilon}_k} \left (\frac{\check{\xi}_k}{4 \check{\epsilon}_k^2}+\frac{\check{\Delta}^2 \check{k}^2}{2 \check{\epsilon}_k^4} \right) $, $ C = \int \dd^3 \check{k} \frac{(-1)}{8 \check{\epsilon}_k^3} $ and $ D = \int \dd^3 \check{k} \frac{\check{\xi}_k}{-4 \check{\epsilon}_k^3} $. There is no linear term in $ \zeta $ in $ \check{M}_{++} (\check{z}, \check{q}) $ because it is an even function of $ \check{z} $. According to relation (A6) of \cite{PRAConcav}, we also have $ B = \int \dd^3 \check{k} \frac{\check{k}^2}{6 \check{\epsilon}_k^3} $ as in equation (23) of reference \cite{sontempnn}.} 
\be
\check{M}_{--}(\check{z},\check{q})\simeq \check{M}_{--}(0,0)\equiv \frac{A}{(2\pi)^3}, \quad \check{M}_{++}(\check{z},\check{q})\simeq \frac{(B+C\zeta^2)}{(2\pi)^3}\check{q}^2, \quad \check{M}_{+-}(\check{z},\check{q})\simeq \frac{D \zeta \check{q}}{(2\pi)^3}
\ee
then replacing $ \sh \Omega $ by $-\ii $ in equations (\ref{eq:fmm}, \ref{eq:fpp}, \ref{eq:fpm}) useful for analytic continuation (\ref{eq:Noz}). 
We assumed in the second step, by checking it numerically, that $ s_2 (z) $ is in the fourth quadrant, just like $ \zeta $. We thus end up with a much simpler eigenenergy equation, 
\be
\label{eq:surs2}
\left[\ii\pi^2\check{\Delta}\,\phi_{--}(s_2)+A\check{q}\right] \left[-\ii\frac{\pi^2\zeta^2}{4\check{\Delta}}\phi_{++}(s_2)+(B+C\zeta^2)\check{q}\right]-\left[-\frac{\pi^2}{2} \zeta\phi_{+-}(s_2)+D\zeta\check{q}\right]^2\!\!\!\simeq 0 \ \ \mbox{where}\ \ 
\left\{
\begin{array}{l}
\phi_{--}(s)=\frac{\ln(1+\sqrt{1-s^2})-\ln(1-\sqrt{1-s^2})}{2} \\
\phi_{++}(s)=\frac{1}{2}  \phi_{--}(s) + \frac{(1-s^2)^{1/2}}{2s^2}\\
\phi_{+-}(s)=\frac{(1-s^2)^{1/2}}{s}
\end{array}
\right.
\ee
To be consistent, we must also approximate equation (\ref{eq:surs}) on $ s_2 (z) $ as follows,
\be
\label{eq:zetas2}
\frac{4}{\zeta^2}(1-\ii s_2\check{\Delta}) \simeq \frac{1-s_2^2}{-s_2^2}
\ee
which allows us to easily eliminate $ \zeta $ and consider (\ref{eq:surs2}) as an equation on $ s_2 (z) $. 

A first type of solution of (\ref{eq:surs2}) corresponds to $ \zeta \to 0 $ therefore $ s_2 \to-\ii/\check{\Delta}+0^+$. In the limit $ \check{q} \to 0 $, we find by trial and error that $ \zeta^2/\check{q} $ has a finite limit $ \gamma $, \footnote{$ \zeta^2/\check{q} $ cannot tend to infinity because $ \phi_{++} \phi_{--}-\phi_{+-}^2 $ does not vanish at $ s =-\ii/\check{\Delta}+0^+$. Suppose that $ \zeta^2/\check{q} = o (1) $ leads to the contradictory result $ \zeta^2 \approx \check{q} $.} and that (\ref{eq:surs2}) is reduced to an affine equation on $ \gamma $. Hence the exact hypoacoustic behavior of the branch: 
\be
\label{eq:hypoacou}
\boxed{
\check{z}_{q} \underset{\check{q}\to 0}{=}\gamma^{1/2} \check{q}^{3/2}[1+O(\check{q})]\quad\mbox{with}\quad \left.\gamma=\frac{-4\ii \pi^{-2} B\check{\Delta}\,\phi_{--}(s)}{(\phi_{++}\phi_{--}-\phi_{+-}^2)(s)}\right\vert_{s=-\ii/\check{\Delta}+0^+}
}
\ee
For $ z_q $ to be in the fourth quadrant, $ \gamma $ must have a negative imaginary part, which imposes $ \check{\Delta} \in ] 0, \check{\Delta}_0 [$ with $ \check{\Delta}_0 \simeq 2, \! 138 $. \footnote{If $ \check{\Delta} = \sh \tau $, $ \phi_{--} = \tau+\ii \pi/2 $, $ \phi_{++} = (\tau-\sh \tau \ch \tau+\ii \pi/2)/2 $ and $ \phi_{+-} = \ii \ch \tau $ in (\ref{eq:hypoacou}). In addition, $ \gamma^{1/2} \sim-2 \ii/[3 \check{\Delta} (\pi^2/8-1)]^{1/2} $ when $ \check{\Delta} \to 0 $.} This explains why the hypoacoustic branch does not appear at low $ \check{q} $ in figure \ref{fig:autres}f. Furthermore, we show that, among the $ Z_{\pm} $ residues of section \ref{subsec:carac}, the one $ Z_+$ in the channel of small phase deviations is dominant, and diverges as $ \check{q}^{-1/2} $; at low $ \check{q} $, the hypoacoustic modes are therefore mainly phase modes of the order parameter. 

A second type of solution of (\ref{eq:surs2}) corresponds to $ | \zeta | \to+\infty $ therefore $ s_2 \to 1-\ii 0^+$. Then, in (\ref{eq:surs2}), we neglect $ B $ in front of $ C \zeta^2 $, we divide the remaining equation by $ (\pi^2 \zeta/2)^2 $, we expand and collect by powers of $ \check{q} $, then we neglect the $ \check{q}^2 $ terms to obtain: 
\be
\label{eq:phiphi}
(\phi_{--}\phi_{++}-\phi_{+-}^2)(s_2)+\frac{4\check{q}}{\pi^2}\left[D\,\phi_{+-}(s_2)+\ii C\check{\Delta}\,\phi_{--}(s_2)-\frac{\ii A}{4\check{\Delta}}\phi_{++}(s_2)\right]\simeq 0
\ee
It remains to replace each contribution by its dominant order in the small parameter $ u = (1-s_2^2)^{1/2} $: in the second term of (\ref{eq:phiphi}), it suffices to know that $ \phi_{--} \sim \phi_{++} \sim \phi_{+-} \sim u $, but in the first, we must expand each function $ \phi $ up to the relative order $ u^4 $ (this is the sub-sub-leading order) to obtain the equivalent $ u^6/45 $. Hence the law $ u \sim \check{q}^{1/5} u_0 $ and the exact hyperacoustic behavior of the branch: 
\be
\label{eq:hyperacou}
\boxed{
\check{z}_{q}\underset{\check{q}\to 0}{=} \beta^{1/2}\check{q}^{4/5}\left[1+O(\check{q}^{2/5})\right]\quad\mbox{with}\quad\beta=\frac{4(\ii\check{\Delta}-1)}{u_0^2} \quad\mbox{and}\quad u_0=\left(\frac{180}{\pi^2}\right)^{1/5}\left(\frac{\ii A}{4\check{\Delta}}-\ii C \check{\Delta}-D\right)^{1/5}}
\ee
In (\ref{eq:hyperacou}), we must choose the value of the fifth root so that $ u_0 $ is in the first quadrant and $ \beta $ has a negative imaginary part, which can be done for any value of $ \check{\Delta}> \check{\Delta}_1 \simeq $ 0.05283, hence the systematic presence of a type IV hyperacoustic branch at low $ \check{q} $ in figure \ref{fig:autres}. \footnote{The continuation of $ \beta $ to $ \check{\Delta} \in] 0, \check{\Delta}_1 [$ has a positive imaginary part and obeys $ \beta^{1/2} \sim (8 \pi \check{\Delta}/45)^{1/5} $ when $ \check{\Delta} \to 0 $.} We verify that, among the residues $ Z_{\pm} $ of section \ref{subsec:carac}, the one $ Z_+$ in the channel of small phase deviations is dominant, and diverges as $ \check{q}^{-4/5} $; at low $ \check{q} $, the hyperacoustic modes are therefore also mainly phase modes of the order parameter. 

It remains to understand physically the type IV branch on the upper wave number interval. We affirm that this connected component is only the continuation at large $ q $, that is to say the result of the irruption in the fourth quadrant, of the hypoacoustic branch or of the hyperacoustic branch. We can clearly see this by mentally extending the curves in figure \ref{fig:autres}d, the large $q$ component would be the return of the hyperacoustic branch after a passage in the first quadrant. Similarly, in figure \ref{fig:autres}e, the large $q$ component seems to be the return of the hypoacoustic branch after a detour in the first quadrant. On the other hand, in figure \ref{fig:autres}f, we interpret the large $q$ component as the first entry of the hypoacoustic branch in the fourth quadrant, this branch pointing from the start at low $ q $ towards the first quadrant, see the discussion below equation (\ref{eq:hypoacou}). To make these interpretations easier to follow, we have shown in figures \ref{fig:autres}g, \ref{fig:autres}h and \ref{fig:autres}i the trajectory of type IV branches for $ \mu> 0 $ in the complex plane. To justify them numerically, we also show the trajectory of the branches for $ \Delta/\mu = 1/2 $ in figure \ref{fig:autres}h: the hypoacoustic branch is a connected set in the fourth quadrant. These figures lead to the following empirical remark: the branches which venture into the third quadrant ($ \re z <0 $ and $ \im z <0 $) never return (at subsequent $ q $) in the fourth quadrant, unlike those that venture into the first quadrant. 

\subsubsection{On the BEC side: $ \mu <0 $} 
\label{subsubsec:autresmuneg} 

Finally, let's look for a continuum collective branch in the case of a negative chemical potential. As we said in section \ref{subsubsec:lpdnae1e2e3}, the spectral densities then have, on the real energy axis, $ \epsilon_3 (q) $ of equation (\ref{eq:om1om3}) as the only point of non-analyticity. The analytic continuation of the eigenenergy equation can only be done through the interval $ [\epsilon_3 (q),+\infty [$, therefore can only be of type IV. The rescaling of the variables must be revised and the calculation of the spectral densities must be redone. We now express the wave numbers in units of $ k_\mu = (2m | \mu |)^{1/2}/\hbar $, for example $ \check{q} = q/k_\mu $, energies in units of $ | \mu | $, for example $ \check{\epsilon} = \epsilon |/| \mu | $ and $ \check{\xi}_{\check{k}} = \check{k}^2+1 $, the spectral densities and the matrix elements of $ M $ in units of $ 2m k_\mu/\hbar^2 $. We find that the reduced spectral densities are always given by form IV of table \ref{tab:rhogen}, provided that the term $ 1-\check{q}^2/4 $ is replaced by $-(1+\check{q}^2/4) $ in cubic equation (\ref{eq:surs}). For $ \check{\epsilon}> \check{\epsilon}_3 (\check{q}) $, the modified equation has indeed, for all $ \check{q}> 0 $, three real roots, $ s_1 (\epsilon) <-1 $, $ s_2 (\epsilon) \in [0,1] $ which appears in the spectral densities, and $ s_3 (\epsilon)> 1 $, as one can show in a graphical discussion. The analytic continuation of $ s_2 $ to complex energies $ z $ is done by continuous mapping as in section \ref{subsubsec:autresmupos}, never passing below its branch point(s), i.e.\  complex roots of the polynomial $ P_8 (X) $ in the fourth quadrant. \footnote{\label{note:muneg} For $ \mu <0 $, you must replace $ \check{q} $ by $ \ii \check{q} $ in the definition (\ref{eq:P8}) of $ P_8 (X) $. Then $ P_8 (X) $ has for $ \check{q} < \check{q}_* (\check{\Delta}) $ a single root $ \alpha $ strictly inside the fourth quadrant, and has two, $ \alpha $ and $ \alpha'$, otherwise. Here, $ \check{q}_* (\check{\Delta}) $ is the real positive root of the Cardan discriminant $ \delta $ of footnote \ref{note:Lagrange} considered as a polynomial in $ \check{q} $ and also subject to the substitution $ \check{q} \to \ii \check{q} $.}

We have studied numerically the case $ \check{\Delta} = \Delta/| \mu | = 1/5 $, see figure \ref{fig:autres}c. At low $ \check{q} $, we find a hyperacoustic branch whose start is well described by (\ref{eq:hyperacou}) if we think of changing the factor $ \ii \check{\Delta}-$ 1 in $ \ii \check{\Delta}+$ 1 in $ \beta $, but which ends up leaving the fourth quadrant; equation (\ref{eq:hyperacou}) thus modified also makes it possible to show that this hyperacoustic branch exists in the vicinity of $ q = 0 $ for any value of $ \check{\Delta} $. We also find a second branch, which does not reach the large $ \check{q} $ but tends to the value $ 2 \Delta $ at the origin of the wave numbers by following the real axis closely (its imaginary part tends to zero as $ q^3 $). We analytically predict a quadratic start \footnote{If we have that $ z_q-2 \Delta \approx q^2 $, we see that $ \sh \Omega \approx q \to 0 $ and $ \ch \Omega \to $ 1 in (\ref{eq:fmm}, \ref{eq:fpp}, \ref{eq:fpm}) so that $ f_{--} (\psi) \sim \frac{\sh^2 \Omega}{2} (\psi+\sin \psi \cos \psi) $, $ f_{++} (\psi) \sim \psi $, $ f_{+-} (\psi) \sim \sh \Omega \sin \psi $. In addition, the lower edge of the broken pair continuum is $ 2 (\Delta^2+\mu^2)^{1/2}> 2 \Delta $ for $ \mu <0 $ and $ q = 0 $; therefore, $ M_{++} (z_q, q) $ and $ M_{+-} (z_q, q) $ have a finite limit $ M_{++} (2 \Delta, 0) $ and $ M_{+-} (2 \Delta, 0) $ when $ q \to 0 $, while $ M_{--} (z_q, q) = O (q^2) $ since $ M_{--} (2 \Delta, 0) = 0 $ as in (\ref{eq:zeroexact}) and $ M_{--} (z, q) $ varies quadratically in $ q $ and linearly in $ z-2 \Delta $ around $ (z, q) = (2 \Delta, 0) $. Given the analytic continuation procedure (\ref{eq:Noz}) and the factor $ 1/q $ in the spectral density, see table \ref{tab:rhogen}, the contribution of $ M_{++} $ and $ M_{--} $ is negligible in the analytically continued matrix element while that of $ M_{+-} $ is not. Finally, by taking the limit $ q \to 0 $ in equation (\ref{eq:surs}) modified for $ \mu <0 $, knowing that $ z_q $ is in the fourth quadrant and that $ \sin \psi = [1-s_2^2 (z_q)]^{1/2} $, we find that $ \sh \Omega \sim-\ii \check{q}/(\check{\Delta} \sin \psi) $ and we end up with (\ref{eq:quadra}). Expression (\ref{eq:valM+-}) of $ \check{M}_{+-} (2 \check{\Delta}, 0) $ and its sign still apply if we set $ \tau = \argsh (\mu/\Delta) <0 $ and if we replace the now purely imaginary factor $-(\eee^{2 \tau}-1)^{1/2} $ by $ (1-\eee^{2 \tau})^{1/2} $.} 
\be
\label{eq:quadra}
\boxed{
z_q \underset{q\to 0}{=} 2\Delta - \frac{1}{\check{\Delta}\sin^2\psi}\frac{\hbar^2q^2}{2m} + O(q^3) \quad\mbox{with}\quad
\frac{\psi(\psi+\sin\psi\cos\psi)}{2\sin^2\psi}=\left[1-8\pi\,\check{M}_{+-}(2\check{\Delta},0)\right]^2
}
\ee
shown as a solid orange line in figure \ref{fig:autres}c and in good agreement with the numerical results; the transcendental equation on $ \psi = \pi/2-\asin s_2 $ has a unique solution in the interval $ [0, \pi] $. Result (\ref{eq:quadra}) is in fact only valid for $ \psi <\pi/2 $, that is to say $ \check{\Delta} <0.222 $; for $ \check{\Delta}> $ 0.222, the branch leaves the fourth quadrant towards the first quadrant (its imaginary part vanishes) before reaching $ \check{q} = 0 $. \footnote{In the limit $ \check{q} \to 0 $, the root $ s_2 (z) $ obtained by continuous mapping is positive for $ \check{z} = 2 \check{\Delta}+(C-\ii 0^+) \check{q}^2/\check{\Delta} $ as soon as the real coefficient $ C $ is $ <-1 $, as it is the case in (\ref{eq:quadra}). Using (\ref{eq:surs}) modified for $ \mu <0 $, we find that $ s_2^2 (z) \to (1+C-\ii 0^+)/(C-\ii 0^+) $ therefore $ s_2 (z) \to [(1+C-\ii 0^+)/(C-\ii 0^+)]^{1/2} $ since $ s_2 $ must be close to $ 1 $ when $ C \gg 1 $. Now we have $ s_2 = \cos \psi $. The condition $ s_2> 0 $ therefore imposes $ \psi <\pi/2 $. In addition, we have analytically calculated the coefficient of the term $ q^3 $ in expansion (\ref{eq:quadra}). We find that it is pure imaginary, that its imaginary part is $ <0 $ for $ \psi <\pi/2 $ and vanishes by changing sign at $ \psi = \pi/2 $. It is also at this threshold value $ \psi = \pi/2 $ ($ \check{\Delta} \simeq 0.222 $) that $ \check{z}_q $ merges to second order in $ \check{q} $ with the real root $ \check{\epsilon}_2 (\check{q}) $ of the polynomial $ P_8 (X) $ of (\ref{eq:P8}) transposed to the case $ \mu <0 $, so with a point of non-analyticity of two of the roots $ s_i (\check{z}) $: for $ \mu <0 $, it is indeed necessary to replace $ \check{q} $ by $ \ii \check{q} $ in (\ref{eq:P8}), and therefore in (\ref{eq:eps2fq}), as footnote \ref{note:muneg} says.} The quadratic start (\ref{eq:quadra}) is reminiscent of that (\ref{eq:quadzq}) of the usual continuum branch; the reasoning that led to (\ref{eq:equivzmp}) here also gives zero limit residues in $ q = 0 $, $ \check{Z}_-\approx \check{q} $ and $ \check{Z}_+\approx \check{q}^3 $. However, there is no hypoacoustic branch. Moreover, the reasoning leading to (\ref{eq:hypoacou}) is not transposed to the case $ \mu <0 $: for $ z = o (\check{q}) $, we find numerically that $ s_2 (z) = o (1) $, and that the root $ s_3 (z) $ has the expected limit $ \ii/\check{\Delta} $, in the continuous mapping passing above the $ \alpha $ branch point.

\section{Conclusion} 
\label{sec:conclusion}

In a spatially homogeneous three-dimensional unpolarized gas of spin $ 1/2 $ fermions with attractive interaction, prepared in the thermodynamic limit at zero temperature, therefore fully paired and condensed in the form of $ \uparrow \downarrow $ bound pairs, we have analytically studied the continuum branch corresponding to the collective pair-breaking excitation modes of the gas. As shown in reference \cite{PRL2019}, under certain conditions, such modes can manifest themselves experimentally by a Lorentzian peak $ \omega \mapsto | Z_\qq/(\hbar \omega-z_\qq) |^2 $ in the intensity of a frequency response function on the collective variable of the perturbed gas given by the complex order parameter $ \Delta (\rr, t) $. We thus analyzed the complex energy $ z_\qq $ of the mode as well as its residue $ Z_\qq $ for an arbitrary very weak non-polarizing and bounded in time excitation, according to wave vector $ \qq $, in order to obtain a complete description of the branch. Mathematically, we use time-dependent BCS theory linearized around the minimal energy stationary solution, to obtain an equation on the eigenenergies of the modes, which one must always analytically continue to the lower complex half-plane across the branch cut resulting from the broken pair continuum at fixed $ \qq $, if we want to find $ z_\qq $. Physically, the fact that $ \im z_\qq $ is negative reflects the damping of the mode by breaking of pairs and emission of free fermions of wave vectors $ \qq/2 \pm \kk $ in the continuum, $ \kk \in \mathbb{R}^3 $. Our work analytically studies the branch at any wave number $ q $, and for any interaction strength; it therefore goes significantly beyond previous work, which obtains analytical results only in the limit of low wave numbers $ q \to 0 $, that is to say only on the quadratic start of the branch $ z_\qq-2 | \Delta | \propto q^2 $, within the weak coupling limit $ | \Delta |/\mu \to 0^+$ \cite{AndrianovPopov} or for an arbitrary coupling $ | \Delta |/\mu $ but a positive chemical potential $ \mu $ \cite{PRL2019}.

In the weak coupling limit $ | \Delta |/\mu \to 0^+$, the real part of the dispersion relation exhibits three scales of variation with wave numbers. $ (i) $ The first one $ q_1 = k_0 | \Delta |/\mu $, where $ k_0 $ is the wave number minimizing the energy $ \epsilon_\kk $ of a BCS quasiparticle ($ \epsilon_{k_0} = | \Delta | $), unsurprisingly corresponds to $ q \approx 1/\xi $ where $ \xi $ is the size of a $ \uparrow \downarrow $ bound pair. The branch has a universal limiting law when $ z_\qq $ is expressed in units of $ | \Delta | $ and when $ | \Delta |/\mu $ tends to zero at fixed $ q/q_1 $: this law  applies as well for fermions with attractive contact interaction in the $ s $ wave (case of cold atoms) as for charged fermions with a long-range repulsive Coulomb interaction (case of a BCS superconductor in the sense of reference \cite{AndrianovPopov} that is to say without real inclusion of the phonons of the crystal lattice). If we start from $ q = 0^+$, the real part increases, exhibits a maximum $ \simeq 2.16 \, | \Delta | $ at $ q/q_1 \simeq 1.70 $, then decreases and tends to zero as $ | \Delta | q_1/q $, up to a logarithmic factor in $ q/q_1 $; the imaginary part is decreasing and diverges linearly as $-\mu q/k_0 $ up to logarithmic corrections in $ q/q_1 $. At the dominant order, the residue is carried by the channel of small modulus deviations of the order parameter from its equilibrium value. The continuum mode is a modulus mode and one can neglect the phase-modulus coupling in the linearized BCS equations. $ (ii) $ The second wave number scale is $ q_2 = k_0 (| \Delta |/\mu)^{2/3} $: on the real part of the branch, the decrease $ 1/q $ of the previous scale is gradually interrupted, $ \re z_\qq $ goes through a minimum of the order of $ \mu (| \Delta |/\mu)^{4/3} $ and located at $ q_{\rm min} \approx q_2 $, up to logarithmic factors in $ | \Delta |/\mu $, then goes up; this is due to the contribution of the phase-modulus coupling. The imaginary part continues to decrease in the same approximately linear way in $ q $. Even if the residue in the modulus channel still prevails, that in the phase channel is no longer negligible. The continuum mode is in phase-modulus hybridization. $ (iii) $ The third scale is $ q_3 = k_0 $; the imaginary part of $ z_\qq $ is approximately proportional to $-\mu q/k_0 $, with a logarithmic coefficient in $ | \Delta |/\mu $, while the real part is increasing, approximately proportional to $ \mu (q/k_0)^{3/2} $, again with a logarithmic coefficient in $ | \Delta |/\mu $; the residues in the phase and modulus channels are almost equal: the mode lives with almost equal amplitudes in the two channels. At $ q = 2k_0 $, the interval between branch points of the eigenenergy equation, through which the analytic continuation was carried out, closes and the branch ceases to exist. The results at the scales $ q_2 $ and $ q_3 $ are valid for a short-range interaction, as in the cold atom gases, to which we limit ourselves hereinafter; we do not know if they survive at least qualitatively the Coulomb interaction.

In the right vicinity of the zero chemical potential, $ \mu/| \Delta | \to 0^+$, which is in the strong interaction regime, on the positive scattering length side of the BEC-BCS crossover, the continuum collective branch $ z_\qq $ is, to leading order in $ \mu/| \Delta | $, purely real and quadratic in $ q $, $ z_\qq-2 | \Delta | \simeq-128 [\Gamma (5/4)]^4 \pi^{-5} \hbar^2 q^2/2m $, over its entire domain of existence $] 0,2k_0 [$, domain which becomes more and more narrow since here $ k_0 = (2m \mu)^{1/2}/\hbar $, where $ m $ is the mass of a fermion, tends to zero. At the next order appears a nonzero imaginary part of $ z_{\qq} $, which reveals the existence of two wave number scales. The first scale, $ Q_1 = k_0 (\mu/| \Delta |)^{1/2} $, corresponds to the limit of validity of the quadratic approximation on $ z_\qq $ (this requires $ q \ll Q_1 $, which reference \cite{PRL2019} did not say) but has not yet received any other physical interpretation. At this scale, the imaginary part is of order of magnitude $ \mu (\mu/| \Delta |)^{3/2} $, and goes monotonously from a quadratic start to an asymptotic semiquintic plunge in $ q $. The second scale is $ Q_2 = k_0 $: at the dominant order, the imaginary part of $ z_\qq $ is of the order of $ \mu (\mu/| \Delta |)^{1/4} $, is decreasing, and varies with the same power law $ q^{5/2} $ as in the asymptotic part $ q \gg Q_1 $ of the previous scale. On both scales, the dominant residue is carried by small modulus deviations of the order parameter from its equilibrium value: the continuum mode remains from this point of view a modulus mode, even if it is absolutely necessary to include phase-modulus coupling in the calculation of its energy to have the right result, unlike what happens in the weak coupling limit at $ q = O (q_1) $. 

The previous results, obtained for $ \mu> 0 $ and $ q <2 k_0 $, result from an analytic continuation of the eigenenergy equation across the branch cut joining its first two branch points $ \epsilon_a (q) $ and $ \epsilon_b (q) $ on the positive real energy half-axis. They always correspond to a continuum collective branch of limit $ 2 | \Delta | $ and of quadratic start at $ q = 0 $. For a not too large wave number ($q <q_0(\Delta, \mu)\in]2 k_0/\sqrt{3}, 2k_0 [$), there is a third branch point $ \epsilon_c (q) $. The analytic continuation through the interval $ [\epsilon_b (q), \epsilon_c (q)] $ gives rise to a new branch of excitation in the lower right quadrant of the complex plane, at least for $ | \Delta |/\mu $ not too large and $ q $ not too small; at weak coupling, it almost coincides with the continuum collective branch studied previously. 

Finally, it remains to perform the analytic continuation across the non-compact interval $ [\epsilon_{\rm max} (q),+\infty [$ where $ \epsilon_{\rm max} (q) $ is the largest branch point on the real axis. For $ \mu> 0 $ and $ q <2k_0 $, we have $ \epsilon_{\rm max} (q) = \epsilon_c (q) $ or $ \epsilon_b (q) $ depending on whether $ q $ is lower or greater than $ q_0 (\Delta, \mu) $, as shown in the previous paragraph. For $ \mu> 0 $ at any $ q> 2 k_0 $, or for $ \mu <0 $ at any $ q> 0 $, the eigenenergy equation has only one branch point with real positive energy, it is the lower edge $ 2 \epsilon_{\kk = \qq/2} $ of the broken pair continuum, with which $ \epsilon_{\rm max} (q) $ therefore coincides. Analytically continuing through $ [\epsilon_{\rm max} (q),+\infty [$, we find two new branches. Their wave number domain generally has several connected components, the borders of which mark their entry or exit points in the lower right quadrant of the complex plane. For $ \mu> 0 $, when these branches reach $ q = 0 $, they acquire hypoacoustic behavior at low wave numbers (the complex energy $ z_q \approx q^{3/2} $ tends to zero faster than linearly and the phase velocity tends to zero) or on the contrary hyperacoustic (the energy $ z_q \approx q^{4/5} $ tends to zero more slowly than linearly and the phase velocity diverges), the dominant residue being always carried by the small phase deviations of the order parameter from its equilibrium value. For $ \mu <0 $, we observe a hyperacoustic branch; on the other hand, the hypoacoustic branch is replaced by another branch, of limit $ 2 | \Delta | $ (as reference \cite{Benfatto} hoped, this is twice the order parameter and not twice the binding energy of a pair $ 2 (| \Delta |^2+\mu^2)^{1/2} $) and of a purely real quadratic start at $ q = 0 $ for $ \Delta/| \mu | <0.222 $, and which then resembles the usual continuum branch, except that it is on the bosonic side of the BEC-BCS crossover. 

Finally, let us indicate some possible extensions of this work. BCS theory that we used is no longer quantitative in the strong interaction regime, in particular at the unitary limit often used in cold atom experiments; in this case it would be good to use a more elaborate approach, like that of reference \cite{Zwerger}. Our calculation of the continuum mode dispersion relation does not take into account coupling to the Bogolioubov-Anderson acoustic branch (in neutral particle gases), which has {\sl a priori} no reason to be negligible at any wave number and any interaction strength. Finally, one could ask what is the contribution of the continuum collective branch to the thermodynamic quantities of the fermion gas at low temperature, that is to say for $ k_B T $ small compared to the binding energy of a pair. 

\smallskip 
\section *{Acknowledgments} 
We are indebted to Serghei Klimin, Jacques Tempere, Alice Sinatra, Felix Werner, Ludovic Pricoupenko and Pascal Naidon for useful discussions and suggestions on the continuum collective branch. We thank Serghei Klimin for his comments on the text of the article. 

\appendix 
\section{On the analytical computation of the spectral densities} 
\subsection{In the weak coupling limit} 
\label{subapp:lcf} 

Let us show how to pass from expression (\ref{eq:rho--0}) of the spectral density $ \rho_{--}^{(0)} $ to expression (\ref{eq:firho--0}), integrating on the reduced wave number $ \bar{K} $ with fixed value $ u $ of the cosine of the polar angle. You just have to find the zeros of the argument of Dirac's $ \delta $ distribution in the integrand. To simplify, we can immediately impose $ \bar{\epsilon} \equiv \epsilon/\Delta \geq 2 $ (the spectral density is zero otherwise) and restrict to the integration domain $ u \geq 0 $ and $ \bar{K} \geq 0 $ due to integrand even parity. We set $a = \bar{q} u \geq 0 $ so that $ x_{\pm} = \bar{K} \pm a/2 $ and $ e_\pm = [1+(\bar{K} \pm a/2)^2]^{1/2} $ in (\ref{eq:defxpmepm}). We must solve the equation $ \bar{\epsilon} = e_++e_-$ on the variable $ \bar{K} $. By cleverly squaring and collecting terms to make the square root disappear, it comes: 
\be
\label{eq:chaine}
\bar{\epsilon}=e_+ + e_- \Leftrightarrow \bar{\epsilon}^2=e_+^2+e_-^2+2 e_+e_- \Leftrightarrow 2 e_+e_- =\bar{\epsilon}^2-(e_+^2+e_-^2)
\Leftrightarrow  (2 e_+e_-)^2-(e_+^2+e_-^2-\bar{\epsilon}^2)^2=0 \ \mbox{and}\ \bar{\epsilon}^2\geq e_+^2+e_-^2
\ee
Let's introduce $ \bar{R} = [\bar{\epsilon}^2-(4+a^2)]/(\bar{\epsilon}^2-a^2) $ and make each term of (\ref{eq:chaine}) explicit, to get the root $ \bar{K}_0 $ and its condition of existence: 
\be
\bar{\epsilon}=e_+ + e_- \Leftrightarrow \bar{K}^2=\frac{\bar{\epsilon}^2}{4}\bar{R} \ \mbox{and}\ \frac{(\bar{\epsilon}^2-a^2)^2+4 a^2}{2(\bar{\epsilon}^2-a^2)}\geq 0 \Leftrightarrow \bar{K}=\bar{K}_0\equiv\frac{\bar{\epsilon}}{2} \bar{R}^{1/2}\ \mbox{and}\ \bar{\epsilon}\geq (4+a^2)^{1/2}
\ee
Indeed, the first inequality imposes $ \bar{\epsilon}^2-a^2 \geq 0 $ (the fraction it contains has a numerator $ \geq 0 $), and you have to have $ \bar{R} \geq 0 $. 
With the same effort, we obtain the value of $ e_+-e_-$ at $ \bar{K} = \bar{K}_0 $ by a clever rewriting, and therefore the values of $ e_\pm $ and of $ x_\pm $: 
\be
\label{eq:astuce}
e_+-e_-=\frac{e_+^2-e_-^2}{e_+ +e_-}=\frac{2a\bar{K}_0}{\bar{\epsilon}}=a \bar{R}^{1/2} \Longrightarrow
e_+=\frac{\bar{\epsilon}+a \bar{R}^{1/2}}{2},\  e_- = \frac{\bar{\epsilon}-a \bar{R}^{1/2}}{2},\  x_+ =\frac{\bar{\epsilon}\bar{R}^{1/2}+a}{2},\ x_-=\frac{\bar{\epsilon}\bar{R}^{1/2}-a}{2}
\ee
The resulting identity below makes it possible to integrate (\ref{eq:rho--0}) on wave number to obtain (\ref{eq:firho--0}): 
\be
\delta(e_++e_--\bar{\epsilon})= J^{-1}\Theta(\bar{\epsilon}-(4+a^2)^{1/2})\,\delta(\bar{K}-\bar{K}_0)
\quad\mbox{with}\quad J=\frac{\dd}{\dd\bar{K}}(e_++e_-)=\frac{x_+}{e_+} + \frac{x_-}{e_-}\geq 0
\ee

\subsection{In the general case} 
\label{subapp:gen} 

In the continuous limit of the lattice model, we show how to express the spectral densities $ \rho_{ss'} (\epsilon, \qq) $, $ s, s' \in \{-,+\} $, in terms of elliptical integrals as in table \ref{tab:rhogen}, starting from their definition (\ref{eq:defdensspec}) and limiting ourselves to the case $ \mu> 0 $ and $ \epsilon> 2 \Delta $. First, we rescale as in (\ref{eq:varadim}, \ref{eq:rhoadim}) then we integrate (\ref{eq:defdensspec}) in spherical coordinates of polar axis the direction of $ \qq $. The integration on the azimuthal angle is straightforward, since the integrand does not depend on it. We then integrate on $ u $, cosine of the polar angle, after having reduced to $ u \geq 0 $ due to even parity of the integrand. It is necessary for that to find the zeros of the argument of the Dirac $ \delta $ distribution in (\ref{eq:defdensspec}), therefore to solve the equation $ \check{\epsilon} = \check{\epsilon}_++\check{\epsilon}_-$ on $ u $, with $ \check{\epsilon}_\pm = \check{\epsilon}_{\kk \pm \qq/2} $ (similarly we use the notation $ \check{\xi}_\pm = \check{\xi}_{\kk \pm \qq/2} $). Let $ a \equiv \check{k}^2+\frac{1}{4} \check{q}^2-1 $ and $ b \equiv \check{k} \check{q} $, so 
\be
\check{\xi}_{\pm}=a\pm bu \quad\mbox{and}\quad\check{\epsilon}_{\pm}=[(a\pm bu)^2+\check{\Delta}^2]^{1/2}
\ee
It is a clever reparameterization, in the sense that $ a = 0 $ at the place of the minimum of $ k \mapsto \min_u \epsilon_++\epsilon_-$ (this function was introduced in \cite{PRL2019} to determine the extreme values of $ k $ accessible for fixed $ \epsilon_++\epsilon_-$). Let's take advantage of some of the work done in the weak interaction limit (section \ref{subapp:lcf}) by directly using the two ends of the chain (\ref{eq:chaine}): 
\be
\label{eq:chq}
\check{\epsilon}=\check{\epsilon}_+ + \check{\epsilon}_- \Leftrightarrow u^2=\frac{\check{\epsilon}^2}{4b^2} \check{R}\ \mbox{and}\
\frac{(\check{\epsilon}^2-4a^2)^2+16 a^2\check{\Delta}^2}{\check{\epsilon}^2-4 a^2}\geq 0 \Leftrightarrow u=u_0\equiv\frac{\check{\epsilon}}{2b}\check{R}^{1/2}\ \mbox{and}\  -\frac{1}{2}(\check{\epsilon}^2-4\check{\Delta}^2)^{1/2} \leq a \leq \frac{1}{2}(\check{\epsilon}^2-4\check{\Delta}^2)^{1/2}
\ee
now with $ \check{R} = [\check{\epsilon}^2-4 (a^2+\check{\Delta}^2)]/(\check{\epsilon}^2-4 a^2) $, then by proceeding as in (\ref{eq:astuce}) to calculate $ \check{\epsilon}_+-\check{\epsilon}_-$ at $ u = u_0 $: 
\be
\check{\epsilon}_+-\check{\epsilon}_-=\frac{\check{\epsilon}_+^2-\check{\epsilon}_-^2}{\check{\epsilon}_++\check{\epsilon}_-} = \frac{4abu_0}{\check{\epsilon}} = 2 a\check{R}^{1/2}
\Longrightarrow
\check{\epsilon}_+ = \frac{\check{\epsilon}}{2} + a \check{R}^{1/2},\quad
\check{\epsilon}_- = \frac{\check{\epsilon}}{2} - a \check{R}^{1/2},\quad
\check{\xi}_+ = a + \frac{\check{\epsilon}}{2} \check{R}^{1/2},\quad
\check{\xi}_- = a - \frac{\check{\epsilon}}{2} \check{R}^{1/2}
\ee
We thus obtain the Dirac $ \delta $ expression adapted to polar integration: 
\be
\label{eq:deltau}
\delta(\check{\epsilon}-\check{\epsilon}_+-\check{\epsilon}_-) = \frac{\Theta(1-u_0)}{J}\delta(u-u_0)
\quad\mbox{with}\quad J=\frac{\dd}{\dd u} (\check{\epsilon}_++\check{\epsilon}_-)= b\left(\frac{\check{\xi}_+}{\check{\epsilon}_+}-\frac{\check{\xi}_-}{\check{\epsilon}_-}\right)\geq 0
\ee
As we have taken into account in (\ref{eq:deltau}) the geometric constraint $ u_0 \leq 1 $ by means of a Heaviside function $ \Theta $, we can integrate on $ u \in [0,+\infty [$. It remains to integrate on the radial variable $ \check{k} $, which, due to function $ \Theta $, is not that simple, the domain of integration splitting into various sub-intervals according to the number of roots of the equation $ u_0 (\check{k}) = 1 $. Let's explain how to do this using the example of $ \check{\rho}_{+-} $. First, we use $ a $ rather than $ \check{k} $ as the integration variable. Then, taking into account the range of variation of $ a $ in (\ref{eq:chq}) and the hyperbolic parametrization (\ref{eq:chgtvar}) of the energy $ \check{\epsilon} $ by $ \Omega $, we put $ a = \frac{1}{2} (\check{\epsilon}^2-4 \check{\Delta}^2)^{1/2} \sin \theta = \check{\Delta} \sh \Omega \sin \theta $ where $ \theta \in [-\pi/2, \pi/2] $. Then, using (\ref{eq:Wexpli}), 
\be
\label{eq:expmapp}
\check{\rho}_{+-}(\check{\epsilon},\check{q})=\int_{-\frac{1}{2}(\check{\epsilon}^2-4\check{\Delta}^2)^{1/2}}^{\frac{1}{2}(\check{\epsilon}^2-4\check{\Delta}^2)^{1/2}} \dd a \frac{4\pi\check{\epsilon}\check{\Delta}^2a\,\Theta(1-u_0)/\check{q}}{(\check{\epsilon}^2-4a^2)^{3/2}(\check{\epsilon}^2-4(a^2+\check{\Delta}^2))^{1/2}}=\frac{\pi\check{\Delta}}{2\check{q}}\sh\Omega\ch\Omega \int_{-\pi/2}^{\pi/2}\dd\theta\frac{\Theta(1-u_0)\sin\theta}{(1+\sh^2\Omega\cos^2\theta)^{3/2}}
\ee
To know the sign of the argument of $ \Theta $ according to the value of $ s \equiv \sin \theta $, we must by continuity find its zeros therefore solve the cubic equation $ u_0^2 = 1 $ on $ s $, that is to say (\ref{eq:surs}) except for a trivial rearrangement. The graphic discussion of (\ref{eq:surs}) is done in figure \ref{fig:discugraph}c and in section \ref{subsubsec:exdesp}, let us recall the conclusions here: $ (i) $ if $ \check{\epsilon} <\check{\epsilon}_2 (\check{q}) $, the equation has only one real root $ s_3 $, which is $> 1 $, so that $ u_0 (s) <1 $ for all $ s \in [-1,1] $; $ (ii) $ if $ \check{\epsilon}_2 (\check{q}) <\check{\epsilon} <\check{\epsilon}_3 (\check{q}) $ (this can happen only if $ \check{q} <\check{q}_0 $), the equation has three real roots, the first two $ s_1 $ and $ s_2 $ in $]-1,1 [$, the third $ s_3 > 1 $, so that $ u_0 (s) <1 $ on $ [-1, s_1 [\cup] s_2,1] $; $ (iii) $ if $ \check{\epsilon}_3 (\check{q}) <\check{\epsilon} $, the equation has three real roots, $ s_1 <-1 $, $ s_2 \in]-1,1 [$ and $ s_3> 1 $, so that $ u_0 (s) <$ 1 on $] s_2,1] $. The critical energies $ \check{\epsilon}_{2,3} (\check{q}) $ and the critical wave number $ \check{q}_0 $ are defined in section \ref{subsubsec:lpdnae1e2e3}. The integral in the right-hand side of (\ref{eq:expmapp}) is therefore written 
\be
\label{eq:rhopmcas}
\int_{-\pi/2}^{\pi/2}\dd\theta\frac{\Theta(1-u_0)\sin\theta}{(1+\sh^2\Omega\cos^2\theta)^{3/2}}=
\left\{
\begin{array}{ll}
\displaystyle\int_{-\pi/2}^{\pi/2}\frac{\dd\theta\sin\theta}{(1+\sh^2\Omega\cos^2\theta)^{3/2}} & \mbox{if}\,2\check{\Delta}<\check{\epsilon}<\check{\epsilon}_2(\check{q}) \\
&\\
\displaystyle\left(\int_{-\pi/2}^{\asin s_1} +\int_{\asin s_2}^{\pi/2}\right) \frac{\dd\theta\sin\theta}{(1+\sh^2\Omega\cos^2\theta)^{3/2}} & \mbox{if}\,\check{\epsilon}_2(\check{q})<\check{\epsilon}<\check{\epsilon}_3(\check{q}) \\
&\\
\displaystyle\int_{\asin s_2}^{\pi/2} \frac{\dd\theta\sin\theta}{(1+\sh^2\Omega\cos^2\theta)^{3/2}} & \mbox{if}\,\check{\epsilon}_3(\check{q})<\check{\epsilon}
\end{array}
\right.
\ee
The integral of the case $ (i) $ in (\ref{eq:rhopmcas}) is zero due to odd parity of the integrand. In the other cases, it remains to make the change of variable $ \theta = \alpha-\pi/2 $ in the integral of lower bound $-\pi/2 $ and $ \theta = \pi/2-\alpha $ in the integral of upper bound $ \pi/2 $; one thus gets the function $ f_{+-} (\psi) $ of (\ref{eq:fpm}), expressible analytically in a simple way. For the spectral densities $ \check{\rho}_{++} $ and $ \check{\rho}_{--} $, we do the same, except that we have to recognize elliptical integrals, that are complete in the case $ (i) $ and incomplete as in (\ref{eq:fmm}, \ref{eq:fpp}) otherwise. We find table \ref{tab:rhogen}. 

\section{On the limit $ \Delta/\mu \to+\infty $} 
\subsection{At the wave number scale $ \check{q} \approx \check{\Delta}^0 $} 
\label{subapp:gdd0} 

This is to explain how the following expansions were calculated, useful for obtaining result (\ref{eq:Zqdevloin}) on the continuum branch of type II (here, $ \check{q} $ and $ \check{Z} = \check{z}-2 \check{\Delta} $ are fixed, with $ \im \check{Z} <0 $): 
\bea
\label{eq:devm++pex}
\check{M}_{++}(\check{z},\check{q}) &\stackrel{\check{q},\,\check{Z}\scriptsize\,\mbox{fixed}}{\underset{\check{\Delta}\to +\infty}{=}}&
\!\!\!\!\!\!\check{\Delta}^{3/4}\!\!\!\int_0^{+\infty}\!\!\!\!\dd\check{K} \frac{(2\pi)^{-2}\check{K}^2}{\check{Z}-\check{K}^4} +\check{\Delta}^{1/2}\!\!\!\int_0^{+\infty}\!\!\!\!\dd\check{K}
\frac{(2\pi)^{-2}}{\check{K}^2} \left(1-\frac{1}{\sqrt{1+\check{K}^4}}\right)+\check{\Delta}^{1/4}\!\!\!\int_0^{+\infty}\!\!\!\!\dd\check{K}
\frac{(2\pi)^{-2}(5\check{q}^2-12)\check{K}^4}{6(\check{Z}-\check{K}^4)^2}+O(\check{\Delta}^{-1/4}) \nonumber\\
&=& -\frac{(-\check{Z})^{-1/4}}{8\pi\sqrt{2}} \check{\Delta}^{3/4} + \frac{[\Gamma(3/4)]^2}{4\pi^{5/2}}\check{\Delta}^{1/2}
+\frac{(5\check{q}^2-12)}{192\pi\sqrt{2}(-\check{Z})^{3/4}}\check{\Delta}^{1/4}+O(\check{\Delta}^{-1/4})\\
\check{M}_{--}(\check{z},\check{q}) &\stackrel{\check{q},\,\check{Z}\scriptsize\,\mbox{fixed}}{\underset{\check{\Delta}\to +\infty}{=}}&
\frac{(-\check{Z})^{3/4}}{8\pi\sqrt{2}}\check{\Delta}^{-1/4}+\frac{3[\Gamma(3/4)]^2\check{Z}+4[\Gamma(5/4)]^2\check{q}^2}{12\pi^{5/2}}\check{\Delta}^{-1/2} +\frac{(12-13\check{q}^2)(-\check{Z})^{1/4}}{192\pi\sqrt{2}}\check{\Delta}^{-3/4}+O(\check{\Delta}^{-5/4})\\
\check{M}_{+-}(\check{z},\check{q}) &\stackrel{\check{q},\,\check{Z}\scriptsize\,\mbox{fixed}}{\underset{\check{\Delta}\to +\infty}{=}}&
-\frac{[\Gamma(5/4)]^2}{\pi^{5/2}}\check{\Delta}^{1/2}+\frac{(-\check{Z})^{1/4}}{8\pi\sqrt{2}}\check{\Delta}^{1/4}+\frac{(3\check{q}^2-4)(-\check{Z})^{-1/4}}{64\pi\sqrt{2}}\check{\Delta}^{-1/4}+O(\check{\Delta}^{-1/2})
\eea
For that, let's start from integral expressions (\ref{eq:defSigma}) of the matrix elements, let us rescale them and formally integrate on the polar and azimuthal angles of axis the direction of $ \qq $. Integrals remain on the relative wave number $ \check{k} $ of certain functions $ f (\check{k}, \Delta) $. Let's apply the multi-scale analysis explained in the text above (\ref{eq:Zqdevloin}), which amounts to splitting the integration domain into three sub-intervals: 
\begin{multline}
\int_0^{+\infty}\dd\check{k}\,f(\check{k},\check{\Delta}) = \int_{0}^{\eta\check{\Delta}^{3/8}}\dd\check{k}\,f(\check{k},\check{\Delta})+\int_{\eta\check{\Delta}^{3/8}}^{A\check{\Delta}^{3/8}}\dd\check{k}\,f(\check{k},\check{\Delta})+\int_{A\check{\Delta}^{3/8}}^{+\infty} \dd\check{k}\,f(\check{k},\check{\Delta}) \\
= \int_0^{\eta/\varepsilon}\dd\check{K}_{\rm b}\ \underset{\equiv f_{\rm b}(\check{K}_{\rm b},\varepsilon)}{\underbrace{\check{\Delta}^{1/4} f(\check{\Delta}^{1/4}\check{K}_{\rm b},\check{\Delta})}}
+\int_{\eta}^{A} \dd\check{K}_{\rm bc} \ \underset{\equiv f_{\rm bc}(\check{K}_{\rm bc},\varepsilon)}{\underbrace{\check{\Delta}^{3/8} f(\check{\Delta}^{3/8}\check{K}_{\rm bc},\check{\Delta})}}
+\int_{A\varepsilon}^{+\infty} \dd\check{K}_{\rm c} \ \underset{\equiv f_{\rm c}(\check{K}_{\rm c},\varepsilon)}{\underbrace{\check{\Delta}^{1/2} f(\check{\Delta}^{1/2}\check{K}_{\rm c},\check{\Delta})}}
\end{multline}
where the small parameter is $ \varepsilon = \check{\Delta}^{-1/8} $. The cut-off parameters $ \eta $ and $ A $ are arbitrary; at the end of Taylor's expansion in $ \varepsilon $, it will be necessary to take the limits $ A \to+\infty $ and $ \eta \to 0 $ in the coefficient of each power $ \varepsilon^n $. \footnote{Taking the limit is actually only necessary for $ \check{M}_{--} $. For $ \check{M}_{++} $ and $ \check{M}_{+-} $, the dependence on $ A $ and $ \eta $ disappears by itself in the coefficients of $ \varepsilon^n $ useful here; we could therefore take $ \eta = A = 1 $ which would amount to keeping the scales (b) and (c) fully but introducing the scale (bc) only as a cut-off.} It remains to expand the functions $ f_{\rm b} $, $ f_{\rm c} $ and $ f_{\rm bc} $ in powers of $ \varepsilon $ at fixed $ \check{K}_{\rm b} $, $ \check{K}_{\rm c} $ and $ \check{K}_{\rm bc} $, with coefficients $ f_{\rm b}^{(n)} $, $ f_{\rm c}^{(n)} $ and $ f_{\rm bc}^{(n)} $, then integrate. If the integral of a $ f_{\rm b}^{(n)} $ is UV divergent or the integral of a $ f_{\rm c}^{(n)} $ is IR divergent when $ \varepsilon \to 0 $, you have to take out a simple equivalent of the integrand (typically, a sum of power laws in the integration variable) which will leave behind a convergent integral. We then reexpand everything in powers of $ \varepsilon $. 

Let us give the example of the matrix element $ \check{M}_{++} $, to be determined with a relative error $ O (\varepsilon^8) $ therefore an absolute error $ O (\varepsilon^2) $. Let's start with scale (c). Taylor's expansion of the integrand is written 
\be
\label{eq:devfc}
f_{\rm c}(\check{K},\varepsilon)\underset{\varepsilon\to 0}{=}-\frac{(2\pi)^{-2}\varepsilon^{-4}}{\check{K}^2\sqrt{1+\check{K}^4}}+
f_{\rm c}^{(4)} (\check{K})\,\varepsilon^4 +f_{\rm c}^{(12)}(\check{K})\,\varepsilon^{12} +O(\varepsilon^{20})
\ee
The integral of the first term is IR divergent when $ \varepsilon \to 0 $; we go through a succession of elementary manipulations to reduce it to a sum of power laws: 
\begin{multline}
\int_{A\varepsilon}^{+\infty}\!\! \frac{\dd\check{K}}{\check{K}^2\sqrt{1+\check{K}^4}}=\int_{A\varepsilon}^{+\infty}\!\!\frac{\dd\check{K}}{\check{K}^2}
\left(\frac{1}{\sqrt{1+\check{K}^4}}-1\right) +\int_{A\varepsilon}^{+\infty}\!\! \frac{\dd\check{K}}{\check{K}^2}
= \int_0^{+\infty}\!\! \frac{\dd\check{K}}{\check{K}^2}\left(\frac{1}{\sqrt{1+\check{K}^4}}-1\right)
-\int_0^{A\varepsilon}\!\!\dd\check{K}[-\frac{1}{2}\check{K}^2+O(\check{K}^6)]+\frac{\varepsilon^{-1}}{A} \\
 = -\frac{[\Gamma(3/4)]^2}{\pi^{1/2}}+\frac{A^3}{6}\varepsilon^3 +\frac{\varepsilon^{-1}}{A}+O(\varepsilon^7)
\end{multline}
The integral of the second and third terms of (\ref{eq:devfc}) is UV convergent but is IR divergent; the same manipulations are necessary. However, it suffices to know that 
\be
f_{\rm c}^{(4)}(\check{K}) \underset{\check{K}\to 0}{=} -\frac{(2\pi)^{-2}\check{Z}}{\check{K}^6} +\frac{(2\pi)^{-2}(5\check{q}^2-12)}{6\check{K}^4}
-\frac{(2\pi)^{-2}\check{Z}}{2\check{K}^2}+O(1)\quad\mbox{and}\quad f_{\rm c}^{(12)}(\check{K})\underset{\check{K}\to 0}{=}O(1/\check{K}^{10})
\ee
At the order of the calculation, the two most divergent terms of $ f_{\rm c}^{(4)} (\check{K}) $ contribute, and $ f_{\rm c}^{(12)} (\check{K}) $ is negligible. The same procedure is to be repeated for scales (bc) and (b). It suffices here to know that 
\bea
\!\!\!\!\!\!\!\!\!f_{\rm bc}(\check{K},\varepsilon)&\!\!\!\!\!\underset{\varepsilon\to 0}{=}\!\!\!\!\!& -\frac{(2\pi)^{-2}}{\check{K}^2}\varepsilon^{-5}+
(2\pi)^{-2}\left(-\frac{\check{Z}}{\check{K}^6}+\frac{1}{2}\check{K}^2\right)\varepsilon^{-1} +
\frac{(2\pi)^{-2}(5\check{q}^2-12)}{6\check{K}^4}\varepsilon+O(\varepsilon^3)\\
\!\!\!\!\!\!\!\!\!f_{\rm b}(\check{K},\varepsilon)&\!\!\!\!\!\underset{\varepsilon\to 0}{=}\!\!\!\!\!& \frac{(2\pi)^{-2}\check{K}^2}{\check{Z}-\check{K}^4}\varepsilon^{-6}
+\frac{(2\pi)^{-2}(5\check{q}^2-12)\check{K}^4}{6(\check{Z}-\check{K}^4)^2}\varepsilon^{-2}+f_{\rm b}^{(2)}(\check{K})\varepsilon^2+O(\varepsilon^6)
\ \mbox{where}\ f_{\rm b}^{(2)}(\check{K})\!\!\!\!\underset{\check{K}\to +\infty}{=}\!\!\!\! (2\pi)^{-2}\frac{\check{K}^2}{2}+ O(\check{K}^{-2})
\eea
We thus end up with the middle part of (\ref{eq:devm++pex}), the elements of which the reader will recognize above. 

\subsection{At the wave number scale $ \check{q} \approx \check{\Delta}^{-1/2} $} 
\label{subapp:gddmundemi} 

To establish equation (\ref{eq:sdomm+-}) on the phase-modulus coupling $ M_{+-} $, we proceed as it is explained in the text above it. After formal angular average of (\ref{eq:defSigma}) in spherical coordinates of polar axis the direction of $ \qq $, $ M_{+-} $ is written as an integral on the wave number $ \check{k} $ of a certain function $ F (\check{k}, \check{\Delta}) $. Let us take as small parameter $ \varepsilon = \check{\Delta}^{-1/8} $, let us split the integration domain in two around the cut-off $ \Lambda \check{\Delta}^{1/4} $ ($ \Lambda $ is a constant) and make the changes of variable (\ref{eq:cvabc}) of type (a) and (c): 
\be
\check{M}_{+-}(\check{z},\check{q})=\int_0^{\Lambda\check{\Delta}^{1/4}}\dd\check{k}\, F(\check{k},\check{\Delta})
+\int_{\Lambda\check{\Delta}^{1/4}}^{+\infty} \dd\check{k}\, F(\check{k},\check{\Delta})
=\int_0^{\Lambda\varepsilon^{-2}} \dd\check{K}_{\rm a}\, \underset{\equiv F_{\rm a}(\check{K}_{\rm a},\varepsilon)}{\underbrace{F(\check{K}_{\rm a},\check{\Delta})}}
+
\int_{\Lambda\varepsilon^{2}}^{+\infty} \dd\check{K}_{\rm c}\,\underset{\equiv F_{\rm c}(\check{K}_{\rm c},\varepsilon)}{\underbrace{\check{\Delta}^{1/2} F(\check{\Delta}^{1/2}\check{K}_{\rm c},\check{\Delta})}}
\ee
It remains to expand in powers of $ \varepsilon $ under the integral sign at fixed $ \check{K}_a $ or $ \check{K}_c $, with $ \check{Q} = \check{\Delta}^{1/2} \check{q} $ and $ \check{\zeta} $ defined by (\ref{eq:defzc}) fixed, 
\be
\label{eq:devF}
F_{\rm a}(\check{K}_{\rm a},\varepsilon)\underset{\varepsilon\to 0}{=} \frac{(2\pi)^{-2}\check{K}_{\rm a}^2(\check{K}_{\rm a}^2-1)}
{\check{\zeta}-(\check{K}_{\rm a}^2-1)^2} + O\left(\frac{\varepsilon^8}{1+\check{K}_{\rm a}^2}\right)\quad\mbox{and}\quad
F_{\rm c}(\check{K}_{\rm c},\varepsilon)\underset{\varepsilon\to 0}{=}\frac{-(2\pi)^{-2}\varepsilon^{-4}}{(1+\check{K}_{\rm c}^4)^{1/2}}+O\left(\frac{\varepsilon^4}{\check{K}_{\rm c}^2}\right)
\ee
then to integrate: 
\bea
\label{eq:intdeva}
\int_0^{\Lambda\varepsilon^{-2}} \dd\check{K}_{\rm a}\, F_{\rm a}^{(0)}(\check{K}_{\rm a}) &\!\!\!\!=\!\!\!\!&
F_{\rm a}^{(0)}(+\infty)\Lambda\varepsilon^{-2}+\int_0^{+\infty} \dd\check{K}_{\rm a}[F_{\rm a}^{(0)}(\check{K}_{\rm a})-F_{\rm a}^{(0)}(+\infty)]+O(\varepsilon^2) \\
\label{eq:intdevb}
\int_{\Lambda\varepsilon^{2}}^{+\infty} \dd\check{K}_{\rm c}\, F_{\rm c}^{(-4)}(\check{K}_{\rm c}) &\!\!\!\!=\!\!\!\!&
\int_0^{+\infty} \dd\check{K}_{\rm c}\, F_{\rm c}^{(-4)}(\check{K}_{\rm c}) -\int_0^{\Lambda\varepsilon^{2}}\dd\check{K}_{\rm c}
F_{\rm c}^{(-4)}(\check{K}_{\rm c})=-\frac{[\Gamma(5/4)]^2}{\pi^{5/2}}-\Lambda\varepsilon^{2}F_{\rm c}^{(-4)}(0) +O(\varepsilon^{10})
\eea
Here, the functions $ F^{(n)} $ are the coefficients of the terms of order $ \varepsilon^n $ in (\ref{eq:devF}). In (\ref{eq:intdeva}), we used a minus-plus trick by subtracting from the integrand its limit at infinity. In the remaining integral, the integrand is now $O(1/\check{K}_a^2) $; we can therefore replace the upper bound by $+\infty $, introducing a negligible error $ O (\varepsilon^2) $. In (\ref{eq:intdevb}), to go from the middle part to the right-hand side, we first carried out the first integral then, in the second integral, we approximated the integrand by its value at the origin; as the latter varies quartically near $ \check{K}_{\rm c} = 0 $, the error made is indeed $O(\varepsilon^{10}) $. By collecting the contributions (\ref{eq:intdeva}) and (\ref{eq:intdevb}) in $ \check{M}_{+-} $ with weights $ 1 $ and $ \varepsilon^{-4} $, we see that the linear terms in $ \Lambda $ in the  right-hand sides of these equations cancel each other. It remains (\ref{eq:sdomm+-}).


\begin{thebibliography}{99}
\bibitem{CKS}
R. Combescot, M. Yu. Kagan, S. Stringari, ``Collective mode of homogeneous superfluid Fermi gases in the BEC-BCS crossover", Phys. Rev. A {\bf 74}, 042717 (2006).

\bibitem{AndrianovPopov}
V.A. Andrianov, V.N. Popov, ``Gidrodinami\v{c}eskoe dejstvie i Boze-spektr sverhteku\v{c}ih Fermi-sistem", Teor.  Mat. Fiz. {\bf 28}, 341 (1976) [Theor. Math. Phys. {\bf 28}, 829 (1976)].

\bibitem{PRL2019} 
H. Kurkjian, S.N. Klimin, J. Tempere, Y. Castin, ``Pair-Breaking Collective Branch in BCS Superconductors and Superfluid Fermi Gases",
Phys. Rev. Lett. {\bf 122}, 093403 (2019).

\bibitem{Thomas2002}
K. M. O'Hara, S. L. Hemmer, M. E. Gehm, S. R. Granade, J. E. Thomas, ``Observation of a strongly interacting degenerate Fermi gas of atoms", Science {\bf 298}, 2179 (2002).

\bibitem{Salomon2003}
T. Bourdel, J. Cubizolles, L. Khaykovich, K. M. Magalh\~aes, S. J. J. M. F. Kokkelmans, G. V. Shlyapnikov, C. Salomon,
``Measurement of the interaction energy near a Feshbach resonance in a $^{6}\mathrm{L}\mathrm{i}$ Fermi gas", Phys. Rev. Lett. {\bf 91}, 020402 (2003).

\bibitem{Grimm2004b}
M. Bartenstein, A. Altmeyer, S. Riedl, S. Jochim, C. Chin, J. H. Denschlag, R. Grimm, ``Collective excitations of a degenerate gas at the BEC-BCS crossover", Phys. Rev. Lett. {\bf 92}, 203201 (2004).

\bibitem{Ketterle2004}
M. W. Zwierlein, C. A. Stan, C. H. Schunck, S. M. F. Raupach, A. J. Kerman, W. Ketterle, ``Condensation of pairs of fermionic atoms near a Feshbach resonance", Phys. Rev. Lett. {\bf 92}, 120403 (2004).

\bibitem{Salomon2010}
S. Nascimb{\`e}ne, N. Navon, K. J. Jiang, F. Chevy, C. Salomon, ``Exploring the thermodynamics of a universal Fermi gas", Nature {\bf 463}, 1057 (2010).

\bibitem{Zwierlein2012}
M.J.H. Ku, A.T. Sommer, L.W. Cheuk, M.W. Zwierlein, ``Revealing the superfluid lambda transition in the universal thermodynamics of a unitary Fermi gas", Science {\bf 335}, 563 (2012).

\bibitem{Anderson1958}
P. W. Anderson, ``Random-phase approximation in the theory of superconductivity", Phys. Rev. {\bf 112}, 1900 (1958).

\bibitem{theseHadrien}
H. Kurkjian, \href{https://hal.archives-ouvertes.fr/tel-01341982}{\it Coh\'erence, brouillage et dynamique de phase dans un condensat de paires de fermions}, PhD Thesis, \'Ecole Normale Sup\'erieure, Paris (2016).

\bibitem{Annalen}
H. Kurkjian, Y. Castin, A. Sinatra, \href{https://hal.archives-ouvertes.fr/hal-01392846}{``Three-phonon and four-phonon interaction processes in a pair-condensed Fermi gas"}, Annalen der Physik {\bf 529}, 1600352 (2017).

\bibitem{Houchesdimred}
Y. Castin, ``Simple theoretical tools for low dimension Bose gases", Lecture notes of the 2003 Les Houches Spring School {\sl Quantum Gases in Low Dimensions}, edited by M. Olshanii, H. Perrin, L. Pricoupenko,  J.  Phys. IV France {\bf 116}, 89 (2004).

\bibitem{PRLGurarie}
V. Gurarie, ``Nonequilibrium Dynamics of Weakly and Strongly Paired Superconductors", Phys. Rev. Lett. {\bf 103}, 075301 (2009). 

\bibitem{CCTlivre}
C. Cohen-Tannoudji, J. Dupont-Roc, G. Grynberg, {\sl Processus d'interaction entre photons et atomes} (InterEditions and \'Editions du CNRS, Paris, 1988).

\bibitem{livreNozieres}
P. Nozi\`eres, {\sl Le probl\`eme \`a $N$ corps~: Propri\'et\'es g\'en\'erales des gaz de fermions} (Dunod, Paris, 1963).

\bibitem{Varenna06}
 Y. Castin, ``Basic tools for degenerate Fermi gases", {\sl Lecture notes of the 2006 Varenna Enrico Fermi School on Fermi gases},
edited by M. Inguscio, W. Ketterle, C. Salomon (SIF, 2007).

\bibitem{Ketterlegap}
A. Schirotzek, Y. Shin, C. H. Schunck, W. Ketterle, ``Determination of the superfluid gap in atomic Fermi gases by quasiparticle spectroscopy",
Phys. Rev. Lett.  {\bf 101}, 140403 (2008).

\bibitem{Marini1998}
M. Marini, F. Pistolesi, G.C. Strinati, ``Evolution from BCS superconductivity to Bose condensation: Analytic results for the crossover in three dimensions", 
Eur. Phys. J. B {\bf 1}, 151 (1998).

\bibitem{GR}
I.S. Gradshteyn, I.M. Ryzhik, {\sl Tables of integrals, series, and products}, edited par A. Jeffrey, 5th edition (Academic Press, San Diego, 1994).

\bibitem{PRAConcav}
H. Kurkjian, Y. Castin, A. Sinatra, ``Concavity of the collective excitation branch of a Fermi gas in the BEC-BCS crossover", Phys. Rev. A {\bf 93}, 013623 (2016).

\bibitem{JLTP}
S.N. Klimin, H. Kurkjian, J. Tempere, ``Anderson-Bogoliubov collective excitations in superfluid Fermi gases at nonzero temperatures", Journal of Low Temperature Physics {\bf 196}, 102 (2019).

\bibitem{sontempnn}
S.N. Klimin, J. Tempere, H. Kurkjian, ``Phononic collective excitations in superfluid Fermi gases at nonzero temperatures", Phys. Rev. A {\bf 100}, 063634 (2019).

\bibitem{KetterleRecord}
Z. Hadzibabic, S. Gupta, C.A. Stan, C.H. Schunck, M.W. Zwierlein, K. Dieckmann, W. Ketterle, ``Fiftyfold improvement in the Number of Quantum Degenerate Fermionic Atoms", Phys. Rev.  Lett. {\bf 91}, 160401 (2003).

\bibitem{DPGSCS}
D.S. Petrov, C. Salomon, G.V. Shlyapnikov, ``Weakly Bound Dimers of Fermionic Atoms", Phys. Rev. Lett. {\bf 93}, 090404 (2004).

\bibitem{Leggett1980}
A.J. Leggett, ``Cooper Pairing in Spin-Polarized Fermi systems", Journal de physique Colloq. {\bf 41}, C7-19 (1980).

\bibitem{Engelbrecht}
J.R. Engelbrecht, M. Randeria, C.A.R. S\'{a} de Melo, ``BCS to Bose crossover: Broken-symmetry state", Phys. Rev.  B {\bf 55}, 15153 (1997).

\bibitem{livreZwerger}
{\sl The BCS-BEC Crossover and the Unitary Fermi Gas}, 
Lecture Notes in Physics 836, edited by W. Zwerger (Springer, Berlin, 2012).

\bibitem{vcLandau}
Y. Castin, I. Ferrier-Barbut, C. Salomon, \href{https://hal.archives-ouvertes.fr/hal-01053706}{``La vitesse critique de Landau
d'une particule dans un superfluide de fermions"}, Comptes Rendus Physique {\bf 16}, 241 (2015).

\bibitem{Benfatto} 
T. Cea, C. Castellani, G. Seibold, L. Benfatto, ``Nonrelativistic Dynamics of the Amplitude (Higgs) Mode in Superconductors", Phys. Rev. Lett. {\bf 115}, 157002 (2015).

\bibitem{Zwerger}
R. Haussmann, M. Punk, W. Zwerger, ``Spectral functions and rf response of ultracold fermionic atoms", Phys. Rev. A {\bf 80}, 063612 (2009).

\end{thebibliography}
\end{document}